\documentclass[a4paper,11pt]{book}

\usepackage[all]{xy}
\usepackage{mathrsfs}
\usepackage{graphicx}  % needed for figures
\usepackage{bm}        % for math
\usepackage[english]{babel}
\usepackage{amsfonts,amsmath,amssymb,mathrsfs}
\usepackage[T1]{fontenc}

\usepackage{latexsym}

\newtheorem{defi}{Definition}

\newcommand{\be}{\begin{equation}}
\newcommand{\ee}{\end{equation}}
\newcommand{\bea}{\begin{eqnarray}}
\newcommand{\nn}{\nonumber}
\newcommand{\eea}{\end{eqnarray}}

\newcommand{\gn}{\bar{\nabla}}
\def\two{\hbox{$_{2}$}}
\def\three{\hbox{$_{3}$}}
\def\four{\hbox{$_{4}$}}

\begin{document}
\frontmatter
%%%%%%%%%%%%%%%%%%%%%%%
%%%%%%%%%%%%%%%%%%%%%%%
\begin{titlepage}
\begin{centering}
%\vspace*{2cm}
  
%{\Large \bf MODIFIED ACTIONS FOR GRAVITY: THEORY AND PHENOMENOLOGY}

%\vspace*{5cm}
%\large
%Thesis submitted for the degree of\\
%``Doctor Philosophi\ae"\\

%\vspace*{1cm}
%October 2007

%\vspace*{7cm}

%Thomas P.~Sotiriou\\
%International School for Advanced Studies\\
%Via Beirut 2-4, 34014 Trieste, Italy.\\
%E-mail: {\sl sotiriou@sissa.it} 
\vspace*{2cm}
  
  {\Large \bf MODIFIED ACTIONS FOR GRAVITY: THEORY AND PHENOMENOLOGY}

\vspace*{3cm}
\large
Thesis submitted for the degree of\\
``Doctor Philosophi\ae"

\vspace*{1cm}
October 2007
\vspace*{4cm}

\begin{tabular*}{400pt}{l @{\extracolsep\fill} r}

CANDIDATE & SUPERVISORS\\
&\\
Thomas P. Sotiriou & Stefano Liberati and John Miller
\end{tabular*}

\vspace*{4cm}

International School for Advanced Studies\\
Via Beirut 2-4, 34014 Trieste, Italy.\\
E-mail: {\sl sotiriou@sissa.it} 

\end{centering}
\end{titlepage}
\cleardoublepage
\vspace*{7cm}
\begin{quote}

\em
\raggedleft
The important thing is not to stop questioning.

   {\bf Albert Einstein}
\end{quote}

\newpage
%%%%%%%%%%%%%%%%%%%%%%%%%%%%%%%%%%%%%%

\chapter*{Abstract}

This thesis is devoted to the study of gravitational theories which
can be seen as modifications or generalisations of General Relativity.
The motivation for considering such theories, stemming from Cosmology,
High Energy Physics and Astrophysics is thoroughly discussed
(cosmological problems, dark energy and dark matter problems, the lack
of success so far in obtaining a successful formulation for Quantum
Gravity). The basic principles which a gravitational theory should
follow, and their geometrical interpretation, are analysed in a broad
perspective which highlights the basic assumptions of General
Relativity and suggests possible modifications which might be made. A
number of such possible modifications are presented, focusing on
certain specific classes of theories: scalar-tensor theories, metric
$f(R)$ theories, Palatini $f(R)$ theories, metric-affine $f(R)$
theories and Gauss--Bonnet theories. The characteristics of these
theories are fully explored and attention is payed to issues of
dynamical equivalence between them. Also, cosmological phenomenology
within the realm of each of the theories is discussed and it is shown
that they can potentially address the well-known cosmological
problems. A number of viability criteria are presented: cosmological
observations, Solar System tests, stability criteria, existence of
exact solutions for common vacuum or matter configurations {\em etc}.
Finally, future perspectives in the field of modified gravity are
discussed and the possibility for going beyond a trial-and-error
approach to modified gravity is explored.

\chapter*{Collaborations}

The research presented in this thesis was mainly conducted in SISSA-International School for Advanced Studies between November 2004 and October 2007. This thesis is the result of the authors own work, as well as the outcome of scientific collaborations stated below, except where explicit reference is made to the results of others.

The content of this thesis is based on the following research papers published in refereed Journals or refereed conference proceedings:

\begin{enumerate}
\small{
%\cite{Sotiriou:2005xe}
\item%{Sotiriou:2005xe}
``The nearly Newtonian regime in Non-Linear Theories of Gravity''
  \\{}T.~P.~Sotiriou
  \\{}Gen.~Rel.~Grav.~{\bf 38} 1407 (2006)
[arXiv:gr-qc/0507027]}
%\href{http://www.slac.stanford.edu/spires/find/hep/www?eprint=gr-qc\%2F0507027}{SPIRES entry}

%\cite{Sotiriou:2005hu}
\item%{Sotiriou:2005hu}
{\bf ``Unification of inflation and cosmic acceleration in the Palatini formalism''}
  \\{}T.~P.~Sotiriou
  \\{}Phys.\ Rev.\ D {\bf 73}, 063515 (2006)
  [arXiv:gr-qc/0509029]
%\href{http://www.slac.stanford.edu/spires/find/hep/www?eprint=gr-qc\%2F0509029}{SPIRES entry}

%\cite{Sotiriou:2005cd}
\item%{Sotiriou:2005cd}
{\bf ``Constraining $f(R)$ gravity in the Palatini formalism''}
  \\{}T.~P.~Sotiriou
  \\{}Class.\ Quant.\ Grav.\  {\bf 23}, 1253 (2006)
  [arXiv:gr-qc/0512017]
%\href{http://www.slac.stanford.edu/spires/find/hep/www?eprint=gr-qc\%2F0512017}{SPIRES entry}

%\cite{Sotiriou:2006qn}
\item%{Sotiriou:2006qn}
{\bf ``Metric-affine $f(R)$ theories of gravity''}
  \\{}T.~P.~Sotiriou and S.~Liberati
  \\{}  Ann.~Phys.~{\bf 322}, 935 (2007)
 [arXiv:gr-qc/0604006]
%\href{http://www.slac.stanford.edu/spires/find/hep/www?eprint=gr-qc\%2F0604006}{SPIRES entry}

%\cite{Sotiriou:2006hs}
\item%{Sotiriou:2006hs}
{\bf ``$f(R)$ gravity and scalar-tensor theory''}
  \\{}T.~P.~Sotiriou
  \\{} Class.~Quant.~Grav.~{\bf 23}, 5117 (2006)
[arXiv:gr-qc/0604028]
%\href{http://www.slac.stanford.edu/spires/find/hep/www?eprint=gr-qc\%2F0604028}{SPIRES entry}

\item {\bf ``The metric-affine formalism of $f(R)$ gravity''}\\
T.~P.~Sotiriou and S.~Liberati
\\{} J.~Phys.~Conf.~Ser.~{\bf 68}, 012022 (2007) [arXiv:gr-qc/0611040]\\
{\it Talk given (by T.~P.~S.) at the 12th Conference on Recent Developments in Gravity (NEB XII), Nafplio, Greece, 29 Jun-2 Jul 2006}

%\cite{Sotiriou:2006sf}
\item %{Sotiriou:2006sf}
 {\bf ``Curvature scalar instability in $f(R)$ gravity.''}
  \\{}T.~P.~Sotiriou,
   \\{} Phys.~Lett.~B {\bf 645}, 389 (2007)
	[arXiv:gr-qc/0611107]
  %%CITATION = GR-QC 0611107;%%

\item {\bf ``The significance of matter coupling in $f(R)$ gravity''}\\
T.~P.~Sotiriou, Proceedings of the 11th Marcel Grossman Meeting in press\\
{\it Talk given at the 11th Marcel Grossman Meeting, Berlin, Germany, 23-29 Jul 2006}

%\cite{Sotiriou:2006pq}
\item %{Sotiriou:2006pq}
 {\bf ``Post-Newtonian expansion for Gauss-Bonnet gravity.''}
\\{} T.~P.~Sotiriou and E.~Barausse,
  \\{} Phys.~Rev.~D {\bf 75}, 084007 (2007)
	[arxiv:gr-qc/0612065]
  %%CITATION = GR-QC 0612065;%%

\item %{Barausse:2007pn}
  {\bf ``A no-go theorem for polytropic spheres in Palatini $f(R)$ gravity.''}
\\{} E.~Barausse, T.~P.~Sotiriou and J.~C.~Miller,
  \\{} Submitted for publication
  [arXiv:gr-qc/0703132]
  %%CITATION = GR-QC/0703132;%%

\item
{\bf ``Theory of gravitation theories: a no-progress report.''}
\\{} T.~P.~Sotiriou, V.~Faraoni and S.~Liberati,
  \\{} Submitted for publication [arXiv:0707.2748 [gr-qc] ]
\\{} {\it Invited paper in the Special Issue: Invited Papers and Selected Essays from the Annual Essay Competition of the Gravity Research Foundation for the Year 2007} 
\end{enumerate}

\chapter*{Notation}

An attempt has been made to keep the basic notation as standard as possible. However, the use of non-metric connections did require the use of some non-standard notation. The following list will hopefully be a useful tool for clarifying these non-standard notation. In general, the notation, standard or not, is always defined at its first occurence in the text and in all places that ambiguities may arise, irrespectivelly of whether it has been included in this guide. The signature of the metric is assumed to be $(-,+,+,+)$ and the speed of light $c$ is taken to be equal to $1$ throughout this thesis. In order to lighten the notation, in some cases  a coordinate system is used in which $G=c=1$, where $G$ is Newton's gravitational constant. However, for clarity $G$ is not set to be equal to $1$ throught the text.

\begin{tabbing}
{\bf Spoken Languages} mla: \=sotiriou@sissa.it; tsotiri@phys.uoa.gr \kill
$g^{\mu\nu}$: \> Lorentzian metric\\
$g$: \> Determinant of $g^{\mu\nu}$\\
$\Gamma^{\lambda}_{\phantom{a}\mu\nu}$:  \> General Affine Connection\\
$\{^{\lambda}_{\phantom{a}\mu\nu}\}$:  \> Levi--Civita Connection \\
$\nabla_\mu$:  \> Covariant derivative with respect to $\{^{\lambda}_{\phantom{a}\mu\nu}\}$\\
$\gn_\mu$:  \> Covariant derivative with respect to $\Gamma^{\lambda}_{\phantom{a}\mu\nu}$\\
$(\mu\nu)$:  \> Symmetrization over the indices $\mu$ and $\nu$\\
$[\mu\nu]$:  \> Anti-symmetrization over the indices $\mu$ and $\nu$\\
$Q_{\mu\nu\lambda}$:  \> Non-metricity ($\equiv-\gn_\mu g_{\nu\lambda}$) \\
$S_{\mu\nu}^{\phantom{ab}\lambda}$:  \> Cartan torsion tensor ($\equiv \Gamma^{\lambda}_{\phantom{a}[\mu\nu]}$)\\
$R^{\lambda}_{\phantom{a}\sigma\mu\nu}$: \> Riemann tensor of $g_{\mu\nu}$\\
$R_{\mu\nu}$: \> Ricci tensor of $g_{\mu\nu}$ ($\equiv R^{\sigma}_{\phantom{a}\mu\sigma\nu}$)\\
$R$: \> Ricci scalar of $g_{\mu\nu}$ ($\equiv g^{\mu\nu}R_{\mu\nu}$)\\
${\cal R}^{\lambda}_{\phantom{a}\sigma\mu\nu}$: \> Riemann tensor constructed with  $\Gamma^{\lambda}_{\phantom{a}\mu\nu}$\\
${\cal R}_{\mu\nu}$: \> $\equiv {\cal R}^{\sigma}_{\phantom{a}\mu\sigma\nu}$\\
${\cal R}$: \> $\equiv g^{\mu\nu}{\cal R}_{\mu\nu}$\\
$S_M$: \> Matter action \\
$T_{\mu\nu}$: \> Stress-energy tensor $\left(\equiv -\frac{2}{\sqrt{-g}}\frac{\delta S_M}{\delta g^{\mu\nu}}\right)$\\
$\Delta^\lambda_{\phantom{a}\mu\nu}$: \> Hypermomentum $\left(\equiv -\frac{2}{\sqrt{-g}}\frac{\delta S_M}{\delta \Gamma^\lambda_{\phantom{a}\mu\nu}}\right)$\\ 
$\phi$: \> Scalar field (generic)\\
$\psi$: \> Matter fields (collectively)
\end{tabbing}

\chapter{Preface}

The terms ``modified gravity'' and ``alternative theory of gravity''
have become standard terminology for theories proposed for describing
the gravitational interaction which differ from the most conventional
one, General Relativity. Modified or alternative theories of gravity
have a long history. The first attempts date back to the 1920s, soon
after the introduction of Einstein's theory. Interest in this research
field, which was initially driven by curiosity or a desire to
challenge the then newly introduced General Theory of Relativity, has
subsequently varied depending on circumstances, responding to the
appearance of new motivations. However, there has been more or less
continuous activity in this subject over the last 85 years.

When the research presented in this thesis began, interest in modified
gravity was already at a high point and it has continued increasing
further until the present day. This recent stimulus has mostly been
due to combined motivation coming from the well-known cosmological
problems related to the accelerated expansion of the universe and the feedback
from High Energy Physics.

Due to the above, and even though the main scope of this thesis is to
present the research conducted by the author during the period
November 2004 - October 2007, a significant effort has been made so
that this thesis can also serve as a guide for readers who have
recently developed an interest in this field. To this end, special
attention has been paid to giving a coherent presentation of the
motivation for considering alternative theories of gravity as well as
to giving a very general analysis of the foundations of gravitation
theory. Also, an effort has been made to present the theories
discussed thoroughly, so that readers less familiar with this subject
can be introduced to them before gradually moving on to their more
complicated characteristics and applications. 

The outline of this thesis is as follows: In the Introduction, several
open issues related to gravity are discussed, including the
cosmological problems related to dark matter and dark energy, and the
search for a theory of Quantum Gravity. Through the presentation of a
historical timeline of the passage from Newtonian gravity to General
Relativity, and a comparison with the current status of the latter in
the light of the problems just mentioned, the motivations for
considering alternative theories of gravity are introduced. Chapter
\ref{foundations} is devoted to the basic principles which gravitation
theories should follow. The Dicke framework, the various forms of the
Equivalence Principle and the so-called metric postulates are
critically reviewed and the assumptions that lead to General
Relativity are examined. Additionally, the ways of relaxing these
assumptions are explored together with the resulting theories. In
Chapter \ref{modeltheo}, we focus on specific theories: scalar-tensor
theory, metric, Palatini and metric-affine $f(R)$ gravity and
Gauss--Bonnet gravity, and their theoretical characteristics are
thoroughly presented. Chapter \ref{equivtheor} contains a discussion
about the possible dynamical equivalence between these theories, while
in Chapter \ref{cosmology} their cosmological phenomenology is
presented. Attention is paid to their ability to address the
well-known cosmological problems and to their cosmological viability.
Chapter \ref{weakstrong} is devoted to the study of the weak and
strong gravity regimes in these modified theories of gravity. The
Newtonian and post-Newtonian limits, stability issues, non-vacuum
solutions, {\em etc.} are discussed as criteria for the viability of
these theories. Finally, Chapter \ref{concl} contains the conclusions
of this work, as well as suggestions and remarks about future work in
the field of modified gravity.

A number of people have contributed in this thesis in various ways.
First and foremost, I would like to thank my PhD advisors, Stefano
Liberati and John Miller, for their constant support during the course
of this work. It is difficult for me to imagine having better advisors
than Stefano and John, to whom I am truly grateful, not only for their
guidance but also for standing by me in all my choices and for the
impressive amount of patience they have exhibited during the course of
these three years. Special thanks to John for his untiring correction
of my spelling, grammar and (ab)use of the English language.

I cannot thank enough my collaborators Enrico Barausse and Valerio
Faraoni, not only for their hard work on our common projects, but also
for numerous hours of conversation and debate mostly, but definitely
not exclusively, on scientific issues. It has really been a pleasure
for me to collaborate with them. I am also very grateful to Matt
Visser and Salvatore Capozziello, my thesis examiners, for undertaking
the task of reviewing this manuscript and for their invaluable
suggestions. 

During the course of this research I have benefited from systematic or
occasional but always stimulating discussions with a number of people,
besides those already mentioned. Aware of the fact that I am running
the risk of forgetting a few --- and apologising in advance for that
--- I will attempt to name them here and express my gratitude:
Sebastiano Sonego, Tomi Koivisto, Gonzalo Olmo, Mihalis Dafermos,
Sergei Odintsov, Mauro Francaviglia, Gianluca Allemandi, Carlo
Baccigalupi, Francesca Perrotta and Urbano Franca.

SISSA has provided an ideal environment for conducting my research
over the last three years. I would like to thank all of my colleagues
for contributing to that. I will have numerous lunches, dinners and
unforgettable get-togethers in Trieste to remember. Special thanks to
my office mate Federico Stivoli, for all the fun we had while sharing
a room in SISSA. 

I am also grateful to the friends and loved ones that visited me for
shorter or longer periods during my stay in Trieste. Their presence
definitely made these three years more memorable. Last, but definitely
not least, I would like to thank my family for their love and their
help in finding my way through life. If it was not for their
continuous and untiring support it would have been impossible for me
to start, let alone finish, this PhD.

\vspace{1cm}
{\bf Trieste, 23/10/2007}  \hspace{4.5cm} {\bf Thomas P.~Sotiriou}

\tableofcontents
%%%%%%%%%%%%%%%%%%%%%%
%%%%%%%%%%%%%%%%%%%%%%
\mainmatter
%%%%%%%%%%%%%%%%%%%%%%
%%%%%%%%%%%%%%%%%%%%%%
%%%%%%%%%%%%%%%%%%%%%%%%
%%%%%%%%%%%%%%%%%%%%%%%%
\chapter{Introduction}
\label{intro}
%%%%%%%%%%%%%%%%%%%%%%%%
%%%%%%%%%%%%%%%%%%%%%%%%

%%%%%%%%%%%%%%%%%%%%%%%%%%%%%%%%%%%%%%%%%%%%%%%%%%%%%%%%%%%%%%%%%
\section{General Relativity is the theory of gravity, isn't it?}
%%%%%%%%%%%%%%%%%%%%%%%%%%%%%%%%%%%%%%%%%%%%%%%%%%%%%%%%%%%%%%%%%
\label{newtoneinstein}

It is remarkable that gravity is probably the fundamental interaction
which still remains the most enigmatic, even though it is so related
with phenomena experienced in everyday life and is the one most easily
conceived of without any sophisticated knowledge.  As a matter of
fact, the gravitational interaction was the first one to be put under
the microscope of experimental investigation, obviously due to exactly
the simplicity of constructing a suitable experimental apparatus. 

Galileo Galilei was the first to introduce pendulums and inclined
planes to the study of terrestrial gravity at the end of the 16th
century. It seems that gravity played an important role in the
development of Galileo's ideas about the necessity of experiment in
the study of science, which had a great impact on modern scientific
thinking. However, it was not until 1665, when Sir Isaac Newton
introduced the now renowned ``inverse-square gravitational force
law'', that terrestrial gravity was actually united with celestial
gravity in a single theory. Newton's theory made correct predictions
for a variety of phenomena at different scales, including both
terrestrial experiments and planetary motion. 

Obviously, Newton's contribution to gravity --- quite apart from his
enormous contribution to physics overall --- is not restricted to the
expression of the inverse square law. Much attention should be paid to
the conceptual basis of his gravitational theory, which incorporates
two key ideas: i) The idea of absolute space, {\em i.e.~}the view of
space as a fixed, unaffected structure; a rigid arena in which
physical phenomena take place. ii) The idea of what was later called
the Weak Equivalence Principle which, expressed in the language of
Newtonian theory, states that the inertial and the gravitational mass
coincide.

Asking whether Newton's theory, or any other physical theory for that
matter, is right or wrong, would be ill-posed to begin with, since any
consistent theory is apparently ``right''. A more appropriate way to
pose the question would be to ask how suitable is this theory for
describing the physical world or, even better, how large a portion of
the physical world is sufficiently described by this theory. Also, one
could ask how unique the specific theory is for the description of the
relevant phenomena. It was obvious in the first 20 years after the
introduction of Newtonian gravity that it did manage to explain all of
the aspects of gravity known at that time. However, all of the
questions above were posed sooner or later.

In 1855, Urbain Le Verrier observed a 35 arc-second excess precession
of Mercury's orbit and later on, in 1882, Simon Newcomb measured this
precession more accurately to be 43 arc-seconds. This experimental
fact was not predicted by Newton's theory. It should be noted that Le
Verrier initially tried to explain the precession within the context
of Newtonian gravity, attributing it to the existence of another, yet
unobserved, planet whose orbit lies within that of Mercury. He was
apparently influenced by the fact that examining the distortion of the
planetary orbit of Uranus in 1846 had led him, and, independently, John Couch Adams, to the discovery of Neptune and the accurate
prediction of its position and momenta. However, this innermost planet
was never found.

On the other hand, in 1893 Ernst Mach stated what was later called by
Albert Einstein ``Mach's principle''. This is the first constructive
attack on Newton's idea of absolute space after the 17th century debate between Gottfried Wilhelm Leibniz and Samuel Clarke (Clarke was acting as Newton's spokesman) on the same subject, known as the Leibniz--Clarke Correspondence. Mach's idea can be
considered as rather vague in its initial formulation and it was
essentially brought into mainstream physics later on by Einstein along
the following lines: ``...inertia originates in a kind of interaction
between bodies...''. This is obviously in contradiction with Newton's
ideas, according to which inertia was always relative to the absolute
frame of space. There exists also a later, probably clearer
interpretation of Mach's Principle, which, however, also differs in substance. This was given by Dicke: ``The
gravitational constant should be a function of the mass distribution
in the universe''. This is different from Newton's idea of the
gravitational constant as being universal and unchanging. Now Newton's
basic axioms were being reconsidered.

But it was not until 1905, when Albert Einstein completed Special
Relativity, that Newtonian gravity would have to face a serious
challenge. Einstein's new theory, which managed to explain a series of
phenomena related to non-gravitational physics, appeared to be
incompatible with Newtonian gravity. Relative motion and all the
linked concepts had gone well beyond the ideas of Galileo and Newton
and it seemed that Special Relativity should somehow be generalised to
include non-inertial frames. In 1907, Einstein introduced the
equivalence between gravitation and inertia and successfully used it
to predict the gravitational redshift. Finally, in 1915, he completed
the theory of General Relativity, a generalisation of Special
Relativity which included gravity. Remarkably, the theory matched
perfectly the experimental result for the precession of Mercury's
orbit, as well as other experimental findings like the Lense-Thirring
gravitomagnetic precession (1918) and the gravitational deflection of
light by the Sun, as measured in 1919 during a Solar eclipse by Arthur
Eddington.

General Relativity overthrew Newtonian gravity and continues to be up
to now an extremely successful and well-accepted theory for
gravitational phenomena. As mentioned before, and as often happens
with physical theories, Newtonian gravity did not lose its appeal to
scientists. It was realised, of course, that it is of limited validity
compared to General Relativity, but it is still sufficient for most
applications related to gravity. What is more, at a certain limit of
gravitational field strength and velocities, General Relativity
inevitably reduces to Newtonian gravity. Newton's equations for
gravity might have been generalised and some of the axioms of his
theory may have been abandoned, like the notion of an absolute frame,
but some of the cornerstones of his theory still exist in the
foundations of General Relativity, the most prominent example being
the Equivalence Principle, in a more suitable formulation of course.

This brief chronological review, besides its historical interest, is
outlined here also for a practical reason. General Relativity is bound
to face the same questions as were faced by Newtonian gravity and many
would agree that it is actually facing them now. In the forthcoming
sections, experimental facts and theoretical problems will be
presented which justify that this is indeed the case. Remarkably,
there exists a striking similarity to the problems which Newtonian
gravity faced, {\em i.e.~}difficulty in explaining particular
observations, incompatibility with other well established theories and
lack of uniqueness. This is the reason behind the question mark in the
title of this section.

%%%%%%%%%%%%%%%%%%%%%%%%%%%%%%%%%%%%%%%%%%%%
\section{A high-energy theory of gravity?}
%%%%%%%%%%%%%%%%%%%%%%%%%%%%%%%%%%%%%%%%%%%%
\label{qg}

Many will agree that modern physics is based on two great pillars:
General Relativity and Quantum Field Theory. Each of these two
theories has been very successful in its own arena of physical
phenomena: General Relativity in describing gravitating systems and
non-inertial frames from a classical viewpoint or on large enough
scales, and Quantum Field Theory in revealing the mysteries of high
energy or small scale regimes where a classical description breaks
down.  However, Quantum Field Theory assumes that spacetime is flat
and even its extensions, such as Quantum Field Theory in curved space
time, consider spacetime as a rigid arena inhabited by quantum fields.
General Relativity, on the other hand, does not take into account the
quantum nature of matter. Therefore, it comes naturally to ask what
happens if a strong gravitational field is present at small,
essentially quantum, scales? How do quantum fields behave in the
presence of gravity? To what extent are these amazing theories
compatible?

Let us try to pose the problem more rigorously. Firstly, what needs to
be clarified is that there is no precise proof that gravity should
have some quantum representation at high energies or small scales, or
even that it will retain its nature as an interaction. The
gravitational interaction is so weak compared with other interactions
that the characteristic scale under which one would expect to
experience non-classical effects relevant to gravity, the Planck
scale, is $10^{-33}$ cm. Such a scale is not of course accessible by
any current experiment and it is doubtful whether it will ever be
accessible to future experiments either\footnote{This does not imply, of course, that imprints of Quantum Gravity phenomenology cannot be found in lower energy experiments.}. However, there are a number
of reasons for which one would prefer to fit together General
Relativity and Quantum Field Theory \cite{brill, isham}. Let us list
some of the most prominent ones here and leave the discussion about
how to address them for the next section.

\subsection{Searching for the unknown}

Curiosity is probably the motivation leading scientific research. From
this perspective it would be at least unusual if the gravity research
community was so easily willing to abandon any attempt to describe the
regime where both quantum and gravitational effects are important. The
fact that the Planck scale seems currently experimentally inaccessible
does not, in any way, imply that it is physically irrelevant. On the
contrary, one can easily name some very important open issues of
contemporary physics that are related to the Planck scale.

A particular example is the Big Bang scenario in which the universe
inevitably goes though an era in which its dimensions are smaller than
the Planck scale (Planck era). On the other hand, spacetime in General
Relativity is a continuum and so in principle all scales are relevant.
From this perspective, in order to derive conclusions about the nature
of spacetime one has to answer the question of what happens on very
small scales. 

\subsection{Intrinsic limits in General Relativity and Quantum Field 
Theory}

The predictions of a theory can place limits on the extent of its
ability to describe the physical world. General Relativity is believed
by some to be no exception to this rule. Surprisingly, this problem is
related to one of the most standard processes in a gravitational
theory: gravitational collapse. Studying gravitational collapse is not
easy since generating solutions to Einstein's field equations can be a
tedious procedure. We only have a few exact solutions to hand and
numerical or approximate solutions are often the only resort. However,
fortunately, this does not prevent one from making general arguments
about the ultimate fate of a collapsing object. 

This was made possible after the proof of the Penrose--Hawking
singularity theorems \cite{hawk,hawkellis}. These theorems state that
a generic spacetime cannot remain regular beyond a finite proper time,
since gravitational collapse (or time reversal of cosmological expansion) will inevitably lead to spacetime
singularities. In a strict interpretation, the presence of a
singularity is inferred by geodesic incompleteness, {\em i.e.~}the
inability of an observer travelling along a geodesic to extend this
geodesic for an infinite time as measured by his clock. In practical
terms this can be loosely interpreted to mean that an observer
free-falling in a gravitational field will ``hit'' a singularity in a
finite time and Einstein's equation cannot then predict what happens
next. Such singularities seem to be present in the centre of black
holes. In the Big Bang scenario, the universe itself emerges out of
such a singularity.

Wheeler has compared the problem of gravitational collapse in General
Relativity with the collapse of the classical Rutherford atom due to
radiation \cite{wheeler}. This raises hopes that principles of quantum
mechanics may resolve the problem of singularities in General
Relativity, as happened for the Rutherford model. In a more general
perspective, it is reasonable to hope that quantization can help to
overcome these intrinsic limits of General Relativity.

On the other hand, it is not only General Relativity that has an
intrinsic limit. Quantum Field Theory presents some disturbing
ultraviolet divergences. Such divergences, caused by the fact that
integrals corresponding to the Feynman diagrams diverge due to very
high energy contributions --- hence the name ultraviolet --- are
discretely removed by a process called renormalization. These
divergences are attributed to the perturbative nature of the
quantization process and the renormalization procedure is somehow
unappealing and probably not so fundamental, since it appears to cure
them in a way that can easily be considered as non-rigorous from a
mathematical viewpoint. A non-perturbative approach is believed to be
free of such divergences and there is hope that Quantum Gravity may
allow that (for early results see
\cite{pauli,deser,khrip,dewitt,isham2}).

\subsection{A conceptual clash}

Every theory is based on a series of conceptual assumption and General
Relativity and Quantum Field Theory are no exceptions. On the other
hand, for two theories to work in a complementary way to each other
and fit well together, one would expect an agreement between their
conceptual bases.  This is not necessarily the case here. 

There are two main points of tension between General Relativity and
Quantum Field Theory. The first has to do with the concept of time:
Time is given and not dynamical in Quantum Field Theory and this is
closely related to the fact that spacetime is considered as a fixed
arena where phenomena take place, much like Newtonian mechanics. On
the other hand, General Relativity considers spacetime as being
dynamical, with time alone not being such a relevant concept. It is
more of a theory describing relations between different events in
spacetime than a theory that describes evolution over some running
parameter. One could go further and seek for the connection between
what is mentioned here and the differences between gauge invariance
as a symmetry of Quantum Field Theory and diffeomorphism invariance as
a symmetry of General Relativity.

The second conceptual issue has to do with Heisenberg's uncertainty
principle in Quantum Theory which is absent in General Relativity as a
classical theory. It is interesting to note that General Relativity, a
theory in which background independence is a key concept, actually
introduces spacetime as an exact and fully detailed record of the
past, the present and the future. Everything would be fixed for a
super-observer that could look at this $4$-dimensional space from a
fifth dimension. On the other hand, Quantum Field Theory, a background
dependent theory, manages to include a degree of uncertainty for the
position of any event in spacetime. 

Having a precise mathematical structure for a physical theory is
always important, but getting answers to conceptual issues is always
the main motivation for studying physics in the first place. Trying to
attain a quantum theory of gravity could lead to such answers.

\subsection{The vision for unification}

Apart from strictly scientific reasons for trying to make a match
between Quantum Field Theory and General Relativity, there is also a
long-standing intellectual desire, maybe of a philosophical nature or
stemming from physical intuition, to bring the fundamental
interactions to a unification. This was the vision of Einstein himself
in his late years. His perspective was that a geometric description
might be the solution. Nowdays most of the scientists active in this
field would disagree with this approach to unification and there is
much debate about whether the geometric interpretation or a field
theory interpretation of General Relativity is actually preferable ---
Steven Weinberg for example even claimed in \cite{weinberg} that
``no-one'' takes a geometric viewpoint of gravity ``seriously''.
However, very few would argue that such a unification should not be
one of the major goals of modern physics. An elegant theory leading to
a much deeper understanding of both gravity and the quantum world
could be the reward for achieving this.

%%%%%%%%%%%%%%%%%%%%%%%%%%%%%%%%%%%%%%%%%%%%%%%%%%%%%%%
\section{The Cosmological and Astrophysical riddles}
%%%%%%%%%%%%%%%%%%%%%%%%%%%%%%%%%%%%%%%%%%%%%%%%%%%%%%%
\subsection{Cosmology in a nutshell}
\label{introcosm}

Taking things in chronological order, we started by discussing the
possible shortcomings of General Relativity on very small scales, as
those were the first to appear in the literature. However, if there is
one scale for which gravity is by far of the utmost importance, this
is surely the cosmic scale. Given the fact that other interactions are
short-range and that at cosmological scales we expect matter
characteristics related to them to have ``averaged out'' --- for
example we do not expect that the universe has an overall charge ---
gravity should be the force which rules cosmic evolution. Let us see
briefly how this comes about by considering Einstein's equations
combined with our more obvious assumptions about the main
characteristics of the observable universe. 

Even though matter is not equally distributed through space and by
simple browsing through the sky one can observe distinct structures
such as stars and galaxies, if attention is focused on larger scales
the universe appears as if it was made by patching together multiple
copies of the same pattern, {\it i.e.~}a suitably large elementary
volume around the Earth and another elementary volume of the same size
elsewhere will have little difference. This suitable scale is actually
$\approx 10^8$ light years, slightly larger than the typical size of a
cluster of galaxies. In Cosmology one wants to deal with scales larger
than that and to describe the universe as a whole. Therefore, as far
as Cosmology is concerned the universe can be very well described as
homogeneous and isotropic. 

To make the above statement useful from a quantitative point of view, we have 
to turn it into an idealized assumption about the matter and geometry of the 
Universe. Note that the universe is assumed to be spatially homogeneous and 
isotropic at each instant of cosmic time. In more rigorous terms, we are 
talking about homogeneity on each one of a set of 3-dimensional space-like 
hypersurfaces. For the matter, we assume a perfect fluid description and these 
spacelike hypersurfaces are defined in terms of a family of fundamental 
observers who are comoving with this perfect fluid and who can synchronise 
their comoving clocks so as to measure the universal cosmic time. The matter 
content of the universe is then just described by two parameters, a uniform 
density $\rho$ and a uniform pressure $p$, as if the matter in stars and atoms 
is scattered through space. For the geometry we idealize the curvature of space 
to be everywhere the same.

Let us proceed by imposing these assumption on the equation describing
gravity and very briefly review the derivation of the equations
governing the dynamics of the universe, namely the Friedmann
equations. We refer the reader to standard textbooks for a more
detailed discussion of the precise geometric definitions of
homogeneity and isotropy and their implications for the form of the
metric ({\it e.g.~} \cite{weinberg}). Additionally, for what comes
next, the reader is assumed to be acquainted with the basics of
General Relativity, some of which will also be reviewed in the
next chapter.

Einstein's equation has the following form
\be
\label{einstein}
G_{\mu\nu}= 8\,\pi\,G\, T_{\mu\nu},
\ee
where
\be
G_{\mu\nu}\equiv R_{\mu\nu}-\frac{1}{2}\,R\,g_{\mu\nu}
\ee
 is the Einstein tensor and $R_{\mu\nu}$ and $R$ are the Ricci tensor
and Ricci scalar of the metric $g_{\mu\nu}$. $G$ is the gravitational
constant and $T_{\mu\nu}$ is the matter stress-energy tensor. Under
the assumptions of homogeneity and isotropy, the metric can take the
form
 \be
\label{flrw}
ds^2=-dt^2+a^2(t)\left[\frac{dr^2}{1-k r^2}+r^2 d\theta^2+r^2 \sin^2(\theta)d\phi^2\right],
\ee
known as the Friedmann-Lema\^itre-Robertson-Walker metric (FLRW). 
$k=-1,0,1$ according to whether the universe is hyperspherical 
(``closed''), spatially flat, or hyperbolic (``open'') and $a(t)$ is called 
the scale factor\footnote{The traditional cosmological language of ``closed''/``flat''/``open''  
is inaccurate and quite misleading and, therefore, should be avoided. Even if one  
ignores the possibility of nonstandard topologies, the $k=0$  
spatially flat 3-manifold is, in any sensible use of the word,  
``open''. If one allows nonstandard topologies (by modding out by a  
suitable symmetry group) then there are, in any sensible use of the  
word,  ``closed'' $k=0$ spatially flat 3-manifolds (tori), and also  
``closed'' $k=-1$ hyperbolic 3-manifolds. Finally the distinction  
between flat and spatially flat is important,  and obscuring this  
distinction is dangerous.}. Inserting this metric into eq.~(\ref{einstein}) and 
taking into account that for a perfect fluid
\be
T^{\mu\nu}=(\rho+p)u^\mu u^\nu+p\, g^{\mu\nu},
\ee
where $u^\mu$ denotes the four-velocity of an observer comoving with 
the fluid and $\rho$ and $p$ are the energy density and the pressure 
of the fluid, one gets the following equations
\bea
\label{Friedmann1}
\left(\frac{\dot{a}}{a}\right)^2&=&\frac{8\,\pi\,G\,\rho}{3}-\frac{k}{a^2},\\
\label{Friedmann2}
\frac{\ddot{a}}{a}&=&-\frac{4\,\pi\,G\,}{3}\left(\rho+3p\right),
\eea
where an overdot denotes differentiation with respect to coordinate time $t$.

Eqs.~(\ref{Friedmann1}) and (\ref{Friedmann2}) are called the
Friedmann equations. By imposing homogeneity and isotropy as
characteristics of the universe that remain unchanged with time on
suitably large scales we have implicitly restricted any evolution to
affect only one remaining characteristic: its size. This is the reason
why the Friedmann equations are equations for the scale factor,
$a(t)$, which is a measure of the evolution of the size of any length
scale in the universe. Eq.~(\ref{Friedmann1}), being an equation in
$\dot{a}$, tells us about the velocity of the expansion or
contraction, whereas eq.~(\ref{Friedmann2}), which involves
$\ddot{a}$, tells us about the acceleration of the expansion or the
contraction. According to the Big Bang scenario, the universe starts
expanding with some initial velocity. Setting aside the contribution
of the $k$-term for the moment, eq.~(\ref{Friedmann1}) implies that
the universe will continue to expand as long as there is matter in it.
Let us also take into consideration the contribution of the $k$-term,
which measures the spatial curvature and in which $k$ takes the values $-1,0,1$. If
$k=0$ the spatial part of the metric (\ref{flrw}) reduces to a flat metric expressed in spherical coordinates. Therefore, the universe is spatially flat and eq.~(\ref{Friedmann1})
implies that it has to become infinite, with $\rho$ approaching zero,
in order for the expansion to halt. On the other hand, if $k=1$ the
expansion can halt at a finite density at which the matter
contribution is balanced by the $k$-term. Therefore, at a finite time
the universe will stop expanding and will re-collapse. Finally for
$k=-1$ one can see that even if matter is completely dissolved, the
$k$-term will continue to ``pump'' the expansion which means that the
latter can never halt and the universe will expand forever.

Let us now focus on eq.~(\ref{Friedmann2}) which, as already
mentioned, governs the acceleration of the expansion. Notice that $k$
does not appear in this equation, {\it i.e.~}the acceleration does not
depend on the characteristics of the spatial curvature.
Eq.~(\ref{Friedmann2}) reveals what would be expected by simple
intuition: that gravity is always an attractive force. Let us see this
in detail. The Newtonian analogue of eq.~(\ref{Friedmann2}) would be
 \be
\frac{\ddot{a}}{a}=-\frac{4\,\pi\,G\,}{3}\rho,
\ee
 where $\rho$ denotes the matter density. Due to the minus sign on the
right hand side and the positivity of the density, this equation
implies that the expansion will always be slowed by gravity. 

The presence of the pressure term in eq.~(\ref{Friedmann2}) is simply
due to the fact that in General Relativity, it is not simply matter
that gravitates but actually energy and therefore the pressure should
be included. For what could be called ordinary matter ({\em
e.g.~}radiation, dust, perfect fluids, {\em etc}.) the pressure can be
expected to be positive, as with the density. More precisely, one
could ask that the matter satisfies the four energy conditions
\cite{wald}:
 \begin{enumerate}
\item Null Energy Condition: $\rho + p \ge 0$,
\item Weak Energy Condition: $\rho\ge 0,\quad \rho + p \ge 0$,
\item Strong Energy Condition: $\rho + p \ge 0,\quad \rho + 3 p \ge 0$,
\item Dominant Energy Condition: $\rho \ge |p|$.
\end{enumerate}
 We give these conditions here in terms of the components of the
stress-energy tensor of a perfect fluid but they can be found in a
more generic form in \cite{wald}. Therefore, once positivity of the
pressure or the validity of the Strong Energy Condition is assumed,
gravity remains always an attractive force also in General Relativity
\footnote{When quantum effects are taken into account, one or more of the energy
conditions can be violated, even though a suitably averaged version
may still be satisfied. However, there are even classical fields that
can violate the energy conditions, as we will see latter on.}. 

To sum up, even without attempting to solve the Friedmann equations,
we have already arrived at a well-established conclusion: Once we assume, according to the Big Bang scenario, that the universe is expanding, then, according to General
Relativity and with ordinary matter considerations, this expansion
should always be decelerated. Is this what actually happens though?

\subsection{The first need for acceleration}
\label{inflationaryexpansion}

We derived the Friedmann equations using two assumptions: homogeneity
and isotropy of the universe. Both assumptions seem very reasonable
considering how the universe appears to be today. However, there are
always the questions of why does the universe appear to be this way
and how did it arrive at its present form through its evolution. More
importantly though, one has to consider whether the description of the
universe by the Big Bang model and the Friedmann equations is
self-consistent and agrees not only with a rough picture of the
universe but also with the more precise current picture of it. 

Let us put the problem in more rigorous terms. First of all one needs
to clarify what is meant by ``universe''. Given that the speed of
light (and consequently of any signal carrying information) is finite and adopting the
Big Bang scenario, not every region of spacetime is accessible to us.
The age of the universe sets an upper limit for the largest distance
from which a point in space may have received information. This is
what is called a ``particle horizon'' and its size changes with time.
What we refer to as the universe is the part of the universe causally
connected to us --- the part inside our particle horizon. What happens
outside this region is inaccessible to us but more importantly it does
not affect us, at least not directly. However, it is possible to have
two regions that are both accessible and causally connected to us, or
to some other observer, but are not causally connected with each
other. They just have to be inside our particle horizon without being
inside each other's particle horizons. It is intuitive that regions
that are causally connected can be homogeneous --- they have had the
time to interact. However, homogeneity of regions which are not
causally connected would have to be attributed to some initial
homogeneity of the universe since local interactions cannot be
effective for producing this. 

The picture of the universe that we observe is indeed homogeneous and
isotropic on scales larger than we would expect based on our
calculation regarding its age and causality. This problem was first
posed in the late 1960s and has been known as the horizon problem
\cite{weinberg, misner}. One could look to solve it by assuming that
the universe is perhaps much older and this is why in the past the
horizon problem has also been reformulated in the form of a question:
how did the universe grow to be so old? However, this would require
the age of the universe to differ by orders of magnitude from the
value estimated by observations. So the homogeneity of the universe,
at least at first sight and as long as we believe in the cosmological
model at hand, appears to be built into the initial conditions.

Another problem, which is similar and appeared at the same time, is
the flatness problem. To pose it rigorously let us return to the
Friedmann equations and more specifically to eq.~(\ref{Friedmann1}).
The Hubble parameter $H$ is defined as $H=\dot{a}/a$. We can use it to
define what is called the critical density
 \be
\rho_c=\frac{3\,H^2}{8\,\pi\,G},
\ee
 which is the density which would make the 3-geometry flat. Finally, 
we can use the critical density in order to create the dimensionless 
fractions
\bea
\Omega&=&\frac{\rho}{\rho_c},\\
\Omega_k&=&-\frac{k}{a^2 H^2}.
\eea
It is easy to verify from eq.~(\ref{Friedmann1}) that
\be
\label{omega}
\Omega+\Omega_k=1.
\ee
 As dimensionless quantities, $\Omega$ and $\Omega_k$ are measurable,
and by the 1970s it was already known that the current value of
$\Omega$ appears to be very close to $1$ (see for example
\cite{dickepeebles}).
 Extrapolating into the past reveals that $\Omega$ would have had to
be even closer to $1$, making the contribution of $\Omega_k$, and
consequently of the $k$-term in eq.~(\ref{Friedmann1}), exponentially
small. 

The name ``flatness problem'' can be slightly misleading and therefore
it needs to be clarified that the value of $k$ obviously remains
unaffected by the evolution. To avoid misconceptions it is therefore
better to formulate the flatness problems in terms of $\Omega$ itself.
The fact that $\Omega$ seems to be taking a value so close to the
critical one at early times is not a consequence of the evolution and
once more, as happened with the horizon problem, it appears as a
strange coincidence which can only be attributed to some fine tuning
of the initial conditions.

But is it reasonable to assume that the universe started in such a
homogeneous state, even at scales that where not causally connected,
or that its density was dramatically close to its critical value
without any apparent reason? Even if the universe started with
extremely small inhomogeneities it would still not present such a
homogeneous picture currently. Even if shortcomings like the horizon
and flatness problems do not constitute logical inconsistencies of the
standard cosmological model but rather indicate that the present state
of the universe depends critically on some initial state, this is
definitely a feature that many consider undesirable. 

So, by the 1970s Cosmology was facing new challenges. Early attempts
to address these problems involved implementing a recurring or
oscillatory behaviour for the universe and therefore were departing
from the standard ideas of cosmological evolution \cite{lemaitre,
lemaitre2, tolman}. This problem also triggered Charles W.~Misner to
propose the ``Mixmaster Universe'' (Bianchi type IX metric), in which
a chaotic behaviour was supposed to ultimately lead to statistical
homogeneity and isotropy \cite{misner2}. However, all of these ideas
have proved to be non-viable descriptions of the observed universe. 

A possible solution came in the early 1980s when Alan Guth proposed
that a period of exponential expansion could be the answer
\cite{guth}. The main idea is quite simple: an exponential increase of
the scale factor $a(t)$ implies that the Hubble parameter $H$ remains
constant. On the other hand, one can define the Hubble radius $c/H(t)$
which, roughly speaking, is a measure of the radius of the observable
universe at a certain time $t$. Then, when $a(t)$ increases
exponentially, the Hubble radius remains constant, whereas any
physical length scale increases exponentially in size. This implies
that in a short period of time, any lengthscale which could, for
example, be the distance between two initially causally connected
observers, can become larger than the Hubble radius. So, if the
universe passed through a phase of very rapid expansion, then the part
of it that we can observe today may have been significantly smaller at
early times than what one would naively calculate using the Friedmann
equations. If this period lasted long enough, then the observed
universe could have been small enough to be causally connected at the
very early stage of its evolution. This rapid expansion would also
drive $\Omega_k$ to zero and consequently $\Omega$ to $1$ today, due
to the very large value that the scale factor $a(t)$ would currently
have, compared to its initial value. Additionally, such a procedure is
very efficient in smoothing out inhomogeneities, since the physical
wavelength of a perturbation can rapidly grow to be larger than the
Hubble radius. Thus, both of the problems mentioned above seem to be
effectively addressed. 

Guth was not the only person who proposed the idea of an accelerated
phase and some will argue he was not even the first. Contemporaneously
with him, Alexei Starobinski had proposed that an exponential
expansion could be triggered by quantum corrections to gravity and
provide a mechanism to replace the initial singularity
\cite{starobinski}. There are also earlier proposals whose spirit is
very similar to that of Guth, such as those by Demosthenes Kazanas
\cite{kazanas}, Katsuhiko Sato \cite{satoinf} and Robert Brout {\em et
al.~}\cite{Brout:1979bd}. However, Guth's name is the one most related
with these idea since he was the first to provide a coherent and
complete picture on how an exponential expansion could address the
cosmological problems mentioned above. This period of accelerated
expansion is known as inflation, a terminology borrowed from economics
due to the apparent similarity between the growth of the scale factor
in Cosmology and the growth of prices during an inflationary period.
To be more precise, one defines as inflation any period in the cosmic
evolution for which
 \be
\label{inflation}
\ddot{a}> 0.
\ee

However, a more detailed discussion reveals that an exponential
expansion, or at least quasi-exponential since what is really needed
is that the physical scales increase much more rapidly than the Hubble
radius increases, is not something trivial to achieve. As discussed in
the previous section, it does not appear to be easy to trigger such an
era in the evolution of the universe, since accelerated expansion
seems impossible according to eq.~(\ref{Friedmann2}), as long as both
the density and the pressure remain positive. In other words,
satisfying eq.~(\ref{inflation}) requires
 \be
\label{inflcrit}
(\rho+3p)<0 \Rightarrow \rho <-3 p,
\ee
 and assuming that the energy density cannot be negative, inflation
can only be achieved if the overall pressure of the ideal fluid which
we are using to describe the universe becomes negative. In more
technical terms, eq.~(\ref{inflcrit}) implies the violation of the
Strong Energy Condition \cite{wald}.

It does not seem possible for any kind of baryonic matter to satisfy
eq.~(\ref{inflcrit}), which directly implies that a period of
accelerated expansion in the universe evolution can only be achieved
within the framework of General Relativity if some new form of matter
field with special characteristics is introduced. Before presenting
any scenario of this sort though, let us resort to observations to
convince ourselves about whether such a cosmological era is indeed
necessary.

\subsection{Cosmological and Astronomical Observations}
\label{observ}

In reviewing the early theoretical shortcomings of the Big Bang
evolutionary model of the universe we have seen indications for an
inflationary era. The best way to confirm those indications is
probably to resort to the observational data at hand for having a
verification. Fortunately, there are currently very powerful and
precise observations that allow us to look back to very early times.

A typical example of such observations is the Cosmic Microwave Background Radiation (CMBR).
In the early universe, baryons, photons and electrons formed a hot
plasma, in which the mean free path of a photon was very short due to
constant interactions of the photons with the plasma through Thomson
scattering. However, due to the expansion of the universe and the
subsequent decrease of temperature, it subsequently became
energetically favourable for electrons to combine with protons to form
hydrogen atoms (recombination). This allowed photons to travel freely
through space. This decoupling of photons from matter is believed to
have taken place at a redshift of $z\sim 1088$, when the age of the
universe was about $380,000$ years old or approximately $13.7$ billion
years ago. The photons which left the last scattering surface at that
time, then travelled freely through space and have continued cooling
since then. In 1965 Penzias and Wilson noticed that a Dicke radiometer
which they were intending to use for radio astronomy observations and
satellite communication experiments had an excess $3.5$K antenna
temperature which they could not account for. They had, in fact,
detected the CMBR, which actually had already been theoretically
predicted in 1948 by George Gamow. The measurement of the CMBR, apart
from giving Penzias and Wilson a Nobel prize publication
\cite{penzias}, was also to become the number one verification of the
Big Bang model. 

Later measurements showed that the CMBR has a black body spectrum
corresponding to approximately $2.7$ K and verifies the high degree of
isotropy of the universe. However, it was soon realized that attention
should be focused not on the overall isotropy, but on the small
anisotropies present in the CMBR, which reveal density fluctuations
\cite{anisot1, anisot2}. This triggered a numbered of experiments,
such as COBE, Toco, BOOMERanG and MAXIMA \cite{exper, exper2, exper3,
exper4, exper5, exper6}. The most recent one is the Wilkinson
Microwave Anisotropy Probe (WMAP) \cite{wmap} and there are also new experiments planned for the near future, such as the Planck mission \cite{planckmis}.

The density fluctuations indicated by the small anisotropies in the
temperature of CMBR are believed to act as seeds for gravitational
collapse, leading to gravitationally bound objects which constitute
the large scale matter structures currently present in the universe
\cite{liddle}. This allows us to build a coherent scenario about how
these structures were formed and to explain the current small scale
inhomogeneities and anisotropies. Besides the CMBR, which gives
information about the initial anisotropies, one can resort to galaxy
surveys for complementary information. Current surveys determining the
distribution of galaxies include the 2 degree Field Galaxy Redshift
Survey (2dF GRS) \cite{2df} and the ongoing Sloan Digital Sky Survey
(SDSS) \cite{sdss}. There are also other methods used to measure the
density variations such as gravitational lensing \cite{grl} and X-ray
measurements \cite{xrays}. 

Besides the CMBR and Large Scale Structure surveys, another class of
observations that appears to be of special interest in Cosmology are
those of type Ia supernovae. These exploding stellar objects are
believed to be approximately standard candles, {\em i.e.~}
astronomical objects with known luminosity and absolute magnitude.
Therefore, they can be used to reveal distances, leading to the
possibility of forming a redshift-distance relation and thereby
measuring the expansion of the universe at different redshifts. For
this purpose, there are a number of supernova surveys
\cite{supernovae, supernovae2, supernovae3}.

But let us return to how we can use the outcome of the experimental
measurements mentioned above in order to infer whether a period of
accelerated expansion has occurred. The most recent CMBR dataset is
that of the Three-Year WMAP Observations \cite{wmap3} and results are
derived using combined WMAP data and data from supernova and galaxy
surveys in many cases. To begin with, let us focus on the value of
$\Omega_k$. The WMAP data (combined with Supernova Legacy Survey data \cite{supernovae2})
indicates that
 \be
\label{datak}
\Omega_k=-0.015^{+0.020}_{-0.016},
\ee
 {\em i.e.~}that $\Omega$ is very close to unity and the universe
appears to be spatially flat, while the power spectrum of the CMBR
appears to be consistent with gaussianity and adiabaticity
\cite{komatsu,peiris}. Both of these facts are in perfect agreement
with the predictions of the inflationary paradigm. 

In fact, even though the theoretical issues mentioned in the previous
paragraph ({\em i.e.~}the horizon and the flatness problem) were the
motivations for introducing the inflationary paradigm, it is the
possibility of relating large scale structure formation with initial
quantum fluctuations that appears today as the major advantage of
inflation \cite{mukhanov}. Even if one would choose to dismiss, or
find another way to address, problems related to the initial
conditions, it is very difficult to construct any other theory which
could successfully explain the presence of over-densities with values
suitable for leading to the present picture of our universe at smaller
scales \cite{liddle}. Therefore, even though it might be premature to
claim that the inflationary paradigm has been experimentally verified,
it seems that the evidence for there having been a period of
accelerated expansion of the universe in the past is very compelling.

However, observational data hold more surprises. Even though $\Omega$
is measured to be very close to unity, the contribution of matter to
it, $\Omega_m$, is only of the order of 24\%. Therefore, there seems
to be some unknown form of energy in the universe, often called {\em
dark energy}. What is more, observations indicate that, if one tries
to model dark energy as a perfect fluid with equation of state $p=w
\rho$ then
 \be
\label{wde}
w_{de}=-1.06^{+0.13}_{-0.08},
\ee
 so that dark energy appears to satisfy eq.~(\ref{inflcrit}). Since it
is the dominant energy component today, this implies that the universe
should be undergoing an accelerated expansion currently as well. This
is also what was found earlier using supernova surveys
\cite{supernovae}.

As is well known, between the two periods of acceleration (inflation
and the current era) the other conventional eras of evolutionary
Cosmology should take place. This means that inflation should be
followed by Big Bang Nucleosynthesis (BBN), referring to the production
of nuclei other than hydrogen. There are very strict bounds on the
abundances of primordial light elements, such as deuterium, helium and
lithium, coming from observations \cite{burles} which do not seem to
allow significant deviations from the standard cosmological model
\cite{carrollbbn}. This implies that BBN almost certainly took place
during an era of radiation domination, {\em i.e.~}a period in which
radiation was the most important contribution to the energy density.
On the other hand, the formation of matter structures requires that
the radiation dominated era is followed by a matter dominated era. The
transition, from radiation domination to matter domination, comes
naturally since the matter energy density is inversely proportional to
the volume and, therefore, proportional to $a^{-3}$, whereas the
radiation energy density is proportional to $a^{-4}$ and so it
decreases faster than the matter energy density as the universe
expands. 

To sum up, our current picture of the evolution of the universe as
inferred from observations comprises a pre-inflationary (probably quantum gravitational) era followed
by an inflationary era, a radiation dominated era, a matter dominated
era and then a second era of accelerated expansion which is currently
taking place. Such an evolution departs seriously from the one
expected if one just takes into account General Relativity and
conventional matter and therefore appears to be quite unorthodox. 

But puzzling observations do not seem to stop here. As mentioned
before, $\Omega_m$ accounts for approximately 24\% of the energy
density of the universe. However, one also has to ask how much of this
24\% is actually ordinary baryonic matter. Observations indicate that
the contribution of baryons to that, $\Omega_b$, is of the order of
$\Omega_b\sim 0.04$ leaving some 20\% of the total energy content of
the universe and some 83\% of the matter content to be accounted for
by some unknown unobserved form of matter, called {\em dark matter}. 
Differently from dark energy, dark matter has the gravitational
characteristics of ordinary matter (hence the name) and does not
violate the Strong Energy Condition. However, it is not directly
observed since it appears to interact very weakly if at all.

The first indications for the existence of dark matter did not come
from Cosmology. Historically, it was Fritz Zwicky who first posed the
``missing mass'' question for the Coma cluster of galaxies
\cite{zwicky, zwicky2} in 1933. After applying the virial theorem in
order to compute the mass of the cluster needed to account for the
motion of the galaxies near to its edges, he compared this with the
mass obtained from galaxy counts and the total brightness of the
cluster. The virial mass turned out to be larger by a factor of almost 
400.

Later, in 1959, Kahn and Waljter were the first to propose the
presence of dark matter in individual galaxies \cite{kahn}. However,
it was in the 1970s that the first compelling evidence for the
existence of dark matter came about: the rotation curves of galaxies,
{\em i.e.~}the velocity curves of stars as functions of the radius,
did not appear to have the expected shapes. The velocities, instead of
decreasing at large distances as expected from Keplerian dynamics and
the fact that most of the visible mass of a galaxy is located near to
its centre, appeared to be flat \cite{rubin, rubin2, bosma}. As long
as Keplerian dynamics are considered correct, this implies that there
should be more matter than just the luminous matter, and this
additional matter should have a different distribution within the
galaxy (dark matter halo).

Much work has been done in the last 35 years to analyse the problem of
dark matter in astrophysical environments (for recent reviews see
\cite{dmreview1, dmreview2, dmreview3}) and there are also recent
findings, such as the observations related to the Bullet Cluster, that
deserve a special mention\footnote{Weak lensing observations of the
Bullet cluster (1E0657-558), which is actually a unique cluster
merger, appear to provide direct evidence for the existence of dark
matter \cite{bullet}.}. The main conclusion that can be drawn is that
some form of dark matter is present in galaxies and clusters of
galaxies. What is more, taking also into account the fact that dark
matter appears to greatly dominate over ordinary baryonic matter at
cosmological scales, it is not surprising that current models
of structure formation consider it as a main
ingredient ({\em e.g.~}\cite{millenium}).

\subsection{The Cosmological Constant and its problems}

We have just seen some of the main characteristics of the universe as
inferred from observations. Let us now set aside for the moment the
discussion of the earlier epochs of the universe and inflation and
concentrate on the characteristic of the universe as it appears today:
it is probably spatially flat ($\Omega_k\sim 0$), expanding in a
accelerated manner as confirmed both from supernova surveys and WMAP,
and its matter energy composition consists of approximately 76\% dark
energy, 20\% dark matter and only 4\% ordinary baryonic matter. One
has to admit that this picture is not only surprising but maybe even
embarrassing, since it is not at all close to what one would have
expected based on the standard cosmological model and what is more it
reveals that almost 96\% of the energy content of the universe has a
composition which is unknown to us.

In any case, let us see which is the simplest model that agrees with
the observational data. To begin with, we need to find a simple
explanation for the accelerated expansion. The first physicist to
consider a universe which exhibits an accelerated expansion was
probably Willem de Sitter \cite{desitter}. A de Sitter space is the
maximally symmetric, simply-connected, Lorentzian manifold with
constant positive curvature. It may be regarded as the Lorentzian
analogue of an $n$-sphere in $n$ dimensions. However, the de Sitter
spacetime is not a solution of the Einstein equations, unless one adds
a cosmological constant $\Lambda$ to them, {\em i.e.~}adds on the left
hand side of eq.~(\ref{einstein}) the term $\Lambda g_{\mu\nu}$.

Such a term was not included initially by Einstein, even though this
is technically possible since, according to the reasoning which he
gave for arriving at the gravitational field equations, the left hand
side has to be a second rank tensor constructed from the Ricci tensor
and the metric, which is divergence free. Clearly, the presence of a
cosmological constant does not contradict these requirements. In fact,
Einstein was the first to introduce the cosmological constant,
thinking that it would allow him to derive a solution of the field
equations describing a static universe \cite{einuni}. The idea of a
static universe was then rapidly abandoned however when Hubble
discovered that the universe is expanding and Einstein appears to have
changed his mind about the cosmological constant: Gamow quotes in his
autobiography, My World Line (1970): ``Much later, when I was
discussing cosmological problems with Einstein, he remarked that the
introduction of the cosmological term was the biggest blunder of his
life'' and Pais quotes a 1923 letter of Einstein to Weyl with his
reaction to the discovery of the expansion of the universe: ``If there
is no quasi-static world, then away with the cosmological term!''
\cite{pais}.

In any case, once the cosmological term is included in the Einstein
equations, de Sitter space becomes a solution. Actually, the de Sitter
metric can be brought into the form of the FLRW metric in
eq.~(\ref{flrw}) with the scale factor and the Hubble parameter given
by\footnote{Note that de Sitter space is an example of a manifold that can be sliced in 3 ways --- $k=+1$, $k=0$, $k=-1$ --- with each coordinate patch covering a different portions of spacetime. We are referring here just to the $k=0$ slicing for simplicity.}.
 \bea
a(t)&=&e^{H\,t},\\
H^2&=&\frac{8\,\pi\,G}{3}\,\Lambda.
\eea
 This is sometimes referred to as the de Sitter universe and it can 
be seen that it is expanding exponentially.

The de Sitter solution is a vacuum solution. However, if we allow the
cosmological term to be present in the field equations, the Friedmann
equations (\ref{Friedmann1}) and (\ref{Friedmann2}) will be modified
so as to include the de Sitter spacetime as a solution:
 \bea
\label{LFriedmann1}
\left(\frac{\dot{a}}{a}\right)^2&=&\frac{8\,\pi\,G\,\rho+\Lambda}{3}-\frac{k}{a^2},\\
\label{LFriedmann2}
\frac{\ddot{a}}{a}&=&\frac{\Lambda}{3}-\frac{4\,\pi\,G\,}{3}\left(\rho+3p\right).
\eea
 From eq.~(\ref{LFriedmann2}) one infers that the universe can now
enter a phase of accelerated expansion once the cosmological constant
term dominates over the matter term on the right hand side. This is
bound to happen since the value of the cosmological constant stays
unchanged during the evolution, whereas the matter density decreases
like $a^3$. In other words, the universe is bound to approach a de
Sitter space asymptotically in time. 

On the other hand $\Omega$ in eq.~(\ref{omega}) can now be split in
two different contributions, $\Omega_{\Lambda}=\Lambda/(3\,H^2)$ and
$\Omega_m$, so that eq.~(\ref{omega}) takes the form
 \be
\Omega_m+\Omega_{\Lambda}+\Omega_k=1.
\ee
 In this sense, the observations presented previously can be
interpreted to mean that $\Omega_\Lambda\sim 0.72$ and the
cosmological constant can account for the mysterious dark energy
responsible for the current accelerated expansion. One should not fail
to notice that $\Omega_m$ does not only refer to baryons. As mentioned
before, it also includes dark matter, which is actually the dominant
contribution. Currently, dark matter is mostly treated as being cold
and not baryonic, since these characteristics appear to be in good
accordance with the data. This implies that, apart from the
gravitational interaction, it does not have other interactions --- or
at least that it interacts extremely weakly --- and can be regarded as
collisionless dust, with an effective equation of state $p=0$ (we will
return to the distinction between cold and hot dark matter shortly).

We have sketched our way to what is referred to as the $\Lambda$ Cold
Dark Matter or $\Lambda$CDM model. This is a phenomenological model
which is sometimes also called the concordance model of Big Bang
Cosmology, since it is more of an empirical fit to the data. It is the
simplest model that can fit the cosmic microwave background
observations as well as large scale structure observations and
supernova observations of the accelerating expansion of the universe
with a remarkable agreement (see for instance \cite{wmap3}). As a
phenomenological model, however, it gives no insight about the nature
of dark matter, or the reason for the presence of the cosmological
constant, neither does it justify the value of the latter.

While it seems easy to convince someone that an answer is indeed
required to the question ``what exactly is dark matter and why is it
almost 9 times more abundant than ordinary matter'', the presence of
the cosmological constant in the field equations might not be so
disturbing for some. Therefore, let us for the moment put aside the
dark matter problem --- we will return to it shortly --- and consider
how natural it is to try to explain the dark energy problem by a
cosmological constant (see \cite{weinberg2, carroll1, carroll2,
peeblesratra} for reviews). 

It has already been mentioned that there is absolutely no reason to
discard the presence of a cosmological constant in the field equations
from a gravitational and mathematical perspective. Nonetheless, it is
also reasonable to assume that there should be a theoretical
motivation for including it --- after all there are numerous
modifications that could be made to the left hand side of the
gravitational field equation and still lead to a consistent theory
from a mathematical perspective and we are not aware of any other
theory that includes more than one fundamental constant. On the other
hand, it is easy to see that the cosmological term can be moved to the
right hand side of the field equations with the opposite sign and be
regarded as some sort of matter term. It can then be put into the form
of a stress-energy tensor
$T^{\mu}_{\nu}=\textrm{diag}(\Lambda,-\Lambda,-\Lambda,-\Lambda)$,
{\em i.e.~}resembling a perfect fluid with equation of state $p=-\rho$
or $w=-1$. Notice the very good agreement with the value of $w_{de}$
inferred from observations (eq.~(\ref{wde})), which explains the
success of the $\Lambda$CDM model.

Once the cosmological constant term is considered to be a matter term,
a natural explanation for it seems to arise: the cosmological constant
can represent the vacuum energy associated with the matter fields. One
should not be surprised that empty space has a non-zero energy density
once, apart from General Relativity, field theory is also taken into
consideration. Actually, Local Lorentz Invariance implies that the
expectation value of the stress energy tensor in vacuum is
 \be
\langle T_{\mu\nu}\rangle=-\langle\rho\rangle g_{\mu\nu},
\ee
 and $\langle\rho\rangle$ is generically non-zero. To demonstrate this, we can
take the simple example of a scalar field \cite{carroll3}. Its energy
density will be
 \be
\rho_\phi=\frac{1}{2}\dot{\phi}^2+\frac{1}{2}(\nabla_{\textrm{sp}}\phi)^2+V(\phi),
\ee
 where $\nabla_{\textrm{sp}}$ denotes the spatial gradient and
$V(\phi)$ is the potential. The energy density will become constant
for any constant value $\phi=\phi_0$ and there is no reason to believe
that for $\phi=\phi_0$, $V(\phi_0)$ should be zero. One could in
general assume that there is some principle or symmetry that dictates
it, but nothing like this has been found up to now. So in general one
should expect that matter fields have a non-vanishing vacuum energy,
{\em i.e.~}that $\langle \rho\rangle$ is non-zero.

Within this perspective, effectively there should be a cosmological
constant in the field equations, given by
 \be
\Lambda=8\,\pi\,G \langle\rho\rangle.
\ee
 One could, therefore, think to use the Standard Model of particle
physics in order to estimate its value. Unfortunately, however,
$\langle\rho\rangle$ actually diverges due to the contribution of very
high-frequency modes. No reliable exact calculation can be made but it
is easy to make a rough estimate once a cutoff is considered (see for
instance \cite{weinberg, carroll3}). Taking the cutoff to be the
Planck scale ($M_\textrm{Plank}=10^{18}$ GeV), which is a typical
scale at which the validity of classical gravity is becoming
questionable, the outcome is
 \be
\label{Ltheory}
\rho_{\Lambda}\sim (10^{27}\,\textrm{eV})^4.
\ee
 On the other hand, observations indicate that
 \be
\label{Lobs}
\rho_{\Lambda}\sim (10^{-3}\,\textrm{eV})^4.
\ee
 Obviously the discrepancy between these two estimates is very large
for being attributed to any rough approximation. There is a difference
of $120$-orders-of-magnitude, which is large enough to be considered
embarrassing. One could validly claim that we should not be comparing
energy densities but mass scales by considering a mass scale for the
vacuum implicitly defined through $\rho_\Lambda=M_{\Lambda}^4$.
However, this will not really make a difference, since a
30-orders-of-magnitude discrepancy in mass scale hardly makes a good
estimate. This constitutes the so-called {\em cosmological constant
problem}.

Unfortunately, this is not the only problem related to the
cosmological constant. The other known problem goes under the name of
{\em the coincidence problem}. It is apparent from the data that
$\Omega_{\Lambda}\sim 0.72$ and $\Omega_m \sim 0.28$ have comparable
values today. However, as the universe expands their fractional
contributions change rapidly since
 \be
\label{comp}
\frac{\Omega_{\Lambda}}{\Omega_m}=\frac{\rho_{\Lambda}}{\rho_m}\propto a^3.
\ee
 Since $\Lambda$ is a constant, $\rho_{\Lambda}$ should once have been
negligible compared to the energy densities of both matter and
radiation and, as dictated by eq.~(\ref{comp}), it will come to
dominate completely at some point in the late time universe. However,
the striking fact is that the period of transition between matter
domination and cosmological constant domination is very short compared
to cosmological time scales\footnote{Note that in the presence of a positive cosmological constant there is an infinite future in which $\Lambda$ is dominating.}. The puzzle is, therefore, why we live
precisely in this very special era \cite{carroll3}. Obviously, the
transition from matter domination to cosmological constant domination,
or, alternatively stated, from deceleration to acceleration, would
happen eventually. The question is, why now?

To sum up, including a cosmological constant in the field equations
appears as an easy way to address issues like the late time
accelerated expansion but unfortunately it comes with a price: the
cosmological constant and coincidence problems. We will return to this
discussion from this point later on but for the moment let us close
the present section with an overall comment about the $\Lambda$CDM
model. Its value should definitely not be underestimated. In spite of
any potential problems that it may have, it is still a remarkable fit
to observational data while at the same time being elegantly simple.
One should always bear in mind how useful a simple empirical fit to
the data may be. On the other hand, the $\Lambda$CDM model should also
not be over-estimated. Being a phenomenological model, with poor
theoretical motivation at the moment, one should not necessarily
expect to discover in it some fundamental secrets of nature.

%%%%%%%%%%%%%%%%%%%%%%%%%%%%%%%%
\section{Is there a way out?}
%%%%%%%%%%%%%%%%%%%%%%%%%%%%%%%%

In the previous sections, some of the most prominent problems of
contemporary physics were presented. As one would expect, since these
questions were initially posed, many attempts to address one or more
of them have been pursued. These problems may be viewed as being
unrelated to each other, or grouped in different categories at will.
For instance, one could follow a broad research field grouping, much
like the one attempted in the previous section, dividing them into
problems related with Cosmology and problems related with high energy
physics, or group them according to whether they refer to unexplained
observations or theoretical shortcomings. In any case there is one
common denominator in all of these problems. They are all somehow
related to gravity.

The way in which one chooses to group or divide these problems 
proposes a
natural path to follow for their solution. In this section let us very
briefly review some of the most well-known and conventional solutions
proposed in the literature, which mainly assume that all or at least
most of these issues are unrelated. Then we can proceed to argue why
and how the appearance of so many yet to be explained puzzles related
to gravity and General Relativity may imply that there is something
wrong with our current understanding of the gravitational interaction
even at a classical level, resembling the historically recorded
transition from Newtonian gravity to General Relativity described in
section \ref{newtoneinstein}. With that we will conclude this
introductory chapter.

\subsection{Scalar fields as matter fields in Cosmology}

We have already discussed the need for an inflationary period in the
early universe. However, we have not yet attempted to trace the cause
of such an accelerated expansion. Since the presence of a
cosmological constant could in principle account for that, one is
tempted to explore this possibility, as in the case of late time
acceleration. Unfortunately, this simple solution is bound not to work
for a very simple reason: once the cosmological constant dominates
over matter there is no way for matter to dominate again. Inflation
has to end at some point, as already mentioned, so that Big Bang
Nucleosynthesis and structure formation can take place. Our presence
in the universe is all the evidence one needs for that. Therefore, one
is forced to seek other, dynamical solutions to this problem.

As long as one is convinced that gravity is well described by General
Relativity, the only option left is to assume that it is a matter
field that is responsible for inflation. However, this matter field
should have a rather unusual property: its effective equation of state
should satisfy eq.~(\ref{inflcrit}), {\em i.e.~}it should have a
negative pressure and actually violate the Strong Energy Condition.
Fortunately, matter fields with this property do exist. A typical
simple example is a scalar field $\phi$. 

A scalar field minimally coupled to gravity, satisfies the
Klein--Gordon equation
 \be
\label{KG}
\nabla^2 \phi+V'(\phi)=0,
\ee
 where $\nabla_\mu$ denotes the covariant derivative, $\nabla^2\equiv
\nabla^\mu\nabla_\mu$, $V(\phi)$ is the potential and the prime
denotes partial differentiation with respect to the argument. Assuming
that the scalar field is homogeneous and therefore $\phi\equiv
\phi(t)$ we can write its energy density and pressure as
 \bea
\label{scalardensity}
\rho_\phi&=&\frac{1}{2} \dot{\phi}^2+V(\phi),\\
\label{scalarpressure}
p_\phi&=&\frac{1}{2} \dot{\phi}^2-V(\phi),
\eea
while, in a FLRW spacetime, eq.~(\ref{KG}) takes the following form:
\be
\label{mscalar}
\ddot{\phi}+3 H \dot{\phi}+V'(\phi)=0.
\ee
 It is now apparent that if $\dot{\phi}^2 < V(\phi)$ then the pressure
is indeed negative. In fact $w_\phi=p_\phi/\rho_\phi$ approaches $-1$
when $\dot{\phi}^2 \ll V(\phi)$.

In general a scalar field that leads to inflation is referred to as
the {\em inflaton}. Since we invoked such a field instead of a
cosmological constant, claiming that in this way we can successfully
end inflation, let us see how this is achieved. Assuming that the
scalar dominates over both matter and radiation and neglecting for the
moment the spatial curvature term for simplicity,
eq.~(\ref{Friedmann1}) takes the form
 \be
\label{Friedmanns}
H^2\approx \frac{8\,\pi\,G}{3}\left(\frac{1}{2} \dot{\phi}^2+V(\phi)\right).
\ee
 If, together with the condition $\dot{\phi}^2 < V(\phi)$, we require
that $\ddot{\phi}$ is negligible in eq.~(\ref{mscalar}) then
eqs.~(\ref{Friedmanns}) and (\ref{mscalar}) reduce to
 \bea
H^2&\approx& \frac{8\,\pi\,G}{3}\,V(\phi),\\
\label{sra}
3 H \dot{\phi}&\approx& -V'(\phi).
\eea
 This constitutes the {\em slow-roll approximation} since the
potential terms are dominant with respect to the kinetic terms,
causing the scalar to roll slowly from one value to another. To be
more rigorous, one can define two slow-roll parameters
 \bea
\epsilon(\phi)&=&4\,\pi\,G \left(\frac{V'}{V}\right)^2,\\
\eta(\phi)&=&8\,\pi\,G\frac{V''}{V},
\eea
 for which the conditions $\epsilon(\phi)\ll 1$ and $\eta(\phi)\ll 1$
are necessary in order for the slow-roll approximation to hold
\cite{liddlelyth, liddlelyth2}. Note that these are not sufficient
conditions since they only restrict the form of the potential. One
also has to make sure that eq.~(\ref{sra}) is satisfied. In any case,
what we want to focus on at this point is that one can start with a
scalar that initially satisfies the slow-roll conditions but, after
some period, $\phi$ can be driven to such a value so as to violate
them. A typical example is that of $V(\phi)=m^2 \phi^2/2$, where these
conditions are satisfied as long as $\phi^2>16\,\pi\, G$ but, as
$\phi$ approaches the minimum of the potential, a point will be
reached where $\phi^2>16\,\pi\, G$ will cease to hold. Once the
slow-roll conditions are violated, inflation can be naturally driven
to an end since $\dot{\phi}^2$ can begin to dominate again in
eq.~(\ref{scalarpressure}).

However, just ending inflation is not enough. After such an era the
universe would be a cold and empty place unable to evolve dynamically
to anything close to the picture which we observe today. A viable
model for inflation should include a mechanism that will allow the
universe to return to the standard Big Bang scenario. This mechanism
is called reheating and consists mainly of three processes: a period
of non-inflationary scalar field dynamics, after the slow-roll
approximation has ceased to be valid, the creation and decay of
inflaton particles and the thermalization of the products of this
decay \cite{liddle}. Reheating is an extensive and intricate subject
and analyzing it goes beyond the scope of this introduction. We refer
the reader to \cite{Abbott:1982hn, Albrecht:1982mp, Shtanov:1994ce,
Kofman:1994rk, Kofman:1997yn, linde, kolbturner} for more information.

On the same grounds, we will refrain here from mentioning specific
models for inflation and from discussing subtleties with using
inflation in order to address problems of initial conditions such as
those stated in paragraph \ref{inflationaryexpansion}. We refer the
reader to the literature for further reading \cite{linde, kolbturner,
mukhanovbook, Peebles, lyth}.

Before closing this paragraph, it should be mentioned that scalar
fields can be used to account for the late-time accelerated expansion
of the universe in the same way as the inflaton is used in
inflationary models. Since, however, this subject overlaps with the
subject of dark energy, we will discuss it in the next sub-section
which is dedicated to the dark energy problem.

\subsection{The dark energy problem}

We have already seen that there seems to be compelling observational
evidence that the universe is currently undergoing an accelerated
expansion and we have also discussed the problems that arise if a
cosmological constant is considered to be responsible for this
acceleration within the framework of the $\Lambda$CDM model. Based on
that, one can classify the attempts to address the problem of finding
a mechanism that will account for the late-time accelerated expansion
in two categories: those that try to find direct solutions to the
cosmological constant and the coincidence problems and consequently
attempt to provide an appealing theoretical explanation for the
presence and the value of the cosmological constant, and those that
abandon the idea of the cosmological constant altogether and attempt
to find alternative ways to explain the acceleration.

Let us state two of the main approaches followed to solve the
cosmological constant problem directly:

 The first approach resorts to High Energy Physics. The general idea
is simple and can be summed up in the question: Are we counting
properly? This refers to the quite naive calculation mentioned
previously, according to which the energy density of the cosmological
constant as calculated theoretically should be $10^{120}$ times larger
than its observed value. Even though the question is simple and
reasonable, giving a precise answer to it is actually very complicated
since, as mentioned already, little is known about how to make an
exact calculation of the vacuum energy of quantum fields. There are
indications coming from contemporary particle physics theories, such
as supersymmetry (SUSY), which imply that one can be led to different
values for the energy density of vacuum from the one mentioned before
(eq.~(\ref{Ltheory})). For instance, since no superpartners of known
particles have been discovered in accelerators, one can assume that
supersymmetry was broken at some scale of the order of $10^3$GeV or
higher. If this is the case, one would expect that
 \be
\rho_\Lambda\sim M^4_{\textrm{SUSY}}\ge (10^{12}\textrm{eV})^4.
\ee
 This calculation gives an estimate for the energy density of the
vacuum which is 60 orders of magnitude smaller than the one presented
previously in eq.~(\ref{Ltheory}). However, the value estimated here
is still 60 orders of magnitude larger than the one inferred from
observations (eq.~(\ref{Lobs})). Other estimates with or without a
reference to supersymmetry or based on string theory or loop quantum
gravity exist. One example is the approach of Ref.~\cite{VolovikL} where an attempt is made to use our knowledge from condensed matter systems in order to explain the value of the cosmological constant. We will not, however, list further examples here but refer the reader to
\cite{weinberg2, carroll2} and references therein for more details.
In any case, the general flavour is that it is very difficult to avoid
the cosmological constant problem by following such approaches without
making some fine tuning within the fundamental theory used to perform
the calculation for the energy density of vacuum. Also, such
approaches mostly fail to address the second problem related to the
cosmological constant: the coincidence problem.

The second direct approach for solving problems related to the
cosmological constant has a long history and was given the name
``anthropic principle'' by Brandon Carter \cite{carter, carter2,
barrowtipler}. Unfortunately, the anthropic principle leaves a lot of
room for different formulations or even misinterpretations. Following
\cite{weinberg2} we can identify at least three versions, starting
from a very mild one, that probably no one really disagrees with but
is not very useful for answering questions, stating essentially that
our mere existence can potentially serve as an experimental tool. The
second version on the other hand is a rather strong one, stating that
the laws of nature are by themselves incomplete and become complete
only if one adds the requirement that conditions should allow
intelligent life to arise, for only in the presence of intelligent
life does science become meaningful. It is apparent that such a
formulation places intelligent life or science at the centre of
attention as far as the universe is concerned. From this perspective
one cannot help but notice that the anthropic principle becomes
reminiscent of the Ptolemaic model. Additionally, to quote Weinberg:
``...although science is clearly impossible without scientists, it is
not clear that the universe is impossible without science''. The third
and most moderate version of the anthropic principle, known as the
``weak anthropic principle'' states essentially that observers will
only observe conditions which allow for observers. This version is the
one mostly discussed from a scientific perspective and even though it
might seem tautological, it acquires a meaning if one invokes
probability theory. 

To be more concrete, as opposed to the second stronger formulation,
the weak anthropic principle does not assume some sort of conspiracy
of nature aimed at creating intelligent life. It merely states that,
since the existence of intelligent observes requires certain
conditions, it is not possible for them in practice to observe any
other conditions, something that introduces a bias in any
probabilistic analysis. This, of course, requires one extra
assumption: that parts of the universe, either in space or time, might
indeed be in alternative conditions.  Unfortunately we cannot conclude
at this point whether this last statement is true. Assuming that it
is, one could put constrains on the value of the cosmological constant
by requiring that it should be small enough for galaxies to form as in
\cite{weinberg3} and arrive at the conclusion that the currently
observed value of the cosmological constant is by no means unlikely.
Some modern theories do allow such alternative states of the universe
to co-exist (multiverse), and for this reason it has recently been
argued that the anthropic principle could even be placed on firm
ground by using the ideas of string theory for the ``anthropic or
string landscape'', consisting of a large number of different false
vacua \cite{susskind}. However, admitting that there are limits on our
ability to unambiguously and directly explain the observable universe
inevitably comes with a disappointment. It is for this reason that
many physicists would refrain from using the anthropic principle or at
least they would consider it only as a last resort, when all other
possibilities have failed.

Let us now proceed to the indirect ways of solving problems related
with the cosmological constant. As already mentioned, the main
approach of this kind is to dismiss the cosmological constant
completely and assume that there is some form of dynamical dark
energy. In this sense, dark energy and vacuum energy are not related
and therefore the cosmological constant problem ceases to exist, at
least in the strict formulation given above. However, this comes with
a cost: as mentioned previously, observational data seem to be in very
good agreement with having a cosmological constant, therefore implying
that any form of dynamical dark energy should be able to mimic a
cosmological constant very precisely at present times. This is not
something easy to achieve. In order to be clearer and also to have the
possibility to discuss how well dynamical forms of dark energy can
address the cosmological constant and coincidence problems, let us use
an example.

Given the discussion presented earlier about inflation, it should be
clear by now that if a matter field is to account for accelerated
expansion, it should have a special characteristic: negative pressure
or more precisely $p\le -\rho/3$. Once again, as in the inflationary
paradigm, the obvious candidate is a scalar field. When such a field
is used to represent dark energy it is usually called {\em
quintessence} \cite{peebrat, ratpeeb, Wetterich:1987fm,
Ostriker:1995su, Caldwell:1997ii, Carroll:1998zi, Bahcall:1999xn,
Armendariz-Picon:2000dh, Wang:1999fa}. Quintessence is one of the
simplest and probably the most common alternative to the cosmological
constant. 

If the scalar field is taken to be spatially homogeneous, its equation
of motion in an FLRW spacetime will be given by eq.~(\ref{mscalar})
and its energy density and pressure will be given by
eqs.~(\ref{scalardensity}) and (\ref{scalarpressure}) respectively,
just like the inflaton. As dictated by observations through
eq.~(\ref{wde}), a viable candidate for dark energy should have an
effective equation of state with $w$ very close to minus one. In the
previous section it was mentioned that this can be achieved for a
scalar field if the condition $\dot{\phi}^2 \ll V(\phi)$ holds. This
should not be confused with the slow-roll condition for inflation,
which just requires that $\dot{\phi}^2 < V(\phi)$ and also places a
constraint for $\ddot{\phi}$. However, there is a similarity in the
spirit of the two conditions, namely that in both cases the scalar
field is required, roughly speaking, to be slowly-varying. It is worth
mentioning that the condition $\dot{\phi}^2 \ll V(\phi)$ effectively
restricts the form of the potential $V$. 

Let us see how well quintessence can address the cosmological constant
problem. One has to bear in mind that the value given in
eq.~(\ref{Lobs}) for the energy density of the cosmological constant
now becomes the current value of the energy density of the scalar
$\rho_\phi$. Since we have asked that the potential terms should be
very dominant with respect to the kinetic terms, this value for the
energy density effectively constrains the current value of the
potential. What is more, the equation of motion for the scalar field,
eq.~(\ref{mscalar}) is that of a damped oscillator, $3H\dot{\phi}$
being the friction term. This implies that, for $\phi$ to be rolling
slowly enough so that $\dot{\phi}^2 \ll V(\phi)$ could be satisfied,
then $H\sim \sqrt{V''(\phi)}$. Consequently, this means that the
current value of $V''(\phi)$ should be that of the observed cosmological
constant or, taking also into account that $\sqrt{V''(\phi)}$
represents the effective mass of the scalar $m_\phi$, that
 \be
 \label{scamass}
m_\phi\sim 10^{-33} \,\,\textrm{eV}.
\ee

Such a small value for the mass of the scalar field raises doubts
about whether quintessence really solves the cosmological constant
problem or actually just transfers it from the domain of Cosmology to
the domain of particle physics. The reason for this is that the scalar
fields usually present in quantum field theory have masses many orders
of magnitude larger than that given in eq.~(\ref{scamass}) and, hence, this poses a naturalness question (see
\cite{carroll2} for more details). For instance, one of the well-known
problems in particle physics, the hierarchy problem, concerns
explaining why the Higgs field appears to have a mass of $10^{11}$ eV
which is much smaller that the grand unification/Planck scale,
$10^{25}$-$10^{28}$ eV. As commented in \cite{carroll3}, one can then
imagine how hard it could be to explain the existence of a field with
a mass equal to $10^{-33}$ eV. In all fairness to quintessence,
however, it should be stated that the current value of the energy
density of dark energy (or vacuum, depending on the approach) is an
observational fact, and so it does not seem possible to completely
dismiss this number in some way. All that is left to do, therefore, is
to put the cosmological constant problem on new grounds that will
hopefully be more suitable for explaining it. 

One should not forget, however, also the coincidence problem. There
are attempts to address it within the context of quintessence mainly
based on what is referred to as tracker models \cite{cope, ferreira,
zlatev, liddlescherrer, steinhardt,zlatev2, sahni}. These are specific
models of quintessence whose special characteristic is that the energy
density of the scalar parallels that of matter or radiation for a part
of the evolution which is significant enough so as to remove the
coincidence problem. What is interesting is that these models do not
in general require specific initial conditions, which means that the
coincidence problem is not just turned into an initial conditions
fine-tuning problem. Of course, the dependence of such approaches on
the parameters of the potential remains inevitable.

It is also worth mentioning that $\phi$ should give rise to some
force, which judging from its mass should be long-range, if the scalar
couples to ordinary matter. From a particle physics point of view, one
could expect that this is indeed the case, even if those interactions
would have to be seriously suppressed by powers of the Planck scale \cite{carrollq, horvat}.
However, current limits based on experiments concerning a fifth-force
or time dependence of coupling constants, appear to be several orders
of magnitude lower than this expectation \cite{carrollq, horvat}. This
implies that, if quintessence really exists, then there should be a
mechanism --- probably a symmetry --- that suppresses these couplings.

Yet another possibility for addressing the cosmological constant
problems, or more precisely for dismissing them, comes when one adopts
the approach that the accelerated expansion as inferred by
observations is not due to some new physics but is actually due to a
misinterpretation or an abuse of the underlying model being used. The
Big Bang model is based on certain assumptions, such as homogeneity
and isotropy, and apparently all calculations made rely on these
assumptions. Even though at present one cannot claim that there is
compelling evidence for this, it could be, for example, that the role
of inhomogeneities is underestimated in the standard cosmological
model and a more detailed model may provide a natural solution to the
problem of dark energy, even by changing our interpretation of current
observations (for instance see \cite{buchert} and references therein).

\subsection{The dark matter problem}

As we have already seen, the presence of dark matter is indirectly
inferred from observations through its gravitational interaction.
Therefore, if one accepts that General Relativity describes gravity
correctly, then an explanation for the nature of dark matter as some
form of matter yet to be observed in the universe or in the laboratory
should be given. Note that dark matter is used here generically to
mean matter that does not emit light. So, to begin with, its nature
could be either baryonic and non-baryonic. The candidates for baryonic
dark matter are mostly quite conventional astrophysical objects such
as brown dwarfs, massive black holes and cold diffuse gas. However,
there is precise evidence from observations that only a small fraction
of dark matter can be baryonic (see for example \cite{wmap3} and
\cite{sumner, Bertone:2004pz} for reviews). Therefore, the real puzzle 
regards the nature of non-baryonic dark matter.

One can separate the candidates into two major categories: hot dark
matter, {\em i.e.~}non-baryonic particles which move
(ultra-)relativistically, and cold dark matter {\em i.e.~}non-baryonic
particles which move non-relativistically. The most prominent
candidate for hot dark matter is the the neutrino. However, studies of
the cosmic microwave background, numerical simulations and other
astrophysical observations indicate that dark matter has clumped to
form some structures on rather small scales and therefore it cannot
consist mainly of particles with high velocities, since this clumping
would then have been suppressed (see for example \cite{whiteetal,
wangzal} and references in \cite{Bertone:2004pz}). For this reason,
and because of its simplicity, cold dark matter currently gives the
favoured picture.

There are many cold dark matter candidates and so we will refrain from
listing them all or discussing their properties in detail here and
refer the reader to the literature \cite{Bertone:2004pz}. The most
commonly considered ones are the axion and a number of weakly
interacted massive particles (WIMPs) naturally predicted in
supersymmetry theories, such as the neutralino, the sneutrino, the
gravitino, the axino {\em etc}. There are a number of experiments aiming for
direct and indirect detection of dark matter and some of them, such as
the DAMA/NaI experiment \cite{Bernabei:2003za}, even claim to have
already achieved that (see \cite{dmexp} for a full list of dark matter
detection experiments and \cite{sumner} for a review of experimental
searches for dark matter). Great hope is also being placed on the
Large Hadron Collider (LHC) \cite{lhc}, which is due to start
operating shortly, to constrain the parameter space of particles
arising from supersymmetric theories. Finally, the improvement of
cosmological and astrophysical observations obviously plays a crucial
role. Let us close by saying that the general flavour or expectation
seems to be that one of the proposed candidates will soon be detected
and that the relevant dark matter scenario will be verified. Of course
expectations are not always fulfilled and it is best to be prepared
for surprises.

\subsection{OK, Quantum Gravity, but how?}

In Section \ref{qg} we discussed some of the more prominent
motivations for seeking a high energy theory of gravity which would
allow a matching between General Relativity and Quantum Field Theory.
These triggered research in this direction at a very early stage and
already in the 1950s serious efforts were being made towards what is
referred to as Quantum Gravity. Early attempts followed the
conventional approach of trying to quantize the gravitational field in
ways similar to the quantization of Electromagnetism, which had
resulted in Quantum Electrodynamics (QED). This led to influential
papers about the canonical formulation of General Relativity
\cite{adm1, adm2}. However, it was soon realized that the obvious
quantization techniques could not work, since General Relativity is
not renormalizable as is the case with Quantum Electrodynamics
\cite{uti}. In simple terms, this means that if one attempts to treat
gravity as another particle field and to assign a gravity particle to
it (graviton) then the sum of the interactions of the graviton
diverges. This would not be a problem if these divergences were few
enough to be removable via the technique called renormalization and
this is indeed what happens in Quantum Electrodynamics, as also
mentioned in Section \ref{qg}. Unfortunately, this is not the case for
General Relativity and renormalization cannot lead to sensible and
finite results. 

It was later shown that a renormalizable gravitation theory ---
although not a unitary one --- could be constructed, but only at the
price of admitting corrections to General Relativity \cite{uti,
stelle}. Views on renormalization have changed since then and more
modern ideas have been introduced such as the concept of effective
field theories. These are approximate theories with the following
characteristic: according to the length-scale, they take into account
only the relevant degrees of freedom. Degrees of freedom which are
only relevant to shorter length-scales and higher energies and are,
therefore, responsible for divergences, are ignored. A systematic way
to integrate out short-distance degrees of freedom is given by the
renormalization group (see \cite{Delamotte:2002vw} for an introduction
to these concepts).

In any case, quantizing gravity has proved to be a more difficult task
than initially expected and quantum corrections seem to appear,
introducing deviations away from General Relativity \cite{quant1,
quant2, quant3}. Contemporary research is mainly focused on two
directions: String Theory and Loop Quantum Gravity. Analysing the
basis of either of these two approaches would go beyond the scope of
this introduction and so we will only make a short mention of them. We
refer the reader to \cite{stringbooks1, stringbooks2, Duff:1996aw} and
\cite{rovelli1, Thiemann:2002nj, Ashtekar:2004eh, asht, rovelli2} for
text books and topical reviews in String Theory and Loop Quantum
Gravity respectively.

String Theory attempts to explain fundamental physics and unify all
interactions under the key assumption that the building blocks are not
point particles but one dimensional objects called strings. There are
five different versions of String Theory, namely Type I, Type IIA,
Type IIB and two types of Heterotic String Theory. M-Theory is a
proposed theory under development that attempts to unify all of the
above types. A simplified version of the idea behind String Theory
would be that its fundamental constituents, strings, vibrate at
resonant frequencies. Different strings have different resonances and
this is what determines their nature and results in the discrimination
between different forces. 

Loop Quantum Gravity follows a more direct approach to the
quantization of gravity. It is close to the picture of canonical
quantization and relies on a non-perturbative method called loop
quantization. One of its main disadvantages is that it is not yet
clear whether it can become a theory that can include the description
of matter as well or whether it is just a quantum theory of 
gravitation. 

It is worth mentioning that a common problem with these two approaches
is that, at the moment, they do not make any experimentally testable
predictions which are different from those already know from the
standard model of particle physics. As far as gravity is concerned,
String Theory appears to introduce deviations from General Relativity
(see for example \cite{gaspven, nojodi1, vassi}), whereas, the
classical limit of Loop Quantum gravity is still under investigation.

\subsection{Gravity on the stand}

In this introductory chapter, an attempt has been made to pose clearly
a series of open questions related, in one way or the other, to
gravity and to discuss some of the most common approaches currently
being pursued for their solution. This brings us to the main question
motivating the research presented in this thesis: could all or at
least some of the problems mentioned earlier be somehow related and is
the fact that General Relativity is now facing so many challenges
indicative of a need for some new gravitational physics, even at a
classical level?

Let us be more analytic. In Section \ref{newtoneinstein} we presented
a brief chronological review of some landmarks in the passage from
Newtonian Gravity to General Relativity. One could find striking
similarities with what has happened in the last decades with General
Relativity itself. For instance, the cosmological and astrophysical
observations which are interpreted as indicating the existence of dark
matter and/or dark energy could be compared with Le Verrier's
observation of the excess precession of Mercury's orbit. Remarkably,
the first attempt to explain this phenomenon, was exactly the
suggestion that an extra unseen --- and therefore dark, in a way ---
planet orbited the Sun inside Mercury's orbit. The basic motivation
behind this attempt, much like the contemporary proposals for matter
fields to describe dark matter and dark energy, was to solve the
problem within the context of an otherwise successful theory, instead
of questioning the theory itself. Another example one could give, is
the theoretical problems faced by Newtonian gravity once Special
Relativity was established. The desire for a unified description of
coordinate frames, inertial or not, and the need for a gravitational
theory that is in good accordance with the conceptual basis of Special
Relativity ({\em e.g.~}Lorentz invariance) does not seem to be very
far from the current desire for a unified description of forces and
the need to resolve the conceptual clash between General Relativity
and Quantum Field Theory.

The idea of looking for an alternative theory to describe the
gravitational interaction is obviously not new. We already mentioned
previously that attempts to unify gravity with quantum theory have
included such considerations in the form of making quantum corrections
to the gravitational field equations (or to the action, from a field
theory perspective). Such corrections became effective at small scales
or high energies. Additionally, many attempts have been made to modify
General Relativity on both small and large scales, in order to address
specific problems, such as those discussed earlier. Since we will
refer to such modification extensively in the forthcoming chapters we
will refrain from listing them here to avoid repetition. At present we
will confine ourselves to giving two very early examples of such
attempts which were not triggered so much by a theoretical or
observational need for a new theory, but by another important issue in
our opinion: the desire to test the uniqueness of General Relativity
as the only viable gravitational theory and the need to verify its
conceptual basis.

Sir Arthur Stanley Eddington, the very man who performed the
deflection of light experiment during the Solar eclipse of 1919 which
was one of the early experimental verifications of General Relativity,
was one of first people to question whether Einstein's theory was the
unique theory that could describe gravity \cite{eddin}. Eddington
tried to develop alternative theories sharing the same conceptual
basis with General Relativity, most probably for the sake of
theoretical completeness, since at the time there was no apparent
reason coming from observations. Robert Dicke was also one of the
pioneers in exploring the conceptual basis of General Relativity and
questioning Einstein's equivalence principle. He reformulated Mach's
principle and together with Carl Brans developed an alternative
theory, known as Brans--Dicke theory \cite{dicke, bransdicke}. Part of
the value of Dicke's work lies on the fact that it helped people to
understand that we do not know as much as we thought about the basic
assumptions of General Relativity, a subject that we will discuss
shortly.

Even though the idea of an alternative theory for gravitation is not
new, a new perspective about it has emerged quite recently. The
quantum corrections predicted in the 1960s were expected to appear
only at small scales. On the other hand, Eddington's modification or
Brans--Dicke theory were initially pursued as a conceptual alternative
of General Relativity and had phenomenological effects on large scales
as well. Now, due to both the shortcomings of Quantum Gravity and the
puzzling cosmological and astrophysical observations, these ideas have
stopped being considered unrelated. It seem worthwhile to consider the
possibility of developing a gravitation theory that will be in
agreement with observations and at the same time will be closer to the
theories that emerge as a classical limit of our current approaches to
Quantum Gravity, especially since it has been understood that quantum
corrections might have an effect on large scale phenomenology as well.

Unfortunately, constructing a viable alternative to General Relativity
with the above characteristics is far from being an easy task since
there are numerous theoretical and observational restrictions. Two
main paths have been followed towards achieving this goal: proposing
phenomenological models tailored to fit observations, with the hope
that they will soon gain some theoretical motivation from high energy
physics and current Quantum Gravity candidates, and developing ideas
for Quantum Gravity, with the hope that they will eventually give the
answer in the form of an effective gravitational theory through their
classical limit which will account for unexplained observations. In
this thesis a different approach will be followed in an attempt to
combine and complement these two. At least according to the author's opinion, we seem to be still at 
too early a stage in the development of our ideas about Quantum 
Gravity to be able to give precise answers about the type and form 
of the expected quantum corrections to General Relativity. Current 
observations still leave scope for a wide range of different 
phenomenological models and so it seems a good idea to attempt 
exploring the limits of classical gravity by combining theory and 
observations. In a sense, this approach lies somewhere in the middle 
between the more conventional approaches mentioned before. Instead 
of starting from something known in order to extrapolate to the 
unknown, we attempt here to jump directly into the unknown, hoping 
that we will find an answer.

To this end, we will examine different theories of gravity, trying to
determine how far one can go from General Relativity. These theories
have been chosen in such a way as to present a resemblance with the
low energy effective actions of contemporary candidates for Quantum
Gravity in a quest to study the phenomenology of the induced
corrections. Their choice has also been motivated by a desire to fit
recent unexplained observations. However, it should be stressed that
both of these criteria have been used in a loose manner, since the
main scope of this study is to explore the limits of alternative
theories of gravity and hopefully shed some light on the strength and
validity of the several assumptions underlying General Relativity. In
that sense, many of the theories which we will consider can be
regarded as toy theories or straw-man theories. The main motivation comes from the fear that we may not know as 
much as we think or as much as needed to be known before making 
the key steps pursued in the last 50 years in gravitational physics; 
and from the hope that a better understanding of classical gravity 
might have a lot to offer in this direction.

As a conclusion to this introduction it is worth saying the following:
it is probably too early to conclude whether it is General Relativity
that needs to be modified or replaced by some other gravitational
theory or whether other solutions to the problems presented in this
chapter, such as those mentioned earlier, will eventually give the
required answers. However, in scientific research, pursuing more than
one possible solution to a problem has always been the wisest and most
rewarding choice; not only because there is an already explored
alternative when one of the proposed solutions fails, but also due to
the fact that trial and error is one of the most efficient ways to get
a deeper understanding of a physical theory. Exploring alternative
theories of gravity, although having some disadvantages such as
complexity, also presents a serious advantage: it is bound to be
fruitful even if it leads to the conclusion that General Relativity is
the only correct theory for gravitation, as it will have helped us
both to understand General Relativity better and to secure our faith 
in it.

\chapter{Foundations of Gravitation Theory}
\label{foundations}
 \section[Viability criteria and the Equivalence 
Principle(s)]{Viability criteria and the various forms of the Equivalence Principle}
\label{epetc}

\subsection{Viability and the Dicke framework}

Even though it took only 4 years for having the first experimental
verification of General Relativity to appear --- Eddington's
measurement of light deflection in 1919 --- Einstein's theory did not
become the object of systematic and accurate experimental testing
until the early 1960s. In fact, it was only in 1960 that the
gravitational redshift of light was successfully measured by Pound and
Rebka \cite{pound} even though this test was proposed by Einstein in
1907 and it is considered one of the three classical tests of General
Relativity, together with the perihelion shift of Mercury and light
deflection. After that, a number of new experimental tests were
performed based on effects which were either new or which had been
discovered earlier but their verification was not technologically
possible at the time. Examples range from the Lense--Thirring effect
\cite{Lense}, the Nordtvedt effect \cite{nordteff} and Shapiro time
delay \cite{shapirotd} to the Nobel Prize discovery of the binary
pulsar by Taylor and Hulse \cite{hulse} which led to the first
indirect evidence for the existence of gravitational waves (for a
historical review see Chapter 1 of \cite{willbook}).

However, it was soon realised and first proposed by Schiff and Dicke
referring to the redshift experiments \cite{schiff, dickered}, that
gravitational experiments do not necessarily test General Relativity,
since, instead of testing the validity of specific field equations,
experiments test the validity of principles, such as the equivalence
principle. Contemporarily, a number of alternative theories of
gravitation had been developed, many of which shared some of the
principles of General Relativity and were therefore indistinguishable
as far as some of the tests were concerned. This triggered the
development of powerful tools for distinguishing and testing theories,
the most commonly used of which is the Parametrized Post-Newtonian
(PPN) expansion, pioneered by Nordtvedt and extended by Nordtvedt and
Will \cite{nordtppn, willppn1, nordtwillppn, willppn2}. 

The idea that experiments actually test principles and not specific
theories highlights the importance of exploring the conceptual basis
of a gravitation theory. In fact, it would be very helpful to
provide a framework for analysing gravitation theories and
experiments and deriving general conclusions about the viability
criteria of the theories. This would provide a starting point for
constructing gravitation theories which are not obviously non-viable
for theoretical or experimental reasons. Dicke was one of the pioneers
in this direction and presented what was later known as the Dicke
Framework \cite{dickeframe}. We will focus on this for the rest of
the present section.

Following \cite{willbook} we identify the two mathematical assumptions
of the Dicke Framework as being:
 \begin{itemize}
 \item Spacetime is a $4$-dimensional manifold, with each point in the 
manifold corresponding to a physical event.
 \item The equations of gravity and the mathematical entities in them
are to be expressed in a form that is independent of the coordinates
used, {\em i.e.} in covariant form.
 \end{itemize}
 A comment is due for each of these statements. Regarding the first
one, it should be stressed that it does not presuppose that the
manifold has either a metric or an affine connection, since one would
prefer to arrive at this as a conclusion from experiments. Regarding
the second one, it is important to bear in mind that non-covariant
equations can in many cases be written in a covariant form if a number
of covariant constraints are imposed. Such constraints introduce
absolute structures into the theory ({\em e.g.~}preferred coordinate
frames) and therefore, even though coordinate invariance is justified,
the theory does not really become background independent (see
\cite{giulini} for an interesting discussion). From this viewpoint,
requiring only covariance of the field equations is not very
restrictive.

Dicke also proposed two further assumptions related to those just
presented: that gravity should be associated with one or more fields
of tensorial character (scalars, vectors, tensors) and that the field
equation governing the dynamics of gravity should be derivable from an
invariant action via a stationary action principle. The first of these
two assumptions appears as an almost direct consequence of the two
previous ones, whereas the second seems less fundamental, at least at
a classical level, and should not be imposed lightheartedly because it
may lead to unnecessary confinement of acceptable theories.

The assumptions of Dicke's framework are probably the minimal unbiased
assumptions that one can start with in order to develop a
gravitation theory. There are also other fundamental criteria which
a gravitation theory should satisfy in order to be viable. From a
theoretical viewpoint there are the two basic requirements of all
theories, {\em i.e.}
 \begin{enumerate}
 \item completeness: The theory should be able to analyse from ``first 
principles'' the outcome of any experiment,
 \item self-consistency: predictions should be unique and independent 
of the calculation method.
\end{enumerate}
 From an experimental viewpoint there are two more very basic 
requirements:
\begin{enumerate}
 \item The theory should be relativistic, {\em i.e.} should reduce to 
Special Relativity when gravity is ``turned off'' (and at low energies).
 \item The theory should have the correct Newtonian limit, {\em i.e.}
in the the limit of weak gravitational fields and slow motion it
should reproduce Newton's laws.
 \end{enumerate}
 Both of these requirements are based on the fact that Special
Relativity and Newtonian Gravity are extremely well tested theories
--- at least in regimes in which we theoretically expect them to be
valid --- and therefore any gravitation theory should be able to
reproduce them in the suitable limit (see also \cite{willbook} for
more details).

One would like to combine with the above requirements also experiments
that aim directly at testing gravity in its full glory in order to
confine acceptable theories. We intend to do so in what follows.

\subsection{Equivalence Principle(s)}
\label{ep}

In an abstract (and loose) sense a theory can usually be thought of as
a set of axioms from which one can derive logical statements. When it
comes to a physical theory one should also add that the statements of
the theory should be able to predict the outcome of experiments that
fall with its purview. However, it is common to think of General
Relativity or other gravitation theories as a set of field equations
(or an action). A complete and coherent axiomatic formulation of
Einstein's theory, or any other gravitation theory, is still pending;
the viability criteria presented above are a step in this direction,
but when referring to an axiomatic formulation one needs to go
further. What is needed here is to formulate ``principles''. One of
these is the general covariance principle included in the Dicke
Framework\footnote{We will return to the issue of the axiomatic
formulation of gravitation theories in Chapter \ref{concl}.}. 

A principle that has triggered much more discussion and is probably
much less understood is the Equivalence Principle, or more precisely
each of its various formulations. As we have already mentioned in the
Introduction, a formulation of the Equivalence Principle was
already incorporated in Newtonian gravity. Newton pointed out in {\it
Principia} that the ``mass'' of any body --- meaning the quantity that
regulates its response to an applied force --- and the ``weight'' of
the body --- the property regulating its response to gravitation ---
should be equal. The terms ``inertial mass'' and ``passive
gravitational mass'' where later introduced by Bondi \cite{bondi1} to
distinguish the quantities present in Newton's second law of motion
 \be
\vec{F}=m_I \vec{a},
\ee
 where $\vec{F}$ is the force $3$-vector and $\vec{a}$ is the 
$3$-acceleration, and Newton's gravitation law
\be
\vec{F}=m_P \vec{g},
\ee
 where $\vec{g}$ is the gravitational acceleration $3$-vector. In
terms of $m_I$ and $m_P$, the Equivalence Principle in Newtonian
theory can be rigorously expressed as
 \be
\label{inerteqpass}
m_I=m_p.
\ee

Einstein, by the use of gedanken experiments such as the famous free falling elevator one, realised that a free falling observer does
not feel the effects of gravity and saw in a reformulation of the
Equivalence Principle the foundations for generalizing Special
Relativity and developing a theory to describe both non-inertial
frames and gravity. The meaning of mass in such a theory is
questionable and so the Equivalence principle should be expressed in
terms of some more fundamental concept. Free-fall comes to the rescue.
Eq.~(\ref{inerteqpass}) within the framework of Newtonian gravity,
implies in practice that all bodies experience the same acceleration
when they are in free-fall, irrespective of composition. An expression
of some kind of universality of free fall should therefore be sought
for when attempting to reformulate the Newtonian version of the
Equivalence Principle.

We will not attempt to review such endeavours historically. We focus
directly on the several current forms of the Equivalence Principle.
These are\footnote{We follow the definitions given in \cite{willbook}.
The reader should be cautious since several different formulations
exist in the literature and the terminology can be misleading ({\em
e.g.~}some authors refer to the EEP as WEP or to the SEP as EEP).}:

{\it Weak Equivalence Principle} (WEP): If an uncharged test body is
placed at an initial event in spacetime and given an initial velocity
there, then its subsequent trajectory will be independent of its
internal structure and composition.

{\it Einstein Equivalence Principle} (EEP): (i) the WEP is valid, (ii)
the outcome of any local non-gravitational test experiment is
independent of the velocity of the freely-falling apparatus (Local
Lorentz Invariance-LLI) and (iii) the outcome of any local
non-gravitational test experiment is independent of where and when in
the universe it is performed (Local Position Invariance-LPI).

{\it Strong Equivalence Principle} (SEP): (i) the WEP is valid for
self-gravitating bodies as well as for test bodies, (ii) the outcome
of any local test experiment is independent of the velocity of the
freely-falling apparatus (Local Lorentz Invariance-LLI) and (iii) the
outcome of any local test experiment is independent of where and when
in the universe it is performed (Local Position Invariance-LPI).

In order for these definitions to be clear, the following
clarifications are needed \cite{willbook}: An uncharged test body is
an electrically neutral body that has negligible self-gravitation as
estimated using Newtonian theory (it does not contribute to the
dynamics of the gravitational field) and it is small enough in size so
that its couplings to inhomogeneities in the external fields can be
ignored. A local non-gravitational test experiment is defined to be
any experiment which is performed in a freely falling laboratory which
is shielded and small enough in size for inhomogeneities in the
external fields to be ignored throughout its volume. Additionally,
self-gravitating effects in this laboratory should be negligible.

Let us now focus on the differences between the WEP, EEP and SEP. The
WEP implies that spacetime is endowed with a family of preferred
trajectories which are the world lines of freely falling test bodies.
Note that the existence of a metric is not suggested by the WEP. Even
if an external assumption for the existence of the metric is made
though, the geodesics of this metric do not necessarily coincide with
the free-fall trajectories as far the WEP is concerned.

The EEP adds two more statement to the WEP: Local Lorentz Invariance
and Local Position Invariance. A freely-falling observer carries a
local frame in which test bodies have unaccelerated motions. According
to the requirements of the LLI, the outcomes of non-gravitational
experiments are independent of the velocity of the freely-falling
frame and therefore if two such frames located at the same event
${\cal P}$ have different velocities, this should not affect the
predictions for identical non-gravitational experiments. Local
Position Invariance requires that the above should hold for all
spacetime points. Therefore, roughly speaking, in local freely falling
frames the theory should reduce to Special Relativity. 

This implies that there should be at least one second rank tensor
field which reduces, in the local freely falling frame, to a metric
conformal with the Minkowski one. The freedom of having an arbitrary
conformal factor is due to the fact that the EEP does not forbid a
conformal rescaling in order to arrive at special-relativistic
expressions for the physical laws in the local freely-falling frame.
Note however, that while one could think of allowing each specific
matter field to be coupled to a different one of these conformally
related second rank tensors, the conformal factors relating these
tensors can at most differ by a multiplicative constant if the
couplings to different matter fields are to be turned into constants
under a conformal rescaling as the LPI requires (this highlights the
relation between the LPI and varying coupling constants)\footnote{This
does not exclude the possibility of having a second metric tensor in
the theory as long as this metric does not couple to the matter (this
then leads to theories of the bi-metric kind).}. We can then
conclude that rescaling coupling constants and performing a conformal
transformation leads to a metric $g_{\mu\nu}$ which, in every freely
falling local frame reduces (locally) to the Minkowski metric
$\eta_{\mu\nu}$.

It should be stressed that all conformal metrics $\phi g_{\mu\nu}$
($\phi$ being the conformal factor) can be used to write down the
equations or the action of the theory. $g_{\mu\nu}$ is only special in
the following sense: Since at each event ${\cal P}$ there exist local
frames called local Lorentz frames, one can find suitable coordinates
in which at ${\cal P}$
 \be
\label{localmet}
g_{\mu\nu}=\eta_{\mu\nu}+{\cal O}\left(\sum_\alpha |x^\alpha-x^\alpha({\cal P})|^2\right),
\ee
 and $\partial g_{\mu\nu}/\partial x^\alpha=0$.  In local Lorentz
frames, the geodesics of the metric $g_{\mu\nu}$ are straight lines.
Free-fall trajectories are straight lines in a local freely-falling
frame. Identifying the two frames we realize that the geodesics of
$g_{\mu\nu}$ coincide with free-fall trajectories. Put in other words,
the EEP requires the existence of a family of conformal metrics, one
of which should have geodesics which coincide with free-fall
trajectories.

Finally, let as focus on the SEP. The SEP extends the validity of the
WEP to self-gravitating bodies and the validity of the EEP to local
gravitational experiments. Note that the Newtonian Equivalence
principle also did not make a distinction between test bodies and self
gravitating bodies. Extending the validity of LLI and LPI to local
gravitational experiments is also a quite strong requirement. For the
time being there is no theory other than General Relativity that
satisfies the SEP. However, there is no explicit proof that the SEP
leads uniquely to General Relativity.

Let us attempt to argue heuristically that this is indeed the case.
First we have to understand how local gravitational experiments are
influenced by the form of the theory. Following \cite{willbook} we can
consider a local freely-falling frame that is small enough for
inhomogeneities in the external gravitational fields to be neglected
throughout its volume, but is large enough to encompass a system of
gravitating matter and its associated gravitational fields. We do not
assume here that the metric is the only gravitational field. In order
to solve the field equations and determine the behaviour of the
system, we need to impose boundary conditions, {\em i.e.}~determine
the values of the fields, gravitational or not, on the boundary of our
local frame. These values will generically depend on the behaviour of
the fields far from the local frame which we are considering.

Since the EEP is anyway included in the SEP, let as assume that the
EEP is indeed valid. LLI and LPI imply that the outcome of local
non-gravitational experiments should be unaffected by the boundary
values of gravitational fields other than the metric, since these are
sensitive to the position or velocity of the frame, depending on their
nature (see also \cite{willbook}). Therefore, in a representation in
which $g_{\mu\nu}$ is taken to be the metric, any gravitational fields
other than the metric should not be coupled to matter directly due to
the EEP (recall the freedom to use conformal metrics). 

Let us now suppose that the theory indeed includes gravitational
fields other than the metric that are not directly coupled to the
matter. If one tries to solve the field equations and determine the
outcome of gravitational experiments, then the boundary values of
these fields will influence the result. This directly implies that the
SEP cannot be satisfied.

All that is left is to consider theories in which the only
gravitational field is the metric. In the local frame which we are
considering it is always possible to find a coordinate system in which
$g_{\mu\nu}$ reduces to $\eta_{\mu\nu}$ and $\partial
g_{\mu\nu}/\partial x^{\alpha}=0$ at the boundary between the local
system and its surroundings ({\em cf.~}eq.~(\ref{localmet}))
\cite{willbook}. Therefore, if the field equation for the metric
contains derivatives of the metric of not higher than second order,
then the outcome of any experiment, gravitational or
non-gravitational, is independent of the surroundings and is therefore
independent of the position and velocity of the frame. 

However, this is not the case if the field equation for the metric is
of higher differential order. The boundary values of the second or
higher derivatives of the metric cannot be ``trivialized'' in any
coordinate system and the outcome of gravitational experiments becomes
sensitive to the position and velocity of the local frame.
Therefore, theories that include higher order derivatives of the
metric, such as fourth-order gravity, do not satisfy the SEP
\footnote{This contradicts what is claimed in \cite{willbook}.}.

We conclude that theories which can satisfy the SEP should not include
gravitational fields other than the metric and that the differential
order of the field equations should be at most second order. As it
stands, this discussion does not prove that General Relativity is the
only theory that satisfies the SEP. However, if one adds some of the
viability arguments listed in the previous section, then the candidate
list reduces significantly. For instance, if one requires that the
theory should come from an action, it is quite straightforward to
show that the Einstein--Hilbert action, modulo surface terms or
topological invariants (see Section \ref{homet}), is the only
generally covariant action that depends only on the metric and leads
to second order field equations under metric variation. There is
therefore strong evidence to believe that the validity of the SEP
leads to General Relativity. It should be stressed that in principle
it is possible to build up other theories that satisfy the SEP but up
to this point only one is actually known: Nordstr\"om's
conformally-flat scalar theory, which dates back to 1913
\cite{nord1913}. However, even this theory is not viable since it
predicts no deflection of light.

Before closing this section, we return to one of the initial
motivations for discussing the principles which a viable gravitation
theory should satisfy: the fact that experiments do not always test
theories, but more often they test principles. There are specific
tests for each version of the Equivalence Principle. The basis of the
WEP is the universality of free-fall, {\em i.e.}~the requirement that
different (test) bodies should experience the same acceleration in an
external gravitational field irrespective of their composition.
Experiments testing the WEP attempt to measure the fractional
difference in acceleration between two bodies, leading to what is
called the ``E\"otv\"os ratio'', named after the classic torsion
balance measurements of E\"otv\"os \cite{eotvos}. There are many very
sophisticated experiments trying to measure violations of the WEP with
accuracies close to $10^{-13}$ and hoping to reach $10^{-17}$ soon. We
refer the reader to \cite{willliv} and references therein for
details.

To test the EEP one has to test, apart from the WEP, Local Lorentz
Invariance and Local Position Invariance. LLI is a principle already
embodied in Special Relativity. From this perspective, questioning it
would affect not only gravitation theories, but also most of modern
physics in general. However, a violation of LLI would not necessarily
constitute a menace for physics as we know it. It can just be a
manifestation of new ``beyond Einstein'' physics related, for
instance, to Quantum Gravity phenomenology. For more information on
testing LLI, we refer the reader to \cite{willliv} and references
therein, and especially to the thorough review of Mattingly
\cite{mattingly}. As far as LPI is concerned, there are two crucial
tests: gravitational redshift experiments ({\em e.g.~}measurement of
clock frequencies at different spacetime locations) and measurements
of possible variations of non-gravitational coupling constants. One
should stress at this point that LPI also refers to the position in
time. See \cite{willliv} for a thorough presentation of relevant
experiments.

Finally, there are tests related to the SEP. Recall that the SEP
extends the validity of the EEP to gravitational experiments as well.
Amongst the most common experimental tests of the SEP are measurements
of the possible variation of the gravitational constant,
preferred-location and preferred-frame effects in the locally measured
gravitational constant, possible violations of the WEP for gravitating
bodies, {\em etc}. \cite{willliv}.

\subsection{Metric Postulates}
\label{metpost}

So far we have stated a number of principles which it is reasonable to
assume that all viable gravitational theories should satisfy.
Experiments will show the extent to which these assumptions are valid
by placing constraints on the possible violations of the principles
concerned. However, this is not quite the end of the story. From a
practical perspective, it is not at all straightforward to understand
whether a specific theory does satisfy these principles. More
precisely, if one is given an action or a set of field equations,
usually a series of tedious manipulations will have to be performed
before concluding that the EEP, for instance, is valid within the
framework of the theory represented by them. 

The reverse problem is also of interest: Given that we have a series
of principles which our theories have to satisfy, can we turn them
into practical constraints on their general form? Could we identify
some mathematically, and not abstractly, formulated characteristics
which a candidate theory should have in order to comply with the
principles described above? An attempt in this direction was made in
1971 by Thorne and Will with the introduction of the so-called metric
postulates \cite{thornewill}. We have already described, in the
previous chapter, how the validity of the EEP implies the existence of
the metric (a member of a family of conformal metrics) whose geodesics
coincide with the free-fall trajectories. This is encapsulated in
Thorne and Will's metric postulates:
 \begin{enumerate}
\item there exists a metric $g_{\mu\nu}$ (second rank non-degenerate 
tensor),
\item $\nabla_{\mu}T^{\mu\nu}=0$ where $\nabla_{\mu}$ is the covariant 
derivative defined with the Levi--Civita connection of this metric and 
$T_{\mu\nu}$ is the stress-energy tensor of non-gravitational (matter) 
fields.
\end{enumerate}
 Note that geodesic motion can be derived using the second metric
postulate \cite{fock}. Theories that satisfy the metric postulates are
referred to as {\em metric theories}.

The metric postulates have proved to be very useful. They are part of
the foundation for the Parametrized Post-Newtonian expansion which has
been extensively used to constrain alternative theories of gravity.
They do, however, have a major disadvantage. As pointed out also by
the authors of \cite{thornewill}, any metric theory can perfectly well
be given a representation that appears to violate the metric
postulates (recall, for instance, that $g_{\mu\nu}$ is a member of a
family of conformal metrics and that there is no {\it a priori} reason
why this metric should be used to write down the field equations). On
top of that, one can add that there are some ambiguities in the
definition of quantities related to the metric postulates. For
example, what exactly is the precise definition of the stress energy
tensor and which fields are included in it? What exactly is the
difference between gravitational and non-gravitational fields?

Let us not elaborate more on these issues here since they will become
much more apparent once we study some alternative theories of gravity
in the next chapter and discuss the equivalence between theories.
Therefore, it is preferable to return to this issue later on. We close
the present section by pointing out that the metric postulates are, at
the moment, the closest thing we have to a guide about where to start
when constructing alternative theories of gravity. 

\section{Geometric description of spacetime}
\label{spacetimegeo}

It should be clear from the discussion of the previous sections that
there is very strong motivation for assuming that gravity is related
to spacetime geometry and that any reasonable theory for the
gravitational interaction is most likely to include a metric. 
Therefore it is useful before going further to take a moment to recall
the basics of the geometric description of a 4-dimensional manifold.
This is by no means a rigorous introduction to the differential
geometry of 4-dimensional manifolds but merely a collection of some
basic definitions which will prove useful later on and in which a
physicist's perspective is probably apparent. 

Let us start by considering a 4-dimensional manifold with a
connection, $\Gamma_{\phantom{a}\mu\nu}^\lambda$, and a symmetric
metric $g_{\mu\nu}(=g_{\nu\mu})$. By definition the metric allows us
to measure distances. We assume that this metric is non-degenerate and
therefore invertible. Consequently it can be used to raise and lower
indices. The connection is related to parallel transport and therefore
defines the covariant derivative. The definition is the following:
 \be
\gn_\mu A^\nu_{\phantom{a}\sigma}=\partial_\mu A^\nu_{\phantom{a}\sigma}+\Gamma^\nu_{\phantom{a}\mu\alpha} A^\alpha_{\phantom{a}\sigma}-\Gamma^\alpha_{\phantom{a}\mu\sigma} A^\nu_{\phantom{a}\alpha}.
\ee
 We give this here even though it may be considered trivial, since
several different conventions exist in the literature.  Additionally
one has to be careful about the position of the indices when the
connection is not symmetric.

Notice that we use $\gn_\mu$ to denote the covariant derivative here
because we have not yet related $\Gamma_{\phantom{a}\mu\nu}^\lambda$
in any way to the metric. This would be an extra assumption that is
not needed at this stage. This connection is not assumed to be the
Levi--Civita connection of $g_{\mu\nu}$ and the symbol $\nabla_\mu$ is
reserved for the covariant derivative defined with the latter. 

 Using the connection, one can construct the Riemann tensor:
\be
\label{riemann}
{\cal R}^\mu_{\phantom{a}\nu\sigma\lambda}=-\partial_\lambda\Gamma^\mu_{\phantom{a}\nu\sigma}+\partial_\sigma\Gamma^\mu_{\phantom{a}\nu\lambda}+\Gamma^\mu_{\phantom{a}\alpha\sigma}\Gamma^\alpha_{\phantom{a}\nu\lambda}-\Gamma^\mu_{\phantom{a}\alpha\lambda}\Gamma^\alpha_{\phantom{a}\nu\sigma}\, .
\ee
 which has no dependence on the metric. Notice that the Riemann tensor
here has only one obvious symmetry; it is antisymmetric in the last
two indices. The rest of the standard symmetries are not present for
an arbitrary connection \cite{schro}.

Since we do not assume here any relation between the metric and the
connections, the former is not necessarily covariantly conserved. The
failure of the connection to preserve the metric is usually measured
by the non-metricity tensor:
 \be
\label{nonmet}
Q_{\mu\nu\lambda}\equiv-\gn_\mu g_{\nu\lambda}.
\ee
 The trace of the non-metricity tensor with respect to its last two 
(symmetric) indices is called the Weyl vector:
\be
\label{weyl}
Q_\mu\equiv \frac{1}{4}Q_{\mu\nu}^{\phantom{a}\phantom{b}\nu}.
\ee
 At the same time, the connection is not necessarily symmetric. The 
antisymmetric part of the connection is often called the Cartan 
torsion tensor:
\be
\label{cartan}
S_{\mu\nu}^{\phantom{ab}\lambda}\equiv \Gamma^{\lambda}_{\phantom{a}[\mu\nu]}.
\ee

One has to be careful when deriving the Ricci tensor in this case,
since only some of the standard symmetry properties of the Riemann
tensor hold here. A straightforward contraction leads, in fact, to two
Ricci tensors \cite{schro}:
 \be
{\cal R}_{\mu\nu}\equiv {\cal R}^\sigma_{\phantom{a}\mu\sigma\nu}=-{\cal R}^\sigma_{\phantom{a}\mu\nu\sigma}, \qquad {\cal R}'_{\mu\nu}\equiv {\cal R}^\sigma_{\phantom{a}\sigma\mu\nu}.
\ee
The first one is the usual Ricci tensor given by
\be
\label{ricci}
{\cal R}_{\mu\nu}={\cal R}^\lambda_{\phantom{a}\mu\lambda\nu}=\partial_\lambda \Gamma^\lambda_{\phantom{a}\mu\nu}-\partial_\nu \Gamma^\lambda_{\phantom{a}\mu\lambda}+\Gamma^\lambda_{\phantom{a}\sigma\lambda}\Gamma^\sigma_{\phantom{a}\mu\nu}-\Gamma^\lambda_{\phantom{a}\sigma\nu}\Gamma^{\sigma}_{\phantom{a}\mu\lambda}.
\ee 
The second is given by the following equation
\be
{\cal R}'_{\mu\nu}=-\partial_\nu \Gamma^\alpha_{\phantom{a}\alpha\mu}+\partial_\mu \Gamma^\alpha_{\phantom{a}\alpha\nu}.
\ee
 For a symmetric connection, this tensor is equal to the antisymmetric
part of ${\cal R}_{\mu\nu}$. Fully contracting both tensors with the
metric to get a scalar gives, for ${\cal R}_{\mu\nu}$
 \be
\label{rscal}
{\cal R}=g^{\mu\nu}{\cal R}_{\mu\nu}
\ee
which is the Ricci scalar, and for ${\cal R}'_{\mu\nu}$
\be
{\cal R}'=g^{\mu\nu}{\cal R}'_{\mu\nu}=0,
\ee
 since the metric is symmetric and $R'_{\mu\nu}$ is antisymmetric. 
Therefore the Ricci scalar is uniquely defined by eq.~(\ref{rscal}). 

We have considered so far second rank tensors that one gets from a
contraction of the Riemann tensor without using the metric, {\em
i.e.}~tensors independent of the metric. There is a third second rank
tensor which can be built from the Riemann tensor \cite{tsamp}: ${\cal
R}''_{\mu\nu}\equiv {\cal
R}^{\phantom{a}\sigma}_{\mu\phantom{a}\sigma\nu}=g^{\sigma\alpha}g_{\mu\beta}{\cal
R}^{\beta}_{\phantom{a}\alpha\sigma\nu}$. This tensor, however,
depends on the metric. A further contraction with the metric will give
${\cal R}''=g^{\mu\nu}{\cal R}''_{\mu\nu}=-{\cal R}$, and so even if
we consider this tensor, the Ricci scalar is in practice uniquely
defined.

\section{General Relativity through its assumptions}

What is really distinguishing General Relativity from other candidate
theories for gravitation? In Section \ref{epetc} this problem was
approached from a conceptual perspective and the discussion evolved
around several principles that can be formed to describe the key
features of the gravitational interaction. Even though this is indeed
the most fundamental and therefore the most noble way to address this
problem, a rigorous axiomatic formulation of General Relativity is
still pending, as already mentioned. The next best thing that one can
do is to list the assumptions that uniquely lead to Einstein's theory
and distinguish it from alternative theories once the geometrical
nature of gravity is itself assumed.

We have already argued why it is very reasonable to describe gravity
as a geometric phenomenon and why a metric is most likely to be
present in the gravity sector. We have also already presented the
tools needed for such a description in the previous section. However,
even if the metric postulates are adopted, General Relativity is not
the only theory that satisfies them and there are extra restrictions
that should be imposed in order to be led uniquely to this theory. 
Let us present these here as they come about in the derivation of the
field equations.

General Relativity is a classical theory and therefore no reference to
an action is really physically required; one could just stay with the
field equations. However, the Lagrangian formulation of the theory has
its merits. Besides its elegance, there are at least two more reasons
it has now become standard:
 \begin{itemize}
 \item At the quantum level the action indeed acquires a physical
meaning and one expects that a more fundamental theory for gravity (or
including gravity), will give an effective low energy gravitational
action at a suitable limit.
 \item It is much easier to compare alternative gravity theories
through their actions rather than by their field equations, since the
latter are far more complicated. Also, it seems that in many cases we
have a better grasp of the physics as described through the action
(couplings, kinetic and dynamical terms {\em etc}.).
 \end{itemize}

For the above reasons we will follow the Lagrangian formulation here.
However, we will be keeping track of the analogy with the geometric
derivation of the field equations of General Relativity, initially
used by Einstein, and comment on it whenever necessary. In Einstein's
derivation the analogy with the Poisson equation, which describes the
dynamics of Newtonian gravity, plays a significant role. Such an
approach can be found in many textbooks (for instance \cite{schutz}).

Let us start with what is probably the most basic assumption of
General Relativity: that the affine connection
$\Gamma_{\phantom{a}\mu\nu}^\lambda$ is the Levi--Civita connection,
{\em i.e.}
 \be
\label{ass0}
\Gamma_{\phantom{a}\mu\nu}^\lambda=\{_{\phantom{a}\mu\nu}^\lambda\}.
\ee 
 This assumption is actually dual, since it requires both the metric 
to be covariantly conserved, 
\be
\label{ass1}
\gn_\lambda g_{\mu\nu}=0,
\ee
 and the connection to be symmetric,
\be
\label{ass2}
\Gamma_{\phantom{a}\mu\nu}^\lambda=\Gamma_{\phantom{a}\nu\mu}^\lambda.
\ee
 Assumption (\ref{ass1}) can also be written in terms of the
non-metricity as $Q_{\mu\nu\lambda}=0$, while assumption (\ref{ass2})
can be written in terms of the Cartan torsion tensor as
$S_{\mu\nu}^{\phantom{ab}\lambda}=0$. General Relativity assumes that
there is neither torsion nor non-metricity.

Given these assumptions, the Riemann tensor will turn out to be
antisymmetric also with respect to the first two indices as well as
symmetric in an exchange of the first and the second pairs. Therefore,
one can construct only one second rank tensor from straightforward
contraction, {\em i.e.}~without using the metric. This is the 
well-known Ricci tensor, $R_{\mu\nu}$ (we use ${\cal R}_{\mu\nu}$ for the
Ricci tensor constructed with an independent connection). A full
contraction with the metric will then lead to the Ricci scalar, $R$ in
the usual way.

Before writing down an action for General Relativity, we need to refer
to another key assumption. {\em General Relativity assumes that no
fields other than the metric mediate the gravitational interaction}.
Any field other than the metric is considered to be matter and should
be included in the matter action. Therefore the general structure of
the action should include a Lagrangian for gravity which depends only
on the metric and a Lagrangian for the matter which depends on the
matter fields. In terms of the field equations, this requirement can
be put in the following terms: the left hand side should depend only
on the metric and the right hand side should depend only on the matter
fields, at least if we want our equations to have a form similar to the Poisson
equation.

For the matter Lagrangian we have one basic requirement: We want its
variation with respect to the metric to lead to the the matter
stress-energy tensor, since this is what we expect to have on the
right hand side of the field equations. Therefore, we define
 \be
T_{\mu\nu}\equiv -\frac{2}{\sqrt{-g}}\frac{\delta S_M}{\delta g^{\mu\nu}},
\ee
 where $\delta/\delta g^{\mu\nu}$ is a functional derivative with 
respect to the metric,
\be
\label{matteraction}
S_M=\int d^4 x \sqrt{-g} L_M(g_{\mu\nu},\psi),
\ee
 is the matter action, $g$ denotes the determinant of the metric
$g_{\mu\nu}$, $L_M$ is the matter Lagrangian and $\psi$ collectively
denotes the matter fields. In a sense, here we just draw our insight
from the analogy with the Poisson equation and the fact that in
Special Relativity the stress-energy tensor is the analogue of the
matter density in Newtonian theory.

Let us now go one step further and examine the form of the
gravitational Lagrangian. Hilbert, to whom we owe the Lagrangian
formulation of General Relativity, recognised two requirements.
Firstly, the Lagrangian should be a generally covariant scalar if it
is to lead to covariant equations (equations of tensors). This depicts
the requirements that the field equations are to be independent of the
coordinates. Secondly, the Lagrangian should depend only on the metric
and its first derivatives and not on any higher order derivatives, so
that metric variation will lead to a second order differential
equation. This requirement comes from the fact that we do not know of
any other theory which has higher order field equations.

However, there was an obstacle to Hilbert's requirements: There is no
generally covariant scalar that one can construct with only the metric
and its first derivatives. The first derivatives of the metric are not
covariant objects and no combination can be formed using them that
turns out to be covariant. The simplest generally covariant scalar
that one can construct is the Ricci scalar which depends also on the
second derivatives of the metric. This was Hilbert's motivation for
defining the gravitational action for General Relativity as:
 \be
\label{actioneh}
S_{EH}=\frac{1}{16 \pi\, G}\int d^4 x \sqrt{-g} R.
\ee
 The coefficient $(16\pi\,G)^{-1}$ is chosen with some anticipation
since at this stage any constant would do and one has to resort to the
Newtonian limit in order to calculate its value.

Let us now see how one derives the field equations from the action
(\ref{actioneh}). We shall not discuss this procedure in detail
however, since it is a standard text-book calculation (see {\em
e.g.}~\cite{wald}). The variation of the action (\ref{actioneh}) with
respect to the metric gives
 \be
\label{varm}
\delta S_{EH}=\frac{1}{16\pi\,G}\left[\int_U d^4 x \sqrt{-g} G_{\mu\nu}\delta g^{\mu\nu}-2\int_{\delta U} d^3 x \sqrt{
|h|}\, \delta K\right],
\ee
 where $U$ denotes the volume, $\delta U$ denotes the boundary of $U$,
and $K$ is, as usual, the trace of the extrinsic curvature of that
boundary \cite{wald}. 
 \be
G_{\mu\nu}\equiv R_{\mu\nu}-\frac{1}{2}\, R\,g_{\mu\nu}
\ee
is the Einstein tensor.

The second term in eq.~(\ref{varm}) is a surface term. Assuming that
$g_{\mu\nu}$ is fixed on the boundary does not imply, however, that
this term goes to zero. That would require also the first derivatives
of the metric to be fixed on the boundary which is not an option since
the number of degrees of freedom of the metric is all that we are
allowed to fix \cite{wald,haw,york}. Note that ignoring the surface
term is not an alternative here; it just means that we are implicitly
fixing the first derivatives. This implies that trying to apply the stationary action principle to the 
action (\ref{actioneh}), or to the sum of this with action 
(\ref{matteraction}), in order to derive field equations is unfeasible due to 
the presence of the non-vanishing surface term.
 In fact, $S_{EH}$ would not even be functionally differentiable at
the solutions of the field equations even if those were attainable.
This is the price which we pay for having allowed the action to
include second derivatives of the metric in order to maintain the
requirement of general covariance. 

Note that these terms turned out to be of a certain form: they can be
combined to give a surface term which does not really affect the
differential order of the field equations, since $G_{\mu\nu}$ does
indeed include only up to second derivatives of the metric. This is a
special and remarkable characteristic of the Einstein--Hilbert action,
$S_{EH}$, and it is not shared by other actions including second
derivatives of the metric. In fact, even before the variation,
$S_{EH}$ can be split into a bulk part and a surface term (see
\cite{landaubook} for an explicit calculation). The bulk Lagrangian,
however, is not a generally covariant scalar. Therefore, one can find
non-covariant actions which lead to a variation (\ref{varm}) without
the unwanted surface term. A typical example is the action proposed by
Schr\"odinger \cite{schro}:
 \be
\label{scrod}
S_{\rm Scr}=\frac{1}{16\pi G}\int_U d^4 x \sqrt{-g} g^{\mu\nu}\left(\{^{\alpha}_{\phantom{a}\beta\mu}\} \{^{\beta}_{\phantom{a}\alpha\mu}\}-\{^{\alpha}_{\phantom{a}\mu\nu}\}\{^{\beta}_{\phantom{a}\alpha\beta}\}\right),
\ee
 In fact, Einstein was one of the first to realize that the
gravitational action does not necessarily have to be built out of a
generally covariant scalar in order to lead to covariant equations
\cite{einstein2}. Of course this does not mean that a non-covariant
action would be physically meaningful as an object, since it is
reasonable to require than an action carrying some physical meaning
should still be coordinate independent or, better yet, diffeomorphism
invariant.

Therefore in order to properly derive the Einstein equations, one has
to redefine the gravitational action in such a way that no surface
term will be present after the variation and at the same time
covariance is preserved. Note that since the surface term is actually
a total variation of a surface action this is not that hard to do.
Starting from the action
 \be
\label{actionehm}
S'_{EH}=S_{EH}+\frac{1}{8\pi\,G}\int_{\delta U}d^3 x \sqrt{|h|}\, K,
\ee
variation with respect to the metric gives
\be
\label{varm2}
\delta S'_{EH}=\frac{1}{16\pi\,G}\int_U d^4 x \sqrt{-g} G_{\mu\nu}\delta g^{\mu\nu}.
\ee
 Adding the variation of the matter action and applying the stationary 
action principle, one can straightforwardly derive the Einstein 
equations
\be
\label{ees}
G_{\mu\nu}=8\pi\,G\,T_{\mu\nu}.
\ee
 Using $S'_{EH}$, one has a cancellation of the surface term and hence
a clean derivation of the Einstein field equations. This action is
usually referred to as the ``healed'' Einstein--Hilbert action.

It is worth commenting that the surface term in the ``healed''
Einstein--Hilbert action is more than a trick in order to find a way to
combine covariance and well defined variation. It has turned out to
have interesting properties since, for instance, it is related to
black hole entropy (for a detailed discussion of the role and nature
of the surface term see {\em e.g.}~\cite{paddysurf}). One also has to
mention that even though $S'_{EH}$ is manifestly covariant, it is not
foliation independent, since the presence of the surface term requires
the choice of a preferred foliation. Therefore, the action
(\ref{actionehm}) cannot be considered really background independent
(which is the actual physical property usually enforced by requiring
diffeomorphism invariance) (see, for instance, the relevant discussion
in~\cite{giulini}). 

Let us conclude the derivation of the Einstein equations by mentioning
that their left hand side can be derived without reference to an
action principle, based on the following arguments: It has to be a
divergence free second rank tensor in order to match the right hand
side and it has to depend only on the metric and its first and second
derivatives. The Einstein tensor is an obvious choice (even though not
the only one). It is worth mentioning at this point that one could
easily add a cosmological constant $\Lambda$ to the field equation,
either by adding the term $-\Lambda\,g_{\mu\nu}$ on the right hand
side of eq.~(\ref{ees}), or by subtracting $2\Lambda$ from the
Einstein--Hilbert Lagrangian.

Before closing this section we can sum up the assumptions used to
arrive at General Relativity within the framework of metric theories
of gravitation:
 \begin{enumerate}
\item $\Gamma_{\phantom{a}\mu\nu}^\lambda=\Gamma_{\phantom{a}\nu\mu}^\lambda$ 
or $S_{\mu\nu}^{\phantom{ab}\lambda}=0$. Spacetime is torsion-less.
\item $\gn_\lambda g_{\mu\nu}=0$ or $Q_{\mu\nu\lambda}=0$. The 
connection is a metric one.
\item No fields other than the metric mediate the gravitational interaction.
\item The field equations should be second order partial differential equations.
\item The field equations should be covariant (or the action should be 
diffeomorphism invariant).
\end{enumerate}

\section{Relaxing the assumptions}

Having listed the assumptions that lead to General Relativity, one may
wonder what a theory which relaxes one or more of these assumptions
would look like. Before going further, let us clarify that the
assumptions listed in the previous section lead to General Relativity
only once one has already adopted some of the viability criteria
presented in Section \ref{epetc}. For
example, we started the discussion presented in this section presupposing the existence
of a metric and the dynamical nature of spacetime.
 Therefore, one should not overestimate the value of the discussion
presented in the previous section: it sums up some of the key features
of General Relativity but it does not necessarily trace their root.

Relaxing some of the assumptions listed above leads, for instance, to
much more drastic departures from General Relativity than others. It
is easy to argue that covariance of the field equations is not an
assumption that can be cast aside as easily as the absence of any
extra field mediating gravity. Indeed, for the reasons discussed in
Section \ref{epetc}, we will consider covariance as being a very basic
principle and will not attempt the relax this assumption in the rest
of this thesis. Let us, therefore, concern ourselves here with the
relaxation of the following assumptions: those related to the symmetry
and the metricity of the connection, the requirement for second order
field equations and the absence of any extra field mediating the
gravitational interaction.

\subsection{The Palatini formalism} 
\label{palform}
 It is obvious that if one does not specify a relation between the
metric $g^{\mu\nu}$ and the connection
$\Gamma^\lambda_{\phantom{a}\mu\nu}$ then this connection can be
regarded as an independent field. Therefore, any theory with this
characteristic would be drastically different from General Relativity.
There is, however, one more possibility: relaxing the assumptions
related to the connection but at the same time ending up with General
Relativity by deriving them as consequences of the field equations. It
is exactly this possibility that we will explore here. It can be found
in standard text books under the name of the Palatini formalism ({\em
e.g.~}\cite{wald, gravitation}) even though it was Einstein and not 
Palatini who introduced it \cite{ffr}.

Let us assume that the connection is indeed symmetric and
eq.~(\ref{ass2}) holds, but abandon the covariant conservation of the
metric, {\em i.e.} assumption (\ref{ass1}). We therefore start with a
symmetric metric and an independent symmetric connection. We now have
two fields describing gravity and we want to construct an action for
our theory.

In the process of deriving the Einstein--Hilbert action,
eq.~(\ref{actioneh}), we considered only $R$, motivated initially by
wanting the resulting field equations to be second order differential
equations. This requirement comes from the fact that all other
theories besides gravity are described by such field equations. We can
build our action here using the same requirement. We need a generally
covariant scalar that depends only on our fundamental fields, the
metric and the connections, and on their first derivatives at most.
Therefore the obvious choice is the Ricci scalar ${\cal R}$
[eq.~(\ref{ricci})].

This is clearly not the only choice. In fact ${\cal R}$ does not even
include the first derivatives of the metric. It also does not include
terms quadratic in the first derivatives of the connection. In this
sense, our choice just comes from analogy with the standard
Einstein--Hilbert action and is, in practice, a choice of convenience
since, as we are about to find out, it will give the desired result.

As far as the matter action is concerned we do not want to abandon the
metric postulates. This implies that the matter action will depend
only on the metric and not on the independent connection
$\Gamma^{\lambda}_{\phantom{a}\mu\nu}$. However, if our theory is to
be a metric theory of gravity, even though it includes an independent
connection, then this connection is not, by definition, carrying its
usual physical meaning \cite{Sotiriou:2006qn, Sotiriou:2006hs, sotmg}.
It does not define parallel transport or the covariant derivative. The
reader should not be surprised by that. In the matter action there can
be covariant derivatives and the only way to avoid having a matter
action generically independent of $\Gamma^\lambda_{\phantom{a}\mu\nu}$ is to
assume that it is the Levi--Civita connection of the metric that is
used for the definition of the covariant derivative. 
 We will analyse this fact extensively later on. For the moment, let
us stress once more that the underlying geometry is indeed {\it a
priori} pseudo-Riemannian. It is worth noticing that this make our
choice for the gravitational action even more {\it ad hoc} since ${\cal R}$
will now not really be related to the curvature of spacetime from a
geometrical perspective.

In any case the total action will be
\be
\label{action}
S_{\rm p}=\frac{1}{16\pi\, G}\int d^4 x \sqrt{-g}{\cal R}+ S_M(g^{\mu\nu},\psi ).
\ee
 the variation of the action (\ref{action}) should now be performed
with respect to both the metric and the connections (or the covariant
derivatives) separately. An independent variation with respect to the
metric and the connection is called Palatini variation. Note that this
should not be confused with the term Palatini formalism, which refers
not only to the Palatini variation, but also to having the matter
action being independent of the connection. 

The easiest way to proceed with the independent variation is to follow
\cite{wald} and express the $\Gamma$s, as a sum of the Levi--Civita
connections of the metric, $g_{\mu\nu}$, and a tensor field
$C_{\phantom{a}\mu\nu}^\lambda$. Variation with respect to the
$\Gamma$s (or the covariant derivative) will then be equivalent to
the variation of $C_{\phantom{a}\mu\nu}^\lambda$. On the boundary,
$g_{\mu\nu}$ and $C_{\phantom{a}\mu\nu}^\lambda$ will be fixed and we
get the following:
 \bea
\label{varp}
0&=&-\frac{1}{8\pi\,G}\int d^4 x\sqrt{-g} g^{\mu\nu}\nabla_{[\mu}\delta C_{\phantom{a}\lambda]\nu}^\lambda+{}\nn\\
& &{}+\frac{1}{16\pi\,G}\int d^4 x\sqrt{-g} \left(C^{\nu\sigma}_{\phantom{a}\phantom{a}\sigma}\delta^\mu_{\phantom{a}\lambda}+C^\sigma_{\phantom{a}\sigma\lambda}g^{\mu\nu}-2C^{\nu\phantom{a}\mu}_{\phantom{a}\lambda}\right)\delta C^\lambda_{\phantom{a}\mu\nu}+\nn\\
& &{}+\frac{1}{16\pi\,G}\int d^4 x\sqrt{-g} \left({\cal R}_{\mu\nu}-\frac{1}{2}{\cal R}\,g_{\mu\nu}-8\pi\,G\,T_{\mu\nu}\right)\delta g^{\mu\nu}.
\eea
 We see immediately that the first term in eq.~(\ref{varp}) is again a
surface term. This time, however, it is exactly zero since now $\delta
C_{\phantom{a}\mu\nu}^\lambda=0$ on the boundary as
$C_{\phantom{a}\mu\nu}^\lambda$ is fixed there. This is, in a sense,
an advantage with respect to the metric formalism since no ``healing''
of the action is required. 

 Coming back to (\ref{varp}) and considering that the independent
variations with respect to the metric and with respect to
$C_{\phantom{a}\mu\nu}^\lambda$ should vanish separately, we see now
that requiring the second term to vanish corresponds to the condition
 \be
C_{\phantom{a}\mu\nu}^\lambda=0,
\ee
or
\be
\label{f1}
\Gamma_{\phantom{a}\mu\nu}^\lambda=\{_{\phantom{a}\mu\nu}^\lambda\},
\ee
 {\em i.e.}, the $\Gamma$s have to be the Levi--Civita connections of
the metric.  So, in the end, the last term leads to the standard
Einstein equations given that now, due to eq.~(\ref{f1}), ${\cal
R}_{\mu\nu}=R_{\mu\nu}$. Note that the above results remain unchanged
if a cosmological constant is added to the action as the resulting
equations will then be just the standard Einstein equations with a
non-vanishing cosmological constant.

It should be stressed that eq.~(\ref{f1}) is now a dynamical equation
and therefore not a definition, so the Palatini formalism leads to
General Relativity without the metricity condition being an external
assumption. However, this comes at a price. Our choice for the action
is much more {\it ad hoc} and the physical meaning of the independent
connection is obscure since, as we argued, it is not present in the
matter action and it is not the one defining parallel transport. 

One might decide to allow $\Gamma_{\phantom{a}\mu\nu}^\lambda$ to be
present in the matter action and to define the covariant derivative. 
Even if we start from the same gravitational action, the resulting
theory in this case will not be General Relativity \cite{hehl,
Sotiriou:2006qn, Sotiriou:2006hs, sotmg}. We will return and fully
analyse these issues in the next chapter.

\subsection{Higher order field equations}

There is yet another way to deviate from General Relativity without
including gravitational fields other than the metric: one can abandon
the assumption of having second order field equations and allow the
action to depend on higher derivatives of the metric. Taking into
account the general covariance requirement, what one does in order to
raise the differential degree of the field equations is to add higher
order curvature invariants in the gravitational action, for instance
$R_{\mu\nu}R^{\mu\nu}$. 

Higher order theories of this short are not new. In fact, they date
back to 1919 \cite{weyl, eddin} and there have been many periods in
which they have received increased interest, including in the last few
years. Since we intend to refer to such theories extensively in the
forthcoming chapters, we will refrain from saying more here, hoping
for the reader's patience. 

\subsection{Extra fields mediating gravity}

Up to this point we have only referred to theories where the metric is
the only gravitational field\footnote{Even in the Palatini formalism
presented in Section \ref{palform}, the final outcome was General
Relativity.}. One can consider having other fields mediating the
gravitational interaction in some way. Let us stress once more that
the terms ``gravitational'' and ``non-gravitational'' field are quite
ambiguous. Even though we will attempt to clarify this issue at a
later stage (Chapter \ref{concl}), it is important to state what is
meant here when we refer to extra fields describing gravity. The term
is used, in a loose sense, to refer to any field that can somehow
participate in the dynamics of gravity. This could be a field directly
describing part of the spacetime geometry, or a field that intervenes
passively in the generation of the spacetime geometry by the matter
fields. In a Lagrangian formalism such a field is mostly expected to
be coupled non-minimally to the metric (otherwise the standard lore is
to consider it as a matter field). 

Several theories including fields other than the metric have been
proposed. Most of them have been ruled out by experiments and can now
be considered obsolete. We will avoid referring to such theories
unless they constituted a crucial step towards a more modern theory or
may seriously contribute to a better understanding of some subtle
issues of contemporary gravitation research. We refer the reader
interested in the history of such theories to \cite{willbook} for a
more complete list of reference.

We will proceed with our discussion, classifying theories according to
the nature of the extra gravitational field (scalar field, vector
field {\em etc}.). However, it should be mentioned that one could also
perform a classification according to the dynamics of the field. Note
that a non-dynamical field can introduce preferred frame effects in a
theory, leading to violation of LLI and/or LPI without necessarily
violating general covariance\footnote{Non covariant expressions can
easily be brought into a covariant form by imposing a list of
covariantly expressed constraints via, for example, a Lagrange
multiplier (see also \cite{giulini}).}.

\subsubsection{Scalar fields}

In Newtonian gravity the gravitational field is represented by a
scalar. Therefore, it should not come as a surprise that one of the
early attempts to create a relativistic gravitation theory is indeed a
generalisation of Newtonian gravity which preserves the scalar
gravitational field as the key field related to gravity. This is
Nordst\"om's theory, which is actually a predecessor of General
Relativity as it was first introduced in 1913 \cite{nord1913}. Apart
from its pedagogical value, Nordstr\"om's theory can now be considered
obsolete. Additionally, it is not a theory that besides the metric
includes also a scalar, but more of a scalar theory of gravity. 

The study of theories which in addition to the metric include also a
scalar field was mainly stimulated by the works of Jordan in 1955
\cite{jordan} and Brans and Dicke in 1961, leading to the development
of what was later called (Jordan--)Brans--Dicke theory. Generalisations
of this theory are now called scalar-tensor theories of gravity. We
will study such theories in some detail in the next chapter.

\subsubsection{Vector fields}

As in the case of scalars, also here the first gravitation theory
including a vector field came before General Relativity; it was
sketched by Hermann Minkowski in 1908. Details about the general form
and characteristics of a theory which includes a dynamical vector
field in addition to the metric can be found in \cite{willbook,
nordtwillppn, hellings}. We want to concentrate here on two theories
that attract significant attention at present.

The first is Tensor-Vector-Scalar gravity (TeVeS), proposed by Jacob
Bekenstein in 2004 \cite{bekenstein}. Bekenstein's theory includes,
besides the metric, not only a vector, but also a scalar field. This
theory was tailored to be a relativistic extension of Milgrom's
modified Newtonian dynamics (MOND) \cite{milgrom1, milgrom2,
milgrom3}. MOND suggests a modification of Newton's law of universal
gravitation in order to account for the unexpected shape of the
rotational curves of galaxies without the need for dark matter (see
Section \ref{observ}). TeVeS reduces to MOND instead of standard
Newtonian gravity in what is usually called the Newtonian limit.

The second theory which we want to consider is the so called
Einstein-Aether theory, proposed by Jacobson and Mattingly
\cite{aether1, aether2}. This theory includes a dynamical vector field
as well as the metric, but no scalar field. Note that the Lagrangian
of the vector field in TeVeS is a special case of the more general
Lagrangian of Einstein-Aether theory. The word aether in
Einstein--Aether theory refers to some preferred frame. This frame is
to be determined by some yet unknown physics which may lead to Lorentz
symmetry violations.  Such violations can leave an imprint not only on
non-gravitational physics, but also on gravity itself and this is
exactly the gap which Einstein--Aether theory is hoping to fill. The
role of the aether is played by the vector field. Even though the
field is dynamical and the theory is fully covariant, the vector is
set to be of unitary length {\it a priori}. This is an implicit
violation of background independence and introduces preferred frame
effects. It should be noted that Bekenstein's theory also shares this
characteristic, even though the fact that the vector field is not
coupled to the matter prevents detection of the preferred frame (at least classically).

\subsubsection{Tensor fields}

Apart from scalar and vector fields, one could also consider including
tensor fields in the mediation of the gravitational interaction. Most
of the theories developed under this perspective include an extra
second rank tensor field, which actually serves as a second metric.
The most well known of these theories is Rosen's bimetric theory,
which, in addition to the spacetime metric, also includes a flat,
non-dynamical metric \cite{rosen1, rosen2, rosen3}. Clearly, the
presence of the flat, non-dynamical metric implies the existence of
some prior geometry and, therefore, the theory is not background
independent. Most of the current interest in bimetric theories comes
from what is called ``variable speed of light Cosmology'' which is
proposed as an alternative way to approach the problems usually
address by the inflationary paradigm \cite{Clayton:1998hv,
Barrow:1999jq, Barrow:1999is, Barrow:1998df}. In brief, the relation
between a variable speed of light and the existence of a second metric
can be explained as follows: The causal propagation of electromagnetic
waves is determined by the metric present in Maxwell's equations.
Therefore, if one introduces a metric different from that describing
the geometry and uses this metric in Maxwell's equations, the outcome
will be a theory in which the light speed will not be determined by
the spacetime metric.

\subsubsection{Affine connections}
\label{afffields}

Finally, let us consider the case of gravitation theories that include
affine connections that are not necessarily related to the metric.
Before going further, it is worth commenting that even in the early
1920s there was an ongoing discussion about whether it is the metric
or the connection that should be considered as being the principal
field related to gravity (see {\em e.g.~}\cite{weylbook}). In 1924
Eddington presented a purely affine version of General Relativity in
vacuum \cite{eddin}. In Eddington's theory the metric came about as a
derived quantity. Later on, Sch\"odinger generalized Eddington's
theory to include a non-symmetric metric \cite{schro2}, therefore
arriving at a purely affine version of Einstein--Straus theory which
was introduced as a unification of gravity and electromagnetism
\cite{einstr} (see also \cite{poplawski1, poplawski2} for a recent
review). Purely affine theories of gravity do not now receive much
attention, most probably due to the difficulties that arise when one
attempts to add matter (however see \cite{Kijowski:2004bj} for some
proposals).

A more conventional approach is to consider theories where both a
metric and a connection are present but are, at least to some degree,
independent. By far the most well-known theory of this sort is
Einstein--Cartan(--Sciama--Kibble) theory \cite{cartan1, cartan2,
cartan3, sciama1, kibble, hehlrev}. This theory assumes that a
connection and a metric describe the geometry. The metric is symmetric
and covariantly conserved by the connection (vanishing non-metricity).
However, the connection is not necessarily symmetric (and therefore it
is not the Levi--Civita connection of the metric). The spacetime
associated with this theory is called a Riemann-Cartan spacetime. One
of the main advantages of the theory is that it allows torsion and
relates its presence with the spin of matter. In fact, one could argue
that if General Relativity were to be extended to microphysics, spin
angular momentum should somehow become a source of the gravitational
field, much like standard macroscopic angular momentum \cite{hehlrev}.

Einstein--Cartan(--Sciama--Kibble) theory is not the only theory that
includes an independent connection. For instance, one can decide to
abandon the metricity assumption as well and allow the connection to
be completely independent of the metric. This generically leads to
metric-affine theories of gravity. However, as we will devote a large
portion of the next chapter to theories with such characteristics and
to the interaction between spin and gravity, we refrain from
mentioning more here and refer the reader to Section \ref{metaffgrav}.

\chapter[Modified actions and field equations]{Modified 
actions and
field equations for gravity}
 \label{modeltheo}
\section{Introduction}

Having discussed the more general aspects of gravitation theory and
briefly reviewed or mentioned some early proposed alternatives to
General Relativity, we will concentrate now on a number of specific
gravitation theories that have received attention lately. We begin by
devoting this chapter to the exploration of their theoretical aspects.
In the following chapters, their phenomenological aspects will be
studied as well.

The theories considered can come from an action as can many of the
interesting theories of gravity. We concentrate on theories which
include a scalar field as an extra field mediating the gravitational
interaction (such as scalar-tensor theories), theories whose action
includes higher order curvature invariants and some specific
combinations of these two cases ({\em e.g.~}Gauss--Bonnet gravity). We
also extensively consider theories with a connection which is
independent of the metric.

The actions of these theories are presented and in many cases their
resemblance with effective low-energy actions coming from more
fundamental theories is briefly discussed. We also present the
derivation of the field equations through the application of a
suitable variational principle and analyse the basic characteristics
of the theory, as expressed through the field equations.

\section{Scalar-Tensor theory of gravity}
\label{scalartensor}
\subsection{A predecessor: Brans--Dicke theory}

As already discussed in the previous chapters, Dicke has been one of
the pioneers in the discussion of the conceptual basis of gravitation
theories. In 1961, motivated by Mach's Principle --- which, according
to Dicke, can take the clearer formulation ``the gravitational
constant should be a function of the mass distribution of the
universe" --- he introduced, together with his student Carl Brans,
what is now called Brans--Dicke Theory \cite{bransdicke}. This theory
includes, apart from the metric, also a scalar field in the mediation
of the gravitational interaction and it was based on earlier works by
Pascual Jordan \cite{jordan} among others. 

The action for Brans--Dicke theory is
\be
\label{bdaction}
S_{BD}=\frac{1}{16\pi\,G}\int d^4 x\sqrt{-g}\left[\phi R-\frac{\omega_0}{\phi} (\partial_\mu \phi \partial^\mu \phi)\right]+ S_M(g_{\mu\nu},\psi),
\ee 
 where $\phi$ is a scalar field and $\omega_0$ is called the
Brans--Dicke parameter. Note that $\phi$ is not present in the matter
action, {\em i.e.~}it is not coupled to the matter, but it is
non-minimally coupled to gravity. Note also that $G$ is, as usual,
Newton's gravitational constant. 

It is apparent from the action (\ref{bdaction}) why Brans--Dicke theory
can be considered as a theory with a varying gravitational constant,
since one can always define an effective gravitational ``constant'', or
better an effective gravitational coupling
 \be
\label{Geff1}
G_{\rm eff}=\frac{G}{\phi}.
\ee
 Therefore, the theory can indeed be thought as a manifestation of 
Dicke's formulation of Mach's Principle. 

Brans--Dicke theory has only one extra free parameter with respect to
General Relativity, $\omega_0$. This is a characteristic many would
consider as a merit for an alternative theory of gravity, since it
makes it easy to test and constrain or even rule out the theory.
Indeed, using the standard post-Newtonian expansion \cite{willbook}
one can utilize Solar System tests to derive a bound for $\omega_0$
(see for example \cite{bertotti}):
 \be
|\omega_0 |> 40\,000.
\ee
 This unusually large value is hardly appealing since one expects
dimensionless coupling parameters to be of order unity. Thus,
Brans--Dicke theory is no longer considered a viable alternative to
General Relativity but serves as a model theory within a more general
class of theories including a scalar field. 

\subsection{Action and field equations}
\label{actfieldst}

Brans--Dicke theory can be straightforwardly generalised into what is
called a scalar-tensor theory of gravity. A general form for the
action of such theories is
 \be
\label{staction}
S_{ST}=\frac{1}{16\pi\,G}\int d^4 x\sqrt{-g}\left[\phi\, R-\frac{\omega(\phi)}{\phi} (\partial_\mu \phi \partial^\mu \phi)-V(\phi)\right]+ S_M(g_{\mu\nu},\psi),
\ee
 where $V(\phi)$ is the potential of the scalar field $\phi$ and
$\omega(\phi)$ is some function of $\phi$. Note that by setting
$\omega(\phi)=\omega_0$ we derive the action
 \be
\label{bdaction2}
S_{BDV}=\frac{1}{16\pi\,G}\int d^4 x\sqrt{-g}\left[\phi R-\frac{\omega_0}{\phi} (\partial_\mu \phi \partial^\mu \phi)-V(\phi)\right]+ S_M(g_{\mu\nu},\psi).
\ee
 If we also exclude the potential term $V(\phi)$, then we return to
the action (\ref{bdaction}).
 
The theory described by action (\ref{bdaction2}) is a Brans--Dicke
theory with a potential for the scalar and is sometimes referred to in
the literature as a scalar-tensor theory and sometimes simply as
Brans--Dicke theory. Even though, strictly speaking, the theory
introduced by Brans and Dicke did not include a potential, we will
reserve the term scalar-tensor theories for more general theories
described by the action (\ref{staction}) and in what comes next we
will be referring to the action (\ref{bdaction2}) as Brans--Dicke
theory with a potential or simply Brans--Dicke theory.
 
It is worth clarifying here that Brans--Dicke theory and any other
version of scalar-tensor theory are metric theories of gravity: the
scalar field is not coupled directly to the matter and so matter
responds only to the metric. The role of the scalar field is just to
intervene in the generation of the spacetime curvature associated with
the metric \cite{willbook}. 

Note also that the bound on $\omega_0$ mentioned earlier for
Brans--Dicke theory without a potential is still applicable in the
presence of a potential or even for a general scalar-tensor theory in
the form
 \be
|\omega(\phi_0) |> 40\,000,
\ee
 where $\phi_0$ is the present value of the scalar. However, for this
constraint to be applicable, the effective mass of the scalar field
should be low or, as commonly said, the potential should be light
($\partial^2 V/\partial \phi^2$ evaluated at $\phi_0$ plays the role
of an effective mass). If the potential is heavy, then the scalar
field becomes very short-ranged and the bound is not applicable. 

Since scalar-tensor theory is one of the most widely-studied
alternatives to General Relativity and there are standard text books
analysing its characteristics \cite{fuji, valeriobook}, we will not go
much further in our discussion of it here. Before deriving the field
equations from the action, let us just comment that non-minimally
coupled scalar fields are present in the low energy effective action
of more fundamental theories, such as String Theory ({\em
e.g.~}dilaton), and a potential might be expected to be present after
supersymmetry breaking. We refer the reader to the literatures for
more details \cite{fuji, valeriobook}.

We can now proceed to vary the action (\ref{staction}) to derive the
field equations. Independent variation with respect to the metric and
the scalar field gives
 \begin{align}
\label{stf1}
G_{\mu\nu}=\frac{8\pi\,G}{\phi} T_{\mu\nu} +\frac{\omega(\phi)}{\phi^2} & \Big(\nabla_\mu \phi \nabla_\nu \phi 
-\frac{1}{2}g_{\mu\nu} \nabla^\lambda \phi \nabla_\lambda \phi \Big)+\nn\\
+\frac{1}{\phi} & (\nabla_\mu\nabla_\nu \phi-g_{\mu\nu} \Box \phi)-\frac{V}{2\phi}g_{\mu\nu},
\end{align}
\be
\label{stf2'}
\Box\phi=-\frac{\phi}{2\omega}\left(R-V'\right)-\frac{1}{2}\nabla^\mu\phi\nabla_\mu\phi\,\left(\frac{\omega'(\phi)}{\omega(\phi)}-\frac{1}{\phi}\right),
\ee
 where $G_{\mu\nu}=R_{\mu\nu}-\frac{1}{2}R g_{\mu\nu}$ is the Einstein
tensor, $T_{\mu\nu}\equiv\frac{-2}{\sqrt{-g}}\frac{\delta S_M}{\delta
g^{\mu\nu}}$ is the stress-energy tensor, $\nabla$ denotes covariant
differentiation, $\Box\equiv \nabla^\mu\nabla_\mu$ and a prime denotes
differentiation with respect to the argument. One can take the trace
of eq.~(\ref{stf1}) and use the result to replace $R$ in
eq.~(\ref{stf2'}) to derive
 \be
\label{stf2}
(2\omega(\phi)+3) \Box \phi= 8\pi\,G\, T-\omega'(\phi)\nabla^\lambda \phi \nabla_\lambda \phi+\phi V'-2V,
\ee
 where $T\equiv g^{\mu\nu}T_{\mu\nu}$ is the the trace of the stress
energy tensor. Note that eq.~(\ref{stf2'}) implies a coupling between
the scalar field and the metric but no coupling with matter, as
expected, so we should not be misled by the presence of matter in
eq.~(\ref{stf2}): the field $\phi$ acts back on matter only through
the geometry.

By setting $\omega(\phi)=\omega_0$ or by varying the action
(\ref{bdaction2}) directly, we can get the simpler field equations for
Brans--Dicke theory with a potential:
 \bea
\label{bdf1}
G_{\mu\nu}=\frac{8\,\pi\,G}{\phi} T_{\mu\nu}&+&\frac{\omega_0}{\phi^2}\Big(\nabla_\mu \phi \nabla_\nu \phi 
-\frac{1}{2}g_{\mu\nu} \nabla^\lambda \phi \nabla_\lambda \phi \Big)+\nn\\
&+&
\frac{1}{\phi}(\nabla_\mu\nabla_\nu \phi-g_{\mu\nu} \Box \phi)-\frac{V}{2\phi}g_{\mu\nu},
\eea
\be
\label{bdf2'}
\frac{2\omega_0}{\phi}\Box\phi+R-\frac{\omega_0}{\phi^2}\nabla^\mu\phi\nabla_\mu\phi-V'=0.
\ee
In Brans--Dicke theory, eq.~(\ref{stf2}) takes the simpler form
\be
\label{bdf2}
(2\omega_0+3) \Box \phi= 8\pi\,G\, T+\phi V'-2V.
\ee

 Let us close this section with a warning about the effective
gravitational coupling. As we said in the previous section, for
Brans--Dicke theory one can define $G_{\rm eff}$ through
eq.~(\ref{Geff1}). In the same way as in eq.~(\ref{Geff1})
  one can define the effective gravitational coupling for any
scalar-tensor theory. However, it should be stressed that this is not
going to be the coupling as measured by a Cavendish experiment. The
latter would be \cite{nordvedt}
 \be
 G^{(\ast)}_{\rm eff}=\frac{G}{\phi} \frac{2 \omega \, \phi+2 }{2\omega \,\phi+3}.
 \ee
 The reason for this difference is quite straightforward: $G_{\rm
eff}$ is, in practice, the inverse of the coefficient of $R$ as read
from the action, whereas $G^{(\ast)}_{\rm eff}$ is the quantity of
dimensions ${\rm cm}^3\,{\rm g}^{-1}\,{\rm s}^{-2}$ which appears in
Newton's second law in a two body problem, such as a Cavendish
experiment. These two quantities are not generically the same.

\section{$f(R)$ gravity in the metric formalism}
\subsection{The action}

We have already briefly discussed in the previous chapter the
possibility of including higher order curvature invariants in the
gravitational action. Attempts towards this direction were first
examined by Weyl and Eddington in 1919 and 1922 respectively
\cite{weyl, eddin}, mainly on the basis of theoretical completeness.
It is easy to understand that complicating the action, and
consequently the field equations, with no apparent reason is not so
appealing. For instance, the degree of the field equations will become
higher than second and we currently are unaware of any other physical
theory with such characteristics.

However, starting from the early 1960s, there appeared indications
that complicating the gravitational action might indeed have its
merits. As discussed in the Introduction, General Relativity is not
renormalisable and therefore cannot be conventionally quantized. In
1962, Utiyama and De Witt showed that renormalisation at one-loop
demands that the Einstein--Hilbert action should be supplemented by
higher order curvature terms \cite{uti}. Later on, Stelle showed that
higher order actions are indeed renormalisable (but not unitary)
\cite{stelle}. More recent results show that when quantum corrections
or String Theory are taken into account, the effective low energy
gravitational action admits higher order curvature invariants
\cite{quant1, quant2, quant3}.

Even though initially the relevance of such terms in the action was
considered to be restricted to very strong gravity phenomena and they
were expected to be strongly suppressed by small couplings, this
perspective has recently changed as discussed in the Introduction. The
main reason for this was the motivation provided by the cosmological
problems such as the dark energy problem, the late-time accelerated
expansion of the universe, the cosmological constant problems {\em etc}.
(see Chapter \ref{intro}). 

Higher order actions may include various curvature invariants, such as
$R^2$, $R_{\mu\nu}R^{\mu\nu}$ {\em etc}., but for orientation purposes one
can consider an action of the form
 \begin{equation}
S=\frac{1}{16\pi\,G}\int d^4 x \sqrt{-g} f(R).
\end{equation}
 The appealing feature of such an action is that it combines
mathematical simplicity and a fair amount of generality. For example,
viewing $f$ as a series expansion of $f$, {\em i.e.}
 \begin{equation}
f(R)=\cdots +\frac{\alpha_2}{R^2}+\frac{\alpha_1}{R}-2\Lambda+R+\frac{R^2}{\beta_2}+\frac{R^3}{\beta_3}\cdots,
\label{eq:fr}
\end{equation}
 where the $\alpha_i$ and $\beta_j$ coefficients have the appropriate
dimensions, we see that the action includes a number of
phenomenologically interesting terms. 

$f(R)$ actions where first rigorously studied by Buchdahl
\cite{buchdahl}. We will proceed to derive the field equations for
such actions here. We will discuss the cosmological implications and
the way in which such theories can address the cosmological problems
in the next chapter.

\subsection{Field equations}
\label{metricfieldeq}

Adding a matter action, the total action for $f(R)$ gravity takes the form
\be
\label{metaction}
S_{met}=\frac{1}{16\pi\,G}\int d^4 x \sqrt{-g} f(R) +S_M(g_{\mu\nu},\psi).
\ee
Variation with respect to the metric gives, after some manipulations,
\bea
\delta S_{met}&=&\frac{1}{16\pi\,G}\int d^4 x \sqrt{-g}  \Big[f'(R)R_{\mu\nu}-\frac{1}{2}f(R)g_{\mu\nu}-\nn\\& &-\nabla_\mu\nabla_\nu f'(R)+g_{\mu\nu}\Box f'-8\pi\,G\,T_{\mu\nu}\Big]\delta g^{\mu\nu}-\nn\\
& &-\frac{1}{8\pi\,G}\int_{\delta U} d^3 x \sqrt{
|h|}\, f'(R)\,\delta K.
\eea
 The integral in the last line represents a surface term. However,
unlike the variation of the Einstein--Hilbert action, this surface term
is not the total variation of a quantity, due to the presence of
$f'(R)$. This implies that it is not possible to ``heal'' the action
just by subtracting some surface term before making the variation. 

Formally speaking, we cannot derive the field equations from this
variation by applying the stationary action principle before finding a
way to treat the surface term. However, the action includes higher
order derivatives of the metric and therefore it is possible to fix
more degrees of freedom on the boundary than those of the metric
itself. It has to be stressed at this point that there are several 
auxiliary variables which one can fix in order to set the surface
term to zero. Additionally, the choice of the auxiliary variable is
not void of physical meaning, even though this might not be obvious at
a classical level, since it will be relevant for the Hamiltonian
formulation of the theory. 

There is no unique prescription for making the fixing in the
literature so far. The situation gets even more complicated if one
takes into account that arbitrary surface terms could also be added
into the action in order to allow different fixings to lead to a well
define variation (see also \cite{barrowsurf} for a discussion on the
surface term of $f(R)$ gravity). Therefore, non-rigorous as it may be,
the standard approach at this stage is to neglect the surface term,
silently assuming that a suitable fixing has been chosen, and go
directly to the field equations
 \be
\label{metf}
f'(R)R_{\mu\nu}-\frac{1}{2}f(R)g_{\mu\nu}-\nabla_\mu\nabla_\nu f'(R)+g_{\mu\nu}\Box f'=8\pi\,G\,T_{\mu\nu}.
\ee
 This mathematical jump might seem worrying and certainly gives no
insight for the choice of auxiliary variables, which would be
necessary for a Hamiltonian formulation or a canonical quantisation.
However, the field equations $(\ref{metf})$ would be unaffected by the
fixing chosen and from a purely classical perspective the field
equations are all that one needs. 

Eqs.~(\ref{metf}) are obviously fourth order partial differential
equations in the metric. Notice, however, that the fourth order terms
--- the last two on the left hand side --- vanish when $f'(R)$ is a
constant, {\em i.e.~}for an action which is linear in $R$. Thus, it is
straightforward for these equations to reduce to the Einstein equation
once $f(R)=R$.

It is also worth noticing that the trace of eq.~(\ref{metf})
 \be
\label{metftrace}
f'(R)R-2f(R)+3\Box f'=8\pi\,G\,T,
\ee
 where $T=g^{\mu\nu}T_{\mu\nu}$, relates $R$ with $T$ differentially
and not algebraically as in General Relativity, where $R=-8\pi\,G\,T$.
This is already an indication that the field equations of $f(R)$
theories will admit more solutions than Einstein's theory. As an
example, we can mention here that Birkhoff's theorem, stating that the
Schwarzschild solution is the unique spherically symmetric vacuum
solution, no longer holds in metric $f(R)$ gravity. Without going into
the details of the calculation, let us stress that $T=0$ no longer
implies that $R=0$, or is even constant.

Another important aspect of such theories has to do with their
maximally symmetric solutions. The functional form of $f$ is what will
affect whether the maximally symmetric solution will be Minkowski, de
Sitter or anti-de Sitter. To see this, let us recall that maximally
symmetric solutions lead to a constant Ricci scalar. For $R={\rm
constant}$ and $T_{\mu\nu}=0$ eq.~(\ref{metftrace}) reduces to
 \be
\label{metftr}
f'(R)R-2f(R)=0,
\ee
 which, for a given $f$, is an algebraic equation in $R$. If $R=0$ is
a root of this equation and one takes this root, then eq.~(\ref{metf})
reduces to $R_{\mu\nu}=0$ and the maximally symmetric solution is
Minkowski spacetime. On the other hand, if the root of
eq.~(\ref{metftr}) is $R=C$, where $C$ is a constant, then
eq.~(\ref{metf}) reduces to $R_{\mu\nu}=C/4 g_{\mu\nu}$ and the
maximally symmetric solution is de Sitter or anti-de Sitter depending
on the sign of $C$, just as in General Relativity with a cosmological
constant.

\section{$f(R)$ gravity in the Palatini formalism}
\subsection{The action}

In Section \ref{palform}, we showed how Einstein's equation can be
derived using, instead of the standard metric variation of the
Einstein--Hilbert action, the Palatini formalism, {\em i.e.~}an
independent variation with respect to the metric and an independent
connection (Palatini variation) of an action with gravitational
Lagrangian ${\cal R}=g^{\mu\nu}{\cal R}_{\mu\nu}$, where ${\cal
R}_{\mu\nu}$ is the Ricci tensor constructed with the independent
connection, and a matter action independent of the connection. Recall
the importance of this last assumption, of the independence of the
matter action and the connection, as it is crucial for the derivation
and is a main characteristic of the Palatini formalism, which as we
argued in Section \ref{palform} has consequences for the physical
meaning of the independent connection: namely, this connection does
not define parallel transport and the geometry is actually
pseudo-Riemannian. 

One can generalise the action in exactly the same way that the
Einstein--Hilbert action was generalised in the previous section:
 \be
\label{palaction}
S_{pal}=\frac{1}{16\pi\,G}\int d^4 x \sqrt{-g} f({\cal R}) +S_M(g_{\mu\nu}, \psi).
\ee
 The motivation for studying such actions is, in practice, the same as
in metric $f(R)$ gravity, and so we will not repeat it here. Applying
the Palatini variation to the action (\ref{palaction}) leads to what
is called $f(R)$ gravity in the Palatini formalism or simply Palatini
$f(R)$ gravity. Even though $f$ is really a function of ${\cal R}$ and
not $R$ in this case, the term $f(R)$ gravity is used as a generic
terminology to refer to a theory whose Lagrangian is a general
function of some Ricci scalar.

\subsection{Field equations}

Let us proceed to derive the field equations for Palatini $f(R)$
gravity. The variation with respect to the metric is quite
straightforward, since ${\cal R}_{\mu\nu}$ does not depend on it.
However, the variation with respect to the connection is more
intricate, since it requires $\delta {\cal R}_{\mu\nu}$. Taking into
account the definition and symmetries of ${\cal R}_{\mu\nu}$, and
after some manipulations, it can be shown that \cite{schro}
 \bea
\label{varR}
\delta {\cal R}_{\mu\nu}&=&\gn_\lambda \delta \Gamma^\lambda_{\phantom{a}\mu\nu}-\gn_{\nu}\delta \Gamma^\lambda_{\phantom{a}\mu\lambda}.
\eea
 We remind to the reader $\gn_\lambda$ denotes the covariant
derivative defined with the independent connection.

The variation of the matter action with respect to the independent
connection is zero since we do not allow the matter action to depend
on $\Gamma^\lambda_{\phantom{a}\mu\nu}$. On the other hand, by
definition
 \be
T_{\mu\nu}\equiv -\frac{2}{\sqrt{-g}}\frac{\delta S_M}{\delta g^{\mu\nu}}.
\ee

Using eq.~(\ref{varR}), the variation of the gravitational part of the
action takes the form
 \bea
\label{varg2}
\delta S_{pal}\!\!&=&\!\!\frac{1}{16\pi\,G}\int d^4 x \sqrt{-g}\left(f'({\cal R}) {\cal R}_{(\mu\nu)}-\frac{1}{2}f({\cal R})g_{\mu\nu}-8\pi\,G\,T_{\mu\nu}\right)\delta g^{\mu\nu}+{}\nn\\& &\!\!\!\!+\frac{1}{16\pi\,G}\int d^4 x \sqrt{-g}f'({\cal R})g^{\mu\nu}\left(\gn_\lambda \delta \Gamma^\lambda_{\phantom{a}\mu\nu}-\gn_{\nu}\delta \Gamma^\lambda_{\phantom{a}\mu\lambda}\right).
\eea
 Integrating by parts the terms in the second line and taking into
account that on the boundary $\delta
\Gamma^\lambda_{\phantom{a}\mu\nu}=0$ and therefore surface terms
linear in $\delta \Gamma^\lambda_{\phantom{a}\mu\nu}$ vanish, we get
 \bea
\label{varg3}
\delta S_{pal}\!&=&\!\!\frac{1}{16\pi\,G}\int d^4 x \Bigg\{\!\sqrt{-g}\left(f'({\cal R}) {\cal R}_{(\mu\nu)}-\frac{1}{2}f({\cal R})g_{\mu\nu}-8\pi\,G\,T_{\mu\nu}\right)\delta g^{\mu\nu}+{}\nn\\& &\!\!\!\!+\left[-\gn_\lambda\left(\sqrt{-g}f'({\cal R})g^{\mu\nu}\right)+\gn_\sigma\left(\sqrt{-g}f'({\cal R})g^{\mu\sigma}\right)\delta^\nu_\lambda\right]\delta \Gamma^\lambda_{\phantom{a}\mu\nu}\Bigg\}.\!\!\!\!
\eea
 Applying the stationary action principle, straightforwardly leads to 
the equations
\bea
\label{palf1}
& &f'({\cal R}) {\cal R}_{(\mu\nu)}-\frac{1}{2}f({\cal R})g_{\mu\nu}=8\pi\,G\, T_{\mu\nu},\\
\label{palf2}
& &-\gn_\lambda\left(\sqrt{-g}f'({\cal R})g^{\mu\nu}\right)+\gn_\sigma\left(\sqrt{-g}f'({\cal R})g^{\sigma(\mu}\right)\delta^{\nu)}_\lambda=0,
\eea
 where indices inside parentheses are symmetrised. Taking the trace of 
eq. (\ref{palf2}), it can be easily shown that
\be
\gn_\sigma\left(\sqrt{-g}f'({\cal R})g^{\sigma\mu}\right)=0,
\ee
 which implies that we can bring the field equations into the more 
economic form
\bea
\label{palf12}
& &f'({\cal R}) {\cal R}_{(\mu\nu)}-\frac{1}{2}f({\cal R})g_{\mu\nu}=8\pi\,G\, T_{\mu\nu},\\
\label{palf22}
& &\gn_\lambda\left(\sqrt{-g}f'({\cal R})g^{\mu\nu}\right)=0,
\eea

\subsection{Manipulations of the field equations}
\label{manfield}

 Let us explore the characteristics of eqs.~(\ref{palf12}) and 
(\ref{palf22}). Taking the trace of eq.~(\ref{palf12}), we get
\be
\label{paltrace}
f'({\cal R}) {\cal R}-2f({\cal R})=8\pi\,G\,T.
\ee
 For a given $f$, this is an algebraic equation in ${\cal R}$. For all
cases for which $T=0$, which includes vacuum and electrovacuum, ${\cal
R}$ will therefore be a constant and a root of the equation
 \be
\label{paltracev}
f'({\cal R}) {\cal R}-2f({\cal R})=0.
\ee
 We will not consider cases for which this equation has no roots since
it can be shown that the field equations are then not consistent
\cite{ferr}. Therefore choices of $f$ that lead to this behaviour
should simply be avoided. Eq.~(\ref{paltracev}) can also be
identically satisfied if $f({\cal R})\propto {\cal R}^2$. This very
particular choice for $f$ leads to a conformally invariant theory
\cite{ferr}. As is apparent from eq.~(\ref{paltrace}), if $f({\cal
R})\propto {\cal R}^2$ then only conformally invariant matter, for
which $T=0$ identically, can be coupled to gravity. Matter is not
generically conformally invariant though and so this particular choice
of $f$ is not suitable for a low energy theory of gravity. We will,
therefore, neglect it for now and return to it in a later section.

Let us now consider eq.~(\ref{palf22}). Notice that if we define a
metric conformal to $g_{\mu\nu}$ to be
 \be
 \label{hgconf}
h_{\mu\nu}=f'({\cal R})g_{\mu\nu},
\ee
then it can easily be shown that\footnote{This calculation holds for $4$ dimensions. When the number of dimensions $D$ is different that $4$ then, instead of using eq.~(\ref{hgconf}), the conformal metric $h_{\mu\nu}$ should be introduced as $h_{\mu\nu}=[f'({\cal R}]^{2/(D-2)}g_{\mu\nu}$ in order for eq.~(\ref{midstep}) to still hold.}
\begin{equation}
\label{midstep}
\sqrt{-h}h^{\mu\nu}=\sqrt{-g} f'({\cal R}) g^{\mu\nu}.
\end{equation}
 This implies that eq.~(\ref{palf22}) becomes the definition of the Levi--Civita connection of $h_{\mu\nu}$.
 In this way, one in practice solves
eq.~(\ref{palf22}) and can then express the independent connection as
 \be
\Gamma^\lambda_{\phantom{a}\mu\nu}=h^{\lambda\sigma}\left(\partial_\mu h_{\nu\sigma}+\partial_\nu h_{\mu\sigma}-\partial_\sigma h_{\mu\nu}\right),
\ee
or equivalently in terms of $g_{\mu\nu}$
\be
\label{gammagmn}
\Gamma^\lambda_{\phantom{a}\mu\nu}=\frac{1}{f'({\cal R})}g^{\lambda\sigma}\left(\partial_\mu \left(f'({\cal R})g_{\nu\sigma}\right)+\partial_\nu \left(f'({\cal R})g_{\mu\sigma}\right)-\partial_\sigma \left(f'({\cal R})g_{\mu\nu}\right)\right),
\ee

Given that eq.~(\ref{paltrace}) relates ${\cal R}$ algebraically with
$T$, and since we have an explicit expression for
$\Gamma^\lambda_{\phantom{a}\mu\nu}$ in terms of ${\cal R}$ and
$g^{\mu\nu}$, we can in principle eliminate the independent connection
from the field equations and express them only in terms of the metric
and the matter fields. In fact, taking into account how the Ricci
tensor transforms under conformal transformations, we can write
 \bea
\label{confrel1}
{\cal R}_{\mu\nu}=R_{\mu\nu}&+&\frac{3}{2}\frac{1}{(f'({\cal R}))^2}\left(\nabla_\mu f'({\cal R})\right)\left(\nabla_\nu f'({\cal R})\right)-\nn\\& &-\frac{1}{f'({\cal R})}\left(\nabla_\mu\nabla_\nu-\frac{1}{2}g_{\mu\nu}\Box\right)f'({\cal R}).
\eea
Contracting with $g^{\mu\nu}$ we get,
\be
\label{confrel2}
{\cal R}=R+\frac{3}{2(f'({\cal R}))^2}\left(\nabla_\mu f'({\cal R})\right)\left(\nabla^\mu f'({\cal R})\right)+\frac{3}{f'({\cal R})}\Box f'({\cal R}).
\ee
 Note the difference between ${\cal R}$ and the Ricci scalar of
$h_{\mu\nu}$ due to the fact that $g_{\mu\nu}$ is used here for the
contraction of ${\cal R}_{\mu\nu}$.

Replacing eqs.~(\ref{confrel1}) and (\ref{confrel2}) in
eq.~(\ref{palf12}), and after some easy manipulations, we get
 \begin{align}
\label{eq:field}
G_{\mu \nu} &= \frac{8\pi\, G}{f'}T_{\mu \nu}- \frac{1}{2}g_{\mu \nu} \left({\cal R} - \frac{f}{f'} \right) + \frac{1}{f'} \left(
			\nabla_{\mu} \nabla_{\nu}
			- g_{\mu \nu} \Box
		\right) f'-\nn\\
& \quad- \frac{3}{2}\frac{1}{f'^2} \left(
			(\nabla_{\mu}f')(\nabla_{\nu}f')
			- \frac{1}{2}g_{\mu \nu} (\nabla f')^2
		\right).
\end{align}
 Notice that, assuming we know the root of eq.~(\ref{paltrace}),
${\cal R}={\cal R}(T)$ and we have completely eliminated the presence
of the independent connection. Therefore, we have successfully reduced
the number of field equations to one and at the same time both side of
eq.~(\ref{eq:field}) depend only on the metric and the matter fields.
In a sense the theory has been brought to the form of General
Relativity with a modified source.

We can now straightforwardly deduce the following:
\begin{itemize}
 \item When $f({\cal R})={\cal R}$, the theory reduces to General 
Relativity, as discussed in Section \ref{palform}.
 \item For matter fields for which $T=0$, due to eq.~(\ref{paltracev})
${\cal R}$ and consequently $f({\cal R})$ and $f'({\cal R})$ are
constants and the theory reduces to General Relativity with a
cosmological constant and a modified coupling constant $G/f'$. If we
denote the value of ${\cal R}$ when $T=0$ as ${\cal R}_0$, then the
value of the cosmological constant is
 \be
\frac{1}{2}\left({\cal R}_0 - \frac{f({\cal R}_0)}{f'({\cal R}_0)} \right)=\frac{{\cal R}_0}{4},
\ee
 where we have used eq.~(\ref{paltracev}). Besides vacuum, $T=0$ also
for electromagnetic fields, radiation, and any other conformally
invariant type of matter.
 \item In the general case $T\neq 0$, the modified source on the right
hand side includes derivatives of the stress-energy tensor, unlike in
General Relativity. These are implicit in the last two terms of eq.~(\ref{eq:field}), since
$f'$ is in practice a function of $T$, given that $f'=f'({\cal R})$
and ${\cal R}={\cal R}(T)$\footnote{Note that, apart from special
cases such as a perfect fluid, $T_{\mu\nu}$ and consequently $T$
already include first derivatives of the matter fields, given that the
matter action has such a dependence. This implies that the right hand
side of eq.~(\ref{eq:field}) will include at least second derivatives
of the matter fields, and possibly up to third derivatives.}.
 \end{itemize}

The last observation is a crucial characteristic of Palatini $f(R)$
gravity. We will return to this later on and discuss its implications.
We will also reconsider the possible representations of the field
equations and the action of Palatini $f(R)$ gravity in Chapter
\ref{equivtheor}.

\section[Higher-order actions]{Other actions which include higher-order
curvature invariants}

\subsection{Metric formalism}
\label{homet}

Generalising the Einstein--Hilbert action into an $f(R)$ action is a
minimal modification that one can pursue in order to include higher
curvature invariants. In fact, there are a number of invariants that
one can construct from the metric, which are not included in an $f(R)$
action. One can, for instance, contract the Ricci or the Riemann
tensor with itself to form $R_{\mu\nu}R^{\mu\nu}$ and
$R_{\mu\nu\lambda\sigma}R^{\mu\nu\lambda\sigma}$. Other combinations
are also allowed, such as
$R^{\mu\nu\lambda\sigma}R_{\mu\nu}R_{\lambda\sigma}$ or invariants
formed with other tensors, such as the Weyl tensor. All of these
invariants can be considered as combinations of contractions of the
Riemann tensor one or more times with itself and the metric.

A specific choice is the Gauss--Bonnet invariant
\be
\label{GB}
{\cal G}=R^2-4 R^{\mu\nu}R_{\mu\nu}+R^{\mu\nu\kappa\lambda}R_{\mu\nu\kappa\lambda}.
\ee
 ${\cal G}$ apart from being an invariant in the sense used here, {\em
i.e.~}being a generally covariant scalar, is also a topological
invariant in four dimensions. This means that it is related through
the Gauss--Bonnet formula to the Euler characteristic of the
$4$-dimensional manifold, which characterises the topology. Also, from
Gauss's theorem, the variation of the scalar density $\sqrt{-g} {\cal
G}$ with respect the metric is a total divergence. Therefore, adding
${\cal G}$ to the Einstein--Hilbert action will not contribute to the
field equations and additionally a suitable surface term can be found
to eliminate the total divergence \cite{bunch}.

Due to the above, it is possible to write the most general action
which is linear in second order curvature invariants as
\cite{lanczos}:
 \be
S=\frac{1}{16\pi\,G}\int d^4 x \sqrt{-g} \left(R+a R^2+b R^{\mu\nu}R_{\mu\nu}\right),
\ee
 where the coefficients $a$ and $b$ should have suitable dimensions.
Including an $R_{\mu\nu\lambda\sigma}R^{\mu\nu\lambda\sigma}$ term is
equivalent to altering those coefficients, since one can always add a
Gauss--Bonnet term with a suitable coefficient in order to eliminate
$R_{\mu\nu\lambda\sigma}R^{\mu\nu\lambda\sigma}$. The theory described
by this action is referred to as fourth-order gravity, since it leads
to fourth order equations. Numerous papers have been devoted to the
study of fourth-order gravity. Instead of listing them here, we refer
the reader to some historical reviews \cite{schmidt1, schmidt2}.

Notice that one can also choose to include invariants involving
derivatives of the curvature terms, such as $R\Box R$. The
differential order of the field equations is increased as one adds
higher derivative terms in the action. The rule of the thumb is that
for every one order increase in the action one gets a two order
increase in the field equations. Thus, the $R$ term leads to second
order equations, the $R^2$ term or more general $f(R)$ actions lead to
fourth order equations and the $R \Box R$ and $R\Box^2 R$ terms lead
to sixth and eighth order equations respectively \cite{ruzm, gott,
amend, batt}. 

Following the example of $f(R)$ gravity, one can also choose to
include arbitrary functions of some of the above invariants in the
action. For instance, actions of the form $f(R, R^{\mu\nu}R_{\mu\nu})$
can be considered. A comment is due at this point: even though ${\cal
G}$ is a topological invariant and does not contribute to the field
equations if included in the action, the presence of functions of
${\cal G}$ in the action will influence the dynamics. For example the
action
 \be
\label{fg}
S=\int d^4 x \sqrt{-g} \left(\frac{R}{16\pi\,G}+f({\cal G})\right),
\ee
 does not lead to the Einstein equations \cite{odigb}. We will discuss
actions that include the Gauss--Bonnet invariant more extensively
towards the end of this chapter.

\subsection{Palatini formalism}

As in the metric formalism, one can generalise the action to include
higher order curvature invariants also in the Palatini formalism. Not
much work has been done in this direction. In this section we shall
focus mainly on two aspects of such generalisations: the role of
${\cal G}$ and the effect of such generalisations on the field
equations.

As we have mentioned, in the Palatini formalism the geometry of
spacetime is pseudo-Riemannian, due to the fact that the independent
connection $\Gamma^{\lambda}_{\phantom{a}\mu\nu}$ is not present in
the matter action and does not define parallel transport. This implies
that ${\cal G}$, as defined in eq.~(\ref{GB}), is still the topological invariant related to the
Euler characteristic. To make this discussion clearer, let us consider
what happens if ${\cal G}$ is added to action (\ref{action}) of
Section \ref{palform}. The variation with respect to the metric will
remain unchanged, since $\delta (\sqrt{-g} {\cal G})$ contributes only
by a surface term that can be removed as mentioned earlier. Variation
with respect to the connection will also remain unchanged, since
${\cal G}$ does not depend on $\Gamma^{\lambda}_{\phantom{a}\mu\nu}$
but is constructed only using the metric.

One should not confuse ${\cal G}=R^2-4 R^{\mu\nu}R_{\mu\nu}+ R^{\mu\nu\kappa\lambda}R_{\mu\nu\kappa\lambda}
$ with the combination ${\cal R}^2-4
{\cal R}^{\mu\nu}{\cal R}_{\mu\nu}+{\cal R}^{\mu\nu\kappa\lambda}{\cal
R}_{\mu\nu\kappa\lambda}$ which is not a topological invariant. This
implies that ${\cal R}^{\mu\nu\kappa\lambda}{\cal
R}_{\mu\nu\kappa\lambda}$ in the action cannot be eliminated in favour
of ${\cal R}^2$ and $ {\cal R}^{\mu\nu}{\cal R}_{\mu\nu}$ terms as in
the metric formalism. In the Palatini formalism one is more interested
in including in the action invariants such as $ {\cal R}^{\mu\nu}{\cal
R}_{\mu\nu}$ and ${\cal R}^{\mu\nu\kappa\lambda}{\cal
R}_{\mu\nu\kappa\lambda}$ which are constructed using the independent
connection as well and not terms like $R^{\mu\nu}R_{\mu\nu}$ which
depend only on the metric.

Let us see how the presence of such invariants will affect the field
equations. Consider the action
 \be
\label{action2}
S_{\rm p}=\frac{1}{16\pi\, G}\int d^4 x \sqrt{-g}\left({\cal R}+a {\cal R}^{\mu\nu}{\cal R}_{\mu\nu}\right)+ S_M(g^{\mu\nu},\psi )
\ee
 where $a$ should be chosen so as to have proper dimensions. Since we
have already computed the variation of the rest of the action, let us
focus on the $\sqrt{-g}{\cal R}^{\mu\nu}{\cal R}_{\mu\nu}$ part. This
gives
 \bea
\delta (\sqrt{-g}{\cal R}^{\mu\nu}{\cal R}_{\mu\nu})&=& -\frac{1}{2}\sqrt{-g}g_{\alpha\beta}{\cal R}^{\mu\nu}{\cal R}_{\mu\nu}\delta g^{\alpha\beta}+\sqrt{-g}\delta({\cal R}^{\mu\nu}{\cal R}_{\mu\nu})=\nn\\
&=&-\frac{1}{2}\sqrt{-g}g_{\alpha\beta}{\cal R}^{\mu\nu}{\cal R}_{\mu\nu}\delta g^{\alpha\beta}+2\sqrt{-g}{\cal R}^{\mu}_\alpha{\cal R}_{\mu\beta}\delta g^{\alpha\beta}\nn\\
& &+2\sqrt{-g}{\cal R}^{\mu\nu} \delta{\cal R}_{\mu\nu}.
\eea
 Using eq.~(\ref{varR}) and the variations (\ref{varg2}) and
(\ref{varg3}) for $f({\cal R})={\cal R}$ as a guide, we can
straightforwardly derive the field equations
 \bea
\label{palfg1}
& & {\cal R}_{(\mu\nu)}+2a {\cal R}^{\sigma}_\mu{\cal R}_{\sigma\nu}-\frac{1}{2}\left({\cal R}+a {\cal R}^{\sigma\lambda}{\cal R}_{\sigma\lambda}\right) g_{\mu\nu}=8\pi\,G\, T_{\mu\nu},\\
\label{palfg2}
& &\gn_\lambda\left(\sqrt{-g}\left(g^{\mu\nu}+2a {\cal R}^{\mu\nu}\right)\right)=0.
\eea

Comparing these equations with eqs.~(\ref{palf12}) and (\ref{palf22}),
the following comment is due: Eq.~(\ref{palf22}) is in practice an
algebraic equation in $\Gamma^{\lambda}_{\phantom{a}\mu\nu}$ since
$\gn$ is linear in the connection and no derivatives of
$\Gamma^{\lambda}_{\phantom{a}\mu\nu}$ are present. This is why we
were able to solve for $\Gamma^{\lambda}_{\phantom{a}\mu\nu}$,
eliminate it and rewrite the equations easily in the form of
eq.~(\ref{eq:field}). This is not the case here because ${\cal
R}^{\mu\nu}$ depends on the derivatives of the connection. Therefore,
eq.~(\ref{palfg2}) is a differential equation relating
$\Gamma^{\lambda}_{\phantom{a}\mu\nu}$ and $g_{\mu\nu}$ and we can
conclude that including a higher order term, such as ${\cal
R}^{\mu\nu}{\cal R}_{\mu\nu}$, induces more dynamics in the theory.

\section{Metric-affine gravity}
\label{metaffgrav}
\subsection{The significance of coupling the connection to matter}

We have mentioned several times that in the Palatini formalism the
independent connection $\Gamma^{\lambda}_{\phantom{a}\mu\nu}$ is not
present in the matter action and that this makes the theory a metric
theory of gravity and the geometry pseudo-Riemannian. In fact,
Palatini $f(R)$ gravity satisfies the metric postulates, since it can
be shown that the stress energy tensor of matter is indeed
divergence-free with respect to the Levi--Civita connection of the
metric \cite{tomi}. This should have been expected from the fact that
the only field coupled to matter is the metric $g_{\mu\nu}$. 

How physical is it though to include an independent connection in the
theory without coupling it to the matter fields? Usually the affine
connection defines parallel transport and the covariant derivative.
The matter action includes covariant derivatives of the matter fields
and consequently couplings between the fields and the connection in
the general cases. Some known exceptions to this rule are scalar
fields (since in the case of a scalar, a covariant derivative reduces
to a partial one) and the Electromagnetic field, due to the specific
structure of its action having its roots in gauge invariance (we will
discuss this in detail shortly). Therefore, the assumption
 \be
\frac{\delta S_M}{\delta \Gamma^{\lambda}_{\phantom{a}\mu\nu}}=0
\ee
 has physical implications \cite{sotmg}. It either implies that the
matter action includes only specific matter fields --- an implausibly
limiting option for a gravitation theory --- or that
$\Gamma^{\lambda}_{\phantom{a}\mu\nu}$ is not the affine connection
with which we define parallel transport and the covariant derivative,
as we have been stressing in the previous sections. 

Of course, one is allowed to add an affinity as an extra field, even
if this affinity does not have the usual geometric interpretation and
it is the Levi--Civita connection of the metric that plays this role. 
This is what happens in the Palatini formalism. However, it is
interesting to explore what actually happens if the independent
connection is given its usual geometric characteristics, {\em i.e.~}if it is $\Gamma^{\lambda}_{\phantom{a}\mu\nu}$ that defines parallel
transport and therefore is coupled to the matter. The matter action
will then be $S_M(g^{\mu\nu}, \Gamma^{\lambda}_{\phantom{a}\mu\nu},
\psi)$ and its variation with respect to the connection will no longer
vanish.

Such a theory is a metric-affine theory of gravity. Besides the
standard motivation for alternative theories of gravity, from High
Energy Physics and Cosmology (mentioned in the Introduction and
discussed previously in this chapter for other theories),
metric-affine gravity has one more appealing characteristic: the
connection can be left non-symmetric and the theory can naturally
include torsion. This implies that the theory can be coupled in a more
natural way to some matter fields, such as fermions (Dirac fields).
Note that the stress energy tensor of a Dirac field is not symmetric
by definition and this is something that poses an extra difficulty
when one attempts to couple such fields to General Relativity. In
fact, one might expect that at some intermediate or high energy
regime, the spin of particles might interact with the geometry and
torsion can naturally arise \cite{hehlrev2} [{\em c.f.} with Section
\ref{afffields} and Einstein--Cartan Theory \cite{hehlrev}].
Metric-affine gravity, unlike General Re\-la\-ti\-vity, allows for this to
happen.

There are a number of early works in which the metric and the parallel
transport defining connection are considered as being, to some degree,
independent (see for instance \cite{pap, kun, hehl, hehl2} and
references therein). In many cases, including Einstein--Cartan theory,
some part of the connection is related to the metric ({\em e.g.~}the
non-metricity) \cite{hehlrev}. We will consider the case where
$\Gamma^{\lambda}_{\phantom{a}\mu\nu}$ is left completely
unconstrained and is determined by the field equations. This approach
was first presented in \cite{hehl} for an action linear in ${\cal R}$.
We will generalize it here for $f({\cal R})$ actions
\cite{Sotiriou:2006qn, sotlib2}. Before going any further, it should
be noted that the metric-affine approach has also been widely used in
order to interpret gravity as a gauge theory (see, for example,
\cite{rub} for a study of $f(R)$ actions and \cite{hehlrev2} for a
thorough review).

\subsection{The action}

Let us construct the action which we will be using step by step. To
begin with, we have already specified that the matter action will have
the general form $S_M=S_M\,(g^{\mu\nu},
\Gamma^{\lambda}_{\phantom{a}\mu\nu}, \psi)$. We can then concentrate
on the gravitational action. We can once more use the requirement for
having second order differential field equations, as with the
Einstein--Hilbert action, and combine it with that of having a
Lagrangian which is a generally covariant scalar. Again ${\cal R}$ is
an obvious choice but not the only one, unlike in purely metric
theories. Remember that in the case of the Palatini formalism, we
commented that the choice of the action was to a large extent ad hoc.

For instance, besides invariants built combining the metric and the
independent connection, one might be tempted to use also invariants
that depend only on the metric. Using $R$, {\em i.e.}~the scalar
curvature related to the metric alone, would still lead to second
order field equations. Another option can arise if the connections are
of such a form that one can define a second metric, $h_{\mu\nu}$, that
is covariantly conserved, {\em i.e.}~the metric of which the $\Gamma$s
are the Levi--Civita connections (note that this is not necessarily
true for a general connection~\cite{Gm}, and so it would lead to a
less general theory). Then we could use this metric to contract the
Riemann tensor and derive the Ricci scalar $R(h)$, which is actually
the scalar curvature of the metric $h_{\mu\nu}$ . Going even further
we could even use one of the two metrics, $g_{\mu\nu}$ or
$h_{\mu\nu}$, to go from the Riemann tensor to the Ricci tensor and
the other to derive the Ricci scalar from the Ricci tensor. The
question that arises is whether not using these other scalar
quantities in the action constitutes a further assumption, which is
not needed in the purely metric formulation. 

From the mathematical point of view, we could use any of the Ricci
scalars defined above. However we think that for any possible choice
other than ${\cal R}$, there are good physical reasons for discarding
it. In fact, when constructing a metric-affine theory, one assumes
that the spacetime is fully described by two independent geometrical
objects, the metric and the connection. The metric defines the
chronological structure, the connection defines the affine structure
of the manifold. This manifold is not chosen to be pseudo-Riemannian
(at least initially). One can always mathematically consider a manifold 
on which two different pseudo-Reimannian geometries are imposed, one 
described by the metric $g_{\mu\nu}$ and the other by the metric $h_{\mu\nu}$ 
(if it exists), but these separate geometries are not relevant for the spacetime in which a metric-affine theory acts.
 Therefore, quantities related to them, such as their
scalar curvatures, should not be used in the action of a theory living
on the non-Riemannian manifold under consideration. Also, using
quantities derived by contracting once with one metric and once with
the other, should also be avoided. There is only one metric that
determines how distances are measured in our spacetime and this is
$g_{\mu\nu}$. This is the metric that is used to evaluate inner
products and therefore it is the one that should be used to raise or
lower indices and perform contractions.

Since ${\cal R}$ does not depend on derivatives higher than first
order in either the metric or the connection, as already mentioned in section
\ref{palform}, there is no reason {\it
a priori} to restrict ourselves to an action linear in ${\cal R}$. 
Therefore, it is equally ``natural'' to consider an $f({\cal R})$
action:
 \be
\label{maaction}
S_{ma}=\frac{1}{16\pi\,G}\int d^4 x \sqrt{-g} f({\cal R}) +S_M(g_{\mu\nu}, \Gamma^{\lambda}_{\phantom{a}\mu\nu}, \psi).
\ee
 Choosing an action linear in $R$, like (\ref{action}), must be
considered as a simplifying choice in metric-affine gravity, unlike in
purely metric theories where an action linear in $R$ is the only one
that leads to second order equations \footnote{We are confining
ourselves to Lagrangians that are functions of the Ricci scalar only.
In a more general setting, one should mention that Gauss--Bonnet type
Lagrangians lead to second order field equations as well.}. 

Even the $f({\cal R})$ action is a simplicity choice in metric-affine
gravity and it is not the most general action that would lead to
second order equations. Apart from not including the first
derivatives of the metric, an $f({\cal R})$ action also does not include terms quadratic in
the first derivative of the connection. Moreover, if the
connection is not symmetric there is an extra tensor available for
constructing invariants: the Cartan torsion tensor
(eq.~(\ref{cartan})). We will comment on possible generalisations of
the action in the next section, as this issue will prove to be crucial
in metric-affine gravity.

\subsection{Field equations}

Since we assume that the metric and the connection are fully
independent, we do not intend to make any assumptions about
non-metricity and torsion. Therefore, the connection will not be taken
to be symmetric or covariantly conserved by the connection. The
definitions presented in section \ref{spacetimegeo} will be used
extensively here. Let us attempt to derive field equations from the
action (\ref{maaction}).

\subsubsection{The variation} 

If we denote the gravitational part of the action as 
\be
S_{\rm grav}=\frac{1}{16\pi\,G}\int d^4 x \sqrt{-g} f({\cal R})
\ee
the least action principle gives
\be
\label{var}
0=\delta S_{ma}=\delta S_{\rm grav}+\delta S_M,
\ee
and the variation of the gravitational part gives
\bea
\label{varg}
\delta S_{\rm grav}&=&\frac{1}{16\pi\,G}\int d^4 x \,\delta\left(\sqrt{-g}f({\cal R})\right)=\nn\\
&=&\frac{1}{16\pi\,G}\int d^4 x \left(f({\cal R})\delta\sqrt{-g}+\sqrt{-g}f'({\cal R})\delta {\cal R}\right)\nn\\
&=&\frac{1}{16\pi\,G}\int d^4 x \left(f({\cal R})\delta\sqrt{-g}+\sqrt{-g}f'({\cal R})\delta \left(g^{\mu\nu}{\cal R}_{\mu\nu}\right)\right)\nn\\
&=&\frac{1}{16\pi\,G}\int d^4 x \sqrt{-g}\left(f'({\cal R}) {\cal R}_{(\mu\nu)}-\frac{1}{2}f({\cal R})g_{\mu\nu}\right)\delta g^{\mu\nu}+\nn\\& &+\frac{1}{16\pi\,G}\int d^4 x \sqrt{-g}f'({\cal R})g^{\mu\nu}\delta {\cal R}_{\mu\nu},
\eea
 where we have used the symmetry of the metric ($\delta
g^{\mu\nu}{\cal R}_{\mu\nu}=\delta g^{\mu\nu}{\cal R}_{(\mu\nu)}$). 

To complete this variation, we need to evaluate the quantity $\delta
{\cal R}_{(\mu\nu)}$. ${\cal R}_{\mu\nu}$ depends only on the
connections and so we can already see that the second term of the last
line of eq.~(\ref{varg}) will be the one related to the variation with
respect to $\Gamma^\lambda_{\phantom{a}\mu\nu}$. We cannot use
eq.~(\ref{varR}) here since this was derived under the assumption that
$\Gamma^\lambda_{\phantom{a}\mu\nu}$ is symmetric.  Taking into
account the definition of the Ricci tensor, eq.~(\ref{ricci}), one can
generalise eq.~(\ref{varR}) for a non-symmetric connection:
 \bea
\label{varR2}
\delta {\cal R}_{\mu\nu}&=&\gn_\lambda \delta \Gamma^\lambda_{\phantom{a}\mu\nu}-\gn_{\nu}\delta \Gamma^\lambda_{\phantom{a}\mu\lambda}+2\Gamma^\sigma_{\phantom{a}[\nu\lambda]}\delta\Gamma^\lambda_{\phantom{a}\mu\sigma}.
\eea
 Using eq.~(\ref{varR2}), the variation of the gravitational part of 
the action takes the form
\bea
\label{varg2ma}
\delta S_G&=&\frac{1}{2\kappa}\int d^4 x \sqrt{-g}\left(f'({\cal R}) {\cal R}_{(\mu\nu)}-\frac{1}{2}f({\cal R})g_{\mu\nu}\right)\delta g^{\mu\nu}+{}\nn\\& &+\frac{1}{2\kappa}\int d^4 x \sqrt{-g}f'({\cal R})g^{\mu\nu}\left(\gn_\lambda \delta \Gamma^\lambda_{\phantom{a}\mu\nu}-\gn_{\nu}\delta \Gamma^\lambda_{\phantom{a}\mu\lambda}\right)+\nn\\& &+\frac{1}{2\kappa}\int d^4 x \,2\sqrt{-g}f'({\cal R})g^{\mu\sigma}\Gamma^\nu_{\phantom{a}[\sigma\lambda]}\delta\Gamma^\lambda_{\phantom{a}\mu\nu}.
\eea
Integrating the terms in the second line by parts, we get
\bea
\label{varg3ma}
\delta S_G&=&\frac{1}{2\kappa}\int d^4 x \sqrt{-g}\left(f'({\cal R}) {\cal R}_{(\mu\nu)}-\frac{1}{2}f({\cal R})g_{\mu\nu}\right)\delta g^{\mu\nu}+{}\\& &+\frac{1}{2\kappa}\int d^4 x \bigg[-\gn_\lambda\left(\sqrt{-g}f'({\cal R})g^{\mu\nu}\right)+\gn_\sigma\left(\sqrt{-g}f'({\cal R})g^{\mu\sigma}\right)\delta^\nu_\lambda\nn\\& &+ 2\sqrt{-g}f'({\cal R})\left(g^{\mu\nu}\Gamma^\sigma_{\phantom{a}[\lambda\sigma]}-g^{\mu\rho}\Gamma^\sigma_{\phantom{a}[\rho\sigma]}\delta^\nu_\lambda+g^{\mu\sigma}\Gamma^\nu_{\phantom{a}[\sigma\lambda]}\right)\bigg]\delta\Gamma^\lambda_{\phantom{a}\mu\nu}+\textrm{ST},\nn
\eea
 where $\textrm{ST}$ stands for ``Surface Terms''. These terms are
total divergences linear in $\delta
\Gamma^\lambda_{\phantom{a}\mu\nu}$. Being total divergences, we can
turn their integral over the volume into an integral over the boundary
surface. Since $\delta \Gamma^\lambda_{\phantom{a}\mu\nu}=0$ on the
boundary, they will then vanish. [Note that the first two terms in the
last line of eq.~(\ref{varg3ma}) came from the integration by parts of
the second line of (\ref{varg2ma}). This is because differentiation by
parts and integration of covariant derivatives becomes non-trivial in
the presence of a non-symmetric connection (for more information on
this, see chapter 2 and p.~109 of Ref.~\cite{schro}).] This concludes
the variation of the gravitational part of the action.

We now have to consider the variation of the matter action. Since
\be
S_M=S_M(g_{\mu\nu},\Gamma^\lambda_{\phantom{a}\mu\nu},\psi),
\ee
 we have
 \be
\delta S_M=\frac{\delta S_M}{\delta g^{\mu\nu}}\delta g^{\mu\nu}+\frac{\delta S_M}{\delta \Gamma^\lambda_{\phantom{a}\mu\nu}}\delta \Gamma^\lambda_{\phantom{a}\mu\nu}.
\ee
We can define the stress-energy tensor in the usual way
\be
\label{set}
T_{\mu\nu}\equiv-\frac{2}{\sqrt{-g}}\frac{\delta {S}_M}{\delta g^{\mu\nu}}.
\ee
 We also define a new tensor, which we shall call (following the 
nomenclature of \cite{hehl}) the ``hypermomentum'', as
\be
\label{defD}
\Delta_{\lambda}^{\phantom{a}\mu\nu}\equiv-\frac{2}{\sqrt{-g}}\frac{\delta {S}_M}{\delta \Gamma^\lambda_{\phantom{a}\mu\nu}},
\ee
 {\em i.e.}~the variation of the matter action with respect to the 
connections. Therefore, the variation of the matter action will be
\be
\label{varmat}
\delta S_M=-\frac{1}{2}\int d^4 x\sqrt{-g}\left[T_{\mu\nu}\delta g^{\mu\nu}+\Delta_{\lambda}^{\phantom{a}\mu\nu}\delta \Gamma^\lambda_{\phantom{a}\mu\nu}\right].
\ee
 Even though $\Gamma^\lambda_{\phantom{a}\mu\nu}$ is not a tensor,
this does not mean that $\Delta_{\lambda}^{\phantom{a}\mu\nu}$ is not
a tensor. $\delta \Gamma^\lambda_{\phantom{a}\mu\nu}$ is a tensor and,
therefore, so is $\Delta_{\lambda}^{\phantom{a}\mu\nu}$.

Note also that the vanishing of $\Delta_{\lambda}^{\phantom{a}\mu\nu}$
would imply independence of the matter action from the connections. As
we discussed, this would be contrary to the spirit of metric-affine
gravity if it happened for any field and the theory would reduce to
$f(R)$ gravity in the Palatini formalism. There are, however, specific
fields that have this attribute; the most common example is the scalar
field. There will therefore be certain sorts of matter field, as we
will see later on, where metric-affine $f(R)$ gravity and $f(R)$
gravity in the Palatini formalism will give equivalent physical
predictions, without of course being equivalent theories overall.  For
instance, if we consider a Dirac field, the
matter action is no longer independent of the connection and
$\Delta_{\lambda}^{\phantom{a}\mu\nu}$ does not vanish. 

\subsubsection{Projective invariance and consistent field equations}
\label{fieldeq}
 We are now ready to derive the field equations using the variation of
the gravitational and matter actions. This can be achieved simply by
summing the variations (\ref{varg3ma}) and (\ref{varmat}) and applying
the least action principle. We obtain
 \be
\label{field1}
f'({\cal R}) {\cal R}_{(\mu\nu)}-\frac{1}{2}f({\cal R})g_{\mu\nu}=\kappa T_{\mu\nu},
\ee
and
\begin{align}
\label{field2}
\frac{1}{\sqrt{-g}}\bigg[&-\gn_\lambda\left(\sqrt{-g}f'({\cal R})g^{\mu\nu}\right)+\gn_\sigma\left(\sqrt{-g}f'({\cal R})g^{\mu\sigma}\right){\delta^\nu}_\lambda\bigg]+{}\\ &+2f'({\cal R})\left(g^{\mu\nu}\Gamma^\sigma_{\phantom{a}[\lambda\sigma]}-g^{\mu\rho}\Gamma^\sigma_{\phantom{a}[\rho\sigma]}{\delta^\nu}_\lambda+g^{\mu\sigma}\Gamma^\nu_{\phantom{a}[\sigma\lambda]}\right)=\kappa\Delta_{\lambda}^{\phantom{a}\mu\nu}.\nn
\end{align}
We can also use the Cartan torsion tensor, eq.~(\ref{cartan}),
to re-express eq.~(\ref{field2}) and highlight the presence of torsion:
\begin{align}
\label{field22}
\frac{1}{\sqrt{-g}}\bigg[&-\gn_\lambda\left(\sqrt{-g}f'({\cal R})g^{\mu\nu}\right)+\gn_\sigma\left(\sqrt{-g}f'({\cal R})g^{\mu\sigma}\right){\delta^\nu}_\lambda\bigg]+{}\\ &+2f'({\cal R})\left(g^{\mu\nu}S^{\phantom{ab}\sigma}_{\lambda\sigma}-g^{\mu\rho}S^{\phantom{ab}\sigma}_{\rho\sigma}{\delta^\nu}_\lambda+g^{\mu\sigma}S^{\phantom{ab}\nu}_{\sigma\lambda}\right)=\kappa\Delta_{\lambda}^{\phantom{a}\mu\nu}.\nn
\end{align}
 A careful look at the above equation reveals that if we take the 
trace on $\lambda$ and $\mu$ we get
\be
\label{contr}
0=\kappa \Delta_{\mu}^{\phantom{a}\mu\nu},
\ee
 since the left hand side is traceless. One can interpret this as a
constraint on the form of $\Delta_{\lambda}^{\phantom{a}\mu\nu}$,
meaning that the matter action has to be chosen in such a way that its
variation with respect to the connections leads to a traceless tensor.
However, it is easy to understand that this is not satisfactory since
there exist common forms of matter which do not have this attribute.
Therefore the field equations which we have derived are inconsistent. 
This problem is not new; it was pointed out for the simple case of the
Einstein--Hilbert action long ago \cite{hehl,schro,sand}. Its roots
can be traced in the form of the action itself and in the fact that in
metric-affine gravity $\Gamma^{\lambda}_{\phantom{a}\mu\nu}$ has no
{\em a priori} dependence on the metric.

Let us consider the projective transformation
\be
\label{proj}
\Gamma^{\lambda}_{\phantom{a}\mu\nu}\rightarrow \Gamma^{\lambda}_{\phantom{a}\mu\nu}+{\delta^\lambda}_\mu\xi_\nu,
\ee
 where $\xi_\nu$ is an arbitrary covariant vector field. One can
easily show that the {\cal R}icci tensor will correspondingly
transform like
 \be
\label{projRicci}
{\cal R}_{\mu\nu}\rightarrow {\cal R}_{\mu\nu}-2\partial_{[\mu}\xi_{\nu]}.
\ee
 However, given that the metric is symmetric, this implies that the 
curvature scalar does not change
\be
{\cal R}\rightarrow {\cal R},
\ee
 {\em i.e.}~${\cal R}$ is invariant under projective transformations. 
Hence the Einstein--Hilbert action or any other action built from a
function of ${\cal R}$, such as the one used here, is projective
invariant in metric-affine gravity.  However, the matter action is not
generically projective invariant and this is the cause of the
inconsistency in the field equations.

The conclusion that we have to draw is that when we want to consider a
theory with a symmetric metric and an independent general connection,
an action that depends only on the scalar curvature is not suitable. 
The way to bypass this problem is then obvious: we have to drop one of
the assumptions just listed. The first option is to abandon the
requirement of having a symmetric metric, since in this case ${\cal
R}$, and consequently the gravitational action, would not be
projectively invariant (see eq.~(\ref{projRicci})).  For the
Einstein--Hilbert Lagrangian this would lead to the well known
Einstein--Straus theory \cite{schro}, and using an $f(R)$ Lagrangian
would lead to a generalisation of it. This theory, even though it
leads to fully consistent field equations, is characterised by the
fact that, in vacuum, neither non-metricity nor torsion
vanish~\cite{schro}.  In particular, this implies that torsion in the
Einstein--Strauss theory is not just introduced by matter fields but
is intrinsic to gravity and can propagate. Although logically
possible, such an option does not seem very well motivated from a
physical point of view, as one would more naturally expect any
``twirling" of spacetime to be somehow directly induced by the
interaction with matter. Additionally, there is no experimental
evidence so far of propagating torsion. Note that the effects of
non-propagating torsion appear only in the presence of the matter
inducing it and therefore they are significantly harder to detect. We
shall therefore not pursue a route that allows for propagating torsion
any further. Instead we will consider the alternative solutions to our
problem.

The second path towards a consistent theory is to modify the action by
adding some extra terms. These terms should be chosen in such a way so
as to break projective invariance. There were proposals in this
direction in the past, based on the study of an action linear in $R$
(see \cite{hehl2} and references therein). As an example, we can
mention the proposal of \cite{pap}: adding to the Lagrangian the term
$g^{\mu\nu}\partial_\mu \Gamma^\sigma_{\phantom{a}[\nu\sigma]}$. Such
a choice leads to a fully consistent theory and is mathematically very
interesting. However, we find it difficult to physically motivate the
presence of this term in the gravitational action. Much more
physically justified, instead, are corrections of the type ${\cal
R}^{\mu\nu}{\cal R}_{\mu\nu}$, ${\cal R}^{\alpha\beta\mu\nu}{\cal
R}_{\alpha\beta\mu\nu}$ {\em etc}. In fact, as we have already mentioned,
such terms might very naturally be present in the gravitational action
if we consider it as an effective, low energy, classical action coming
from a more fundamental theory \cite{quant1, quant2, quant3, gaspven,
nojodi1, vassi}. We shall not discuss such modifications in detail
here, since this goes beyond the scope of this study; however, we will
make some comments. It is easy to verify, working for example with the
simplest term ${\cal R}^{\mu\nu}{\cal R}_{\mu\nu}$, that such
modifications will in general lead to consistent field equations. One
should also mention that from a field theory point of view one could
choose to include all of the terms of the same order in some variable. 
As we have already mentioned, an $f({\cal R})$ action does not include
first derivatives of the metric and, what is more, there are a number
of terms which one could consider that can be constructed with
combinations of the derivatives of the connection, especially now that
the latter is not symmetric.

However, any of the additions discussed above will generically lead to
a theory with the same attribute as Einstein--Straus theory, {\em
i.e.}~in vacuum, torsion will not generically vanish. One might
imagine that a certain combination of higher order curvature
invariants would lead to a theory with vanishing torsion in vacuum. To
find such a theory would certainly be very interesting but is beyond
the scope of the present investigation\footnote{ One could even
imagine proposing the absence of torsion in vacuum as a possible
criterion in order to select a suitable combination of high energy
(strong gravity) corrections to our $f(R)$ action.}. In conclusion,
this route generically leads to theories where again the presence of
torsion seems to be an unmotivated complication rather than a physical
feature. 

With no prescription for how to form a more general gravitational
action which can lead to a physically attractive theory, we are left
with only one alternative: to find a way of deriving consistent field
equations with the action at hand. To understand how this is possible,
we should re-examine the meaning of projective invariance. This is
very similar to gauge invariance in Electromagnetism (EM). It tells us
that the corresponding field, in this case the connections
$\Gamma^\lambda_{\phantom{a}\mu\nu}$, can be determined from the field
equations up to a projective transformation (eq.~(\ref{proj})).
Breaking this invariance can therefore come by fixing some degrees of
freedom of the field, similarly to gauge fixing. The number of degrees
of freedom which we need to fix is obviously the number of the
components of the four-vector used for the transformation, {\em
i.e.}~simply four. In practice, this means that we should start by
assuming that the connection is not the most general which one can
construct, but satisfies some constraints. Instead of placing an
unphysical constraint on the action of the matter fields, as dictated
by eqs.~(\ref{field22}) and (\ref{contr}), we can actually make a
statement about spacetime properties. This is equivalent to saying
that the matter fields can have all of the possible degrees of freedom
but that the spacetime has some rigidity and cannot respond to some of
them. (We shall come back to this point again later on. Let us just
say that this is, for example, what happens in General Relativity when
one assumes that there is no torsion and no non-metricity.)

We now have to choose the degrees of freedom of the connections that
we need to fix. Since there are four of these, our procedure will be
equivalent to fixing a four-vector. We can again let the studies of
the Einstein--Hilbert action \cite{hehl2} lead the way. The proposal
of Hehl {\em et al.}~\cite{hehl2} was to fix part of the
non-metricity, namely the Weyl vector $Q_\mu$ (eq.~(\ref{weyl})). The
easiest way to do this is by adding to the action a term containing a
Lagrange multiplier $A^\mu$, which has the form
 \be
S_{LM}=\int d^4 x \sqrt{-g} A^\mu Q_{\mu}.
\ee
 This way, one does not need to redo the variation of the rest of the
action, but instead, only to evaluate the variation of the extra term.
Varying with respect to the metric, the connections and $A$
respectively, we get the new field equations
 \bea
\label{fieldQ1}
& &  f'({\cal R}) {\cal R}_{(\mu\nu)}-\frac{1}{2}f({\cal R})g_{\mu\nu}=\kappa T_{\mu\nu}+\frac{\kappa}{4\sqrt{-g}}\partial_\sigma(\sqrt{-g}A^\sigma)g_{\mu\nu},\\
\label{fieldQ2}
& &\frac{1}{\sqrt{-g}}\bigg[-\gn_\lambda\left(\sqrt{-g}f'({\cal R})g^{\mu\nu}\right)+\gn_\sigma\left(\sqrt{-g}f'({\cal R})g^{\mu\sigma}\right){\delta^\nu}_\lambda\bigg]+\nn\\& &\qquad\qquad+2f'({\cal R})\left(g^{\mu\nu}S^{\phantom{ab}\sigma}_{\lambda\sigma}-g^{\mu\rho}S^{\phantom{ab}\sigma}_{\rho\sigma}{\delta^\nu}_\lambda+g^{\mu\sigma}S^{\phantom{ab}\nu}_{\sigma\lambda}\right)=
\nn\\& &\qquad\qquad\qquad\qquad\qquad\qquad\qquad\quad=
\kappa\left(\Delta_{\lambda}^{\phantom{a}\mu\nu}-\frac{1}{4}{\delta^\mu}_\lambda A^\nu\right),\\
\label{Q0}
& & Q_\mu=0.
\eea
Taking the trace of eq.~(\ref{fieldQ2}) gives
\be
\label{DA}
A^\nu=\Delta_{\mu}^{\phantom{a}\mu\nu},
\ee
 which is the consistency criterion, {\em i.e.}~it gives the value
which we should choose for $A^\nu$ so that the equations are
consistent. This procedure obviously works when $f(R)$ is a linear
function as shown in \cite{hehl2}. However, we will demonstrate here
that it is not equally appealing in any other case. 

Consider the simple case where no matter is present and let us search
for the solution of the field equations for which the torsion
vanishes, {\em i.e.}
 \be
S^{\phantom{ab}\nu}_{\sigma\lambda}=0.
\ee
In this case eqs.~(\ref{fieldQ2}) and (\ref{DA}) give
\be
\label{fieldQ22}
\frac{1}{\sqrt{-g}}\bigg[-\gn_\lambda\left(\sqrt{-g}f'({\cal R})g^{\mu\nu}\right)+\gn_\sigma\left(\sqrt{-g}f'({\cal R})g^{\mu\sigma}\right){\delta^\nu}_\lambda\bigg]=0,
\ee
 which is no different from eq.~(\ref{palf2}) which we derived for
Palatini $f(R)$ gravity. Therefore, once more by contracting the
indices $\nu$ and $\lambda$ and replacing the result back in the
equation, we get
 \be
\label{fieldQ24}
\gn_\lambda\left(\sqrt{-g}f'({\cal R})g^{\mu\nu}\right)=0.
\ee
This equation implies that one can define a metric $h_{\mu\nu}$ such that
\be
h_{\mu\nu}=f'({\cal R})g_{\mu\nu},
\ee
 which is covariantly conserved by the connections
$\Gamma_{\phantom{a}\mu\nu}^\lambda$ [see section \ref{manfield} and discussion after eq.~(\ref{hgconf})].  Now notice the following:
$h_{\mu\nu}$ has zero non-metricity by definition, leading to
 \be
\gn_\lambda h_{\mu\nu}=0.
\ee
A contraction with the metric will give
\be
4\frac{1}{f'({\cal R})}\partial_\lambda f'({\cal R})+g^{\mu\nu}f'({\cal R})\gn_\lambda g_{\mu\nu}=0
\ee
 Now remember that eq.~(\ref{Q0}) forces the vanishing of the Weyl
vector $Q_{\lambda}\equiv g^{\mu\nu}\gn_\lambda g_{\mu\nu}$. Therefore
the above equation implies that
 \be
\frac{1}{f'({\cal R})}\partial_\lambda f'({\cal R})=0,
\ee
 {\em i.e.}~that $f'({\cal R})$ is just a constant. If $f({\cal R})$
is taken to be linear in ${\cal R}$, everything is consistent, but
this is not the case if one considers a more general $f({\cal R})$
action\footnote{In \cite{Sotiriou:2006qn} a miscalculation (eq.~(57))
led to an erroneous claim that torsion vanishes in vacuum in this
version of the theory. This is not true, but the result concerning
whether a non metricity condition should be forced still holds as
shown by the current discussion.}. 

The above exercise clearly shows that there exist no solutions of the
field equations under our assumptions whenever $f({\cal R})$ is
non-linear, {\em i.e.~}there is no vacuum solution with vanishing
torsion. The reason for this is simply that part of the non-metricity
in our theory is due to the form of the action. Therefore,
constraining the non-metricity in any way turns out to be a constraint
on the form of the Lagrangian itself, unless the rest of the
unconstrained part of the connection, torsion, can help to cancel out
the non-metricity induced by $f({\cal R})$. This indicates that if we
want to consider an action more general than the Einstein--Hilbert
one, we should definitely avoid placing such kinds of constraint.

One could add that in a true metric-affine theory of gravity, the
connection and the metric are assumed to be completely independent
fields, related only by the field equations. Therefore, imposing a
constraint that includes both the metric and the connection, such as a
metricity condition, seems to be contradicting the very spirit of the
theory, since it gives an {\it a priori} relation between the two
quantities. 

 The above not only demonstrate the unappealing features of the
procedure adopted in \cite{hehl2} but also makes it clear that the
four degrees of freedom which we have to fix are related to torsion.
This implies that the torsionless version of the theory should be
fully consistent without fixing any degrees of freedom. Let us now
verify that. We can go back to the variation of the action in
eq.~(\ref{varg3ma}) and force the connection to be symmetric. This
gives
 \bea
\label{varg3sym}
\delta S_G&=&\frac{1}{2\kappa}\int d^4 x\Bigg[\sqrt{-g}\left(f'({\cal R}) {\cal R}_{(\mu\nu)}-\frac{1}{2}f({\cal R})g_{\mu\nu}\right)\delta g^{\mu\nu}+{}\\& &+\left[-\gn_\lambda\left(\sqrt{-g}f'({\cal R})g^{\mu\nu}\right)+\gn_\sigma\left(\sqrt{-g}f'({\cal R})g^{\sigma(\mu}\right){\delta^{\nu)}}_\lambda\right]\delta\Gamma^\lambda_{\phantom{a}\mu\nu}\Bigg],\nn
\eea
and so the corresponding field equations are
\be
\label{field1sym}
f'({\cal R}) {\cal R}_{(\mu\nu)}-\frac{1}{2}f({\cal R})g_{\mu\nu}=\kappa T_{\mu\nu},\ee\be
\label{field2sym}
\frac{1}{\sqrt{-g}}\bigg[-\gn_\lambda\left(\sqrt{-g}f'({\cal R})g^{\mu\nu}\right)+\gn_\sigma\left(\sqrt{-g}f'({\cal R})g^{\sigma(\mu}\right){\delta^{\nu)}}_\lambda\bigg]=\kappa\Delta_{\lambda}^{\phantom{a}(\mu\nu)}.
\ee
 where $\Delta_{\lambda}^{\phantom{a}\mu\nu}$ is also symmetrized due
to the symmetry of the connection. One can easily verify that these
equations are fully consistent. They are the field equations of $f(R)$
metric-affine gravity without torsion.

Turning back to our problem, we need to fix four degrees of freedom of
the torsion tensor in order to make the version of the theory with
torsion physically meaningful. A prescription has been given in
\cite{sand} for a linear action and we shall see that it will work for
our more general Lagrangian too. This prescription is to set the
vector $S_\mu=S_{\sigma\mu}^{\phantom{ab}\sigma}$ equal to zero. Note
that this does not mean that $\Gamma^{\phantom{ab}\sigma}_{\mu\sigma}$
should vanish but merely that
$\Gamma^{\phantom{ab}\sigma}_{\mu\sigma}=\Gamma^{\phantom{ab}\sigma}_{\sigma\mu}$.
We shall again use a Lagrange multiplier, $B^\mu$, for this purpose. 
The additional term in the action will be
 \be
\label{lm2}
S_{LM}=\int d^4 x \sqrt{-g} B^\mu S_{\mu}.
\ee
 It should be clear that the addition of this term does not imply that
we are changing the action, since it is simply a mathematical trick to
avoid doing the variation of the initial action under the assumption
that $S_{\mu}=0$. The new field equations which we get from the
variation with respect to the metric, the connections and $B^\mu$ are
 \bea
\label{field1t1}
& &f'({\cal R}) {\cal R}_{(\mu\nu)}-\frac{1}{2}f({\cal R})g_{\mu\nu}=\kappa T_{\mu\nu},\\
\label{field2t1}
& &\frac{1}{\sqrt{-g}}\bigg[-\gn_\lambda\left(\sqrt{-g}f'({\cal R})g^{\mu\nu}\right)+\gn_\sigma\left(\sqrt{-g}f'({\cal R})g^{\mu\sigma}\right){\delta^\nu}_\lambda\bigg]+{}\nn\\ & &\qquad\qquad+2f'({\cal R})\left(g^{\mu\nu}S^{\phantom{ab}\sigma}_{\lambda\sigma}-g^{\mu\rho}S^{\phantom{ab}\sigma}_{\rho\sigma}{\delta^\nu}_\lambda+g^{\mu\sigma}S^{\phantom{ab}\nu}_{\sigma\lambda}\right)=\nn\\& &\qquad\qquad\qquad\qquad\qquad\qquad\qquad\qquad=\kappa(\Delta_{\lambda}^{\phantom{a}\mu\nu}-B^{[\mu}{\delta^{\nu]}}_{\lambda}),\\
& & S_{\mu\sigma}^{\phantom{ab}\sigma}=0,
\eea
 respectively. Using the third equation, we can simplify the second
one to become
 \bea
\frac{1}{\sqrt{-g}}\bigg[&-&\gn_\lambda\left(\sqrt{-g}f'({\cal R})g^{\mu\nu}\right)+\gn_\sigma\left(\sqrt{-g}f'({\cal R})g^{\mu\sigma}\right){\delta^\nu}_\lambda\bigg]+{}\nn\\ & &+2f'({\cal R})g^{\mu\sigma}S^{\phantom{ab}\nu}_{\sigma\lambda}=\kappa(\Delta_{\lambda}^{\phantom{a}\mu\nu}-B^{[\nu}{\delta^{\mu]}}_{\lambda}).
\eea
Taking the trace over $\mu$ and $\lambda$ gives
\be
B^\mu=\frac{2}{3}\Delta_{\sigma}^{\phantom{a}\sigma\mu}.
\ee
Therefore the final form of the field equations is
\bea
\label{field1t}
& &f'({\cal R}) {\cal R}_{(\mu\nu)}-\frac{1}{2}f({\cal R})g_{\mu\nu}=\kappa T_{\mu\nu},\\
\label{field2t}
& &\frac{1}{\sqrt{-g}}\bigg[-\gn_\lambda\left(\sqrt{-g}f'({\cal R})g^{\mu\nu}\right)+\gn_\sigma\left(\sqrt{-g}f'({\cal R})g^{\mu\sigma}\right){\delta^\nu}_\lambda\bigg]+{}\nn\\ & &\qquad\qquad+2f'({\cal R})g^{\mu\sigma}S^{\phantom{ab}\nu}_{\sigma\lambda}=\kappa(\Delta_{\lambda}^{\phantom{a}\mu\nu}-\frac{2}{3}\Delta_{\sigma}^{\phantom{a}\sigma[\nu}{\delta^{\mu]}}_{\lambda}),\\
\label{field3t}
& & S_{\mu\sigma}^{\phantom{ab}\sigma}=0.
\eea
 These equations have no consistency problems and are the ones which 
we will be using from now on.

So, in the end, we see that we can solve the inconsistency problem of
the unconstrained field equations by imposing a certain rigidness on
spacetime, in the sense that spacetime is allowed to twirl due to its
interaction with the matter fields but only in a way that keeps
$S_\mu=0$. This is not, of course, the most general case that one can
think of but as we demonstrated here, it is indeed the most general
within the framework of $f(R)$ gravity.

We are now ready to investigate further the role of matter in
determining the properties of spacetime. In particular, we shall
investigate the physical meaning of the hypermomentum
$\Delta_{\lambda}^{\phantom{a}\mu\nu}$ and discuss specific examples
of matter actions so as to gain a better understanding of the
gravity-matter relation in the theories under scrutiny here.

\subsection{Matter actions}

In the previous section, we derived the field equations for the
gravitational field in the presence of matter. We considered both the
case where torsion was allowed (eqs.~(\ref{field1t}), (\ref{field2t})
and (\ref{field3t})) and the torsionless version of the same theory
(eqs.~(\ref{field1sym}) and (\ref{field2sym})). Observe that the first
equation in both sets is the same, namely eqs.~(\ref{field1sym}) and
(\ref{field1t}). The second one in each set is the one that has an
explicit dependence on $\Delta_{\lambda}^{\phantom{a}\mu\nu}$, the
quantity that is derived when varying the matter action with
respect to the connection, which has no analogue in General
Relativity. We shall now consider separately more specific forms of
the matter action.

\subsubsection{Matter action independent of the connection}
\label{sec:noD}

Let us start by examining the simple case where the quantity
$\Delta_{\lambda}^{\phantom{a}\mu\nu}$ is zero, {\em i.e.}~$S_M$ is
independent of the connection. In this case eq.~(\ref{field2t}) takes
the form
 \begin{align}
\label{0}
\frac{1}{\sqrt{-g}}\bigg[-&\gn_\lambda\left(\sqrt{-g}f'({\cal R})g^{\mu\nu}\right)+\gn_\sigma\left(\sqrt{-g}f'({\cal R})g^{\mu\sigma}\right){\delta^\nu}_\lambda\bigg]+{}\nn\\ &+2f'({\cal R})g^{\mu\sigma}S^{\phantom{ab}\nu}_{\sigma\lambda}=0.
\end{align}
 Contracting the indices $\nu$ and $\lambda$ and using 
eq.~(\ref{field3t}), this gives
\be
\label{new1}
\gn_\sigma\left(\sqrt{-g}f'({\cal R})g^{\mu\sigma}\right)=0.
\ee
Using this result, eq.~(\ref{0}) takes the form
\bea
\label{anti0}
& &-\frac{1}{\sqrt{-g}}\gn_\lambda\left(\sqrt{-g}f'({\cal R})g^{\mu\nu}\right)+2f'({\cal R})g^{\mu\sigma}S^{\phantom{ab}\nu}_{\sigma\lambda}=0.
\eea
 Taking the antisymmetric part of this equation with respect to the 
indices $\mu$ and $\nu$ leads to 
\be
\label{notorsion}
g^{\sigma[\mu}{S_{\sigma\lambda}}^{\nu]}=0,
\ee
which can be written as
\be
\label{not2}
S_{\mu\lambda\nu} = S_{\nu\lambda\mu}.
\ee
 This indicates that the Cartan torsion tensor must be symmetric with
respect to the first and third indices. However, by definition, it is
also antisymmetric in the first two indices. 

It is easy to prove that any third rank tensor with symmetric and
antisymmetric pairs of indices, vanishes: Take the tensor
$M_{\mu\nu\lambda}$ which is symmetric in its first and third index
($M_{\mu\nu\lambda}=M_{\lambda\nu\mu}$) and antisymmetric in the first
and second index ($M_{\mu\nu\lambda}=-M_{\nu\mu\lambda}$). 
Exploiting these symmetries we can write
 \bea
M_{\mu\nu\lambda}=M_{\lambda\nu\mu}=-M_{\nu\lambda\mu}=-M_{\mu\lambda\nu}=M_{\lambda\mu\nu}=M_{\nu\mu\lambda}=-M_{\mu\nu\lambda}\nonumber.
\eea
Therefore, $M_{\mu\nu\lambda}=0$. 

Consequently, eq.~(\ref{not2}) leads to
\be
S^{\phantom{ab}\nu}_{\sigma\lambda}=0,
\ee
 and torsion vanishes. The connection is now fully symmetric and the 
field equations are
\bea
\label{d01}
& &f'({\cal R}) {\cal R}_{(\mu\nu)}-\frac{1}{2}f({\cal R})g_{\mu\nu}=\kappa T_{\mu\nu},\\
\label{d02}
& &\gn_\lambda\left(\sqrt{-g}f'({\cal R})g^{\mu\nu}\right)=0.
\eea
 Note that these are the same equations that one derives for a theory
in which the matter action is assumed {\it a priori} to be independent
of the connection, {\em i.e.}~for Palatini $f(R)$ gravity and
eqs.~(\ref{palf12}) and (\ref{palf22}). It should be stressed,
however, that here the independence of the matter action from the
connection is due to the fact that we have chosen to consider matter
fields with this property and not to a general characteristic of the
theory, as in Palatini $f(R)$ gravity. We will discuss shortly which
matter fields have this property and what is the form of the field
equations when matter fields without this property are present.

Returning to the field equations, we see that eq.~(\ref{d02}) implies
that one can define a metric $h_{\mu\nu}$ such that
 \be
h_{\mu\nu}=f'({\cal R})g_{\mu\nu},
\ee
 which is covariantly conserved by the connections
$\Gamma_{\phantom{a}\mu\nu}^\lambda$ [see section \ref{manfield} and discussion after eq.~(\ref{hgconf})]. This metric is, of course,
symmetric since it is conformal to $g_{\mu\nu}$, and so the
connections should be symmetric as well. In other words, it has been
shown that $\Delta_{\lambda}^{\phantom{a}\mu\nu}=0$ leads to a
symmetric connection, which means that there is no torsion when the
matter action does not depend on the connection. This is an important
aspect of this class of metric-affine theories of gravity. It shows
that {\it metric-affine $f(R)$ gravity allows the presence of torsion
but does not force it}. {\it Torsion is merely introduced by specific
forms of matter}, those for which the matter action has a dependence
on the connections. Therefore, as ``matter tells spacetime how to
curve'', matter will also tell spacetime how to twirl. Notice also
that the non-metricity does not vanish. This is because, as we also
saw previously, part of the non-metricity is introduced by the form of
the Lagrangian, {\em i.e.}~$f({\cal R})$ actions lead generically to
theories with intrinsic non-metricity.

It is interesting to note the special nature of the particular case in
which the $f({\cal R})$ Lagrangian is actually linear in ${\cal R}$,
{\em i.e.}
 \be
f({\cal R})={\cal R}-2\Lambda.
\ee
Then eq.~(\ref{d01}) gives
\be
{\cal R}_{(\mu\nu)}-\frac{1}{2}{\cal R} g_{\mu\nu}+\Lambda g_{\mu\nu}=\kappa T_{\mu\nu},
\ee
and eq.~(\ref{d02}) gives
\be
\Gamma_{\phantom{a}\mu\nu}^\lambda=\{_{\phantom{a}\mu\nu}^\lambda\},
\ee
 {\em i.e.}~the $\Gamma$s turn out to be the Levi--Civita connections
of the metric and so the theory actually reduces to standard General
Relativity which, from this point of view, can now be considered as a
sub-case of a metric-affine theory. 

\subsubsection{Vacuum}

Having explored the case where
$\Delta_{\lambda}^{\phantom{a}\mu\nu}=0$, it is easy to consider the
vacuum case, where also $T_{\mu\nu}=0$. The field equations in this
case take the form
 \bea
\label{d01v}
& &f'({\cal R}) {\cal R}_{(\mu\nu)}-\frac{1}{2}f({\cal R})g_{\mu\nu}=0,\\
\label{d02v}
& &\gn_\lambda\left(\sqrt{-g}f'({\cal R})g^{\mu\nu}\right)=0.
\eea
 We do not need to make any manipulations to investigate the nature of
these equations. They coincide with the equations of Palatini $f(R)$
gravity in vacuum and therefore we can just follow the step of section
\ref{manfield} setting $T_{\mu\nu}$ to zero in order to realize that
the theory reduces to General Relativity with a cosmological constant. 

However, for the sake of clarity, let us repeat some of the steps.
Contracting eq.~(\ref{d01v}) we get
 \be
\label{scalar}
f'({\cal R}) {\cal R}-2 f({\cal R})=0.
\ee
 This is an algebraic equation for ${\cal R}$ once $f({\cal R})$ has
been specified. In general, we expect this equation to have a number
of solutions,
 \be
\label{sol}
R=c_i,\quad i=1,2,\dots
\ee
 where the $c_i$ are constants. As already mentioned in section
\ref{manfield}, there is also a possibility that eq.~(\ref{scalar})
(eq.~(\ref{paltrace}) in section \ref{manfield}) will have no real
solutions or will be satisfied for any ${\cal R}$ (which happens for
$f({\cal R})=a {\cal R}^2$, where $a$ is an arbitrary constant) but
since such cases mainly seem uninteresting or are burdened with
serious difficulties when matter is also considered, we shall not
study them here (see section \ref{manfield} and \cite{ferr}).

Let us, therefore, return to the case where eq.~(\ref{scalar}) has the
solutions given in eq.~(\ref{sol}). In this case, since ${\cal R}$ is
a constant, $f'({\cal R})$ is also a constant and eq.~(\ref{d02v}) 
becomes
 \be
\label{e3}
\gn_\lambda\left(\sqrt{-g}g^{\mu\nu}\right)=0.
\ee
 This is the metricity condition for the affine connections,
$\Gamma^\lambda_{\phantom{a}\mu\nu}$. Therefore, the affine
connections now become the Levi--Civita connections of the metric,
$g_{\mu\nu}$,
 \be
\label{e5}
\Gamma_{\phantom{a}\mu\nu}^\lambda=\{_{\phantom{a}\mu\nu}^\lambda\},
\ee
 and ${\cal R}_{\mu\nu}=R_{\mu\nu}$. Eq.~(\ref{d01v}) can be 
re-written in the form
\be
\label{e4}
R_{\mu\nu}-\frac{1}{4}c_i g_{\mu\nu}=0,
\ee
which is exactly the Einstein field equation with a cosmological constant.

Therefore, in the end we see that a general $f(R)$ theory of gravity
in vacuum, studied within the framework of metric-affine variation,
will lead to the Einstein equation with a cosmological constant.  This
is not the case if one uses the metric variational principle as, in
this case, one ends up with fourth order field equations, {\em
i.e.}~with a significant departure from the standard Einstein
equations (see for example section \ref{metricfieldeq} or
\cite{buchdahl}). Another important feature that deserves to be
commented upon is the following: Contrary to the spirit of General
Relativity where the cosmological constant has a unique value, here
the cosmological constant is also allowed to have different values,
$c_i$, corresponding to different solutions of eq.~(\ref{scalar}). 
So, in vacuum, the action (\ref{maaction}) is in a sense equivalent to
a whole set of Einstein--Hilbert actions \cite{mang} (or, more
precisely, actions of the form (\ref{actionehm}) plus a cosmological
constant).

\subsubsection{Matter action dependent on the connection}
\label{yesD}

We now focus on the more general case in which
$\Delta_{\lambda}^{\phantom{a}\mu\nu}\neq 0$ and therefore the matter
action includes matter fields coupled to the connection. We can find
two interesting sub-cases here. These are when
$\Delta_{\lambda}^{\phantom{a}\mu\nu}$ is either fully symmetric or
fully antisymmetric in the indices $\mu$ and $\nu$. As before, the
equation under investigation will be eq.~(\ref{field2t}). We shall
split it here into its symmetric and antisymmetric parts in the
indices $\mu$ and $\nu$:
 \bea
\label{2sym}
& &\frac{1}{\sqrt{-g}}\bigg[-\gn_\lambda\left(\sqrt{-g}f'({\cal R})g^{\mu\nu}\right)+\gn_\sigma\left(\sqrt{-g}f'({\cal R})g^{\sigma(\mu}\right){\delta^{\nu)}}_\lambda\bigg]+{}\nn\\ & &\qquad\qquad+2f'({\cal R})g^{\sigma(\mu}{S_{\sigma\lambda}}^{\nu)}=\kappa\Delta_{\lambda}^{\phantom{a}(\mu\nu)},\\
\label{2anti}
& &\frac{1}{\sqrt{-g}}\gn_\sigma\left(\sqrt{-g}f'({\cal R})g^{\sigma[\mu}\right){\delta^{\nu]}}_\lambda+2f'({\cal R})g^{\sigma[\mu}{S_{\sigma\lambda}}^{\nu]}=\nn\\
& &\qquad\qquad=\kappa(\Delta_{\lambda}^{\phantom{a}[\mu\nu]}-\frac{2}{3}\Delta_{\sigma}^{\phantom{a}\sigma[\nu}{\delta^{\mu]}}_{\lambda}).
\eea
Let us assume now that
\be
\label{dsym}
\Delta_{\lambda}^{\phantom{a}[\mu\nu]}=0,
\ee
and take the trace of either of the above equations. This leads to
\be
3\gn_\sigma\left(\sqrt{-g}f'({\cal R})g^{\sigma\mu}\right)=2\sqrt{-g}\kappa\Delta_{\sigma}^{\phantom{a}\sigma\mu}.
\ee
Using this and eq.~(\ref{dsym}), eq.~(\ref{2anti}) takes the form
\be
g^{\sigma[\mu}S^{\phantom{ab}\nu]}_{\sigma\lambda}=0,
\ee
which is the same as eq.~(\ref{notorsion}) which we have shown leads to
\be
S^{\phantom{ab}\nu}_{\sigma\lambda}=0.
\ee
 Then, once again, the torsion tensor vanishes and we drop to the 
system of equations
\be
f'({\cal R}) {\cal R}_{(\mu\nu)}-\frac{1}{2}f({\cal R})g_{\mu\nu}=\kappa T_{\mu\nu},\ee\be
\frac{1}{\sqrt{-g}}\bigg[-\gn_\lambda\left(\sqrt{-g}f'({\cal R})g^{\mu\nu}\right)+\gn_\sigma\left(\sqrt{-g}f'({\cal R})g^{\sigma(\mu}\right){\delta^{\nu)}}_\lambda\bigg]=\kappa\Delta_{\lambda}^{\phantom{a}(\mu\nu)}.
\ee
 which are the same as eqs.~(\ref{field1sym}) and (\ref{field2sym})
{\em i.e.}~the equations for the torsionless version of the theory.
This indicates that any torsion is actually introduced by the
antisymmetric part of $\Delta_{\lambda}^{\phantom{a}\mu\nu}$.

 We can now examine the opposite case where it is the symmetric part 
of $\Delta_{\lambda}^{\phantom{a}\mu\nu}$ that vanishes. Then
\be
\label{danti}
\Delta_{\lambda}^{\phantom{a}(\mu\nu)}=0,
\ee
 and taking the trace of either eq.~(\ref{2sym}) or eq.~(\ref{2anti}) 
straightforwardly gives
\be
\gn_\sigma\left(\sqrt{-g}f'({\cal R})g^{\sigma\mu}\right)=0.
\ee
Therefore, eqs.~(\ref{2sym}) and (\ref{2anti}) take the form
\bea
\label{2symanti}
& &-\frac{1}{\sqrt{-g}}\gn_\lambda\left(\sqrt{-g}f'({\cal R})g^{\mu\nu}\right)+2f'({\cal R})g^{\sigma(\mu}{S_{\sigma\lambda}}^{\nu)}=0,\\
\label{2antianti}
& &2f'({\cal R})g^{\sigma[\mu}{S_{\sigma\lambda}}^{\nu]}=\kappa(\Delta_{\lambda}^{\phantom{a}[\mu\nu]}-\frac{2}{3}\Delta_{\sigma}^{\phantom{a}\sigma[\nu}{\delta^{\mu]}}_{\lambda}).
\eea
 Taking into account the general expression for the covariant
derivative of a tensor density
 \be
\gn_\lambda(\sqrt{-g} J^{\alpha\dots}_{\:\:\beta\dots})=\sqrt{-g}\gn_\lambda( J^{\alpha\dots}_{\:\:\beta\dots})-\sqrt{-g}\Gamma^{\sigma}_{\phantom{a}\sigma\lambda}J^{\alpha\dots}_{\:\:\beta\dots},
\ee
 and the fact that
$\Gamma^{\sigma}_{\phantom{a}\sigma\lambda}=\Gamma^{\sigma}_{\phantom{a}\lambda\sigma}$
by eq.~(\ref{field3t}), one can easily show that eq.~(\ref{2symanti})
can be written as
 \be
\hat{\nabla}_\lambda\left(\sqrt{-g}f'({\cal R})g^{\mu\nu}\right)=0,
\ee
 where $\hat{\nabla}_\lambda$ denotes the covariant derivative defined
with the symmetric part of the connection. This equation tells us
that, as before, we can define a symmetric metric
 \be
h_{\mu\nu}=f'({\cal R})g_{\mu\nu},
\ee
 which is now covariantly conserved by the symmetric part of
connections, $\Gamma_{\phantom{a}(\mu\nu)}^\lambda$. If $f({\cal R})$
is linear in ${\cal R}$, $h_{\mu\nu}$ and $g_{\mu\nu}$ coincide, of
course. Additionally, eq.~(\ref{2antianti}) shows that the torsion is
fully introduced by the matter fields. Therefore we can conclude that
when $\Delta_{\lambda}^{\phantom{a}\mu\nu}$ is fully antisymmetric,
there is torsion, but the only non-metricity present is that
introduced by the form of the gravitational Lagrangian, {\em
i.e.}~matter introduces no extra non-metricity.

 We can then conclude that, in the metric-affine framework discussed
here, matter can induce both non-metricity and torsion: the symmetric
part of $\Delta_{\lambda}^{\phantom{a}\mu\nu}$ introduces
non-metricity, the antisymmetric part is instead responsible for
introducing torsion. While some non-metricity is generically induced
also by the $f({\cal R})$ Lagrangian (with the relevant exception of
the linear case), torsion is only a product of the presence of matter.

\subsubsection{Specific matter fields}
\label{spmfields}

 Having studied the implications of a vanishing or non vanishing
$\Delta_{\lambda}^{\phantom{a}\mu\nu}$, we now want to discuss these
properties in terms of specific fields. Since
$\Delta_{\lambda}^{\phantom{a}\mu\nu}$ is the result of the variation
of the matter action with respect to the connection, we will need the
matter actions of the fields in curved spacetime for this purpose. In
purely metric theories one knows that any covariant equation, and
hence also the action, can be written in a local inertial frame by
assuming that the metric is flat and the connections vanish, turning
the covariant derivatives into partial ones. Therefore, one can expect
that the inverse procedure, which is called the minimal coupling
principle, should hold as well and can be used to provide us with the
matter action in curved spacetime starting from its expression in a
local inertial frame. This expectation is based on the following
conjecture: {\it The components of the gravitational field should be
used in the matter action on an ``only as  necessary'' basis}. The root of this
conjecture can be traced to requiring minimal coupling between the
gravitational field and the matter fields (hence the name ``minimal
coupling principle''). In General Relativity this conjecture can be
stated for practical purposes in the following form: {\it the metric
should be used in the matter action only for contracting indices and
constructing the terms that need to be added in order to write a
viable covariant matter action}. This implies that the connections
should appear in this action only inside covariant derivatives and
never alone which is, of course, perfectly reasonable since, first of
all, they are not independent fields and, secondly, they are not
tensors themselves and so they have no place in a covariant
expression. At the same time, other terms that would vanish in flat
spacetime like, for example, contractions of the curvature tensor
with the fields or their derivatives, should be avoided. 

The previous statements are not applicable in metric affine gravity
for several reasons: the connections now are independent fields and,
what is more, if they are not symmetric, there is a tensor that one
can construct via their linear combination: the Cartan torsion tensor. 
Additionally, going to some local inertial frame in metric-affine
gravity is a two-step procedure in which one has to separately impose
that the metric is flat and that the connections vanish. However, the
critical point is that when inverting this procedure one should keep
in mind that there might be dependences on the connections in the
equations other than those in the covariant derivatives. {\it The
standard minimal coupling principle will therefore not, in general,
give the correct answer in metric-affine gravity theories.}

The above discussion can be well understood through a simple example,
using the electromagnetic field.  In order to compute the
hypermomentum $\Delta_{\lambda}^{\phantom{a}\mu\nu}$ of the
electromagnetic field, we need to start from the action
 \be
\label{elaction}
S_{EM}=-\frac{1}{4}\int d^4 x \sqrt{-g} F^{\mu\nu}F_{\mu\nu},
\ee
 where $F^{\mu\nu}$ is the electromagnetic field tensor. As we know, 
in the absence of gravity this tensor is defined as
\be
\label{elf}
F_{\mu\nu}\equiv \partial_\mu A_{\nu}-\partial_\nu A_{\mu},
\ee
 where $A_{\mu}$ is the electromagnetic four-potential. If we naively
followed the minimal coupling principle and simply replaced the
partial derivatives with covariant ones, the definition of the 
electromagnetic field tensor would take the form:
 \be
\label{wrong}
F_{\mu\nu}\equiv \gn_\mu A_{\nu}-\gn_\nu A_{\mu}=\partial_\mu A_{\nu}-\partial_\nu A_{\mu}-2\,\Gamma^\sigma_{\phantom{a}[\mu\nu]}A_\sigma,
\ee
 and one can easily verify that it would then no longer be gauge
invariant, {\em i.e.}~invariant under redefinition of the four
potential of the form $A_{\mu}\rightarrow A_{\mu}+\partial_\mu \phi$,
where $\phi$ is a scalar quantity. Gauge invariance, however, is a
critical aspect of the electromagnetic field since it is related to
the conservation of charge and the fact that the electric and magnetic
fields are actually measurable quantities. Therefore breaking gauge
invariance cannot lead to a viable theory.  One could assume that the
problem lies in the fact that the connection is not symmetric, {\em
i.e.}~torsion is allowed, since it is the antisymmetric part of the
connection that prevents gauge invariance of eq.~(\ref{wrong}), and
hence it might seem that standard electromagnetism is incompatible with
torsion. This explanation was given for example in~\cite{maj} (see
also references therein for other discussions following the same
line). We do not agree with either this approach or its conclusion: As
we said, the problem is actually much simpler but also more
fundamental and lies in the assumption that the minimal coupling
principle still holds in metric-affine gravity.

In order to demonstrate this point, let us turn our attention to the
definition of the electromagnetic field tensor in the language of
differential forms. This is
 \be
\label{defdf}
{\bf F}\equiv{\bf d}{\bf A},
\ee
 where ${\bf d}$ is the standard exterior derivative
\cite{gravitation}. Remember that the exterior derivative is related
to Gauss's theorem which allows us to go from an integral over the
volume to an integral over the boundary surface of this volume. Now
notice that the volume element has no dependence on the connection and
is the same as that of General Relativity, $\sqrt{-g}\, d^4 x$. This
implies that the definition of the exterior derivative should remain
unchanged when expressed in terms of partial derivatives. Partial
derivatives on the other hand are defined in the same way in this
theory as in General Relativity. Therefore, from the definition
(\ref{defdf}) we understand that $F_{\mu\nu}$ should be given in terms
of the partial derivatives by the following equation
 \be
\label{elf2}
F_{\mu\nu}\equiv [{\rm d}A]_{\mu\nu}=\partial_\mu A_{\nu}-\partial_\nu A_{\mu},
\ee
 which is the same as eq.~(\ref{elf}) and respects gauge invariance. 
The expression in terms of the partial derivatives may not look
covariant but can easily be written in a manifestly covariant form:
 \bea
\label{right}
F_{\mu\nu}\equiv \partial_\mu A_{\nu}-\partial_\nu A_{\mu}&=&\gn_\mu A_{\nu}-\gn_\nu A_{\mu}+2\,\Gamma^\sigma_{\phantom{a}[\mu\nu]}A_\sigma\nn\\&=&\gn_\mu A_{\nu}-\gn_\nu A_{\mu}+2\,S^{\phantom{ab}\sigma}_{\mu\nu}A_\sigma.
\eea
 Besides, the expression for ${\bf F}$ in terms of the exterior
derivative is covariant anyway.

It is now obvious that the minimal coupling principle was leading us
to the wrong expression, causing a series of misconceptions. However,
we are still in need of a prescription that will allow us to derive
the matter actions in curved spacetime. Notice that if we require
gravity and matter to be minimally coupled, then the physical basis of
the conjecture that {\it the components of the gravitational field
should be used in the matter action on an ``only as necessary'' basis} still holds,
since its validity is not related to any of the assumptions of General
Relativity. Thus, we can use it to express a metric-affine minimal
coupling principle: {\it The metric should be used in the matter
action only for contracting indices and the connection should be used
only in order to construct the extra terms that must be added in order
to write a viable covariant matter action}. The analogy with the
statement used in General Relativity is obvious, and differences lie
in the different character of the connections in the two theories. One
can easily verify that the matter action of the electromagnetic field
which we derived earlier can be straightforwardly constructed using
this metric--affine minimal coupling principle. 

We would like to stress once more that both the metric--affine minimal
coupling principle presented above and the standard one, are
based on the requirement that the gravitational field should be
minimally coupled to the matter. One could, of course, choose to
construct a theory without such a requirement and allow non-minimal
coupling\footnote{Note that if one considers the possible actions
for classical gravity as effective ones --- obtained as the low energy
limit of some more fundamental high energy theory --- then it is
natural to imagine that the form of the coupling (minimal or some
specific type of non-minimal) might cease to be a free choice (see
{\em e.g.}~Chapter 7 of \cite{valeriobook} for an enlightening
discussion). However, one could still expect that non-minimal coupling
terms will be suppressed at low energies by appropriate powers of the
scale associated with the fundamental theory (Planck scale, string
scale, {\em etc}.) and in this sense the use of a minimal coupling principle
at low energies could be justified.}. 
 This can be done both in metric-affine gravity and in General
Relativity. Clearly, in metric-affine gravity one has more options
when it comes to non-minimal coupling, since besides curvature terms,
also terms containing the Cartan torsion tensor can be used. However,
it is easy to see that the number of viable coupling terms is strongly
reduced by the symmetry of the metric (which also implies symmetry of
the stress-energy tensor) and by the constraints of the theory, {\em
e.g.}~the vanishing of the trace of $S^{\phantom{ab}\sigma}_{\mu\nu}$
when considering $f({\cal R})$ actions.

Allowing non-minimal coupling between gravity and matter in a
gravitation theory drastically changes the corresponding
phenomenology and there might be interesting prospects for such
attempts in metric-affine gravity. For the rest of this thesis,
however, we will continue to assume minimal coupling between gravity
and matter, since this is the most conventional option.

Let us now return to the electromagnetic field. Now that we have a
suitable expression for the electromagnetic field tensor, we can
proceed to derive the field equations for electrovacuum. Notice that
$F_{\mu\nu}$ has no real dependence on the connections and so we can
straightforwardly write
 \be
\label{Del}
\Delta_{\lambda}^{\phantom{a}\mu\nu}=0.
\ee
 The stress-energy tensor $T_{\mu\nu}$ can be evaluated using 
eq.~(\ref{set}) and has the standard form
\be
T_{\mu\nu}=F_\mu^{\phantom{a}\sigma}F_{\sigma\nu}-\frac{1}{4}g_{\mu\nu}F^{\alpha\beta}F_{\alpha\beta}.
\ee
 With the use of eqs.~(\ref{d01}) and(\ref{d02}), we can write the 
field equations as
\bea
\label{field1el}
& &f'({\cal R}) {\cal R}_{(\mu\nu)}-\frac{1}{2}f({\cal R})g_{\mu\nu}=\kappa F_\mu^{\phantom{a}\sigma}F_{\sigma\nu}-\frac{\kappa}{4}g_{\mu\nu}F^{\alpha\beta}F_{\alpha\beta},\\
\label{field2el}
& &\gn_\lambda\left(\sqrt{-g}f'({\cal R})g^{\mu\nu}\right)=0.
\eea
 We can use, however, the fact that the stress-energy tensor of the
electromagnetic field is traceless. If we take the trace of
eq.~(\ref{field1el}) we get
 \be
f'({\cal R}){\cal R}-2 f({\cal R})=0,
\ee
 which, as we discussed previously for the vacuum case, is an
algebraic equation in ${\cal R}$ once $f({\cal R})$ has been
specified. Solving it will give a number of roots (see also the
discussion after eq.~(\ref{sol}))
 \be
{\cal R}=c_i,\quad i=1,2,\dots
\ee
 and $f(c_i)$ and $f'(c_i)$ will be constants. Therefore
eq.~(\ref{field2el}) implies that the metric is covariantly conserved
by the covariant derivative defined using the connection and so
 \be
\Gamma_{\phantom{a}\mu\nu}^\lambda=\{_{\phantom{a}\mu\nu}^\lambda\},
\ee
and we are left with the following field equation:
\be
\label{field1eleh}
{\cal R}_{\mu\nu}-\frac{1}{4}c_i g_{\mu\nu}=\kappa' F_\mu^{\phantom{a}\sigma}F_{\sigma\nu}-\frac{\kappa'}{4}g_{\mu\nu}F^{\alpha\beta}F_{\alpha\beta},
\ee
 which is the Einstein equation for electrovacuum with a cosmological
constant and a modified ``coupling constant''
$\kappa'=\kappa/f'(c_i)$. The rescaling of $\kappa$ should not mislead
us into thinking that either the gravitational constant, $G$, or the
fine structure constant, $\alpha$, change in any way. It just affects
the strength of the ``coupling'' between gravity and the
electromagnetic field, {\em i.e.}~how much curvature is induced per
unit energy of the electromagnetic field. The values of the
cosmological constant and $\kappa'$ depend on the functional form of
$f({\cal R})$ and therefore they are fixed once one selects an action.
For example, $f({\cal R})={\cal R}$ or $f({\cal R})=a {\cal R}^2+{\cal
R}$ both lead to $c_i=0$ and $\kappa'=\kappa$ and the resulting theory
will be indistinguishable from General Relativity. For more general
forms of $f({\cal R})$, the theory is still formally equivalent to
General Relativity but note that the modification of $\kappa$ should,
at least theoretically, be subject to experiment. If such an
experiment is technically possible it might help us place bounds on
the form of the action. 

As already mentioned, a vanishing
$\Delta_{\lambda}^{\phantom{a}\mu\nu}$ implies that there is no
dependence of the matter action on the connections, or equivalently on
the covariant derivative. As we just saw, the electromagnetic field,
and consequently any other gauge field, has this attribute. The same
is true for a scalar field, as the covariant derivatives of a scalar
are reduced to partial derivatives. Therefore, neither of these fields
will introduce torsion or extra non-metricity. For the electromagnetic
field specifically, the fact that the trace of its stress energy
tensor is zero leads to the Einstein field equations, since the
non-metricity introduced by the form of the Lagrangian has to vanish
as well. For the scalar field, whose stress energy tensor does not
have a vanishing trace, this will not happen. The field equations can
be derived straightforwardly by replacing the usual stress energy
tensor of a scalar field in eqs.~(\ref{d01}) and (\ref{d02}).

Let us now turn to matter fields for which
$\Delta_{\lambda}^{\phantom{a}\mu\nu}$ does not vanish. In principle,
any a tensor field should have an action with an
explicit dependence on the connection, leading to a non vanishing
$\Delta_{\lambda}^{\phantom{a}\mu\nu}$. For a massive vector field, which is in general described by the Proca lagrangian, one can choose to define the field strength in terms of the exterior derivative, as it was done before for the electromagnetic field (massless vector field). In this case, there will be only partial derivatives in the matter action and torsion will not couple to this field. This way, massive vector fields will not produce or feel torsion. This generalization of the special relativistic Proca lagrangian to a covariant one, suitable for curved space time, seems to be in accordance to the metric-affine minimal coupling principle we expressed earlier, as it does not needlessly promote the partial derivatives to covariant ones. However, since in the case of a massive vector field there is no gauge invariance, one could indeed choose to just promote the partial derivatives in the Proca lagrangian to covariant ones (see for instance \cite{hehlrev}). If the latter prescription is followed then torsion couples to massive vector fields as well. 

Another typical example of a field with a non-vanishing $\Delta_{\lambda}^{\phantom{a}\mu\nu}$  would be the
Dirac field. The Dirac Lagrangian has an explicit dependence on the
covariant derivative, and therefore an explicit dependence on the
connections. Additionally, there are no viability criteria, unlike in
the case of the electromagnetic field, that will force us to include
extra terms proportional to the Cartan torsion tensor which will
cancel out the presence of the antisymmetric part of the connection.
Therefore, the procedure for deriving the matter action is
straightforward (see \cite{hehlrev} for the full form of the
action\footnote{Note that the result of \cite{hehlrev} is for a theory
that has, by definition, vanishing non-metricity ($U_4$ theory).
However, the form of the matter action is the same once the proper
covariant derivative is used. For discussions about the matter actions
in theories with torsion see also \cite{benn,carfie}. Note that, even
though the results obtained here are in complete agreement with the
ones presented in those works, in many cases the reasoning differs
since there is no attempt there to formulate a metric-affine minimal
coupling principle. The standard minimal coupling principle is used
there in cases where it provides the correct results, while it is
noted that it does not apply to specific cases, such as the
electromagnetic field. For each of these cases, individual arguments
are used in order to derive the matter action in curved spacetime. The
underlying physics in the two approaches is the same, but we believe
that the idea of a metric-affine minimal coupling principle is an
essential concept since, besides its elegance and the analogy with the
standard minimal coupling principle, it leaves no room for
exceptions.}). We can infer from the above that a Dirac field will
potentially introduce both torsion and non-metricity. Note that the
fields which cannot introduce torsion will also not ``feel'' it, since
they are not coupled to the Cartan tensor, and so photons or scalar
particles will not be affected by torsion even if other matter fields
produce it.

It is also interesting to study matter configurations in which matter
is treated macroscopically, the most common being that of a perfect
fluid with no vorticity. Let us here consider separately the cases where torsion is
allowed in the theory and where it is not included. In the latter
case, the consideration of a perfect fluid with no vorticity is identical to standard
General Relativity. Without going into a long detour regarding the correct Lagrangian  
formulation of general relativistic fluid mechanics, we will quote  
some standard results: Since the matter action can be described by three
scalars, the energy density, the pressure and the velocity potential
(see for example \cite{scha,stone}), the action has no dependence on
the covariant derivative and so $\Delta_{\lambda}^{\phantom{a}\mu\nu}$
will vanish. When torsion is allowed, there are two distinct cases
depending on the microscopic properties of the fluid. If a perfect
fluid is used to effectively describe particles whose corresponding
field description does not introduce torsion, then no difference from
the previous case arises. If, on the other hand, the fluid is composed
of particles whose field description {\em can} introduce torsion, then
their spin has to be taken into account (see \cite{hehlrev} and
references therein). There will however be an averaging over volume of
the quantities describing the matter, and if one assumes that the spin
is randomly oriented and not polarized, then it should average to
zero. This description can be applied in physical situations such as
gravitational collapse or Cosmology. The fact that the expectation value
of the spin will be zero will lead to a vanishing expectation value
for the torsion tensor. However, fluctuations around the expectation
value will affect the geometry leading to corrections to the field
equations which will depend on the energy density of the specific
species of particle. Since the torsion tensor is coupled to the
hypermomentum through the gravitational constant
(eq.~(\ref{field2t})), the effect of these fluctuations will be
suppressed by a Planck mass squared. Therefore we can conclude that
for Cosmology, and especially for late times where the energy density
is small, the standard perfect fluid description might serve as an
adequate approximation. 

It is remarkable that the two matter descriptions most commonly used
in Cosmology, the perfect fluid with no vorticity and the scalar field, lead to a
vanishing $\Delta_{\lambda}^{\phantom{a}\mu\nu}$ for a symmetric
connection. It is also noticeable that in our framework, even if
torsion is allowed, the results remain unchanged for the perfect fluid
case, apart from small corrections which should be negligible. It
would be interesting to consider also the case of a an imperfect fluid
({\it i.e.}~to allow also viscosity, heat flow, {\em etc}.), which is
certainly relevant for some observationally interesting systems in
relativistic Astrophysics.  As in the case of a perfect fluid, if we
consider particles with a spin and allow torsion, the standard
imperfect fluid description will not be exact. Note however, that even
in the simpler case of {\em a priori} symmetric connections, we do not
expect the matter action to be independent of such connections (in
contrast with the perfect fluid case). This could lead to a
non-vanishing $\Delta_\lambda^{\phantom{a}\mu\nu}$ and consequently to
some non-metricity, which might lead to interesting deviations away
from General Relativity results.

\subsubsection{Discussion}

It has been shown that when the variation of the matter action leads
to a tensor symmetric in its last two indices, then torsion vanishes.
When the same tensor is antisymmetric, matter introduces only torsion
and not non-metricity. Matter fields whose matter action is
independent of the connection cannot introduce either torsion or
non-metricity. As already mentioned, since torsion is absent in vacuum
and in some specific matter configurations but present in all other
cases, we can infer that it is actually introduced by matter. By
considering for which kind of fields torsion vanishes and for which it
does not, we can arrive at a very interesting conclusion. Torsion is
zero in vacuum and in the presence of a scalar field or an
electromagnetic field. It does not necessarily vanish, however, in the
presence of a Dirac field or other vector and tensor fields. This
shows a correspondence between torsion and the presence of fields that
describe particles with spin. We are, therefore, led to the idea that
particles with spin seem to be the sources of torsion. Of course a
photon, the particle associated with the electromagnetic field, is a
spin one particle. However, in Quantum Field theory a photon and in general massless particles with spin are not characterized their spin but actually by their helicity. Such particles are described by gauge fields which, just as the electromagnetic field, will not introduce any torsion. It is remarkable that this exceptional nature of the photon and other massless fields seems to be present also here. 

The study of the electromagnetic field turned out to be very helpful,
since it demonstrated that the usual minimal coupling principle does
not hold in metric-affine gravity. However, as we showed, one can
still express a metric-affine minimal coupling principle based on the
spirit of minimal coupling between gravity and matter. 

We have also discussed the case where matter is treated
macroscopically. As already mentioned, a perfect fluid with no vorticity cannot
introduce any extra non-metricity for a symmetric connection. When
torsion is allowed, the concept of a perfect fluid has to be
generalized if one wants to include particles with spin, but also in
this case only small contributions to torsion will be introduced
which will be negligible in most cases. On the other hand, for both symmetric and general connections, we 
suspect that there might be larger deviations from General 
Relativity when a seriously imperfect fluid is considered. However, 
for many applications in Cosmology and Astrophysics, a perfect fluid 
description is taken as being a good approximation.
 Moreover, many of the experimental tests passed by General Relativity
are related to either vacuum or to environments where matter can be
more or less accurately described as a perfect fluid. This means that
a metric-affine theory could be in total accordance with these tests
when the Einstein--Hilbert action and possibly many of its extensions
are used. 

However, the possible relevance of imperfect fluid matter in some 
yet to be accurately observed astrophysical systems (such as 
accretion flows or compact objects \cite{john}) leaves open the 
possibility for future discrimination between the class of theories 
discussed here and standard General Relativity. In physical systems 
where matter cannot necessarily be described accurately enough by a 
perfect fluid, one might hope to see deviations from the standard 
behaviour predicted by General Relativity. Even starting with the 
standard Einstein--Hilbert action, torsion and non-metricity should 
affect the dynamics and might make them deviate noticeably from the 
standard ones. This deviation could persist even in a 
nearly-Newtonian regime. It could be interesting to study this in 
the context of galactic dynamics since in this case the effects may 
be important and may even make some contribution in relation to the 
unexpected behaviour of galactic rotation curves. Of course, until a 
thorough and quantitative study is performed, all of the above 
remains at the level of speculations, even though they seem 
qualitatively interesting.

It is important to note that our attempt to include torsion showed
that this cannot be done in the context of $f(R)$ gravity unless one
fixes some degrees of freedom of the connection as mentioned earlier.
The other possibility that was discussed here was to modify the action
by adding some higher order curvature invariant. As we said, it is
very difficult to find a prescription for an action of this form that
will lead to a physically meaningful theory of gravitation with
torsion since the simple case will have unwanted attributes. This is
the reason why we did not pursue this here. Note however, that we
already know that rotating test particles do not follow geodesics.
Therefore, it would be reasonable to assume that, since macroscopic
angular momentum interacts with the geometry, intrinsic angular
momentum (spin) should interact with the geometry as well. This
property should become more important at small length-scales or high
energies. Therefore, it seems remarkable that an attempt to include
torsion and at the same time avoid placing {\it a priori} constraints
on the connection, leads to the conclusion that the action should be
supplemented with higher order curvature invariants, which is in total
agreement with the predictions coming from quantum corrections, String
Theory and M-theory. 

To conclude this section, we would like to stress once more that
metric-affine $f(R)$ gravity reduces to General Relativity, or a
theory very close to it, in most of the cases relevant to known
experimental tests (vacuum, electrovacuum, {\em etc}.) and yet is
phenomenologically much richer. This may help to address some of the
puzzles of physics related to gravity. 

 \section{Gauss--Bonnet gravity}
\label{gbgrav}
\subsection{The action}

In the course of this chapter we have studied $f(R)$ theories of
gravity extensively and we have briefly considered scalar-tensor
theory and theories whose action includes higher-order curvature
invariants, such as $R_{\mu\nu}R^{\mu\nu}$. Since an action may
include such invariants, one is tempted to consider the option that a
scalar field might not only be coupled to the Ricci scalar, as in
scalar-tensor theory, but also to higher order terms. A theory with a
scalar field and more general couplings would be quite complicated and
difficult to handle though. Therefore, besides the general motivation
for pursuing alternative theories of gravity coming from puzzles
related to Cosmology and Quantum Gravity, one would like to have some
motivation for specific couplings in order to go further.

Indeed there are motivations from String Theory to believe that scalar
fields might be coupled to the Gauss--Bonnet invariant ${\cal G}$, as
defined in eq.~(\ref{GB}). To be more precise, one expects to find two
types of scalar field in the low energy effective action of gravity
coming from heterotic String Theory: moduli, $\phi$, which are related
to the size and shape of the internal compactification manifold, and
the dilaton, $\sigma$, which plays the role of the string loop
expansion parameter. There are reasons to believe that moduli
generally couple to curvature squared terms \cite{strings1, strings12}
but that moduli-dependent higher loop contributions, such as terms
cubic or higher order in the Riemann tensor, vanish leaving a coupling
with a Gauss--Bonnet term to be of specific interest \cite{strings1,
strings12, strings2}. On the other hand, the dilaton usually couples
to the the Ricci scalar, as in scalar-tensor theory\footnote{A
conformal transformation of the metric can be used in order to find a
representation of the theory in which the coupling with the Ricci
scalar is avoided and a coupling to matter is introduced (Jordan to
Einstein frame). We will discuss this extensively in the forthcoming
chapters.}. However, there are claims that the dilaton might evolve in
such a way so as to settle to a constant \cite{strings3, strings32}.
Under these assumptions, the effective low energy gravitational action
takes the form
 \be
\label{gbaction}
S_{GB}=\int d^4 x \sqrt{-g}\left[\frac{R}{16\pi\,G}-\frac{\lambda}{2}\partial_\mu \phi \partial^\mu \phi-V(\phi)+f(\phi) {\cal G}\right]+S_M(g^{\mu\nu},\psi),
\ee
 where $\lambda$ is $+1$ for a canonical scalar field and $-1$ for a 
phantom field. 

It is straightforward to generalize this action in order to include a
kinetic term and a coupling with ${\cal G}$ for the dilaton $\sigma$
\footnote{In scalar-tensor theory, the scalar is in many cases
considered to be the dilaton. Even though we denote the dilaton here
by $\sigma$ and the moduli by $\phi$, in section \ref{scalartensor} we
used $\phi$ for the scalar since this is standard notation for a
general scalar field.}. One can also allow a coupling for the dilaton
to $R$ and/or matter, if, of course, the dilaton is not assumed to settle to a
constant as claimed in \cite{strings3, strings32}.
However, considering these claims and the complications which such
couplings would introduce, we will concern ourselves here with the
action (\ref{gbaction}), which in any case can work as an excellent
starting point for studying couplings between a scalar and the
Gauss--Bonnet invariant. 

A theory described by the action (\ref{gbaction}) is usually called
Gauss--Bonnet gravity. Note, however, that the term Gauss--Bonnet
gravity is sometimes used to refer to other theories in 4 or more
dimensions including in some way the Gauss--Bonnet invariant in the
gravitational action and, therefore, care should be taken to avoid
confusion. We will be using this terminology here strictly referring to
the action (\ref{gbaction}). Before going further and deriving the
field equations, it is also worth mentioning that Gauss--Bonnet gravity
has been shown to have many appealing features when it comes to
singularities and cosmological applications ({\em
e.g.~}\cite{strings2, nojodsas}) and therefore part of the motivation
for its study comes from that. We will discuss such applications of
the theory in the next chapter.

\vfill

\subsection{The field equations}

We proceed here with the variation of the action. Variation with
respect to the metric $g_{\mu\nu}$ quite straightforwardly leads to the
equation
 \bea
\label{GB4}
&&\frac{1}{\kappa^2}G^{\mu\nu}
         -\frac{1}{2}g^{\mu\nu} f(\phi) {\cal G}  +2 f(\phi) R R^{\mu\nu} - 2 \nabla^\mu \nabla^\nu \left(f(\phi)R\right)+\nn\\
 &&     + 2 g^{\mu\nu}\nabla^2\left(f(\phi)R\right) - 8f(\phi)R^\mu_{\ \rho} R^{\nu\rho}   + 4 \nabla_\rho \nabla^\mu \left(f(\phi)R^{\nu\rho}\right)+\nn\\
&&      + 4 \nabla_\rho \nabla^\nu \left(f(\phi)R^{\mu\rho}\right) - 4 \nabla^2 \left( f(\phi) R^{\mu\nu}  \right)
- 4g^{\mu\nu} \nabla_{\rho} \nabla_\sigma \left(f(\phi) R^{\rho\sigma} \right)+\nn\\
& &  + 2 f(\phi) R^{\mu\rho\sigma\tau}R^\nu_{\ \rho\sigma\tau}
- 4 \nabla_\rho \nabla_\sigma \left(f(\phi) R^{\mu\rho\sigma\nu}\right)=T^{\mu\nu}+T^{\mu\nu}_\phi,
\eea
where we have defined
\be
\label{sephi}
T^{\mu\nu}_\phi= \lambda \left(\frac{1}{2}\partial^\mu \phi \partial^\nu \phi
      - \frac{1}{4}g^{\mu\nu} \partial_\rho \phi \partial^\rho \phi \right)
    -\frac{1}{2}g^{\mu\nu}V(\phi).
\ee
 Note that, as stressed in section \ref{homet}, ${\cal G}$ is a
topological invariant and the variation of the term $\sqrt{-g}{\cal
G}$ is a total divergence not contributing to the field equations
(formally a suitable surface term should be added in the action in
order to cancel the total divergence). However, the term
$\phi\sqrt{-g}{\cal G}$ will contribute in the field equations for the
metric since $\phi \delta \sqrt{-g}{\cal G}$ will no longer be a
surface term but can only be turned into one after an integration by
parts.

Following \cite{nojodsas}, we can use the following relations coming
from the Bianchi identities:
 \bea
\label{GB5}
\nabla^\rho R_{\rho\tau\mu\nu}&=& \nabla_\mu R_{\nu\tau} - \nabla_\nu
R_{\mu\tau}\ ,\\
\label{bianchi}
\nabla^\rho R_{\rho\mu} &=& \frac{1}{2} \nabla_\mu R\ , \\
\nabla_\rho \nabla_\sigma R^{\mu\rho\nu\sigma} &=&
\nabla^2 R^{\mu\nu} - \frac{1}{2}\nabla^\mu \nabla^\nu R
+ R^{\mu\rho\nu\sigma} R_{\rho\sigma}
      - R^\mu_{\ \rho} R^{\nu\rho} ,\\
\nabla_\rho \nabla^{(\mu} R^{\nu )\rho}
&=& \frac{1}{2} \nabla^{(\mu} \nabla^{\nu)} R
      - R^{\mu\rho\nu\sigma} R_{\rho\sigma}
+  R^\mu_{\ \rho} R^{\nu\rho} \ ,\\
\nabla_\rho \nabla_\sigma R^{\rho\sigma}&=& \frac{1}{2} \Box R \ ,
\eea
in order to obtain from eq.~(\ref{GB4}) the equation
\bea
\label{GB4b}
&&\frac{1}{\kappa^2}G^{\mu\nu}     
   -\frac{1}{2}g^{\mu\nu}f(\phi) {\cal G} + 2 f(\phi) R R^{\mu\nu} + 4f(\phi)R^\mu_{\ \rho} R^{\nu\rho} +\\&&
+2 f(\phi) R^{\mu\rho\sigma\tau}R^\nu_{\ \rho\sigma\tau}-4 f(\phi) R^{\mu\rho\sigma\nu}R_{\rho\sigma} =T^{\mu\nu}+T^{\mu\nu}_\phi+T^{\mu\nu}_f,\nn
\eea
where
\bea
\label{sef}
T^{\mu\nu}_f&=& 2 \left( \nabla^\mu \nabla^\nu f(\phi)\right)R
   - 2 g^{\mu\nu} \left( \nabla^2f(\phi)\right)R-\nn\\ &&
   - 4 \left( \nabla_\rho \nabla^\mu f(\phi)\right)R^{\nu\rho}
     - 4 \left( \nabla_\rho \nabla^\nu f(\phi)\right)R^{\mu\rho} +\nn\\
&& + 4 \left( \nabla^2 f(\phi) \right)R^{\mu\nu}
+ 4g^{\mu\nu} \left( \nabla_{\rho} \nabla_\sigma f(\phi) \right) R^{\rho\sigma}-\nn\\&&
- 4 \left(\nabla_\rho \nabla_\sigma f(\phi) \right) R^{\mu\rho\nu\sigma}
\eea
 We know that the Gauss--Bonnet term in the action is topologically
invariant and therefore for $f(\phi)=\textrm{constant}$ the field
equations should be unmodified with respect to General Relativity.
Thus, the terms proportional to $f(\phi)$ without derivatives should
cancel out leading to the identity
 \be
\label{gbident}
g^{\mu\nu} {\cal G}=4 R R^{\mu\nu} - 8 R^\mu_{\ \rho} R^{\nu\rho}
   +4 R^{\mu\rho\sigma\tau}R^\nu_{\ \rho\sigma\tau}-8 R^{\mu\rho\sigma\nu}R_{\rho\sigma}.
\ee

Now eq.~(\ref{GB4b}) can be simply written as
\be
\label{gbfield1}
G^{\mu\nu}=\kappa^2 \left[T^{\mu\nu}+T^{\mu\nu}_\phi+T^{\mu\nu}_f \right].
\ee
On the other hand, variation of the action with respect to $\phi$ 
gives
\be
\label{gbfield2}
\lambda \nabla^2 \phi - V'(\phi) + f'(\phi) {\cal G}=0,
\ee
 and eqs.~(\ref{gbfield1}) and (\ref{gbfield2}) constitute the field 
equations of the theory.

A comment is due at this point concerning the conservation of
energy-momentum. The matter action for Gauss--Bonnet gravity is built
out of a generally covariant scalar and the matter is minimally
coupled to the metric and not coupled to the scalar field $\phi$.
Therefore, Gauss--Bonnet gravity is a metric theory of gravity and
$T_{\mu\nu}$ is divergence free. Also, one can add that the action
(\ref{gbaction}) is manifestly diffeomorphism invariant, being
constructed with a generally covariant scalar. It is trivial to use
diffeomorphism invariance to derive that $\nabla_{\mu}T^{\mu\nu}=0$. 

However, in General Relativity the fact that $T^{\mu\nu}$ is
divergence free follows also as a consequence of the field equations
due to the Bianchi identity $\nabla_\mu G^{\mu\nu}=0$
(eq.~(\ref{bianchi})). Therefore, one expects that
$\nabla_{\mu}T^{\mu\nu}=0$ should be derivable also from a combination
of the field equations (\ref{gbfield1}) and (\ref{gbfield2}), the
Bianchi identity and probably some generalization of the Bianchi
identity. As an exercise, we will prove that this is indeed the case.

For $\nabla_{\mu} T^{\mu\nu}=0$ to hold, and given that $\nabla_{\mu} G^{\mu\nu}=0$ is a mathematical identity (eq.~(\ref{bianchi})), one needs
\be
\label{must}
\nabla_\mu T^{\mu\nu}_\phi+\nabla_\mu T^{\mu\nu}_f=0.
\ee
 Let us examine these terms separately. For the first one, a
straightforward calculation together with the use of the identity
$\nabla_\mu \nabla_\nu \psi=\nabla_\nu \nabla_\mu \psi$ for any scalar
$\psi$, gives
 \be
\label{phiid}
\nabla_\mu T^{\mu\nu}_\phi=\frac{1}{2}\left(\lambda \nabla^2 \phi - V'(\phi)\right)\nabla^\nu \phi.
\ee
 Calculating $\nabla_\mu T^{\mu\nu}_f$ is not, unfortunately, equally
straightforward but is rather a tedious calculation, so we will not
present it here in detail. We will, however, sketch the steps so that
the reader can easily reproduce it. The first step is to take into
account that for an arbitrary vector $V^\mu$
 \be
\label{com}
\nabla_\beta \nabla_\alpha V^\mu-\nabla_\alpha\nabla_\beta  V^\mu=R^\mu_{\phantom{a}\nu\beta\alpha}V^\nu,
\ee
and that for an arbitrary scalar $\psi$
\be
(\nabla^2\nabla_\nu-\nabla_\nu \nabla^2)\psi=R_{\mu\nu}\nabla^\mu \psi,
\ee
and one can then deduce, after some manipulations, that
\bea
\label{prel}
\nabla_\mu T^{\mu\nu}_f&=&\left(2 R R^{\mu\nu} - 4 R^\mu_{\ \rho} R^{\nu\rho}
   -4 R^{\mu\rho\sigma\nu}R_{\rho\sigma}\right)\nabla_\mu f(\phi)-\nn\\& &-4(\nabla_\mu \nabla_\rho\nabla_\sigma f) R^{\mu\rho\nu\sigma},
\eea
 where the identity (\ref{bianchi}) has also been used extensively to
replace $\nabla_{\mu}R^{\mu\nu}$ at all occurrences and the symmetries
of the Riemann tensor have been used as well. Now note that from
eq.~(\ref{com}), with a suitable contraction with the Riemann tensor,
one gets
 \be
R^{\mu\rho\sigma\tau} R^{\nu}_{\phantom{a}\rho\sigma\tau}\nabla_\mu f=-R^{\nu\sigma\mu\rho}(\nabla_\mu\nabla_\rho\nabla_\sigma f-\nabla_\rho\nabla_\mu\nabla_\sigma f),
\ee
 where some relabelling of the dummy indices has also taken place.
Since the Riemann tensor is antisymmetric in its last two indices and
symmetric in the exchange of pairs of indices, we can write
 \bea
R^{\mu\rho\sigma\tau} R^{\nu}_{\phantom{a}\rho\sigma\tau}\nabla_\mu f&=&-2 R^{\nu\sigma\mu\rho} \nabla_\mu\nabla_\rho\nabla_\sigma f=\nn\\&=&-2R^{\mu\rho\nu\sigma} \nabla_\mu\nabla_\rho\nabla_\sigma f.
\eea
We can then use this to substitute for the last term in 
eq.~(\ref{prel}), giving
\bea
\nabla_\mu T^{\mu\nu}_f&=&(2 R R^{\mu\nu} - 4 R^\mu_{\ \rho} R^{\nu\rho}+2 R^{\mu\rho\sigma\tau} R^{\nu}_{\phantom{a}\rho\sigma\tau}-\nn\\
  & & -4 R^{\mu\rho\sigma\nu}R_{\rho\sigma})\nabla_\mu f(\phi).
\eea
 Finally, we re-write this equation in a more economical form, taking
advantage of the identity (\ref{gbident}):
 \be
\label{identity}
\nabla_\mu T^{\mu\nu}_f=\frac{1}{2}{\cal G} \nabla^\nu f.
\ee
Note that eq.~(\ref{identity}) is just a mathematical identity.

We can now substitute eqs.~(\ref{phiid}) and (\ref{identity}) into
eq.~(\ref{must}). This gives
 \be
\frac{1}{2}\left(\lambda \nabla^2 \phi - V'(\phi)+f'(\phi) {\cal G}\right)\nabla^\nu \phi=0.
\ee
 Obviously this equation is trivially satisfied if and only if the
scalar field satisfies its field equation, namely eq.~(\ref{field2}).
Therefore, the matter
stress-energy tensor is divergence-free on shell, {\em i.e.~}when
$\phi$ satisfies its field equation. Note that in General
Relativity one can consider the matter stress-energy tensor as being
divergence-free as a consequence of the Bianchi identity, whereas in
this case eq.~(\ref{must}) is not a mathematical identity, as just
demonstrated, but requires knowledge of the dynamics of the scalar
field. Therefore we will avoid calling it a generalised Bianchi
identity, even though this is often done for similar equations in the
literature. In a sense, one could refer to the combination of
eq.~(\ref{identity}) with the Bianchi identity as the generalised
Bianchi identity.

\chapter[Equivalence of theories]{Redefinition of variables and equivalence of theories}
\label{equivtheor}
\section{Dynamical Equivalence}

In the previous chapters, we have presented a number of alternative
theories of gravity. A reasonable question to ask is how different
these theories really are. Indeed, as we will see shortly, some of the
theories which we have considered so far can be cast into the form of
others, once suitable redefinitions of the fields are utilized.

There is no unique prescription for redefining the fields of a theory.
Some of the most common redefinitions are renormalizations and
conformal transformations. Additionally, one can utilize auxiliary
fields in order to re-write the action or the field equations of a
theory. Before getting into this issue though, some clarifying remarks
are needed. 

It is important to mention that, at least within a classical
perspective like the one followed here, two theories are considered to
be dynamically equivalent if, under a suitable redefinition of the
gravitational and matter fields, one can make their field equations
coincide. The same statement can be made at the level of the action.
Dynamically equivalent theories give exactly the same results in 
describing a dynamical system to which the theories are
applicable. There are clearly advantages in exploring the dynamical
equivalence between theories: we can use results already derived for
one theory in another equivalent theory. 

The term dynamical equivalence can be considered misleading in
classical gravity. Within a classical perspective, a theory is fully
described by a set of field equations. When we are referring to
gravitation theories, these equations will be describing the dynamics
of gravitating systems. Therefore, two dynamically equivalent theories
can actually be considered just different representations of the same
theory.

The issue of distinguishing between truly different theories and
different representations of the same theory (or dynamically
equivalent theories) is an intricate one. It has serious implications
and has been the cause of many misconceptions in the past, especially
when conformal transformations are used in order to redefine the
fields ({\em e.g.~}the Jordan and Einstein frames in scalar-tensor
theory). Since many of its aspects can be more easily appreciated once
a complete discussion of the alternative theories of gravity has
already been presented, we have decided to allow this discussion to
extend over two different chapters, the current one and Chapter
\ref{concl}.

In the current chapter we will approach the problem only from an
operational viewpoint and consider only the theories presented in the
previous chapter. We will, therefore, merely analyse specific field
redefinitions that are necessary to show the dynamical equivalence
between some of these theories. The use of conformal redefinitions of
the metric will be avoided in order to simplify the discussion and we
will confine ourselves to deriving results that are needed in the
forthcoming chapters.

We will return to this subject again in Chapter \ref{concl}, where we
intend to discuss the role of conformal transformations and
redefinition of fields in gravitation theories, and to analyse
extensively the implications which these have for our understanding of
the underlying theory and our ability to propose alternative gravity
theories, hopefully clarifying some common and longstanding
misconceptions concerning these issues.

\section{$f(R)$ gravity and Brans--Dicke theory}
\subsection{Redefinition of variables}

The dynamical equivalence between $f(R)$ gravity and scalar-tensor
theory, or more specifically Brans--Dicke theory with a potential for
the scalar, has been considered by many authors (see, for instance,
\cite{higgs, tey, whitt, maeda, Barrow:1988xh, wandseq, sok, flan,
Flanagan, olkomp, gianl3, olmo1, Sotiriou:2006hs}). Let us follow the
lines of \cite{Sotiriou:2006hs} in order to see how it comes
about\footnote{Note that there are minor differences between the
terminology of Ref.~\cite{Sotiriou:2006hs} and the one used here.}. We
will work at the level of the action but the same approach can be used
to work directly at the level of the field equations. We begin with
metric $f(R)$ gravity. For the convenience of the reader, we re-write
here the action (\ref{metaction}):
 \be
S_{met}=\frac{1}{16\pi\,G}\int d^4 x \sqrt{-g} f(R) +S_M(g_{\mu\nu},\psi).
\ee
 One can introduce a new field $\chi$ and write a dynamically 
equivalent action \cite{tey}:
\be
\label{metactionH}
S_{met}=\frac{1}{16\pi\,G}\int d^4 x \sqrt{-g} \left(f(\chi)+f'(\chi)(R-\chi)\right) +S_M(g_{\mu\nu},\psi).
\ee
 Variation with respect to $\chi$ leads to the equation $\chi=R$ if
$f''(\chi)\neq 0$, which reproduces action (\ref{metaction}).
Redefining the field $\chi$ by $\phi=f'(\chi)$ and setting
 \be
\label{defV}
V(\phi)=\chi(\phi)\phi-f(\chi(\phi)),
\ee
 the action takes the form
\be
\label{metactionH2}
S_{met}=\frac{1}{16\pi\,G}\int d^4 x \sqrt{-g} \left(\phi R-V(\phi)\right) +S_M(g_{\mu\nu},\psi).
\ee
 Comparison with the action (\ref{bdaction2}) reveals that action
(\ref{metactionH2}) is the action of a Brans--Dicke theory with
Brans--Dicke parameter $\omega_0=0$ \footnote{Action
(\ref{metactionH2}) is also known as the O'Hanlon action
\cite{ohanlon}.}. Therefore, as has been observed long ago, metric
$f(R)$ theories are dynamically equivalent to a class of Brans--Dicke
theories with vanishing kinetic term \cite{tey,wandseq}. 

Let us set aside Palatini $f(R)$ gravity for the moment, and consider
directly metric-affine $f(R)$ gravity. For simplicity, we will assume
that the independent connection is symmetric (torsion-less theory)
since, as we will argue later on, the results of this section will be
completely unaffected by this choice. Once more, we re-write the
action for this theory here for the convenience of the reader:
 \be
\label{maaction2}
S_{ma}=\frac{1}{16\pi\,G}\int d^4 x \sqrt{-g} f({\cal R}) +S_M(g_{\mu\nu}, \Gamma^{\lambda}_{\phantom{a}\mu\nu}, \psi).
\ee
 Following the same steps as before, we can introduce the scalar field 
$\chi$ and then redefine it in terms of $\phi$. The action takes the form:
\be
\label{palactionH2}
S_{ma}=\frac{1}{16\pi\,G}\int d^4 x \sqrt{-g} \left(\phi {\cal R}-V(\phi)\right) +S_M(g_{\mu\nu}, \Gamma^{\lambda}_{\phantom{a}\mu\nu}, \psi).
\ee
 Even though the gravitational part of this action is formally the
same as that of action (\ref{metactionH2}), this action is not a
Brans--Dicke action with $\omega_0=0$ for two reasons: Firstly, the
matter action depends on the connection, unlike Brans--Dicke theory,
and secondly ${\cal R}$ is not the Ricci scalar of the metric
$g_{\mu\nu}$. Therefore, there is no equivalence between Brans--Dicke
theory and the general case of $f(R)$ theories in which the
connections are independent of the metric. The reason is that, unlike
Brans--Dicke theory, the theory described by the action
(\ref{maaction2}) is not a metric theory. The matter action is coupled
to the connection as well, which in this case is an independent field.
This makes the theory a metric-affine theory of gravity, as has been
discussed extensively in the previous chapter.

Let us examine what will happen if we force the matter action to be
independent of the connection, as is usually done in the literature
\cite{flan,olmo1}. Essentially, by forbidding the coupling between the
matter fields and the connection, we reduced the action to
(\ref{palaction})
 \be
\label{palaction2}
S_{pal}=\frac{1}{16\pi\,G}\int d^4 x \sqrt{-g} f({\cal R}) +S_M(g_{\mu\nu}, \psi)
\ee
 and the theory to Palatini $f(R)$ gravity. The field equations of the
theory are eqs.~(\ref{palf12}) and (\ref{palf22}) and as already
mentioned, the latter implies that the connections are the Levi--Civita
connections of the metric $h_{\mu\nu}=f'({\cal R})g_{\mu\nu}$ (see
Section \ref{manfield}). Using the redefinition which we introduced to
relate the actions (\ref{palaction2}) and (\ref{palactionH2}), we can
express the relation between the two conformal metrics simply as
$h_{\mu\nu}=\phi g_{\mu\nu}$. Then, using eq.~(\ref{confrel2}), we can
express ${\cal R}$ in terms of $R$ and $\phi$:
 \be
{\cal R}=R+\frac{3}{2\phi^2}\nabla_\mu \phi \nabla^\mu \phi-\frac{3}{\phi}\Box \phi.
\ee

Putting this into the action (\ref{palactionH2}), the latter takes the 
form:
\be
\label{palactionH2d0}
S_{pal}=\frac{1}{2\kappa}\int d^4 x \sqrt{-g} \left(\phi R+\frac{3}{2\phi}\partial_\mu \phi \partial^\mu \phi-V(\phi)\right) +S_M(g_{\mu\nu}, \psi),
\ee
 where we have neglected a total divergence. The matter action now has 
no dependence on $\Gamma^{\lambda}_{\phantom{a}\mu\nu}$ since this was
our initial requirement. Therefore, this is indeed the action of a
Brans--Dicke theory with Brans--Dicke parameter $\omega_0=-3/2$. 

The equations that one derives from the action (\ref{palactionH2d0})
are eqs.~(\ref{bdf1}) and (\ref{bdf2'}) once $\omega_0$ is set to be
$-3/2$. Note that, for $\omega_0=-3/2$ and once we set $\phi\equiv
f'({\cal R})$, eqs.~(\ref{bdf1}) and (\ref{bdf2'}) can be combined to
give eq.~(\ref{eq:field}), which we derived in Section \ref{manfield}
after simple mathematical manipulations and without any reference to
Brans--Dicke theory. Additionally, it is worth mentioning that, for
$\omega_0=-3/2$, eq.~(\ref{bdf2}) reduces to
 \be
\label{bdf3}
\kappa T+\phi V'(\phi)-2V(\phi)=0,
\ee
 which is an algebraic equation linking $\phi$ and $T$ for a given
potential.  One can then verify again that in vacuum, where $T=0$,
$\phi$ will have to be a constant and so the theory reduces to
Einstein gravity with a cosmological constant, this time determined by
the value of $\phi$.

\subsection{Physical Implications and special cases}
\label{physimpl}

In order to obtain the equivalence between Brans--Dicke theory with
$\omega_0=-3/2$ and metric-affine $f(R)$ gravity, we had to force the
matter action to be independent of the connections, i.e. to reduce the
theory to Palatini $f(R)$ gravity. This led to the fact that the
connections became the Levi--Civita connections of the metric
$h_{\mu\nu}=\phi g_{\mu\nu}$, which allowed us to eliminate the
dependence of the action on the connections. We can construct a theory
where the matter action would be allowed to depend on the connections,
but the connections would be assumed to be the Levi--Civita connections
of a metric conformal to $g_{\mu\nu}$ {\it a priori}. One may be
misled into thinking that such a theory could be cast into the form of
a Brans--Dicke theory, since in this case the dependence of the action
on the connections can indeed be eliminated. No mathematical
calculations are required to show that this is not so. The
gravitational part of the action would, of course, turn out to be the
same as that of (\ref{palactionH2d0}) if $\phi$ (its square root to
be precise) is used to represent the conformal factor. Notice,
however, that since the matter action initially had a dependence on
the independent connection, after eliminating the connection in favour
of scalar $\phi$, the matter will have a dependence not only on the
metric but also on $\phi$ because the connection will be function of
both the metric and $\phi$. Therefore, the scalar field would be
coupled to matter directly and in a non-trivial way, unlike
Brans--Dicke theory or scalar--tensor theory in general.

The above discussion demonstrates that it is the coupling of the
connections to matter that really prevents the action
(\ref{maaction2}) from being dynamically equivalent to
(\ref{bdaction}). One cannot achieve such equivalence by constraining
the connection. The only exception is if the conformal factor relating
$g_{\mu\nu}$ and the metric that is compatible with the connection, is
a constant.  In this case the theory will just reduce to metric $f(R)$
gravity and, as mentioned before, it will be equivalent to a
Brans--Dicke theory with $\omega_0=0$.

The fact that $f(R)$ gravity in the Palatini formalism is equivalent
to a class of Brans--Dicke theories when the matter action is
independent of the connection, demonstrates clearly that the former is
intrinsically a metric theory. This, as mentioned in the previous
chapter, should have been expected since the matter is coupled only to
the metric. Even though $\Gamma^{\lambda}_{\phantom{a}\mu\nu}$ is not
a scalar, the theory actually has only one extra scalar degree of
freedom with respect to General Relativity \footnote{{\em
cf.~}\cite{tomi} where similar conclusions about the role of the
independent connection in Palatini $f(R)$ gravity are derived by
examining energy conservation.}.  The independent connection
representation of the theory just prevents us from seeing this
directly, because the action is written in this frame in terms of what
turns out to be an unfortunate choice of variables. On the other hand,
if one wants to construct a metric-affine theory of gravity, matter
should be coupled to the connection as also claimed in
\cite{Sotiriou:2006qn}. In this case, any dynamical equivalence with
Brans--Dicke theory breaks down. This clarifies once more why we have
reserved the term ``metric--affine $f(R)$ theory of gravity'' for
these theories, in order to distinguish them from those for which
there is no coupling between the matter and the connection and which
are usually referred to in the literature as $f(R)$ theories of
gravity in the Palatini formalism.

It is also important to mention that $f(R)$ theories of gravity in the
metric formalism and in the Palatini formalism are dynamically
equivalent to different classes of Brans--Dicke theories. This implies
that they cannot be dynamically equivalent to each other, {\em
i.e.~}no redefinition of variables or manipulation can be found that
will bring a Palatini $f(R)$ theory into the form of some metric
$f(R)$ theory. Therefore, these theories will give different physical
predictions.  The same is, of course, true for metric-affine $f(R)$
theories of gravity as well, since they cannot be cast into the form
of a Brans--Dicke theory. There is, however, an exception:
metric--affine $f(R)$ gravity will reduce to Palatini $f(R)$ gravity
in vacuum, or in any other case where the only matter fields present
are by definition independent of the connection; such as scalar fields,
the electromagnetic field or a perfect fluid \cite{Sotiriou:2006qn}.
Therefore, even though there is no equivalence between metric-affine
$f(R)$ gravity and Brans--Dicke theories with $\omega_0=-3/2$, their
phenomenology will be identical in many interesting cases, including
cosmological applications.

\begin{figure}[t]
\begin{center}
\includegraphics[width=13cm,angle=0]{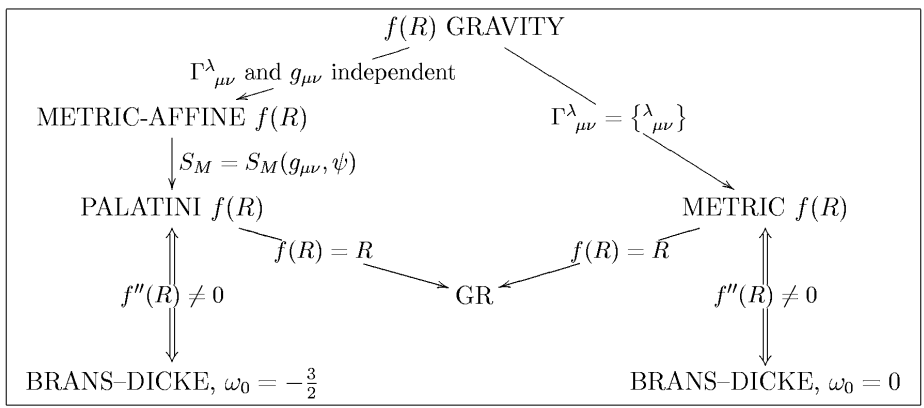}
 \caption{Schematic diagram relating various versions of $f(R)$
gravity and Brans--Dicke theory, stating the various assumptions needed
in passing from one to another.}\label{scem}
 \end{center}
\end{figure}

We have mentioned that the results will remain unchanged if we allow
the connection present in the action (\ref{maaction2}) to be
non-symmetric. Let us now justify this: in the case of $f(R)$ gravity
where the matter is independent of the connection, it is true since
the non-symmetric part of the connection vanishes even if no such
assumption is made {\it a priori}, and the field equations of the
corresponding theory are identical to eqs.~(\ref{palf12}) and
(\ref{palf22}) (see Section \ref{metaffgrav} and
\cite{Sotiriou:2006qn}).  On the other hand, when studying the case
where matter is coupled to the connection, we did not have to use the
symmetry of the connections, neither did we have to use the field
equations, but we worked at the level of the action. We have summed up
the results presented so far in this section in the schematic diagram
of fig.~\ref{scem} \cite{Sotiriou:2006hs}. 

It should be mentioned that Brans--Dicke theory with $\omega_0=-3/2$
has not received very much attention (see however \cite{fabris}).  The
reason is that when Brans--Dicke theory was first introduced, only the
kinetic term of the scalar field was present in the action. Therefore,
choosing $\omega_0=-3/2$ would lead to an ill-posed theory, since only
matter described by a stress-energy tensor with a vanishing trace
could be coupled to the theory. This can be understood by examining
eq.~(\ref{bdf3}) in the absence of terms including the potential.
However, once the potential of the scalar field is considered in the
action, no inconsistency occurs. Note that a Brans--Dicke
gravitational action with $\omega_0=-3/2$ and no potential term is
conformally invariant and dynamically equivalent to Conformal
Relativity (or Hoyle-Narlikar theory) \cite{bla}. The action of
Conformal Relativity \cite{conf1,conf2,conf3} has the form
 \be
\label{craction}
S_{CR}=\frac{1}{2\kappa}\int d^4 x \sqrt{-g} \Psi\left(\frac{1}{6} \Psi R -\Box \Psi\right),
\ee 
which can also be re-written as
 \be
\label{craction1}
S_{CR}=-\frac{1}{2\kappa}\int d^4 x \sqrt{-g} \Psi \left[\Box-\frac{1}{6} R\right]\Psi,
\ee 
in order to highlight the fact that it is conformally invariant.
A field redefinition $\Psi^2=6\phi$ will give
\be
\label{craction2}
S_{CR}=\frac{1}{2\kappa}\int d^4 x \sqrt{-g} \left(\phi R+\frac{3}{2\phi}\nabla_\mu \phi \nabla ^\mu \phi\right),
\ee 
 where a total divergence has been discarded. The dynamical 
equivalence is therefore straightforward. 

The case of a vanishing potential has no analogue in $f(R)$ gravity.
Using eq.~(\ref{defV}) and remembering that $\phi=f'(\chi)$ and on
shell ${\cal R}=\chi$, one can easily verify that setting $V(\phi)=0$
will lead to the the following equation for $f({\cal R})$:
 \be
f'({\cal R}) {\cal R}-f({\cal R})=0.
\ee
 This equation can be identically satisfied only for $f({\cal
R})={\cal R}$. However, to go from the $f(R)$ representation to the
Brans--Dicke representation, one assumes $f''({\cal R})\neq 0$, so no
$f({\cal R})$ Lagrangian can lead to a Brans--Dicke theory with a
vanishing potential. It is remarkable that this ill-posed case does
not exist in Palatini $f(R)$ gravity. There is, however, a conformally
invariant gravitational action in this context as well. One has to
choose $f({\cal R})=a {\cal R}^2$, where $a$ is some constant
\cite{higgs}. In this case the potential of the equivalent
Brans--Dicke theory will be $V(\phi)=\phi^2/4a$. For this potential,
all terms apart from the one containing $T$ in eq.~(\ref{bdf3}) will
again vanish, as would happen for a vanishing potential. The
correspondence can easily be generalised for $n$-dimensional
manifolds, where $n\geq 2$. For the gravitational action to be
conformally invariant in the context of $f(R)$ gravity, one should
choose $f({\cal R})=a {\cal R}^{n/2}$ \cite{ferr}.  The corresponding
potential can be computed using eq.~(\ref{defV}) and has the form
 \be
V(\phi)=\left(\frac{n}{2}-1\right)a \left(\frac{2\phi}{n a}\right)^{n/(n-2)}
\ee 
Eq.~(\ref{bdf2}) will generalize for $n$ dimensions in the following way:
\be
\label{bdf2n}
(n-2)\left(\omega_0+\frac{n-1}{n-2}\right) \Box \phi= \kappa T+\left(\frac{n}{2}-1\right)
\left(\phi V'-\frac{n}{n-2}V\right),
\ee
 implying that for $n$ dimensions the special case which we are
examining corresponds to $\omega_0=-(n-1)/(n-2)$. This indicates that
a Brans--Dicke gravitational action with $\omega_0=-(n-1)/(n-2)$ and a
potential $V(\phi)=b \phi^{n/(n-2)}$, where $b$ is some constant, will
be conformally invariant in an $n$-dimensional manifold. As an example
we can examine the $4$-dimensional action:
 \be
\label{bdactionphi2}
S_{n=4}=\frac{1}{2\kappa}\int d^4 x \sqrt{-g} \left(\phi R+\frac{3}{2\phi}\nabla_\mu \phi \nabla ^\mu \phi-b \phi^2\right).
\ee 
 Using the redefinition $\Psi^2=6\phi$ as before, we can bring the 
action to the following form
\be
\label{cractionpot}
S_{n=4}=\frac{1}{2\kappa}\int d^4 x \sqrt{-g} \Psi\left(\frac{1}{6} \Psi R -\Box \Psi-b\frac{\Psi^3}{36}\right),
\ee 
 which is a generalization of the action (\ref{craction}). It is easy
to verify that this specific potential will not break conformal
invariance. Under the conformal transformation $g_{\mu\nu}\rightarrow
\Omega^2 g_{\mu\nu}$, the root of the determinant will transform as
$\sqrt{-g}\rightarrow \Omega^4 \sqrt{-g}$ and so, with an appropriate
redefinition of the scalar field $\tilde{\Psi}=\Omega^{-1}\Psi$, the
action will return to the form of (\ref{cractionpot}). In fact, eq.~(\ref{cractionpot}) can also be written in the form
\be
%\label{cractionpot}
S_{n=4}=-\frac{1}{2\kappa}\int d^4 x \sqrt{-g} \left(\Psi\left[\Box-\frac{1}{6}  R\right]\Psi+b\frac{\Psi^4}{36}\right),
\ee 
in which the conformal invariance can be more obvious for some and resembles action (\ref{craction1}).

\subsection{Higher-order curvature invariants}

We have seen that both metric and Palatini $f(R)$ gravity acquire a
Brans--Dicke representation. It is interesting to examine whether we
can eliminate higher-order curvature invariants in favour of scalar
fields in the action as well. As has been discussed in Section
\ref{homet}, in the metric formalism, the presence of a general
function of the scalar curvature or of quadratic terms such as
$R^{\mu\nu}R_{\mu\nu}$, leads to fourth order equations, whereas an $R
\Box R$ term or an $R\Box^2 R$ term lead to sixth and eighth order
equations respectively \cite{ruzm, gott, amend, batt}. As can be found
in the literature, theories including any of the above terms can be
rewritten as a multi-scalar-tensor theory, {\em i.e.~}these terms can
be eliminated in favour of one or more scalars. The number of scalar
fields needed is directly related to the order of the field equations:
for fourth order equations, as in metric $f(R)$ gravity, one needs
just one scalar, for sixth order equations one needs two scalars, and
for every further two orders one more scalar has to be introduced
\cite{gott,ritis}.

In Palatini $f(R)$ gravity, however, things are not equally
straightforward. To understand this, one has to recall that, in order
to bring a Palatini $f(R)$ action into the form of a Brans--Dicke
action, we used the solution of the second field equation,
eq.~(\ref{palf22}). More specifically, due to this equation we were
able to introduce the conformal metric $h_{\mu\nu}$ and therefore
relate quantities constructed with the initially independent
connection, such as ${\cal R}$, with purely metric quantities, such as
$R$. Adding a higher order curvature term in the action will
inevitably modify eq.~(\ref{palf22}). In Section \ref{homet}, we have
given the simple example of adding an ${\cal R}^{\mu\nu}{\cal
R}_{\mu\nu}$ term to an otherwise linear action in ${\cal R}$, in
order to demonstrate how the presence of this term introduces more
dynamics for the independent connection. Indeed, this simple example
can be used here as well: eq.~(\ref{palfg1}) is no longer an algebraic
equation in $\Gamma^\lambda_{\phantom{a}\mu\nu}$, as is
eq.~(\ref{palf22}), and therefore it cannot be trivially solved in
order to express this connection, and consequently the quantities
constructed with it, in terms of the metric $g_{\mu\nu}$. 
 Thus, a dynamical equivalence with some scalar-tensor theory is
neither straightforward nor guaranteed.

\section{Why $f(R)$ gravity then?}

Since $f(R)$ gravity in both the metric formalism and the Palatini
formalism can acquire a scalar-tensor theory representation, one might
be led to ask two questions: firstly, why should we consider the
$f(R)$ representation and not just work with the scalar-tensor one,
and secondly, why, since we know quite a lot about scalar-tensor
theory, should we consider $f(R)$ gravity unexplored or interesting? 

The answer to the first question is quite straightforward. There is
actually no reason to prefer either of the two representations {\it a
priori} --- at least as far as classical gravity is concerned. There
can be applications where the $f(R)$ representation can be more
convenient and applications where the scalar-tensor representation is
more convenient. One should probably mention that habit affects
our taste and, therefore, an $f(R)$ representation seems to appear
more appealing to relativists due to its more apparent geometrical
nature, whereas the scalar-tensor representation seems to be more
appealing to particle physicists. This issue can have theoretical
implications. To give an example: if $f(R)$ gravity is considered as a
step towards a more complicated theory, which generalisation would be
more straightforward will depend on the chosen representation. This
issue will be addressed more extensively and in more general terms in
Chapter \ref{concl}.

Whether $f(R)$ theories of gravity are unexplored and interesting or
just an already-studied subcase of scalar tensor theory, is a more
practical question that certainly deserves a direct answer. It is
indeed true that scalar-tensor theory and, more precisely, Brans--Dicke
theory are well-studied theories which have been extensively used in
many applications, including Cosmology. However, the specific choices
$\omega_0=0,-3/2$ for the Brans--Dicke parameter are quite exceptional.
We have already mention in Section \ref{physimpl} why the
$\omega_0=-3/2$ case has not been studied in the past. It is also
worthwhile mentioning that most calculations which are done for a
general value of $\omega_0$ in the literature actually exclude
$\omega_0=-3/2$, mainly because they are done in such a way that the
combination $2\omega_0+3$ appears in a denominator (see Chapter
\ref{weakstrong} for details and examples). As far as the $\omega_0=0$
case is concerned, one can probably speculate that it is the apparent
absence of the kinetic term for the scalar in the action which did not
seem appealing and prevented the study of this theory.  In any case,
the conclusion is that the theories in the Brans--Dicke class that
correspond to metric and Palatini $f(R)$ gravity had not yet been
explored before the recent re-introduction of $f(R)$ gravity and, as
will also become apparent later, several of their special
characteristics when compared with more standard Brans--Dicke theories
were revealed though studies of $f(R)$ gravity.

\section{Gauss--Bonnet gravity and $f({\cal G})$ gravity}
\label{gbequiv}

In Section \ref{gbgrav} we introduced Gauss--Bonnet gravity through the
action (\ref{gbaction}) which we repeat here for convenience
 \be
\label{gbaction2}
S_{GB}=\int d^4 x \sqrt{-g}\left[\frac{R}{16\pi\,G}-\frac{\lambda}{2}\partial_\mu \phi \partial^\mu \phi-V(\phi)+f(\phi) {\cal G}\right]+S_M(g^{\mu\nu},\psi).
\ee
 This theory includes a specific combination of higher-order curvature
invariants: the Gauss--Bonnet term ${\cal G}$ (see also Section
\ref{homet}). As has been discussed, the Gauss--Bonnet term is a
topological invariant and the variation of the density $\sqrt{-g}{\cal
G}$ leads to a total divergence, therefore not contributing to the
field equations. However, in Gauss--Bonnet gravity this term is coupled
to the scalar field $\phi$ and, therefore, does contribute to the
field equations. One could think along the following lines: a
conformal redefinition of the metric, together with a suitable
redefinition of the scalar field could potentially decouple the
transformed Gauss--Bonnet term from the redefined scalar, therefore
allowing us to omit its presence. However, this idea cannot work in
practice, simply because $\sqrt{-g}{\cal G}$ transforms under
conformal redefinition of the metric $g_{\mu\nu}\rightarrow \Omega^2
g_{\mu\nu}$ as
 \bea
\sqrt{-g}{\cal G}\,\rightarrow\, \sqrt{-g}{\cal G} &+&\sqrt{-g}\big[8 R^{\mu\nu}\left(\nabla_\mu \ln{\Omega}\nabla_\nu \ln{\Omega}-\nabla_\mu\nabla_\nu \ln{\Omega}\right)\nn\\
&+&8(\nabla^2\ln{\Omega})^2-8(\nabla_\mu\nabla_\nu \ln{\Omega})^2 \nn\\
&+&8 (\nabla^2 \ln{\Omega})(\nabla_\mu \ln{\Omega})^2-4 R(\nabla^2 \ln{\Omega})\nn\\
&+&16 (\nabla_\mu \ln{\Omega}\nabla_\nu\ln{\Omega})(\nabla^\mu\nabla^\nu \ln{\Omega})\big].
\eea
 Even though extra terms containing derivatives of the conformal
factor will appear after the conformal transformation, no factor
appears in front of the Gauss--Bonnet term. Therefore, it is not
possible to eliminate the coupling with the scalar field, and
consequently the presence of the Gauss--Bonnet term, by means of a
conformal transformation. We conclude that Gauss--Bonnet gravity cannot
be rewritten as a scalar-tensor theory.

However, there is a specific subcase of the action (\ref{gbaction2})
that can be cast into the form of another theory already present in
the literature, namely that given by the action
 \be
\label{fg2}
S=\int d^4 x \sqrt{-g} \left(\frac{R}{16\pi\,G}+f({\cal G})\right),
\ee
 which has already been mentioned in Section \ref{homet} as action
(\ref{fg}) \cite{odigb}. The function $f({\cal G})$ can be a general function of the Gauss--Bonnet term but for the computation which follow we identify its functional form with that of the function $f$ present in eq.~(\ref{gbaction2}). Following Ref.~\cite{odigb}, one
can introduce two auxiliary scalar fields $A$ and $B$, in order to
re-write the action (\ref{fg2}) as
 \be
\label{fg3}
S=\int d^4 x \sqrt{-g} \left(\frac{R}{16\pi\,G}+B\,({\cal G}-A)+f(A)\right).
\ee
 Variation with respect to $B$ leads to $A={\cal G}$ and so action
(\ref{fg2}) is then recovered. Variation with respect to $A$ leads to
 \be
B=f'(A).
\ee
Replacing this back in eq.~(\ref{fg3}) gives
\be
\label{fg4}
S=\int d^4 x \sqrt{-g} \left(\frac{R}{16\pi\,G}+f'(A)\,({\cal G}-A)+f(A)\right).
\ee
Simply redefining
\begin{align}
&\phi=A,\\
& V(\phi)=A f'(A)-f(A),
\end{align}
leads to
\be
\label{gbaction0}
S=\int d^4 x \sqrt{-g}\left[\frac{R}{16\pi\,G}-V(\phi)+f(\phi) {\cal G}\right]+S_M(g^{\mu\nu},\psi).
\ee
 Clearly, this is action (\ref{gbaction2}) for $\lambda=0$, {\em
i.e.~}when the scalar field has no kinetic term. Therefore, for the
specific case of $\lambda=0$, Gauss--Bonnet gravity is dynamically
equivalent to a theory described by action (\ref{fg2}), which includes
a general function of the Gauss--Bonnet term in addition to the
standard Einstein--Hilbert action \cite{odigb}.

\chapter{Cosmology in modified gravity}
\label{cosmology}
\section{Introduction}

During the course of the 90 years that have passed since the
introduction of General Relativity by Einstein, the study and
development of alternative theories of gravity has always been
pursued in parallel, even though there have been periods of intense
effort and periods of slower development, depending on the
contemporary motivation. We have extensively discussed the motivations
for modifying gravity and it would not be wise to attempt to rank them
according to importance. However, one could still observe that the
current stimulus in this subject area is mainly powered by
observational cosmology, simply because the cosmological riddles are
the most recent of the problems which alternative gravity aims to
address. Therefore, it is important to consider the cosmological
phenomenology of the theories presented in the previous chapter and to
explore how well they can address issues such as the late time
accelerated expansion of the universe, the nature of dark energy {\em etc}.

Irrespective of the theory of gravity, the main assumptions of
cosmology remain the same since they are not related to the dynamics
but to the symmetries that we expect the universe to exhibit when one
focuses on large scale evolution and ignores small scale
inhomogeneities. Therefore, the arguments for homogeneity and isotropy
presented in Section \ref{introcosm} are still valid and we can use
the Friedmann-Lema\^itre-Robertson-Walker metric as a local
description of spacetime:
 \be
\label{flrw2}
ds^2=-dt^2+a^2(t)\left[\frac{dr^2}{1-k r^2}+r^2 d\theta^2+r^2 \sin^2(\theta)d\phi^2\right].
\ee
 We remind the reader that $k=-1,0,1$ according to whether the
universe is hyperspherical, spatially flat, or hyperbolic
 and that $a(t)$ is called the scale factor. Part of the
standard approach, which we follow here as well, is to use a perfect
fluid description for the matter for which
 \be
\label{pf}
T^{\mu\nu}=(\rho+p)u^\mu u^\nu+p\, g^{\mu\nu},
\ee
 where $u^\mu$ denotes the four-velocity of an observer comoving with
the fluid and $\rho$ and $p$ are the energy density and the pressure
of the fluid respectively. Once a gravity theory is chosen, one can
insert the FLRW ansatz (\ref{flrw2}) and the stress-energy tensor
(\ref{pf}) into the field equations of the theory and derive equations
governing the evolution of the scale factor $a(t)$. These are
generalizations of the Friedmann equations (\ref{Friedmann1}) and
(\ref{Friedmann2}) and we will occasionally refer to them as the
``generalised Friedmann equations''.

Note that the value of $k$ is an external parameter. As in many other
works in the literature, for what follows we will choose $k=0$, {\em
i.e.~}we focus on a spatially flat universe. This choice in made in
order to simplify the equations and should be viewed sceptically. It
is sometimes claimed in the literature that such a choice is favoured
by the data. However, this is not entirely correct. Even though the
data ({\em e.g.~}\cite{wmap3}) indicate that the current value of
$\Omega_k$ is very close to zero (see eq.~(\ref{datak}) and the
related discussion in Section \ref{observ}) it should be stressed that
this does not really reveal the value of $k$ itself. Since
 \be
\Omega_k=-\frac{k}{a^2 H^2},
\ee
 the current value of $\Omega_k$ is sensitive of the current value of
$a(t)$, {\em i.e.~}the amount of expansion the universe has undergone
after the Big Bang. A significant amount of expansion can easily drive
$\Omega_k$ very close to zero. The success of the inflationary
paradigm is exactly that it explains the flatness problem --- how did
the universe become so flat (see Section \ref{inflationaryexpansion})
--- in a dynamical way, allowing us to avoid having to fine tune the
parameter $k$ (having $k=0$ is statistically very exceptional).

The above having been said, choosing $k=0$ for simplicity is not a
dramatic departure from generality when it come to late time
cosmology. If it is viewed as an approximation and not as a choice of
an initial condition, then one can say that, since $\Omega_k$ as
inferred from observations is very close to zero at current times, the
terms related to $k$ will be subdominant in the Friedmann or
generalised Friedmann equations and therefore one could choose to
discard them by setting $k=0$, without great loss of accuracy. In any
case, results derived under the assumption that $k=0$ should be
considered preliminary until the influence of the spatial curvature is
precisely determined, since there are indications that even a very
small value of $\Omega_k$ may have an effect on them (see, for
instance, Ref.~\cite{Clarkson:2007bc}).

Having set the ground, we are now ready to explore the cosmological
implication of a number of alternative theories of gravitation. Taking
into account that cosmology in scalar-tensor theory has already been
extensively studied \cite{valeriobook}, we focus on the various
versions of $f(R)$ gravity and on Gauss--Bonnet gravity. A study
limited to these theories cannot be considered as exhaustive and many
more modifications of the gravitational actions are possible. However,
from a phenomenological point of view, such theories can work as
useful examples for understanding how modifications of the
gravitational action can help us to address the well-known
cosmological problems, since they include a number of interesting
terms in the gravitational action.

\section{$f(R)$ gravity in the metric formalism}
\subsection{Generalised Friedmann equations}

We start with $f(R)$ gravity in the metric formalism. We present
in this section the modified Friedmann equations for such theories,
which date back to Buchdahl's paper \cite{buchdahl}. The procedure for
deriving these equations is actually quite straightforward since the
components of $R_{\mu\nu}$ for the ansatz (\ref{flrw2}) can be easily
found in textbooks. The time-time component of the field equations
(\ref{metf}) gives the modified Friedmann equation
 \be
\label{metfried}
 \left(\frac{\dot{a}}{a}\right)^2-\frac{1}{3 f'(R)}\left\{\frac{1}{2}\left[f(R)-Rf'(R)\right]
-3\left(\frac{\dot{a}}{a}\right)\dot{R}f''(R) \right\}=\frac{1}{3} \kappa\rho,
\ee
and the space-space components gives
\begin{align}
\label{spacespace}
2\left(\frac{\ddot{a}}{a}\right)+\left(\frac{\dot{a}}{a}\right)^2&+\frac{1}{f'(R)}\Big\{2\left(\frac{\dot{a}}{a}\right)\dot{R}f''(R)+\ddot{R}f''(R)+\nn\\&+\dot{R}^2f'''(R)
-\frac{1}{2}\left[f(R)-Rf'(R)\right] \Big\}=-\kappa p,
\end{align}
where $\kappa=8\pi\,G$. $R$ is given by
\be
R=6\left[\frac{\ddot{a}}{a}+\left(\frac{\dot{a}}{a}\right)^2\right]=6 \left(\dot{H}+2 H^2\right),
\ee
 and $H=\dot{a}/a$ as usual. A combination of eqs.~(\ref{metfried})
and (\ref{spacespace}) gives
 \begin{align}
 \label{metfried2}
 \left(\frac{\ddot{a}}{a}\right)&+\frac{1}{2 f'(R)}\Big\{\left(\frac{\dot{a}}{a}\right)\dot{R}f''(R)+\ddot{R}f''(R)+\nn\\
&+\dot{R}^2f'''(R)
-\frac{1}{3}\left[f(R)-Rf'(R)\right] \Big\}=-\frac{\kappa}{6}\left[\rho+3p\right].
 \end{align}
 Setting $f(R)=R$, one has $f'(R)=1$ and $f''=f'''=0$ and
eqs.~(\ref{metfried}) and (\ref{metfried2}) reduce to the standard
Friedmann equations (\ref{Friedmann1}) and (\ref{Friedmann2}).

\subsection{$1/R$ terms and late time cosmology}
\label{metfde}

The first attempt to consider $f(R)$ theories of gravity as a way to
explain late-time cosmological acceleration was probably
Ref.~\cite{capoz00}. The main objective is to have modified
gravity account for dark energy and consequently explain the late time
accelerated expansion of the universe. One of the easiest ways to see
how this comes about is the following \cite{capozziello}: If we define
the quantities
 \begin{align}
\rho_{de}=\frac{\kappa^{-1}}{f'(R)} & \Bigg\{\frac{1}{2}\left[f(R)-Rf'(R)\right]-3\left(\frac{\dot{a}}{a}\right)\dot{R}f''(R) \Bigg\},\\
p_{de}=\frac{\kappa^{-1}}{f'(R)} & \Bigg\{2\left(\frac{\dot{a}}{a}\right)\dot{R}f''(R)+\ddot{R}f''(R)+\nn\\
& \qquad\qquad +\dot{R}^2f'''(R)-\frac{1}{2}\left[f(R)-Rf'(R)\right] \Bigg\},
\end{align}
and use them to re-write eqs.~(\ref{metfried}) and (\ref{metfried2}) we get
\begin{align}
\left(\frac{\dot{a}}{a}\right)^2 &=\frac{1}{3} \kappa\rho_{\rm tot},\\
 \left(\frac{\ddot{a}}{a}\right)&=-\frac{\kappa}{6}\left[\rho_{\rm tot}+3p_{\rm tot}\right],
\end{align}
where
\begin{align}
\rho_{\rm tot}&=\rho+\rho_{de},\\
p_{\rm tot}&=p+p_{de}.
\end{align}

 Through some simple redefinitions, we have brought the equations
governing the cosmological dynamics into the form of the standard
Friedmann equations. Additionally, the terms related to high order
terms present in the action are now conveniently denoted as
$\rho_{de}$ and $p_{de}$, since these terms are playing here the role
of dark energy. Therefore, the whole theory, for what regards
cosmology, has been brought into the form of General Relativity with
some kind of dark energy, whose nature can actually be attributed to a
modification of gravity.

Now the important question is: What is the effective equation of state
relating $\rho_{de}$ and $p_{de}$? Viewed as functions of $R$, these
quantities are obviously related and the effective equation of state
will depend on the functional form of $f(R)$. Therefore, without going
into more details, one can expect that a convenient choice for $f(R)$
can lead to a suitable value of $w_{de}$ ($p_{de}=w_{de} \rho_{de}$),
so that the modified theory of gravity can account for the
observations indicating late time accelerated expansion. 

We will give two simple examples that can be found in the literature:
Firstly, one can consider the function $f$ to be of the form
$f(R)\propto R^n$. It is quite straightforward to calculate $w_{de}$
as a function of $n$ if the scale factor is assumed to be a generic
power law $a(t)=a_0(t/t_0)^\alpha$ (a general $a(t)$ would lead to a time dependent $w_{de}$) \cite{capozziello}. The result is
 \be
w_{de}=-\frac{6n^2-7n-1}{6n^2-9n+3},
\ee
for $n\neq 1$, and $\alpha$ is given is terms of $n$ as
\be
\alpha=\frac{-2n^2+3n-1}{n-2}.
\ee
 A suitable choice of $n$ can lead to a desired value for $w_{de}$. 
The second example which we will refer to is a model of the form
$f(R)=R-\mu^{2(n+1)}/R^n$, where $\mu$ is a suitably chosen parameter
\cite{cdtt}. In this case, and once again if the scale factor is assumed to be a generic power law, $w_{de}$ can again be written as a function of $n$ \cite{cdtt}:
 \be
w_{de}=-1+\frac{2(n+2)}{3(2n+1)(n+1)}.
\ee
 The most typical model within this class is that with $n=1$
\cite{cdtt}, in which case $w_{de}=-2/3$. Note that in this class of
models, a positive $n$ implies the presence of a term inversely
proportional to $R$ in the action, contrary to the situation for the
$R^n$ models.

We have chosen to discuss metric $f(R)$ gravity and late time
acceleration in terms of its representation as introducing a form of
effective dark energy, since this approach is simple and has a direct
relation with the usual approach to cosmological problems. Obviously,
there is more to say about the cosmological dynamics of metric $f(R)$
gravity, such as making a complete dynamical analysis of the equations
governing the cosmic evolution, or making a precise study of specific
models and their cosmological behaviour. For the moment, however, let
us mention that our goal here is merely to demonstrate how simple
modifications of gravity can address the dark energy problem. Note
also that we have not chosen the examples according to their overall
viability as gravitation theories --- we will discuss these issues
later on --- but according to simplicity.

\subsection{More general models and cosmological constraints}

We have seen qualitatively how simple metric $f(R)$ gravity models,
and especially those including an inverse power of $R$, can be used as an attempt to
solve the cosmological puzzle of the late time accelerated expansion.
Clearly one can consider much more general functions $f$. These do not
have to be polynomials necessarily. However, taking $f$ to be a
polynomial with positive and/or negative powers of $R$ has certain
advantages. Besides simplicity, one can argue that choosing $f$ to be
a polynomial is a practical way to include in the action some
phenomenologically interesting terms which might be of leading order
in the Taylor expansion of an effective Lagrangian [{\em
c.f.~}eq.~(\ref{eq:fr})].

However, before referring to more general models we should mention
that specific models of what is now called metric $f(R)$ gravity were
not initially introduced in cosmology in order to account for the
phenomenology related to the later stages of its evolution: Alexei
Starobinski, who, as mentioned in the introduction, was one of the
pioneers of the idea of inflation, had first proposed a scenario of
this sort for giving a gravity driven inflationary period
\cite{starobinski}. The model, which was actually presented before the
more conventional models based on scalar fields, included an $R^2$
term in the gravitational Lagrangian. The presence of this term is
able to drive the universe to an accelerated expansion which takes
place at early times. We refer the reader to the literature for more
details ({\em e.g.~}\cite{starobinski, mijic, Barrow:1988xh}).

It seems reasonable to expect that a single model, including both
positive and negative powers of $R$, would be able to lead to both an
early time inflationary period and a late time accelerated expansion.
This was indeed shown in \cite{nojodiposneg}. Another interesting
class of models are those containing a $\ln{R}$ term \cite{lnR}. See
also \cite{nojodirev, Capozziello:2007ec} for reviews of metric $f(R)$
gravity and cosmology and \cite{dunsby} for a discussion of the
cosmological dynamics of $R^n$ models.

For a model of $f(R)$ gravity to be considered successful from a
cosmological perspective, however, it is not enough to have the
correct early or late time behaviour in a qualitative sense. There are
a number of precise tests related to cosmological observations that
any gravity theory should pass\footnote{In addition, viable gravity
theories should at the same time pass also tests relevant to other
scales, such as the scales of the Solar System and compact objects. It
is not an easy task to construct a theory that fulfils all of these
requirements simultaneously. We will however discuss this issue
later.}. For instance, there have been studies of the constraints
imposed on specific models of metric $f(R)$ gravity by Big Bang
Nucleosynthesis and local fifth-force experiments
\cite{Brookfield:2006mq} and attempts to explore the details of
cosmological perturbations \cite{bean}. The stability of the de-Sitter
solution, which is supposed to be a late time attractor for models
with late time acceleration, has been considered \cite{Faraoni:2005ie,
Faraoni:2005vk, Cognola:2007vq}, as well as the process of producing
the baryon asymmetry in the universe, Baryogenesis
\cite{Lambiase:2006dq}. To avoid getting into technical details let us
just say that, even though some of these studies, such as the one
related to Baryogenesis, show that metric $f(R)$ gravity does not lead
to significant deviations away from the standard picture, the overall
impression is that simple models are unlikely to produce the
cosmological dynamics and also agree in detail with all of the other
observations. There is, of course, ongoing research on this 
({\em e.g.}~\cite{Song:2007da, De Felice:2007ez}).

An issue that requires a special mention is the question raised in
\cite{amendola} about whether metric $f(R)$ gravity can lead to
cosmological models which include both a standard matter dominated era
and a phase of accelerated expansion. According to \cite{amendola} all
$f(R)$ theories that behave as a power of $R$ at large or small $R$
will have a matter era in which the scale factor will scale like
$t^{1/2}$ instead of the standard law $t^{2/3}$, making the theory
grossly inconsistent with observations. This issue has raised a lively
debate \cite{Capozziello:2006dj,Amendola:2006eh}. The outcome is that
$R^n$ and $R-\mu^{2(n+1)}/R^n$ models do indeed lead to an
unacceptable behaviour during matter domination
\cite{Amendola:2006we}, but there can be more complicated models that
do not have this unappealing characteristic. In fact, a scheme has
been developed to reconstruct the action for metric $f(R)$ gravity
from a desired cosmological evolution as inferred from observation
\cite{Nojiri:2006gh, Nojiri:2006su}.

Let us close this section by stressing that $f(R)$ actions, and
especially $R^n$ and $R-\mu^{2(n+1)}/R^n$ models, should be considered
as toy theories. Even from a dimensional analysis or leading order
point of view, it is hard to consider such actions as exact effective
low energy actions. An action including an $R^2$ term, for example, is
most likely to include an $R_{\mu\nu}R^{\mu\nu}$ term as well. From
this view point, $f(R)$ gravity is just a preliminary step that one
takes in order to explore the possibilities which are offered by
modifications of the gravitational actions.  Attempts to study the
cosmology of more general actions including higher order curvature
invariants have been made ({\em e.g.~}\cite{Carroll:2004de}) and there
is hope that some of the shortcomings of metric $f(R)$ gravity may not
be there for more complete theories.

\section{$f(R)$ gravity in the Palatini formalism}
\subsection{Generalised Friedmann equations}

Let us now consider Palatini $f(R)$ gravity. The action of the theory
is (\ref{palaction}) and the field equations are eqs.~(\ref{palf12})
and (\ref{palf22}). We start by deriving the generalised Friedmann
equation. Using the FLRW metric (\ref{flrw2}) we need to compute the
components of ${\cal R}_{\mu\nu}$, which is the Ricci tensor
constructed with the independent connection (see eq.~(\ref{ricci})).
Since what we know is an ansatz for the metric, it is practical to
work with metric quantities and, therefore, eqs.~(\ref{gammagmn}),
(\ref{confrel1}) and (\ref{confrel2}) can be used to arrive to the
result. The non-vanishing components of ${\cal R}_{\mu\nu}$ are
 \begin{align}
\label{ricci1}
{\cal R}_{00}&=-3\frac{\ddot{a}}{a}+\frac{3}{2}(f')^{-2}(\partial_{0} f')^2-\frac{3}{2}(f')^{-1}\nabla_0\nabla_0 f',\\
\label{ricci2}
{\cal R}_{ij}&=[a \ddot{a}+2 \dot{a}^2+(f')^{-1}\left\{^{\lambda}_{\mu\nu}\right\} \partial_0 f'
%+{}\nn\\& &{}
+\frac{a^2}{2}(f')^{-1}
\nabla_0\nabla_0 f']\delta_{ij},
\end{align}
 where the subscript $0$ denotes the time component and we remind the
reader that ${\nabla}$ is the covariant derivative associated with
$g_{\mu\nu}$. Combining eqs.~(\ref{ricci1}) and (\ref{ricci2}) with
eq.~(\ref{palf12}) one quite straightforwardly arrives at the
generalised Friedmann equation ({\em e.g.~}\cite{mengwang1})
 \be
\label{pfriedmann}
\left(H+\frac{1}{2} \frac{\dot{f'}}{f'}\right)^2=\frac {1}{6} \frac{\kappa (\rho+3p)}{f'}+\frac {1}{6}
\frac{f}{f'},
\ee
 where the overdot denotes differentiation with respect to coordinate
time. Note that when $f$ is linear, $f'=1$ and, therefore,
$\dot{f'}=0$. Taking into account eq.~(\ref{paltrace}), one can easily
show that in this case eq.~(\ref{pfriedmann}) reduces to the standard
Friedmann equation.

\subsection{A toy model as an example}
\label{toy}

Having derived the generalised Friedmann equation, we can now go ahead
and study the cosmological evolution in Palatini $f(R)$ gravity. The
first thing that we would like to check is which cosmological eras can
take place in general. It is required that any model should lead to a
radiation dominated era followed by a matter dominated era. In
addition to this, a model which draws its motivation from late time
cosmology should provide a resolution for the accelerated expansion of
the universe. In fact, it was shown in \cite{vollick} that models
which include in the action a $1/{\cal R}$ term in addition to the
more standard ${\cal R}$ term do indeed have this property. Several
studies of this issue followed \cite{Allemandi:2004ca,
Allemandi:2005qs, Sotiriou:2005hu, Sotiriou:2005cd}.

It is also interesting to consider whether specific choices for the
Lagrangian can lead to early time inflation without the need for a
specific inflaton field introduced for this purpose. In the case of metric $f(R)$
gravity, this was indeed the case. In Palatini $f(R)$ gravity things
are quite different. Models which include an ${\cal R}^2$ term have
been studied and it has been shown that the presence of this term
cannot lead to an inflationary period \cite{mengwang1, Meng:2004yf,
Meng:2004wg}. This, however, as will also become clearer later, is
actually due to the special nature of an ${\cal R}^2$ term within the
framework of Palatini $f(R)$ gravity \cite{Sotiriou:2005hu}. An ${\cal
R}^2$ term, as already mentioned, gives a zero contribution on the
left hand side of eq.~(\ref{paltrace}) and we will see that this makes
this term quite ineffective as far as inflation is concerned.

Let us explore all of the above in more detail through an example.
Consider the specific model in which $f$ is given by
 \be
\label{lagrang}
f({\cal R})=\frac{{\cal R}^3}{\beta^2}+{\cal R}-\frac{\epsilon^2}{3 {\cal R}},
\ee
 where $\epsilon$ and $\beta$ are for the moment some parameters, on
which we will try to put constraints later. Our choice of the form
of the Lagrangian is based on the interesting phenomenology which
it will lead to. When $f$ is chosen to have the form given in eq.
(\ref{lagrang}), in vacuum eq.~(\ref{paltrace}) gives
 \be
\label{alg1}
{\cal R}^4-\beta^2 {\cal R}^2+\epsilon^2\beta^2=0.
\ee
 Note that even if we included an ${\cal R}^2$ term in
eq.~(\ref{lagrang}), eq.~(\ref{alg1}) would remain unchanged due to
the form of eq. (\ref{paltrace}). Thus, even though we have avoided
including this term for the sake of simplicity, there is no reason to
believe that this will seriously affect our results in any way. One
can easily solve eq.~(\ref{alg1}) to get
 \be
{\cal R}^2=\frac{\beta^2}{2} \left[1\pm\sqrt{1-4\left(\epsilon/\beta\right)^2}\right].
\ee
 If $\epsilon\ll\beta$, this corresponds to two de Sitter and two
anti-de Sitter solutions for ${\cal R}$. Here we will consider the two
de Sitter solutions, namely:
 \be
\label{solutions}
{\cal R}_1\sim \beta,\quad {\cal R}_2\sim \epsilon.
\ee
 If we further assume that $\epsilon$ is sufficiently small and
$\beta$ is sufficiently large, then since the expansion rate of the de
Sitter universe scales like the square root of the scalar curvature,
${\cal R}_1$ can act as the seed for an early-time inflation and
${\cal R}_2$ as the seed for a late-time accelerated expansion.

To see this explicitly, let us consider FLRW cosmology in more detail. 
At very early times, we expect the matter to be fully relativistic.
Denoting by $\rho_r$ and $p_r$ the energy density and pressure, the
equation of state will be $p_r=\rho_r /3$. Thus $T=0$ and eq.
(\ref{paltrace}) will reduce to eq.~(\ref{alg1}) and have the solution
given by eqs.~(\ref{solutions}). If we ask for the curvature to be
large, we infer that ${\cal R}={\cal R}_1=\beta$.  Therefore, the universe will
undergo a de Sitter phase which can account for the early-time
inflation. As usual, conservation of energy implies $\rho_r \sim
a^{-4}$. On the other hand, ${\cal R}$ and consequently $f({\cal R})$
are now large constants, whereas $\dot{f'}=0$ since $f'({\cal R})$ is
a constant as well. Therefore, it is easy to verify that the last term
on the right hand side of eq. (\ref{pfriedmann}) will quickly
dominate, with $H$ being given by
 \be
H\sim\sqrt{\beta/12}.
\ee
 In this sense, an inflationary period can occur in Palatini $f(R)$
gravity. Whether this scenario is realistic or not will be explored
shortly.

Let us now consider the matter and radiation dominated eras. When the
temperature is low enough, we expect some matter components to be
non-relativistic. As an idealisation we can assume that the matter has
two components. Radiation, for which $p_r=\rho_r /3$, and
non-relativistic matter, for which the pressure $p_m=0$ (dust) and the
energy density is denoted by $\rho_m$. Eq. (\ref{paltrace}) takes the
following form:
 \be
\label{eq1}
\frac{{\cal R}^3}{\beta^2}-{\cal R}+\frac{\epsilon^2}{{\cal R}}=-\kappa^2\rho_m.
\ee
Energy conservation requires that
\be
\label{eq2}
\dot{\rho}_m+3H\rho_m=0.
\ee
 Using eqs. (\ref{eq1}) and (\ref{eq2}), it is easy to show after some
mathematical manipulations that
 \be
\label{eq3}
\dot{{\cal R}}=\frac{3H {\cal R}\left({\cal R}^2-\frac{{\cal R}^4}{\beta^2}-\epsilon^2\right)}{\left(\frac{3 {\cal R}^4}{\beta^2}-{\cal R}^2-\epsilon^2\right)}.
\ee
The modified Friedmann equation (\ref{pfriedmann}) takes the form
\be
\label{mf2}
H^2=\frac{ 2\kappa^2 \rho+\Lambda_{\rm eff}}
{6\left(\frac{3 {\cal R}^2}{\beta^2}+1+\frac{\epsilon^2}{3 {\cal R}^2}\right) \left(1+\frac{3}{2} A\right)^2}
\ee
where $\rho=\rho_r+\rho_m$, 
\bea
A&=&\frac{\left(\frac{6 {\cal R}^4}{\beta^2}-\frac{2}{3}\epsilon^2\right) \left( {\cal R}^2-\frac{{\cal R}^4}{\beta^2}-\epsilon^2\right)}
{\left(\frac{3 {\cal R}^4}{\beta^2}-{\cal R}^2-\epsilon^2\right)\left(\frac{3 {\cal R}^4}{\beta^2}+{\cal R}^2+\frac{\epsilon^2}{3}\right)},\\
\Lambda_{\rm eff}&=&2\left(\frac{{\cal R}^3}{\beta^2}+\frac{\epsilon^2}{3{\cal R}}\right),
\eea
 and we have used eq. (\ref{eq1}) and the equation of state for the 
relativistic component of the cosmological fluid.

Now ${\cal R}$ is no longer a constant. Its value is given by the root
of eq.~(\ref{eq1}). Assuming that ${\cal R}$ is now less than $\beta$
and significantly larger than $\epsilon$, eqs.~(\ref{eq3}) and
(\ref{mf2}) give
 \be
\dot{{\cal R}}\sim -3 H {\cal R},\qquad H^2\sim \frac{{\cal R}^3}{3\beta^2}
\ee
Thus, it is easy to see that
\be
\label{r1}
{\cal R}\sim t^{-2/3},\qquad a(t)\sim t^{2/9}.
\ee
From eqs. (\ref{r1}), one concludes that
\be
\rho_r\sim t^{-8/9},\qquad
\Lambda_{\rm eff}\sim t^{-2}.
\ee
 {\em i.e.}~$\Lambda_{\rm eff}$ decreases much faster than the energy
density of relativistic matter. Hence, the universe will soon enter a
radiation dominated era characterized by a very low value of ${\cal
{\cal R}}$ (and consequently $\Lambda_{\rm eff}$). 

We next investigate the behaviour of the modified Friedmann equation
(\ref{mf2}). $\Lambda_{\rm eff}$ at this stage of the evolution will
be negligible compared to $\kappa \rho$ and $A\sim 0$ to a very good
approximation. On the other hand, $f'=3{\cal
R}^2/\beta^2+1+\epsilon^2/(3{\cal R}^2)$ will tend to $1$ provided
that $\epsilon$ is small enough.  Therefore, eq. (\ref{mf2}) reduces
to
 \be
\label{stand}
H^2\sim \kappa^2\rho,
\ee
 where, as before, $\rho=\rho_r+\rho_m$. Eq. (\ref{stand}) resembles
standard cosmology. It is reasonable, therefore to assume that
everything can continue as expected, {\em i.e.} radiation dominated
era, Big Bang Nucleosynthesis (BBN), and matter dominated era. 

Finally, we consider what will happen at late times. At some point we
expect matter to become subdominant with respect to $\Lambda_{\rm
eff}$ due to the increase of the scale factor. Note that $\Lambda_{\rm
eff}$ asymptotically reaches $2 \epsilon/3$ as matter dilutes. Thus at
late times we can arrive at the picture where $\rho\sim 0$ and the
universe will therefore again enter a de Sitter phase of accelerated
expansion qualitatively similar to that indicated by current
observations. 

What needs to be stressed here is that the analysis of this section
gives a very rough description of the cosmological dynamics of
Palatini $f(R)$ gravity. Additionally, the model used is chosen {\it
ad hoc}, just because it leads to some interesting phenomenology from
a demonstrative point of view. In the next section, we will proceed to
check the validity of what has been presented here in more detail and
for more generic choices of $f$.

\subsection{Constraining positive and negative powers of ${\cal R}$}
\label{palgencosm}

Even though a simple model like the one described by
eq.~(\ref{lagrang}) was helpful for understanding the basic features
of cosmology in Palatini $f(R)$ gravity, one would like to consider
more generic choices for $f$. At the same time it is important to go
beyond the qualitative results and use the numerous observations to
get quantitative ones.  Such a study was performed in \cite{ama}.
Assuming that the gravitational action includes, besides the standard
linear term, a term inversely proportional to ${\cal R}$, the authors
used four different sets of cosmological data to constrain it. These
are the Supernova Type Ia gold set \cite{supernovae}, the CMBR shift
parameter \cite{bond}, the baryon oscillation length scale
\cite{supernovae3} and the linear growth factor at the 2dF Galaxy
Redshift Survey effective redshift \cite{hawkins,tegmark}. However, as
stated in the conclusions of \cite{ama}, the restricted form of
$f({\cal R})$, including only a term inversely proportional to ${\cal
R}$, prevents the study from being exhaustive.

In \cite{Sotiriou:2005cd} the cosmological behaviour of more general
models of Palatini $f(R)$ gravity was studied and the results of
\cite{ama} where generalised. Following \cite{Sotiriou:2005cd} we
consider here a general model. We leave the function $f$ unspecified
and try to derive results independent of its form as long as this is
possible. Since such a general study is a quite tedious analytical
task given the complexity of the functions involved and the
non-linearity of the equations, we also adopt the following
representation for $f$, which is suitable for our purposes, whenever
needed:
 \be
\label{ans3}
f({\cal R})=\frac{1}{\epsilon_1^{d-1}} {\cal R}^d+{\cal R}-\frac{\epsilon_2^{b+1}}{{\cal R}^b},
\ee
 with $\epsilon_1,\epsilon_2>0$, $d>1$ and $b\geq 0$; $b=0$
corresponds to the $\Lambda$CDM model when $\epsilon_1\rightarrow
\infty$. The dimensions of $\epsilon_1$ and $\epsilon_2$ are
$(\textrm{eV})^2$. Part of our task in this section will be to
constrain the value of $\epsilon_1$, given that for the value of
$\epsilon_2$ no extended discussion is really necessary. In fact, in
order for a model to be able to lead to late-time accelerated
expansion consistent with the current observations, $\epsilon_2$
should by roughly of the order of $10^{-67}$ $(\textrm{eV})^2$
\cite{vollick}.

First of all, let us see how a model with a general function $f$ would
behave in vacuum, or whenever $T=0$ (radiation, {\em etc}.).  If we define
 \be
\label{defF}
{\cal F}({\cal R})\equiv f'({\cal R}){\cal R}-2f({\cal R}),
\ee
then eq.~(\ref{paltrace}) becomes
\be
\label{Fscalar}
{\cal F}({\cal R})=\kappa T,
\ee
and for $T=0$,
\be
\label{Fscalar0}
{\cal F}({\cal R})=0.
\ee
 Eq.~(\ref{Fscalar0}) is an algebraic equation which, in general, will
have a number of roots, ${\cal R}_n$. Our notation implies that ${\cal
R}$ is positive in the presence of ordinary matter and so here we will consider the positive solutions
({\em i.e.}~the positive roots). Each of these solutions corresponds
to a de Sitter expansion, since ${\cal R}$ is constant. If one wants
to explain the late-time accelerated expansion one of these solutions,
say ${\cal R}_2$, will have to be small. If in addition to this we
also want our model to drive an early-time inflation, there should be
a second solution, ${\cal R}_1$ corresponding to a larger value of
${\cal R}$. 

For example, introducing in eq.~(\ref{paltrace}) the ansatz given for
$f$ in eq.~(\ref{ans3}), one gets
 \be
\label{alg}
\frac{d-2}{\epsilon_1^{d-1}}{\cal R}^{d+b}-{\cal R}^{b+1}+(b+2)\epsilon_2^{b+1}=0.
\ee
 If $\epsilon_1\gg \epsilon_2$ and $d>2$ then this equation has two 
obvious solutions
\be
{\cal R}_1\sim \epsilon_1, \qquad {\cal R}_2\sim \epsilon_2.
\ee
 These solutions can act as seeds for a de Sitter expansion, since the
expansion rate of the de Sitter universe scales like the square root
of the scalar curvature. Notice that the rest of the solutions of
eq.~(\ref{Fscalar0}), for ${\cal R}<{\cal R}_1$ and ${\cal R}>{\cal
R}_2$ will not be relevant here. During the evolution we do not
expect, as will become even more obvious later on, that ${\cal R}$
will exceed ${\cal R}_2$ or become smaller than ${\cal R}_1$.

Before going further, it is worth mentioning that one can use
eq.~(\ref{paltrace}) together with the conservation of energy to
express $\dot{{\cal R}}$ as a function of ${\cal R}$:
 \be
\label{rdot}
\dot{{\cal R}}=-\frac{3H ({\cal R} f'-2f)}{{\cal R} f''-f'}.
\ee
 Using eq.~(\ref{rdot}) to re-express $\dot{f'}(=f''\dot{{\cal R}})$
and assuming that the universe is filled with dust ($p=0$) and
radiation ($p=\rho/3$), eq.~(\ref{pfriedmann}) gives, after some
mathematical manipulation,
 \be
\label{HRform}
H^2=
\frac{1}{6 f'}
\frac{2 \kappa \rho+{\cal R} f'-f}
{\left(1-\frac{3}{2}\frac{f''({\cal R}f'-2f)}{f'({\cal R}f''-f')}\right)^2},
\ee
 where again $\rho=\rho_r+\rho_m$.

\subsubsection{Early times} 
\label{early}
 Following the lines of Section \ref{toy}, we can make the following
observation. Since we expect the matter to be fully relativistic at
very early times, in this regime $T=0$ and consequently ${\cal R}$ is
constant. This implies that the second term on the left hand side of
eq.~(\ref{pfriedmann}) vanishes. Additionally, conservation of energy
requires that the first term on the right hand side of the same
equation scales like $a(t)^{-4}$, which means that the second term on
the same side, depending only on the constant curvature, will soon
dominate if $f({\cal R})$ is large enough for this to happen before
matter becomes non relativistic. ${\cal R}$ can either be equal to
${\cal R}_1$ or to ${\cal R}_2$. Since we want ${\cal R}_2$ to be the
value that will provide the late time acceleration, $f$ should be
chosen in such a way that $f({\cal R}_2)$ will become dominant only at
late times when the energy densities of both matter and radiation have
dropped significantly. Therefore, if we want to have an early
inflationary era we have to choose the larger solution ${\cal R}_1$,
and $f$ should have a form that allows $f({\cal R}_1)$ to dominate
with respect to radiation at very early times. The Hubble parameter
will then be given by
 \be
H\sim \sqrt{\frac{f({\cal R}_1)}{6 f'({\cal R}_1)}},
\ee
 As an example, we can use the ansatz given in eq.~(\ref{ans3}). The 
modified Friedmann equation is then
\be
H\sim \sqrt{\frac{\epsilon_1}{3 (d+1)}},
\ee
 and the universe undergoes a de Sitter expansion which can account 
for the early-time inflation.

\begin{figure}[t]
\begin{center}
\includegraphics[width=12cm,angle=0]{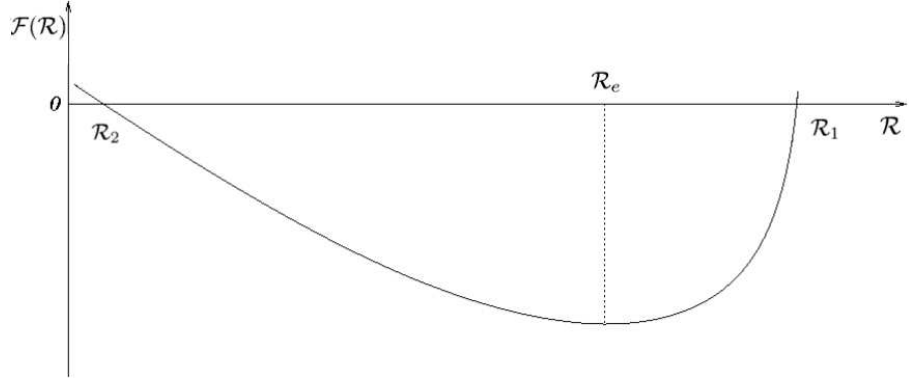}
 \caption{The behaviour of a general function ${\cal F}({\cal R})$
over the interval ${\cal R}_1>{\cal R}>{\cal R}_2$. ${\cal R}_e$
denotes the value of ${\cal R}$ where ${\cal F}$ has a minimum. From
this graph one can easily see that ${\cal F}'$ is positive when ${\cal
R}_e<{\cal R}<{\cal R}_1$ and negative when ${\cal R}_2<{\cal R}<{\cal
R}_e$.}\label{fig1}
 \end{center}
\end{figure}

Sooner or later this inflationary expansion will lead to a decrease of
the temperature and some portion of the matter will become non
relativistic. This straightforwardly implies that ${\cal R}$ will stop
being constant and will have to evolve. Recall that in Palatini $f(R)$
gravity, the field equation for the connection implies the existence
of a metric $h_{\mu\nu}$, which is conformal to $g_{\mu\nu}$ (see
Sections \ref{manfield} and eq.~(\ref{hgconf})). Then $f'({\cal R})$
plays the role of the conformal factor relating these two metrics, and
therefore we do not consider sign changes to be feasible throughout
the evolution of the universe. We also know that in a certain range of
values of ${\cal R}$ it should be close to one. This is the case
because there should be a range of values of ${\cal R}$, for which
$f({\cal R})$ behaves essentially like ${\cal R}$, {\em i.e.}~our
theory should be well approximated by standard General Relativity in
order for us to be able to derive the correct Newtonian limit (see
\cite{Sotiriou:2005xe} and Section \ref{newtf}). Together with
$T\leq0$, the above implies the following:
 \be
\label{constraints1}
f'({\cal R})> 0, \quad {\cal F}<0, \qquad \forall\quad {\cal R}_2<{\cal R}<{\cal R}_1.
\ee
 Since ${\cal F}$ is a continuous function keeping the same sign in
this interval and ${\cal F}({\cal R}_1)={\cal F}({\cal R}_2)=0$, there
should be a value for ${\cal R}$, say ${\cal R}_e$, where ${\cal
F}'({\cal R}_e)$=0, i.e. an extremum. Eq. (\ref{Fscalar}) implies that
the time evolution of ${\cal R}$ is given by
 \be
\label{rdot1}
\dot{{\cal R}}={\kappa \dot{T}}/{{\cal F}'({\cal R})}.
\ee
Differentiating eq.~(\ref{defF}), we get
\be
\label{above}
{\cal F}'=f''{\cal R}-f'.
\ee 
 Using the fact that $\dot{f'}=f''\dot{{\cal R}}$, and using
eq.~(\ref{above}) to express $f''$ in terms of ${\cal F}'$, $f'$ and
${\cal R}$, one can easily show that
 \be
\label{fdot}
\frac{\dot{f'}}{f'}=\frac{{\cal F}'+f'}{{\cal R} f'}\dot{{\cal R}}.
\ee
 The constraints given in eq.~(\ref{constraints1}) imply that for
${\cal R}_e<{\cal R}<{\cal R}_1$, ${\cal F}'>0$. An easy way to
understand this is to remember that ${\cal F}$ is negative in that
interval but zero at ${\cal R}={\cal R}_1$ and so it should be an
increasing function (see also fig. \ref{fig1}). Since, $f'$ and ${\cal
R}$ are also positive, then what determines the sign of
$\dot{f'}/{f'}$ in the neighbourhood of ${\cal R}_e$ is the sign of
$\dot{{\cal R}}$. 

Let us see what will happen if we require ${\cal R}$ to decrease, i.e.
$\dot{{\cal R}}<0$. Eq. (\ref{rdot}) implies that as ${\cal
R}\rightarrow {\cal R}_e$, $\dot{{\cal R}}\rightarrow -\infty$ if
$\dot{T}\neq 0$, since ${\cal F}'({\cal R}_e)=0$. Therefore,
$\dot{f'}/{f'}\rightarrow -\infty$ and using eq.~(\ref{pfriedmann}) we
can infer that $H\rightarrow \infty$. Physically, the above implies
the following: ${\cal R}$ has no way to decrease to a value less than
${\cal R}_e$ without giving the universe an infinite expansion. In
practice, any attempt for ${\cal R}$ to approach ${\cal R}_e$ would
lead to a dramatically rapid expansion until non-relativistic matter
fully dilutes and ${\cal R}$ settles back to ${\cal R}_1$. Thus, once
the curvature terms in the modified Friedmann equation dominate the
evolution, there is no turning back to matter domination through a
continuous process. The two vacuum solutions ${\cal R}_1$ and ${\cal
R}_2$ seem to be somehow disconnected in the evolution and ${\cal R}$
has to remain in the region close to only one of them. This is a
general statement independent of the form of matter that is present
since $T$ was left unspecified in its derivation. So, even though, as
shown in \cite{Sotiriou:2005hu}, including positive powers of ${\cal
R}$ in the action can lead to early-time inflation, there seems to be
no graceful exit from it. The only alternative left would be to
consider that due to some other physical and non-classical process,
the equation presented here ceases to be valid for some time interval,
which, however seems highly implausible.

Since it seems impossible to provide an exit from this gravity driven
inflation it seems reasonable to check whether we can at least totally
avoid it. If we choose as our initial solution ${\cal R}_2$ instead of
${\cal R}_1$ then the curvature terms will not dominate as long as
${\cal R}$ is constant. However, there is still one subtle point. At
some stage during the evolution the energy density of non-relativistic
matter will have to rise sooner or later, forcing ${\cal R}$ to change
its value. It is also reasonable to assume that if inflation is not
driven by curvature, we will have to adopt a more standard approach to
guarantee that it will happen, like an inflaton field. It is obvious,
however, keeping in mind the previous discussion, that one would want
${\cal R}$ to be always less than ${\cal R}_e$ and this will impose 
a
constraint which will depend on the functional form of $f$. For
example, if one assumes that $f$ is described by the ansatz given in
eq.~(\ref{ans3}) then, considering ordinary matter, ${\cal R}<{\cal
R}_e$ at all times implies that $\epsilon_1\gg\kappa\rho_m$ at all
times. Let us also consider the case of a slow-rolling inflaton field,
$\phi$. Then, if we denote its energy density by $\rho_\phi$ and its
pressure by $p_\phi$ we have, as usual,
 \bea
\rho_\phi&=&\frac{1}{2}\dot{\phi}^2+V(\phi),\\
p_\phi&=&\frac{1}{2}\dot{\phi}^2-V(\phi),
\eea
 where $V(\phi)$ is the scalar field potential. During the period when
$\phi$ dominates the evolution, $T=\dot{\phi}^2-4 V(\phi)$ and since
slow-roll implies that $\dot{\phi}^2\ll V(\phi)$, $T\approx -4
V(\phi)$. Therefore, if we want ${\cal R}<{\cal R}_e$, so that
inflation proceeds as usual, then $\epsilon_1\gg V(\phi)$ at all
times.

If ${\cal R}$ is less than ${\cal R}_e$ for all values of $T$ then it
is easy to verify that everything will evolve naturally after the end
of inflation.  For ${\cal R}_2<{\cal R}<{\cal R}_e$, ${\cal F}'<0$ and
as non-relativistic matter dilutes, $\dot{\rho}_m<0$, and so from
eq.~(\ref{rdot1}) we see that $\dot{{\cal R}}<0$ so that ${\cal R}$
will have to decrease to reach the value ${\cal R}_2$ asymptotically.

The above discussion is not relevant, of course, if the only term with
a positive power present in the action, besides ${\cal R}$ itself, is
${\cal R}^2$. In this case, due to the form of ${\cal F}$, this term
does not appear in eq.~(\ref{Fscalar}). This specific case has been
studied in \cite{Meng:2004wg}. One thing that is worth commenting on,
before closing this discussion, is the following. The constraint
$f'>0$ (see eq.~(\ref{constraints1})), which is implied by the fact
that $f'$ plays the role of the conformal factor relating the metrics
$g_{\mu\nu}$ and $h_{\mu\nu}$, can, depending on the form of $f$,
impose a further constraint on the value of the constants in front of
the positive power terms. In \cite{Meng:2004yf} inflation driven by an
inflaton field was studied in the presence of an ${\cal R}^2$ term.
The authors derived a constraint for the constant appearing in front
of the ${\cal R}^2$ term in the action by requiring that the square of
the Hubble parameter should be positive during the kinetic dominated
phase. This constraint is exactly what would one derive by requiring
$f'$ to be always positive.

\subsubsection{Big Bang Nucleosynthesis}

Let us now turn our attention to the next cosmological era, radiation
domination and Big Bang Nucleosynthesis (BBN). Current observations
indicate that the standard cosmological model can fit the data related
to the primordial abundances of light elements. On the other hand, how
a modified gravity model like the one discussed here would fit those
data has not been yet worked out. However, there is little room for
modifying the behaviour of the Friedmann equation during BBN and it
seems reasonable to ask that the model under investigation should
resemble standard cosmology during these cosmological eras
\cite{carrollbbn}. This implies that eq.~(\ref{HRform}) should be
similar to the standard Friedmann equation
 \be
\label{stfriedmann}
H^2=\frac{1}{3}\kappa \rho.
\ee
 By comparing eqs. (\ref{HRform}) and (\ref{stfriedmann}), one sees 
that during BBN
\be
\label{c1}
f'\sim 1
\ee
\be
\label{c2}
1-\frac{3}{2}\frac{f''({\cal R}f'-2f)}{f'({\cal R}f''-f')}\sim 1
\ee
\be
\label{c3}
{\cal R} f'-3f\sim 0
\ee
 To make the picture clearer, we give the explicit expressions for 
$f'$ and $f''$ when $f$ is given by eq.~(\ref{ans3}):
\bea
f'&=&d \frac{{\cal R}^{d-1}}{\epsilon_1^{d-1}}+1+b \frac{\epsilon_2^{b+1}}{{\cal R}^{b+1}},\\
f''&=& d (d-1) \frac{{\cal R}^{d-2}}{\epsilon_1^{d-1}}-b(b+1) \frac{\epsilon_2^{b+1}}{{\cal R}^{b+2}}.
\eea
 Let us for the moment assume that the term inversely proportional to
${\cal R}$ is not present. In order for condition (\ref{c1}) to be
fulfilled $\epsilon_1\gg {\cal R}_{BBN}$. This is the natural
constraint on the value of $\epsilon_1$ imposed when one asks for the
model to have almost identical behaviour to the standard one during
BBN. Once $\epsilon_1$ is chosen to have a large enough value, all
three constraints (\ref{c1}), (\ref{c2}) and (\ref{c3}) are easily
fulfilled and the modified Friedmann equation (\ref{HRform}) becomes
identical to the standard one, eq.~(\ref{stfriedmann}), for the
relevant values of ${\cal R}$. The above constraint can be viewed as a
sufficient constraint for the model to be viable but not as a
necessary one. However, one could also claim that, even if the
modifications in the Friedmann equation do not necessarily have to be
negligible, they should at least lead to second order corrections and
not affect the leading order. This implies that $\epsilon_1$ should
definitely be larger than ${\cal R}_{BBN}$. 

We have, however, neglected the presence of the term inversely
proportional to ${\cal R}$. In the absence of positive powers, this
term should be negligible during BBN since ${\cal R}_{BBN}$ is much
larger than $\epsilon_2$. This picture may change if we consider the
full version of the model. Eq.~(\ref{scalar}) can take the following
form
 \be
\label{red}
{\cal R}f'-2f=-\kappa \rho_m^0 (1+z)^3,
\ee
 where $\rho_m^0$ is the present value of the energy density of
non-relativistic matter and $z$ is the redshift. We have assumed here
that $a_0=1$. Using eq. (\ref{red}) one can derive how ${\cal R}$ will
scale with the redshift. If $\epsilon_1\gg {\cal R}_{BBN}$, then for
all of the evolution of the universe after BBN, ${\cal R}$ scales
almost like $(1+z)^3$. This indicates that, since BBN takes place at a
very high redshift, ${\cal R}_{BBN}$ is indeed much larger than
$\epsilon_2$. If, however, one assumes that $\epsilon_1$ is large
enough to alter the behaviour of eq. (\ref{red}), then ${\cal R}$ will
have a milder scaling with the redshift, meaning that ${\cal R}_{BBN}$
can get very close to $\epsilon_2$. Then the three constraints
(\ref{c1}), (\ref{c2}) and (\ref{c3}) might not be fulfilled not only
due to the presence of the term with the positive power of ${\cal R}$
greater than 1 but also because of the term with the negative power. 
This will be a secondary effect related to the term with a positive
power greater than 1, as shown earlier. 
 It will be avoided if again $\epsilon_1\gg {\cal R}_{BBN}$ and will
be subdominant if $\epsilon_1$ is just smaller than ${\cal R}_{BBN}$.
Unfortunately, since the value of ${\cal R}_{BBN}$ is very model
dependent, it is difficult to turn this constraint into a numerical
one.

\subsection{Late times}

Now let us check the behaviour of the modified Friedmann equation at
late times. The scalar curvature ${\cal R}$ decreases with time to
reach a value close to $\epsilon_2$. Therefore the conditions
(\ref{c1}), (\ref{c2}) and (\ref{c3}) will at some point cease to hold
because of the term involving the negative power of ${\cal R}$. Any
contribution of the term involving the positive power of ${\cal R}$
greater than 1 will be negligible for two reasons. Firstly, since the
value of $\epsilon_1$ should be such that these terms are already
negligible during BBN, it is safe to assume that they will remain so
throughout the rest of the evolution of the universe. The same results
can be inferred by using the constraints derived in Section
\ref{early}. 

As we mentioned earlier, satisfying the present bounds related to 
the primordial abundancies of light elements according to BBN is 
straightforward if the modified Friedmann equation of the model does 
not deviate significantly from the standard one during the BBN 
epoch. This was the condition which we imposed to derive the 
constraints just presented. However, one cannot completely discard 
the possibility that a modified gravity model whose modified 
Friedmann equation does deviate slightly but significantly from the 
standard one during BBN could still be viable: it is not yet clear 
how a modification of gravity will then influence the light element 
abundancies and BBN as a whole. Under this perspective, the 
constraints presented here are suffiecient for a viable model: if 
the modified Friedmann equation reduces to the standard one with 
high precision during BBN, one is assured that the current bounds on 
light element abundancies will be both unaffected and satisfied. 
However, these constraints cannot yet be considered as being an 
absolutely necessary condition for the model to be viable, until a 
more detailed study of the effect of a modification of gravity on 
BBN is performed. In any case, even if one assumes that there is 
some slight contribution from the terms being discussed in the 
modified Friedmann equation during BBN, such a contribution should 
become weaker at later times.

The constraints coming from the early-time behaviour are necessary but
it is difficult to turn them into numerical ones. At the same time one
can always claim that the early time evolution of the universe is not
very well established and there might still be room for new physics
there affecting these constraints. However, let us anticipate that the
Newtonian limit of the theory will also provide constraints which will
turn out to be in agreement with those derived here and actually more
stringent (see Section \ref{newtf}). The range of values of ${\cal R}$
which is of interest for late-time observations is between the value
of ${\cal R}$ at decoupling ${\cal R}_{dec}$ and the current value of
${\cal R}$, ${\cal R}_0$. For these values, it is safe to consider
that \cite{Sotiriou:2005cd}
 \bea
f&\sim& {\cal R}-\frac{\epsilon_2^{b+1}}{{\cal R}^b},\\
f'&\sim&1+b \frac{\epsilon_2^{b+1}}{{\cal R}^{b+1}},\\
f''&\sim& -b(b+1) \frac{\epsilon_2^{b+1}}{{\cal R}^{b+2}},
\eea
 with extremely high accuracy for all times after decoupling. It is
easy to see that the modified Friedmann equation of the model
described in (\ref{ans3}) will be identical at late times to that of a
model with no positive powers of the curvature greater than 1
($\epsilon_1\rightarrow \infty$).

In \cite{ama} the authors consider $f$ to be of the form
\be
\label{ans}
f({\cal R})={\cal R}\left(1+\alpha\left(\frac{{\cal R}}{H_0^2}\right)^{\beta-1}\right),
\ee
 where $\alpha$ and $\beta$ are dimensionless parameters, with
$\beta<1$ (note that in our notation ${\cal R}$ is positive). This
representation of the function $f$ is very useful when one wants to
constrain some dimensionless parameter. Comparing it with our ansatz,
eq.~(\ref{ans3}), we get $d=\delta+1$, $b=-\beta$ and $\epsilon_1
\rightarrow \infty$, since in eq.~(\ref{ans}) there is no positive
power of ${\cal R}$ greater than 1. In order to constrain the values
of $\alpha$ and $\beta$ they use a rather extensive list of
cosmological observations. The first quantity which they consider is
the CMBR shift parameter \cite{bond,melchiorri,odman} which in a
spatially flat universe is given by
 \be
\label{shiftdef}
\mathscr{R} = \sqrt{\Omega_m H_0^2}\int_0^{z_{dec}}\frac{d\tilde{z}}{H(\tilde{z})},
\ee
 where $z_{dec}$ is the redshift at decoupling and $\Omega_m\equiv
\kappa \rho^0_m/(3H_0^2)$. When expressed in terms of the scalar
curvature, eq.~(\ref{shiftdef}) becomes
 \bea
\label{shift}
\mathscr{R} & = & \sqrt{\Omega_m H_0^2}\int_{0}^{z_{dec}}\frac{dz}{H(z)}\nonumber\\
& = & \sqrt{\Omega_m H_0^2}\int_{{\cal R}_{dec}}^{{\cal R}_0}\frac{a'({\cal R})}{a({\cal R})^2}
\frac{d{\cal R}}{H({\cal R})}\\
& = & \frac{1}{3^{4/3}}\left(\Omega_m H_0^2\right)^{1/6}\int_{{\cal R}_0}^{{\cal R}_{dec}}
\frac{{\cal R}f''-f'}{\left({\cal R}f'-2f\right)^{2/3}}\frac{d{\cal R}}{H({\cal R})}\nonumber.
\eea
 Using the values for $z_{dec}$ and $\mathscr{R}$ obtained with WMAP
\cite{wmap3}, namely $z_{dec}=1088^{+1}_{-2}$ and
$\mathscr{R}=1.716\pm 0.062$, they find that the best fit model is
$(\alpha,\beta)=(-8.4,-0.27)$. The ``Gold data set'' of Supernovae is
also used \cite{supernovae}. What is important for this analysis is
the expression for the luminosity distance, which in terms of ${\cal
R}$ is
 \bea
\label{lumdist}
D_L(z) & = & (1+z)\int_{0}^z\frac{d\tilde{z}}{H(\tilde{z})}\nonumber\\
& = & \sqrt{\Omega_m H_0^2}\frac{1}{a({\cal R})}\int_{{\cal R}}^{{\cal R}_0}\frac{a'({\cal R})}{a({\cal R})^2}\\
& = & \frac{1}{3}\sqrt{\Omega_m H_0^2}\left({\cal R}f'-2f\right)^{1/3} \times \nonumber\\&&\int_{{\cal R}_0}^{{\cal R}_{dec}}
\frac{{\cal R}f''-f'}{\left({\cal R}f'-2f\right)^{2/3}}\frac{d{\cal R}}{H({\cal R})}\nonumber.
\eea
 Marginalizing over the Hubble parameter $h$, the authors again
constrain $\alpha$ and $\beta$ and the best fit model is
$(\alpha,\beta)=(-10.0,-0.51)$. Another independent observation which
they use is that of the imprint of the primordial baryon-photon
acoustic oscillations on the matter power spectrum. The dimensionless
quantity $A$ \cite{linder1, linder2,eisen,hu,eisen2},
 \begin{equation}
A= \sqrt{\Omega_m}E(z_1)^{-1/3}\left[ \frac{1}{z_1}\int_{0}^{z_1}\frac{dz}{E(z)}\right]^{2/3},
\end{equation}
 where $E(z)=H(z)/H_0$, can act as a ``standard ruler''. The data from
the Sloan Digital Sky Survey \cite{supernovae3} provide a value for
$A$, namely
 \be
\label{boe}
A=D_v(z=0.35)\frac{\sqrt{\Omega_m H_0^2}}{0.35c}=0.469\pm0.017,
\ee
where 
\be
\label{distance}
D_v(z)=\left[D_M(z)^2\frac{cz}{H(z)}\right]^{1/3},
\ee
 and $D_M(z)$ is the comoving angular diameter distance. The best fit
model using this value is $(\alpha,\beta)=(-1.1,0.57)$. Finally, in
\cite{ama} these three sets of data are combined to give a best fit
for $(\alpha,\beta)=(-3.6,0.09)$.

The above observations are potentially very useful, of course, in
studying the viability of a model like (\ref{ans}). Two comments are
due:

Firstly, as shown here, the modified Friedmann equation of a more
general model like (\ref{ans3}), which also includes positive powers
of ${\cal R}$ greater than 1, is effectively identical to that of
(\ref{ans}) at late times. Therefore, it is expected that the results
of \cite{ama} will remain unaffected by the inclusion of positive
powers of the scalar curvature greater than 1, since these terms have
to satisfy the constraints derived in this section. Additionally, one
can conclude that the late evolution of the universe is {\it not}
affected by the positive powers of the scalar curvature greater than 1
present in the action. This can be rephrased in two interesting ways:
{\it The results of observational tests relevant to the late-time
evolution of the universe are insensitive to the inclusion of
additional positive powers of $R$} or {\it observational tests
relevant to the late-time evolution cannot constrain the presence of
additional positive powers of $R$ in the gravitational action}. The
second expression implies that such tests are not sufficient to judge
the overall form of the gravitational action.

Secondly, it is worth commenting on the result of \cite{ama}.  The
best fit model for the combination of the different data sets suggests
that their exponent $\beta$ is equal $0.09$ (see eq.~(\ref{ans})) and
therefore favours the $\Lambda$CDM model, being well within the
$1\sigma$ contour. However, one gets different values for $\beta$ when
the different data sets are considered individually. For the SNe data
the best fit model has $\beta=-0.51$ and the baryon oscillations
$\beta=0.57$, both disfavouring the $\Lambda$CDM model, but also being
mutually contradictory. The CMBR shift parameter gives $\beta=-0.27$
which again is significantly different from the other two values. Of
course, one might expect that the combination of the data will give
the most trustworthy result. However, this is not necessarily true,
since one could regard the very wide discrepancies in the value of
$\beta$ coming from different observations as an indication that more
accurate data are needed to derive any safe conclusion. It would also
be interesting to study to what extent a model with $\beta=-1$, which
is the model commonly used in the literature, can individually fit the
current data and to compare the results with other models built to
explain the current accelerated expansion, such as quintessence,
scalar-tensor theories, {\em etc}. One should bear in mind that the
$\Lambda$CDM model has always been the best fit so far. However, the
motivation for creating alternative models does not come from
observations but from our inability to solve the theoretical problems
that arise if one adopts the standard picture (coincidence problem,
{\em etc}.).

Before closing this section, let us briefly mention the fourth scheme
used in \cite{ama} in order to obtain constraints: large scale
structure and growth of perturbations. The results derived there
through this scheme do not actually improve the constraints obtained
with the three schemes already mentioned.  As the authors of
\cite{ama} correctly state, a more detailed analysis should be
performed along the lines of \cite{Koivisto:2005yc}, in which a more
standard linearized perturbation analysis is performed. In fact there
are a a number of works which place constraints on Palatini $f(R)$
gravity using perturbation analysis, the matter power spectrum, large
scale structure {\em etc}. The general conclusion is that the models
which are very close to the $\Lambda$CDM model are the only ones that
could satisfy the relevant constraints. It is, therefore, very
difficult to construct viable alternative models.  Simple models such
as those used above are very much disfavoured by the observations. 
 We refer the reader to the literature for more details
\cite{Koivisto:2005yc,Koivisto:2006ie,Li:2006vi,Li:2006ag,Fay:2007gg,Uddin:2007gj}.

\section{Metric-affine $f(R)$ gravity and cosmology}

Instead of presenting here a detailed study of the cosmological
aspects of metric-affine $f(R)$ theories of gravity, we wish to remind
the reader that the main difference from Palatini $f(R)$ gravity is
the following: In the Palatini formalism, the matter action is assumed
to be independent of the connections whereas in the metric-affine
formalism no such assumption is made. More precisely, as we have
argued, such an {\em a priori} assumption is against the spirit of
metric-affine gravity. However, as was explained in Section
\ref{metaffgrav}, whenever the only matter fields considered are
perfect fluids with no vorticity, electromagnetic fields or scalar fields, metric-affine
$f(R)$ gravity does reduce to Palatini $f(R)$ gravity, as the matter
action of those fields are independent of the connection without this
having to be imposed as an external assumption. In cosmology, these
are indeed the only fields considered in most applications. This implies that the main
features of cosmology in metric-affine $f(R)$ gravity will not be
different from those of Palatini $f(R)$ gravity.

Therefore, no detailed study is needed and the reader may refer to the
previous section. However, what is mentioned here has to be approached
with caution. In Section \ref{spmfields} we have already commented on
the difficulties that arise when one attempts to adopt macroscopic
descriptions of matter, such as perfect fluids, when spin and torsion
are taken into account. The definition of a perfect fluid might have
to be generalised and the details of cosmological evolution could be
affected (see \ref{spmfields} for details). A more detailed analysis
of cosmology in metric-affine $f(R)$ gravity is still pending.

\section{Gauss--Bonnet gravity}
\label{gbcosm}

\subsection{Generalised Friedmann equations}

In Section \ref{gbgrav} we introduced Gauss--Bonnet gravity, derived
the field equations and studied some of their characteristics. We have
also made a reference to the motivation from heterotic String Theory
for the study of such actions. As is the case for many alternative
theories of gravity, in Gauss--Bonnet gravity there is also strong
motivation from cosmology. More specifically there have been works
showing that Gauss--Bonnet gravity can address the dark energy problem
without the need for any exotic matter components \cite{nojodsas,
nojodisam, Leith:2007bu}. Additionally, such a theory can have other
interesting characteristics in relation to cosmological phenomenology
including early time inflation \cite{carneu, neup, neup2} and
avoidance of future and past singularities \cite{strings2, nojodsas,
tsuji}.

Let us briefly derive the modified Friedmann equations for
Gauss--Bonnet gravity (see also \cite{nojodsas,carneu}). Recall that
the action of the theory is given by eq.~(\ref{gbaction}) and the
field equations for the metric and the scalar field are given by
eqs.~(\ref{gbfield1}) and (\ref{gbfield2}) respectively. Using the
flat ($k=0$) form of the FLRW metric, eq.~(\ref{flrw2}), and assuming
that the scalar only depends on time, one gets from the time-time and
space-space components of eq.~(\ref{gbfield1}) respectively (after
some manipulations involving also the definitions for the quantities
$T^{\mu\nu}_\phi$ and $T^{\mu\nu}_f$)
 \begin{align}
\label{gbfried1}
H^2=&\frac{1}{3}\kappa \left(\rho+\frac{\lambda}{2} \dot{\phi}^2+V(\phi)-24 \dot{\phi} f'(\phi) H^3\right),\\
\label{gbfried2}
\left(\frac{\ddot{a}}{a}\right)\equiv \dot{H}+H^2 = &-\frac{\kappa}{6}\left(\rho+3p\right)- \frac{\kappa}{6}\left(2\lambda \dot{\phi}^2-2 V(\phi)\right.+\nn\\&\left.+24 H^3 \dot{\phi}f'(\phi)+24 \frac{\partial}{\partial t} \left(H^2 \dot{f}\right)\right).
\end{align}
 The Gauss--Bonnet invariant can easily be expressed in terms of $H$ 
and its time derivative as
\be
{\cal G}=24 H^2 \left(\dot{H}+H^2\right).
\ee
The equation of motion for the scalar takes the form
\be
\label{phiflrw}
\lambda\left(\ddot{\phi}+3 H \dot{\phi}\right)+ V'(\phi)-24f'(\phi)H^2 \left(\dot{H}+H^2\right)=0.
\ee

Eqs.~(\ref{gbfried1}), (\ref{gbfried2}) and (\ref{phiflrw}) govern the
cosmological dynamics. Note that the potential $V(\phi)$ and the
coupling between the scalar and the Gauss--Bonnet term $f(\phi)$ are
left unspecified at this stage. Besides providing initial conditions,
one needs to choose the functional form of $V(\phi)$ and $f(\phi)$ in
order to solve the equations.

\subsection{Gauss--Bonnet gravity as dark energy}

Let us see how Gauss--Bonnet gravity can account for dark energy. We
will proceed as in Section \ref{metfde} and attempt to bring the
modified Friedmann equations into the form for standard matter plus a
dark energy component. It is not difficult to see that if one defines
 \begin{align}
\rho_{de}&=\frac{\lambda}{2} \dot{\phi}^2+V(\phi)-24 \dot{\phi} f'(\phi) H^3,\\
p_{de}&=\frac{\lambda}{2} \dot{\phi}^2-V(\phi)+8\frac{\partial}{\partial t}\left(H^2 \dot{f}\right)+16 H^3 \dot{\phi}f'(\phi),
\end{align}
then eqs.~(\ref{gbfried1}) and (\ref{gbfried2}) can be written in the form
\begin{align}
\label{sfr1}
H^2&=\frac{1}{3}\kappa \left(\rho_{\rm tot}\right),\\
\label{sfr2}
\left(\frac{\ddot{a}}{a}\right)&=-\frac{\kappa}{6}\left(\rho_{\rm tot}+3p_{\rm tot}\right),
\end{align}
where
\begin{align}
\rho_{\rm tot}&=\rho+\rho_{de},\\
p_{\rm tot}&=p+p_{de}.
\end{align}
 Clearly, eqs.~(\ref{sfr1}) and (\ref{sfr2}) are formally the same as
those which one would derive in General Relativity once the presence
of a dark energy component is assumed. As usual, we can define an
effective equation of state parameter $w_{de}\equiv p_{de}/\rho_{de}$.
Recall that, if we assume that the scalar field dominates the
evolution, then
 \be
w_{de}=\frac{2q-1}{3},
\ee
where $q\equiv-a \ddot{a}/\dot{a}^2$ is the deceleration parameter.

Since the role of dark energy is to provide the late-time accelerated
expansion, let us focus on how this can be achieved within the
framework of Gauss--Bonnet gravity. Due to the resemblance of the
theory to General Relativity with a minimally coupled scalar field, we
know that if $\dot{f}=0$ (or $f=0$), the minimal condition for having
acceleration, $\rho_{de}+3p_{de}< 0$, is satisfied when $V(\phi) >
\lambda \dot{\phi}^2$. However, in our case $\dot{f}\neq 0$ and its
sign is important for determining the behaviour of the scale factor.
One can show that when $\dot{f}< 0$ (which will generically hold for a
canonical scalar, $\lambda> 0$), the acceleration condition is indeed
satisfied if $V(\phi) > \lambda \dot{\phi}^2$.

Note that the role of $f$ can actually be more active. In principle
it can even lead to an era of acceleration even if there is no
potential term. The effective equation of state depends on both $f$
and $V$. Choosing these functions appropriately, one can control how
much effect each of them will have on the cosmic evolution.  Common
well motivated choices for $f$ and $V$ are exponentials, {\em
i.e.~}$V=V_0 e^{-a\kappa\phi}$ and $f=f_0 e^{b\kappa\phi}$ where $a$,
$b$ and $f_0$ are of order unity while $V_0$ is as small as the energy
density of the cosmological constant in order to guarantee that the
theory will fit observations related to the late-time cosmological
expansion ({\em e.g.~}\cite{nojodsas}). In such models, the
acceleration is mainly due to the potential terms.

In the same way that Gauss--Bonnet gravity can lead to a late-time
acceleration, it can also lead to an early-time inflationary period.
As already mentioned, the theory is very similar to General Relativity
plus a minimally coupled scalar field, such as the inflaton field
usually used to drive inflation. Of course, the coupling of the scalar
field to the Gauss--Bonnet term does lead to qualitative differences.
For instance, the slow roll variables should be redefined, the
generation of perturbations will differ {\em etc}. \cite{carneu}. We will
not go further in examining specific models or analysing the
cosmological features of Gauss--Bonnet gravity. We refer the reader to
the relevant literature instead \cite{nojodsas, Leith:2007bu,carneu,
neup, neup2}. 

However, before closing it is important to refer to the confrontation
with cosmological observations. Current literature on the subject includes studies of the impact of the Gauss--Bonnet coupling in relation to several constraints: 
those coming from the 
Cosmic Microwave Background, galaxy distributions, large scale 
structure and supernovae as well as from studies of cosmological 
perturbations related to the Cosmic Microwave Background and the 
matter power spectrum \cite{tomigb1,tomigb2}.
 Remarkably, the theory seems to be in good agreement with the data as
far as these studies are concerned. However, as reported in
\cite{tomigb2}, constraints from baryon oscillations and
nucleosynthesis appear to disfavour simple models. A scheme with which
one can reconstruct the action from the expansion history has been
developed in \cite{cogneli}. Finally, note that more general actions
which include couplings between the dilaton and the Gauss--Bonnet term
have also been studied ({\em e.g.~}\cite{neup,neup2}).

\chapter[Weak and strong gravity regimes]{Weak and strong 
gravity
regimes in Modified gravity}
\label{weakstrong}
\section{Introduction}

Up to now we have studied the theoretical basis of several theories of
gravity and have examined their cosmological features. However, we
have not yet referred to a number of other important issues for any
theory of gravitation.

To begin with, we have not considered the Newtonian limit of any of
the theories mentioned. This is obviously a crucial issue, since any
theory of gravity should reduce to Newtonian gravity at a suitable
limit. The validity of Newton's theory in a weak gravity regime and at
certain length scales is hardly questionable. Additionally, deviations
from it as gravity becomes stronger are well constrained by Solar
System tests and the post-Newtonian expansion \cite{willbook} is a
powerful tool for judging the viability of a theory. Constraints
coming from Cosmology are important but in most cases constraints
coming from Solar System tests with the use of the post-Newtonian
approximation are more stringent. One should take into account that
alternative theories of gravity motivated by cosmological problems are
tailored to fit cosmological observations to some extent. Difficulties
start to arise when a theory is required to perform well in Cosmology
and, at the same time, to comply with the bounds imposed by Solar
System tests.

Another aspect of the theories under investigation which has not been
discussed so far regards other solutions of their field equations,
apart from the cosmological ones. Even though many of the
characteristics of a gravitation theory can be inferred from the form
of its action or of its field equations and without any reference to
specific solutions, the study of specific solutions always adds to our
insight. It is important, for instance, that one should check whether
solutions which describe any configuration of physical interest do
exist and do have properties which agree with observations. In
addition to this, it is not only the weak gravity regime that can
provide constraints for alternative theories of gravity. Binary system
tests (see \cite{Damour:2007uf} and references therein) or other
strong gravity tests ({\em e.g.~}\cite{Damour:1993hw}) offer the
opportunity to test gravity beyond the weak field regime.

We will attempt to cover this gap in the present chapter. To this end,
we will study the Newtonian limit and the post-Newtonian expansion of
$f(R)$ gravity in both the metric formalism and the Palatini
formalism, as well as in Gauss--Bonnet gravity. We will also address
other issues related to the weak field regime of $f(R)$ theories of
gravity as well as referring to vacuum solutions in these theories. 
For what regards non-vacuum solutions and the strong gravity regime
the progress in the literature is much smaller. We restrict ourselves
to discussing non-vacuum solutions in Palatini $f(R)$ gravity which,
apart from the interest which they have within the framework of this
theory, also serve as a very good example to demonstrate how studying
matter configurations in the strong gravity regime can help to
constrain or even rule out theories.

It should be mentioned that our discussion of the strong and weak
gravity regimes of alternative theories of gravity in this chapter is
far from exhaustive. One could also consider other theories but, more
importantly, there are aspects of the theories under investigation
that we will not be extensively referring to here. In some cases, such
as vacuum \cite{Clifton:2006ug,Multamaki:2006zb,Capozziello:2007wc}
and non vacuum \cite{Multamaki:2006ym} solutions in metric $f(R)$
gravity, the reader can refer to the literature for more details.
However, studies of most of the subjects which we will not refer to
here are still pending. To name a few: the Newtonian and Post
Newtonian limits in metric-affine $f(R)$ gravity have not yet been
considered and not much attention has yet been paid to vacuum and
non-vacuum solutions in Gauss--Bonnet gravity and metric-affine $f(R)$
gravity, or to strong gravity tests in either of these two theories.

\section{The Nearly Newtonian regime in $f(R)$ gravity}
\label{newtf}

\subsection{Metric $f(R)$ gravity}

Within the context of metric $f(R)$ gravity, the subjects of the
Newtonian limit, the post-Newtonian expansion and confrontation with
Solar System experiments, have long been debated. A large number of
papers have been published and a lot of subtleties have been revealed.
\cite{chiba,Dick:2003dw,sousa,olmo1,olmo2,Sotiriou:2005xe,navaco,capotroisi,Capozziello:2006fa,Erickcek:2006vf,Jin:2006if,Chiba:2006jp,Zhang:2007ne}.
Since it is not possible to extensively review all of the works in
this subject, we will attempt to focus on the major points and guide
the reader through the literature.

As we have seen in Chapter \ref{equivtheor}, metric $f(R)$ gravity is
dynamically equivalent to a Brans--Dicke theory with a potential
$V(\phi)$ and Brans--Dicke parameter $\omega_0=0$. Solar System
constraints on Brans--Dicke theories are quite well known (see Section
\ref{actfieldst} and \cite{bertotti}) and, therefore, it is natural to
exploit this equivalence in order to derive such constraints for
metric $f(R)$ gravity. This is indeed what was done in \cite{chiba}. 

The PPN parameter $\gamma$ is given in terms of $\omega_0$ as 
\cite{willbook}
\be
\gamma=\frac{\omega_0+1}{\omega_0+2}.
\ee
 Thus, in our case where $\omega_0=0$ one gets $\gamma=1/2$. Obviously
this value is far below the current bound, $|\omega_0 |> 40\,000$.
However, this bound only applies for very light scalar fields, {\em
i.e.~}scalars with very small effective masses. A large mass for the
scalar implies that the force mediated by it would be short range and
would not affect the results of Solar System experiments. In
Brans--Dicke theory the square of the effective mass of the scalar is
given by the second derivative of its potential evaluated at the
minimum (extremum). 

When one expresses metric $f(R)$ gravity as a Brans--Dicke theory, the
functional form of the potential depends on $f$. In \cite{chiba}
attention was focused on the model of \cite{cdtt} for which
 \be
\label{cdttf}
f(R)=R-\frac{\mu^{4}}{R},
\ee
 and $\mu\sim 10^{-42}$ GeV. The effective mass of the potential was
evaluated for $R\sim H_0^2\sim \mu$ and it was found to be of the
order of $\mu^2$. This is clearly a very small value and therefore the
conclusion of \cite{chiba} was that this model is ruled out.
Additionally, even though it is possible to construct sophisticated
models in order to make the scalar heavy (see for example
\cite{nojodiposneg}), this requires significant fine tuning of the
parameters and in general models with $1/R$ terms will lead to a very
small mass for the scalar.

As already mentioned, however, the effective mass is given by the
second derivative of the potential at the extremum, {\em i.e.~}at a
point where the first derivative of the potential vanishes. It was
pointed out in \cite{olmo1,olmo2} that, even though this is indeed the
case for a general Brans--Dicke theory
\cite{willbook,wagoner,steinwill}, having the scalar satisfying the
extremum condition cannot be trivially assumed for the $\omega_0=0$
case. The equivalence with metric $f(R)$ gravity requires that the
Jordan frame potential $V(\phi)$ is given by \cite{olmo1}
 \be
\label{Vf}
V(\phi)=R f'-f
\ee
and that 
\be
V'(\phi)=R.
\ee
 Since in a post-Newtonian expansion $R=R_0+\sigma(t,x)$, where $R_0$
is the background value and $\sigma(t,x)$ denotes the local deviation
from this value, the extremum condition is not generically satisfied (neither $R_0$ nor $\sigma(t,x)$ have to vanish).
In this sense, $\omega_0=0$ Brans--Dicke theory constitutes an
exception and standard results related to the post-Newtonian expansion
should not be trusted according to \cite{olmo1}. In the same paper the
post-Newtonian expansion was re-developed. However, the results were
not qualitatively different from those of \cite{chiba} and the
cosmologically interesting $f(R)$ models with $1/R$ terms were again
ruled out.

Contemporarily with \cite{chiba}, Dick considered the Newtonian limit
of metric $f(R)$ gravity without resorting to the equivalent
Brans--Dicke theory \cite{Dick:2003dw}. The approach was based on the
more standard linearized perturbative expansion. However, this
expansion was performed around a de Sitter background, since this is
the generic maximally symmetric solution for metric $f(R)$ gravity.
Again, attention was focused on the $1/R$ models and, again, these 
were ruled out.

 Such a perturbative treatment requires a Taylor expansion to be made
for $f(R)$ and $f'(R)$ around their background values. This is easy to
see, since the field equations, eq.~(\ref{metf}), involve these
functions. As pointed out in \cite{Sotiriou:2005xe}, even
though the results of \cite{Dick:2003dw} might well be correct, the
convergence of these expansions was not checked and relevant higher
orders were truncated {\it ad hoc}. For example, since
 \be
f(R)=f(R_0)+f'(R_0) R_1+\frac{1}{2}f''(R_0) R_1^2+\ldots
\ee
if we use the model of eq.~(\ref{cdttf}) we get
\be
\label{expans1}
f(R)=f(R_0)+\left(1+\frac{\mu^4}{R_0^2}\right) R_1-\frac{1}{2}\frac{2 \mu^4}{R_0^3} R_1^2+\ldots
\ee
 where now $R_0=\mu^2$. It is then easy to see that the second term on
the right hand side of the above equation is of the order of $R_1$,
whereas the third term is of the order of $R_1^2/a$. Therefore, in
order to truncate before the third term, one needs $R_1\gg R_1^2/a$ or
 \be
\label{cond1}
\mu^2\gg R_1. 
\ee
 The evolution of $R$ is governed by the trace of the field equation 
which for this model takes the form
\be
\label{dyn}
-R+\frac{3 \mu^4}{R}-\frac{6 \mu^4}{R^3}\nabla^2 R+\frac{18 \mu^4}{R^4}\nabla^{\mu}R\nabla_{\mu}R=8\,\pi\,G\, T,
\ee
 where we have denoted $8\,\pi\,G$ by $\kappa$. It is therefore not
straightforward to judge whether the condition (\ref{cond1}) is indeed
satisfied.

The same issue is relevant also for the approach of \cite{olmo1} where
the equivalent Brans--Dicke theory is used, since one has to expand the
potential of the scalar field $V(\phi)$ around a background value
$\phi_0$ in order to arrive the post-Newtonian expansion. Since
$V(\phi)$ is given in terms of $f(R)$ by eq.~(\ref{Vf}), it is
reasonable to assume that any problematic behaviour in the expansion
of $f(R)$ and $f'(R)$ might be inherited by the expansion of
$V(\phi)$. Let us stress that this is not to say that the results of
\cite{Dick:2003dw,olmo1} are necessarily incorrect, but merely that a
more detailed and rigorous approach is required.

Another point is that part of the debate about the Newtonian and
post-Newtonian limits of metric $f(R)$ gravity was based on a quite
common misconception: that the existence of the Schwarzschild--de
Sitter solution in vacuum guarantees that the Solar System tests will
be passed (see for instance \cite{rama,Multamaki:2006zb}). To be more
explicit, let us consider the trace of the field equations,
eq.~(\ref{metftrace}):
 \be
\label{metftrace2}
f'(R)R-2f(R)+3\Box f'(R)=8\,\pi\,G\,T,
\ee
 In vacuum $T=0$. If we search for solutions for which the Ricci 
scalar is constant, then $\Box f'(R)=0$ and the equation reduces to
\be
\label{metalg}
f'(R)R-2f(R)=0,
\ee
 where $R$ is now a constant. Eq.~(\ref{metalg}) then becomes an
algebraic equation. We can use this equation to write the field
equation (\ref{metf}) in vacuum as
 \be
\label{metf2}
R_{\mu\nu}-\frac{C}{4} g_{\mu\nu}=0,
\ee
 where $C$ is the constant value of $R$ (see also the last paragraph
of Section \ref{metricfieldeq}). Since eq.~(\ref{metf2}) is formally
the same as the field equation of General Relativity with a
cosmological constant in vacuum, we know that, according to the sign
of $C$, the static spherically symmetric solutions will be
Schwarzschild--de Sitter or Schwarzschild--anti-de Sitter. The mere
existence of these solutions does not, however, have any implication
for the Newtonian and post-Newtonian limits. As was correctly pointed
out in \cite{Erickcek:2006vf}, these are not the unique spherically
symmetric static solutions ($R$ does not have to be a constant) and in
order to find the correct vacuum solution for the exterior of a
spherically symmetric star, one has to search for the solution that
can be properly matched to the interior. The results of
\cite{Erickcek:2006vf} support the findings of \cite{chiba}.

One more issue that was presented as a drawback for the use of the
equivalent Brans--Dicke theory for deriving Solar System constraints
was that of \cite{Faraoni:2006hx}. The claim there was that, since the
equivalence between the two theories enforces the requirement $f''\neq
0$ and, on the other hand, in the weak field regime $f''\rightarrow
0$, the equivalence should break down at this limit and bounds derived
though the equivalent Brans--Dicke theory should not be considered
trustworthy. This claim was later retracted in \cite{Faraoni:2007yn}.
In fact, as pointed out also in \cite{Olmo:2006eh}, the condition
$f''\neq 0$ is not needed if the equivalence between the two theories
is shown at the level of the field equations, instead of using the
action, and it then constitutes a superfluous condition.

Later works appear to resolve some of the issues raised earlier
concerning the validity of the results of
\cite{chiba,Dick:2003dw,olmo1}. In \cite{Jin:2006if}, the approach of
\cite{Erickcek:2006vf} was followed and the results were extended to
more general models. The outcome was that if one properly takes into
account the matching with an interior solution, then only very special
models which are very close to General Relativity with a cosmological
constant can pass the Solar System tests and at the same time give
interesting late time cosmological phenomenology. In
\cite{Chiba:2006jp}, the derivation of constraints from the Solar
System tests by means of using the equivalent Brans--Dicke theory was
considered again and attention was paid to ensuring that the Taylor
expansions of $f(R)$ and $f'(R)$ were well defined and dominated by
terms that are linear in deviations away from $R=R_0$ (as proposed in
\cite{Sotiriou:2005xe}). Again the outcome was that the results of
\cite{chiba, olmo1} are indeed valid. However, different opinions are
still present \cite{Zhang:2007ne}.

To summarise: after a long debate, it seems that most of the models of
metric $f(R)$ gravity that have been proposed as solutions to the dark
energy problem and which therefore include $1/R$ terms, do not have
correct Newtonian and post-Newtonian limits. Exceptions to this do
exist, but significant fine tuning is required, to the extent that it
can be characterised as unnatural. It should be mentioned that this is
not the case for some other models which lead to interesting early
time cosmological phenomenology, such as the Starobinski model
($f(R)=R+\epsilon R^2$) \cite{starobinski}.

\subsection{Palatini $f(R)$ gravity}
\label{postnewtpal}

The Newtonian and post-Newtonian limits of Palatini $f(R)$ gravity
have also been a matter of some debate
\cite{mengwangnewt,barraco,Allemandi:2005tg,Sotiriou:2005xe,Allemandi:2006bm,Ruggiero:2006qv},
similarly to metric $f(R)$ gravity. Again, most of the attention has
focused on models with terms inversely proportional to the scalar
curvature, since these are the cosmologically motivated ones. The two
first results in this direction were in clear contradiction.  In
\cite{mengwangnewt}, Meng and Wang claimed that all models with
inverse powers of the scalar curvature in the action give a correct
Newtonian limit. On the other hand, in \cite{barraco} it was claimed
that this is not true and that there are constraints on the form of
the Lagrangian. However, it was shown in \cite{Sotiriou:2005xe} that
both of these results suffered from a serious problem.

Let us see this in more detail. In \cite{mengwangnewt} and
\cite{barraco}, the authors expand around de Sitter in order to derive
the Newtonian limit.  We can write
 \be
{\cal R}={\cal R}_0+{\cal R}_1,
\ee
 where ${\cal R}_0$ is the scalar curvature of the background de
Sitter spacetime and ${\cal R}_1$ is the correction to ${\cal R}_0$,
including all possible terms, with ${\cal R}_1/{\cal R}_0$ being
considered as being a small quantity. We will need to calculate
$f({\cal R}_0+{\cal R}_1)$ and $f'({\cal R}_0+{\cal R}_1)$. The usual
approach is to Taylor expand around ${\cal R}={\cal R}_0$ and keep
only the leading order terms in ${\cal R}_1$ but we will show that
this cannot be done in the present context because ${\cal R}_1/{\cal
R}_0$ is not small.

Take as an example the model of \cite{cdtt} studied by Vollick
\cite{vollick} in the Palatini formalism. Then
 \be
\label{cdtt2}
f({\cal R})={\cal R}-\frac{\epsilon_2^2}{{\cal R}},
\ee
and $\epsilon_2\sim 10^{-67}(\textrm{eV})^2\sim 10^{-53}\textrm{m}^{-2}$.
Expanding, we get
\be
f({\cal R})=f({\cal R}_0)+f'({\cal R}_0) {\cal R}_1+\frac{1}{2}f''({\cal R}_0) {\cal R}_1^2+\ldots
\ee
and, using (\ref{cdtt2}), we get
\be
\label{expans}
f({\cal R})=f({\cal R}_0)+\left(1+\frac{\epsilon_2^2}{{\cal R}_0^2}\right) {\cal R}_1-\frac{1}{2}\frac{2 \epsilon_2^2}{{\cal R}_0^3} {\cal R}_1^2+\ldots
\ee
 where now ${\cal R}_0=\epsilon_2$. It is then easy to see that the
second term on the right hand side of the above equation is of the
order of ${\cal R}_1$, whereas the third term is of the order of
${\cal R}_1^2/\epsilon_2$. Therefore, in order to truncate before the
third term, one needs ${\cal R}_1\gg {\cal R}_1^2/\epsilon_2$ or
 \be
\label{cond}
\epsilon_2\gg {\cal R}_1. 
\ee
 Note that this is not any exceptional constraint. ${\cal
R}_0\sim \epsilon_2$ and so this is the usual condition for being able to
truncate non linear terms in a Taylor expansion.

Let us now return to eq.~(\ref{paltrace}). For the model of
eq.~(\ref{cdtt2}) this gives
 \be
\label{rt}
{\cal R}=\frac{1}{2}\left(-8\,\pi\,G\,T\pm\sqrt{ (8\pi\,G)^2\, T^2+12 \epsilon_2^2}\right).
\ee

When discussing whether a theory has a good Newtonian limit, we are in
practice checking whether the field equations reduce to a high
precision to the Poisson equation under the following assumptions:
energy densities should be small enough so that there are no strong
gravity effects, and velocities related to the motion of the matter
should be negligible compared to the velocity of light. At the same
time, energy densities should be high enough so that the system under
investigation can be considered gravitationally bound\footnote{For
example, in General Relativity with a cosmological constant one could
consider, even on non-cosmological scales, densities low enough so
that the correction coming from the cosmological constant dominates
with respect to the matter density in the Poisson equation. This, of
course, would not imply that this model does not have a correct
Newtonian limit}. 

It is clear from eq.~(\ref{rt}) that the value of ${\cal R}$, and
consequently ${\cal R}_1$, is algebraicly related to $T$. This
already implies that whether or not the condition (\ref{cond}) is
satisfied will critically depend on the value of the energy density.
To demonstrate this, let us pick a typical example of a density
satisfying the weak field limit criteria: the mean density of the
Solar System, $\rho\sim 10^{-11} \textrm{gr/cm}^3$. For this value
$\left|{\epsilon_2/8\pi\,G\,T}\right|\sim 10^{-21}$, where $T\sim
-\rho$. The ``physical'' branch of the solution given in
eq.~(\ref{rt}) seems to be the one with the plus sign in front of the
square root. In fact, given that $T<0$, on this branch it is ensured
that the matter leads to a standard positive curvature in a strong
gravity regime. Then
 \be
\label{r}
{\cal R}\sim -8\pi\,G\,T-\frac{3\epsilon_2^2}{8\pi\,G\, T}
\ee
 and ${\cal R}_1\sim -8\pi\,G\,T\sim 8\pi\,G\,\rho$.  Thus
$\epsilon_2/{\cal R}_1\sim 10^{-21}$ and it is now evident that
condition (9) does not hold for some typical densities that could be
related to the Newtonian limit.

Note that the situation does not improve even if we choose the
``unphysical'' branch of eq.~(\ref{rt}) which has a minus sign in
front of the square root. In fact, in this case ${\cal R}_1\sim
\epsilon_2 [3 \epsilon_2/(8\pi\,G\,T)+\sqrt{3}]$ and so the correction to
the background curvature is of the order $\epsilon_2$ and not much smaller than
that, as required in order to truncate before the higher order terms
in the expansion eq.~(\ref{expans}).

In \cite{barraco}, this fact was overlooked and only linear terms in
${\cal R}_1$ were kept in the expansion of $f({\cal R})$ and $f'({\cal
R})$ around ${\cal R}_0$. In \cite{mengwangnewt} even though it is
noticed in the final stages of the analysis and is actually used in
order to neglect some terms, the authors do not take it into account
properly from the beginning, keeping again only first order terms
(eq.~(11) of \cite{mengwangnewt} for example).

An alternative way to see the dependence of the weak field limit on
the energy density is the following. We already know (see Section
\ref{manfield}) that in Palatini $f(R)$ gravity the connection is the
Levi--Civita connection of the metric
 \be
h_{\mu\nu}=f'({\cal R})g_{\mu\nu}.
\ee
 For the model given by eq.~(\ref{cdtt2}) then, and if we define 
$\epsilon=\epsilon_2^2/{\cal R}^2$, eq.~(\ref{palf12}) takes the form
\be
\label{ein2}
(1+\epsilon){\cal R}_{\mu\nu}-\frac{1}{2}(1-\epsilon){\cal R} g_{\mu\nu}=8\pi\,G\,T_{\mu\nu},
\ee
and
\be
\label{hg}
h_{\mu\nu}=(1+\epsilon)g_{\mu\nu}.
\ee
Due to eq.~(\ref{paltrace}), $\epsilon$ depends only on $T$.
Combining eqs.~(\ref{hg}) and (\ref{ein2}) we get
\be
\label{ein}
{\cal R}_{\mu\nu}-\frac{1}{2}{\cal R} h_{\mu\nu}+\epsilon\left(\frac{{\cal R}}{1+\epsilon}h_{\mu\nu}+{\cal R}_{\mu\nu}\right)=8\pi\,G\,T_{\mu\nu}.
\ee
 Note that up to this point no approximation or truncation has been
used. We have merely expressed the left hand side of the field
equation for the metric in terms of quantities depending only on the
$h_{\mu\nu}$ metric, which is conformal to $g_{\mu\nu}$. However,
using eq.~(\ref{rt}) and (\ref{r}) we see that $\epsilon \sim
10^{-42}$ if we consider the mean density of the Solar System as
before and is even smaller for higher densities. Therefore the two
metrics are practically indistinguishable in such cases, due to
eq.~(\ref{hg}). Thus we can use the $h$ metric to derive the Newtonian
limit.

As usual, it is expected that a suitable coordinate system can be
found in which
 \be
h_{\mu\nu}=\eta_{\mu\nu}+h^1_{\mu\nu},\qquad |h^1|\ll 1,
\ee
 where $h^1_{\mu\nu}$ denotes the correction with respect to the
Minkowski metric. Then the first two terms of eq.~(\ref{ein}) will
give the standard Newtonian limit and the last two terms will give a
negligible contribution, since they are suppressed by the $\epsilon$
coefficient.  A deviation of the order of $10^{-42}$ is far below the
accuracy of any known experiment. In fact, one can consider densities
several orders of magnitude smaller and still get corrections which
will be far below experimental accuracies.

A critical point is that we assumed here that the metric is flat plus
a small correction instead of de Sitter plus a small correction. Note,
however, that we are not claiming that we are expanding around the
background or any corresponding maximally symmetric spacetime. We are
merely asking for the matter to account for the deviation from
flatness, which is the basic concept related to the Newtonian limit. 
In any case, de Sitter is essentially identical to Minkowski for the
densities discussed, and the important corrections to the metric come
from the local matter, not from considerations of the universe as a
whole.

According to the above, the Lagrangian of eq.~(\ref{cdtt2}) can give a
perfectly good Newtonian limit for some typical weak-field-limit
densities. The approach can be extended to more general Lagrangians.
Indeed, for a general function $f$, eq.~(\ref{ein}) will be
 \be
\label{eingen}
{\cal R}_{\mu\nu}-\frac{1}{2}{\cal R} h_{\mu\nu}+(f'-1)\left({\cal R}_{\mu\nu}-\frac{{\cal R}}{2f'}h_{\mu\nu}\right)=8\pi\,G\,T_{\mu\nu}.
\ee
 Since due to eq.~(\ref{paltrace}) ${\cal R}$ and consequently
$f'({\cal R})$ are functions of the energy density, the deviation of
$f'$ from $1$ will always depend on it. This dependence of the weak
field limit on the energy density is a novel characteristic of
Palatini $f(R)$ gravity \footnote{This discussion clarifies why in
Section \ref{palgencosm} we required that at least in some regime
$f'\rightarrow 1$. Additionally, it is now apparent that if one
selects the model of eq.~(\ref{ans3}) and assumes a typical value for
the density, then stringent constraints on the value of $\epsilon_1$
can be placed in the spirit of \cite{Sotiriou:2005cd}.}. 

Similar things can be said if the problem is approach via the
equivalent Brans--Dicke theory. This was studied in \cite{olmo1}. Note
that the usual bounds coming from Solar System experiments do not
apply in the $\omega_0=-3/2$ case, which is equivalent to Palatini
$f(R)$ gravity. This is because the standard treatment of the
post-Newtonian expansion of Brans--Dicke theory, which one uses to
arrive at such bounds, is critically based on the assumption that
$\omega_0\neq -3/2$ and the term $(2\omega_0+3)$ frequently appears as
a denominator. Making this assumption is not necessary, of course, in
order to derive a post-Newtonian expansion, but is a convenience
choice, which allows for this otherwise general treatment. Therefore,
a different approach, such as the one followed in \cite{olmo1}, was
indeed required for the $\omega_0=-3/2$ case . Following the standard
assumptions of a post-Newtonian expansion around a background
specified by a cosmological solution \cite{willbook}, the following
relations were derived for the post-Newtonian limit
 \bea
\label{olmo11}
-\frac{1}{2}\nabla^2\left[h^1_{00}-\Omega(T)\right]&=&\frac{8\pi\,G\,\rho-V(\phi)}{2\phi},\\
\label{olmo21}
-\frac{1}{2}\nabla^2\left[h^1_{ij}+\delta_{ij}\Omega(T)\right]&=&\left[\frac{8\pi\,G\,\rho+V(\phi)}{2\phi}\right],
\eea
 where $V$ is the potential of the scalar field $\phi$ and
$\Omega(T)\equiv \log[\phi/\phi_0]$. The subscript $0$ in $\phi_0$,
and in any other quantity from here on, denotes that it is evaluated
at $T=0$. Note at this point that normalization by $\phi_0$ in this
definition is not required. In \cite{olmo1}, the constant
$\log(\phi_0)$ was just added inside the brackets on the left hand
side of eq. (\ref{olmo11}) (and subtracted in eq.  (\ref{olmo21}))
using the fact that the latter remains unchanged. Thus we are not
going to use it here and will refer to $\Omega(T)$ just as
$\Omega(T)=\log[\phi]$. 

The solutions of eqs. (\ref{olmo11}) and (\ref{olmo21}) are
\bea
\label{olmo1s}
h^1_{00}(t,x)&=&2 G_{\rm eff} \frac {M_0}{r}+\frac{V_0}{6\phi_0}r^2+\Omega(T),\\
\label{olmo2s}
h^1_{ij}(t,x)&=&\left[2\gamma G_{\rm eff} \frac {M_0}{r}-\frac{V_0}{6\phi_0}r^2-\Omega(T)\right]\delta_{ij},
\eea
 where $M_0 \equiv \phi_0 \int d^3 x' \rho(t,x')/\phi$. The
effective Newton constant $G_{\rm eff}$ and the post-Newtonian
parameter $\gamma$ are defined as
 \bea
G_{\rm eff}&=&\frac{G}{\phi_0}\left(1+\frac{M_V}{M_0}\right),\\
\gamma&=&\frac{M_0-M_V}{M_0+M_V},
\eea
 where $M_V\equiv (8\pi\,G\,)^{-1} \phi_0 \int d^3 
x'\left[V_0/\phi_0-V(\phi)/\phi\right]$. 

Even though we agree with the approach followed to derive eqs.
(\ref{olmo1s}) and (\ref{olmo2s}) and on their validity, we disagree
with the line of reasoning used by the author to argue that models
with inverse powers of the scalar curvature do not have a good
Newtonian limit. We will demonstrate this using, once again, the model
of eq.~ (\ref{cdtt2}).

As stated in different words in \cite{olmo1}, if we define the
Newtonian mass as $M_N\equiv \int d^3 x' \rho(t,x')$, the requirement
for a theory to have a good Newtonian limit is that $G_{\rm eff}
M_0$ is equal to $G M_N$, where $N$ denotes Newtonian and
$\gamma\sim 1$ to very high precision. Additionally, the second
term on the right hand side of both eq.~(\ref{olmo1s}) and
eq.~(\ref{olmo2s}) should be negligible, since it acts as a term
coming from a cosmological constant. $\Omega(T)$ should also be small
and have a negligible dependence on $T$. The above have to be true for
the range of densities relevant to the Newtonian limit, as discussed
before. Using the equation that related $V$ and $\phi$ with ${\cal
R}$ (see Chapter \ref{equivtheor})
 \begin{align}
\phi&=f',\\
V(\phi)&={\cal R}f'-f,
\end{align}
 one can easily show that
 \bea
\label{eq1ol}
\phi&=&1+\frac{\epsilon_2^2}{{\cal R}^2},\\
V(\phi)&=& 16\pi\,G\, \epsilon_2\sqrt{\phi-1}.
\eea
Additionally, for $T=0$, ${\cal R}=\sqrt{3} a$ and so
\bea
\label{eq2ol}
\phi_0&=&4/3,\\
\label{eq3ol}
V_0&=& 16\pi\,G\,\epsilon_2/\sqrt{3}.
\eea
 For the densities which we are considering, we can use the parameter 
$\epsilon$ defined above. Then
\be
\label{eq4ol}
V(\phi)=16\pi\,G\,\frac{\epsilon_2^2}{{\cal R}}=16\pi\,G\,\epsilon_2\sqrt{\epsilon},
\ee
 and $M_V\sim \epsilon_2$. It is easy to see, using eq. 
(\ref{eq1ol}), (\ref{eq2ol}), (\ref{eq3ol}) and (\ref{eq4ol}), that
 \bea
G_{\rm eff}&\approx&\frac{G}{\phi_0},\\
\gamma&\approx&1,
\eea
 and $\phi\approx 1$ plus corrections of order $\epsilon_2$ or smaller, which 
is well beyond the limit of any experiment.

Additionally
\be
\Omega(T)\equiv\log[\phi]=\log\left[1+\epsilon\right]\approx\log\left[1+\frac{\epsilon_2^2}{(8\,\pi\,G)^2  T^2}\right].
\ee
 $V_0$ is of the order of $\epsilon_2$, which is a perfectly
acceptable value, and $\Omega(T)$ is negligible at the densities being
considered and decreases even more when the density increases.
Therefore, our previous results are valid and theories including
inverse powers of the scalar curvature can have a correct Newtonian
limit in the Palatini formalism for a specific density range.

This result contradicts that reported in \cite{olmo1}, even though the
approach followed there seems to be satisfactory. The main reason for
this problem seems to be the following. In \cite{olmo1} the fact that
$\Omega(T)$ should have a mild dependence on $T$ is used to obtain a
constraint for the dependence of $\phi$ on $T$ (eq.~(26) of
\cite{olmo1}). Following a number of steps, this constraint is turned
into a constraint on the functional form of $f(R)$ (eq.~(37) of
\cite{olmo1}) and from this a conclusion is derived about its possible
nonlinearity. We disagree with this line of thought. Such inequalities
constrain merely the value of the relevant quantity at the point where
it is evaluated and not its true functional form. One could probably
use them to make some assumptions about the leading order term but not
to exclude any terms of a different form, as long as they are
negligible with respect to the leading order for the relevant values
of $R$. This, for example, is the case for the model discussed above.
Any constraint placed by the Newtonian limit has to hold over a
certain range of relevant densities (and consequently curvatures), and
not for all densities as implied in \cite{olmo1}.

However, the dependence of the outcome of the Newtonian limit on the
energy density is not only surprising but also problematic. Even
though, according to the above, we can expect that inside or outside a
cloud of matter of a typical weak-field density, gravity may behave in
the same way as in Newtonian gravity, this is definitely not the end
of the story. As correctly pointed out in \cite{olmo1} the dependence
on the energy density, and especially that coming from $\Omega(T)$,
signals a problem. One has to take into account that matter can also
come as a perturbation. Indeed this is the case for Solar System tests
(light deflection, Shapiro time delay, {\em etc}.) which do not necessarily
examine gravitationally bound systems but are essentially vacuum tests
in which the presence of matter ({\it e.g.~}Solar winds) has to be
taken into account as a correction \cite{willbook}. Therefore the
relevant densities can be many orders of magnitude smaller than those
discussed above. In addition to this, in eqs.~(\ref{olmo1s}) and
(\ref{olmo2s}) $\Omega(T)$ is algebraically related to the metric,
which implies that the metric depends directly on the density and not
on some integral over it, as would be expected. This creates doubts
about how the theory would behave if a very weak point source
(approximated by a delta function) is taken into account as a
perturbation. 

Due to the above, we can conclude the following: Even though it can be
shown that, for some typical energy densities, an acceptable weak
field limit can be recovered from Palatini $f(R)$ gravity, this
provides no guarantee that the theory passes Solar System tests.
Additionally, the direct dependence of the outcome on the density,
signals the existence of a deeper problem. In Section \ref{nogo} this
problem will become apparent through a completely different approach,
so we will refrain from saying more here.

Before closing, it should be mentioned that, similarly to the metric
formalism, also in Palatini $f(R)$ gravity the existence of the
Schwarzschild--de Sitter solution as a vacuum spherically symmetric
solution has triggered some confusion concerning Solar System tests.
As shown in Section \ref{manfield}, Palatini $f(R)$ gravity reduces in
vacuum to General Relativity with a cosmological constant. This
implies that this theory retains a useful characteristic of GR: the
exterior spherically symmetric solution is unique (\textit{Birkhoff's
theorem}) \footnote{This does not hold for metric $f(R)$ gravity, as
discussed in the previous section.}. Depending to the sign of the
effective cosmological constant, the solutions are either
Schwarzschild--de Sitter or Schwarzschild--anti-de Sitter. This was
interpreted in \cite{Allemandi:2006bm,Ruggiero:2006qv} as an
indication that the only parameter that can be constrained is the
effective cosmological constant and therefore models that are
cosmologically interesting, for which this parameter is very small,
trivially satisfy Solar System tests. However, even though the
uniqueness of the solution implies that here we will not face problems
like those discussed in the previous section for metric $f(R)$ gravity
(concerning which exterior solution can properly match an interior
one, {\em etc}.), this claim is still incorrect. It should be clarified that
the existence of a spherically symmetric vacuum solution solution,
irrespective of its uniqueness, is not enough to guarantee a good
Newtonian limit for the theory. For instance, the Schwarzschild--de
Sitter solution has two free parameters. One of them can be associated
with the effective cosmological constant in a straightforward manner.
However, it is not clear how the other parameter, which in General
Relativity is identified as the mass of the object in the Newtonian
regime, is related to the internal structure of the object in Palatini
$f(R)$ gravity. Assuming that it represents the mass defined in the
usual way is not enough of course. The essence of deriving the
Newtonian limit of the theory is exactly in deriving an explicit
relation for this quantity and showing that it agrees with the
Newtonian expression.

\section{Curvature scalar instability in $f(R)$ gravity}
\label{palinst}

Besides the post-Newtonian limit, there is another problem related to
the weak field regime of metric $f(R)$ gravity which was pointed out
soon after the introduction of the model with a $1/R$ term
\cite{cdtt}: an instability in the equation governing the dynamics of
the scalar curvature $R$ was discovered by Dolgov and Kawasaki
\cite{DolgovKawasaki} in the presence of matter for the specific model
where $f(R)=R-\mu^4/R$, with $\mu$ being a constant. This instability
is not just a special characteristic of this model but occurs in a
more general class of models \cite{valerio1}.

Let us briefly review the results of \cite{DolgovKawasaki,valerio1}.
By contracting eq.~(\ref{metf}) one gets
 \be
\label{inst}
3\Box f'(R)+f'(R)R-2f(R)=8\pi\,G\, T,
\ee
 where $T=g^{\mu\nu}T_{\mu\nu}$. Following \cite{valerio1}, we can 
write $f(R)=R+\epsilon \varphi(R)$, where $\epsilon$ is a constant.
%. The eq.~(\ref{inst}) takes the form
%\be
%\Box R+\frac{\varphi '''}{\varphi ''}\nabla^\mu R \nabla_\mu R+\frac{(\epsilon \varphi '-1)}{3\epsilon \varphi ''} R=\frac{8\pi\,G\, T}{3\epsilon \varphi ''}+\frac{\varphi}{3\epsilon \varphi ''}.
%\ee
%Finally, 
 If we consider a small region in a weak field regime within matter,
we can assume that 
%\begin{equation}\label{4}
$g_{ab}=\eta_{ab}+h_{ab}$ and $R=-8\pi\,G\,T +R_1$, 
%\end{equation}
where $\eta_{ab}$ is the Minkowski metric and $\left| R_1/(8\pi\,G\,
T)\right| \ll 1$. In this approximation, and to first order in $R_1$,
eq.~(\ref{inst}) gives
\begin{align}
\label{veq}
\ddot{R}_1 &-\nabla^2 R_1 -\frac{16\pi\,G\,\varphi '''}{\varphi ''}\, (\dot{T}\dot{R}_1-\vec{\nabla}T \cdot \vec{\nabla}R_1)\\&+\frac{1}{3\varphi ''} \left( \frac{1}{\epsilon}-\varphi' \right) R_1=8\pi\,G\,\ddot{T}-8\pi\,G\,\nabla^2 T -\, \frac{\left(8\pi\,G\,T\varphi '+\varphi \right)}{3\varphi ''}\nn,
\end{align} 
 where an over-dot denotes differentiation with respect to time, while
$\vec{\nabla}$ and $\nabla^2$ denote the gradient and Laplacian
operators respectively in Euclidean three-dimensional space.

The instability occurs if $\varphi ''=f''(R)<0$ and $\epsilon$ is very
small, since the coefficient of the last term on the left hand side of
eq.~(\ref{veq}) is the square of an effective mass (notice the
resemblence with a damped harmonic oscilator). As already mentioned in
\cite{valerio1}, it can be considered as an instability in the gravity
sector. Because of this, and since it appears in the equations
governing the dynamics of the curvature scalar, we refer to it as the
``curvature scalar instability''. Theories with $f''(R)>0$ will be
stable irrespective of the value of $\epsilon$. However, for several
models that lead to the desired cosmological dynamics at late times,
$\epsilon$ is indeed very small and $f''(R)$ is indeed negative. A
typical example is the model of \cite{cdtt}, where
$\varphi(R)=-\mu^4/R$, with $\mu\sim 10^{-33}$eV, and the time-scale
for the instability to occur is of the order of $10^{-26}$ s
\cite{DolgovKawasaki}.

All of the above is with reference to the metric formalism. Let us now
consider the Palatini formalism. Following the lines of
\cite{Sotiriou:2006sf}, we will argue that such an instability cannot
occur in this case irrespective of the form of the Lagrangian.
Contracting eq.~(\ref{palf12}) gives eq.~(\ref{paltrace}), which we
repeat here for the convenience of the reader:
 \be
\label{stru}
f'({\cal R}){\cal R}-2f({\cal R})=8\pi\,G\,T.
\ee
 Recall that ${\cal R}$ is not the Ricci scalar of the metric. In
Section \ref{manfield}, we derived eq.~(\ref{confrel2}) in which
$R$ in expressed in terms of ${\cal R}$:
 \be
\label{R}
R={\cal R}-\frac{3}{2[f'({\cal R})]^2}\nabla_\mu f'({\cal R}) \nabla^\mu  f'({\cal R})+\frac{3}{ f'({\cal R})}\Box  f'({\cal R}).
\ee
 Now notice that eq.~(\ref{stru}) is an algebraic equation in ${\cal 
R}$ for a given $f({\cal R})$, which will have solutions of the form 
 %\be
%\label{theta}
${\cal R}=\theta(T)$,
%\ee
 where $\theta$ is some function. As has been mentioned several times
before, we are not interested in cases in which eq.~(\ref{stru}) has
no solutions or is identically satisfied ($f({\cal R})\propto {\cal
R}^2$), since these do not constitute viable choices for a low-energy
gravitational theory \cite{ffr,Sotiriou:2006qn}. 

We can now write eqs.~(\ref{R}) as
\begin{align}
\label{matter}
R=\theta(T)&-\frac{3}{2[f'(\theta(T))]^2}\nabla_\mu f'(\theta(T)) \nabla^\mu  f'(\theta(T))+\nn\\&+\frac{3}{ f'(\theta(T))}\Box  f'(\theta(T)),
\end{align}
or alternatively
%\be
%\label{Rnew}
$R=W(T)$,
%\ee
 where $W(T)$ is a function of $T$. This clearly demonstrates that the
Ricci scalar of the metric can be expressed directly as a function of
the trace of the stress-energy tensor. In fact, eq.~(\ref{matter}) is
a straightforward generalization of the contracted Einstein equation,
$R=-8\pi\,G\,T$. From the form of eq.~(\ref{matter}), it is clear that
no instability can occur in this case, since $R$ carries no dynamics
in eq.~(\ref{matter}), unlike eq.~(\ref{inst}).

Let us now analyse where this difference between the two formalisms
stems from. By generalizing the Lagrangian from $R$ or ${\cal R}$ one
inevitably adds a scalar degree of freedom \cite{Sotiriou:2006hs}.
However, as mentioned in Chapter \ref{equivtheor}, this degree of
freedom seems to be of a different nature in the two versions of the
theory. In the metric version, it is dynamical and therefore care
should be taken to ensure stability, whereas in the Palatini version
it is non-dynamical. This is related to the fact that the Palatini
formalism leads to second order field equations in the metric whereas
the metric formalism leads to fourth order field equations, but it can
also be easily seen by using the equivalence of $f(R)$ gravity and
scalar-tensor theory (see Chapter \ref{equivtheor} and references
therein). 

 As we have seen, the Brans--Dicke action equivalent to metric $f(R)$ 
gravity is
\be
\label{metactionH2new}
S_{}=\frac{1}{16\pi\,G}\int d^4 x \sqrt{-g} \left(\phi R-V(\phi)\right) +S_M(g_{\mu\nu},\psi),
\ee
 with $\omega_0=0$. In the Palatini formalism, the action will be 
formally the same apart from the fact the $R$ will become ${\cal R}$,
%\be
%\label{palactionH2}
%S=\frac{1}{16\pi\,G}\int d^4 x \sqrt{-g} \left(\phi {\cal R}-V(\phi)\right) +S_M(g_{\mu\nu}, \Gamma^{\lambda}_{\phantom{a}\mu\nu}, \psi),
%\ee
 but in this case it will not be a scalar-tensor theory with
$\omega_0=0$ since ${\cal R}$ is not the Ricci scalar of the metric
\cite{Sotiriou:2006hs}. However, if we use eq.~(\ref{R}) and 
$\phi=f'({\cal R})$, we get
 \be
\label{palactionH2d0new}
S_{pal}=\frac{1}{16\pi\,G}\int d^4 x \sqrt{-g} \left(\phi R+\frac{3}{2\phi}\partial_\mu \phi \partial^\mu \phi-V(\phi)\right) +S_M(g_{\mu\nu}, \psi),
\ee
which is indeed a scalar-tensor theory, but with $\omega_0=-3/2$.

The field equation of the scalar field in scalar-tensor theory is 
\be
\label{bdf22}
(2\omega_0+3) \Box \phi= 8\pi\,G\,T+\phi V'-2V.
\ee
 Note that $\phi$ is the extra degree of freedom of $f(R)$ gravity,
with respect to General Relativity. Using eq.~(\ref{bdf22}), it is
obvious that $\phi$ satisfies the field equations
 \bea
3\Box \phi+2V(\phi)-\phi V'(\phi)&=& 8\pi\,G\,T,\\
2V(\phi)-\phi V(\phi) &=& 8\pi\,G\,T,
\eea 
 in the metric and Palatini formalisms respectively. This demonstrates
that $\phi$ is indeed dynamical in the metric formalism, whereas it is
not dynamical in the Palatini formalism, as mentioned above. At this
point, it is worth mentioning that one should not be misled into
judging the dynamics of a non-minimally coupled field by the presence
or absence of a kinetic term in the action. There are no kinetic terms
for $\phi$ in action (\ref{metactionH2new}) but it is still dynamical.
Exactly the opposite holds for the Palatini formalism. The reason for
this is that both fields are coupled not only to the metric, but also
to its derivatives. Therefore, when varying the action with respect to
the metric and then integrating by parts in order to ``free'' $\delta
g^{\mu\nu}$, terms including derivatives of the scalar field are bound
to appear. Therefore, in the metric formalism, even though there are
no apparent kinetic terms for $\phi$ in the action, there will be
kinetic terms in the field equations. For Palatini $f(R)$ gravity,
$\omega_0=-3/2$ and this is the remarkable case where these kinetic
terms exactly cancel out the ones coming from the kinetic part of the
action.

To conclude, the curvature scalar instability discovered by Dolgov and
Kawasaki for metric $f(R)$ gravity places an additional constraint on
the form of the Lagrangian, whereas it is not present in the Palatini
formalism, irrespective of the functional form of $f$. It should be
stressed, however, that even though this instability does not occur in
Palatini $f(R)$ gravity, other types of instability might well be
present. For example, judging from the form of eq.~(\ref{matter}), it
is not difficult to imagine that specific forms of $f$ could lead to a
blow-up of the scalar curvature for small density perturbations around
a stable matter configuration. Such instabilities would be, of course, of a different
nature. This issue seems to be directly related to the problems with
the weak field limit of the theory discussed in the previous section
and it will be fully clarified in Section \ref{nogo}.

\section{Post-Newtonian expansion of Gauss--Bonnet\\ \mbox{gravity}}

In Section \ref{gbgrav} we presented the action and field equations of
Gauss--Bonnet gravity and in Section \ref{gbcosm} we studied its
cosmological applications. In order to confront the theory with Solar
System observations, as we have already done for metric and Palatini
$f(R)$ gravity, one needs the Post-Newtonian Parametrized expansion of
the theory.  This is the issue that will concern us in this section
and we will approach it along the lines of
Ref.~\cite{Sotiriou:2006pq}. We will not consider the exceptional
case, where $\lambda=0$ and the scalar field has no kinetic term in
the action. Such actions are dynamically equivalent to an action with
a general function of ${\cal G}$ added to the Ricci scalar (see
Section \ref{gbequiv}) and their Newtonian limit has been considered
in \cite{odi}.

We begin by bringing the field equations of the theory, namely
eqs.~(\ref{gbfield1}) and (\ref{gbfield2}), into a form more suitable
for our purposes. Taking the trace of eq.~(\ref{gbfield1}) and using
the definitions for the quantities $T^{\mu\nu}_\phi$ and
$T^{\mu\nu}_f$ given in eqs.~(\ref{sephi}) and (\ref{sef}), we get
 \be
\label{tracegb}
R= 8\,\pi\,G  \left[-T-T^\phi+2 (\Box f(\phi))R-4(\nabla^\rho\nabla^\sigma f(\phi))R_{\rho\sigma}\right],
\ee
 where $T=g^{\mu\nu}T_{\mu\nu}$ and
$T^\phi=g^{\mu\nu}T_{\mu\nu}^\phi$. Replacing eq.~(\ref{tracegb}) back
in eq.~(\ref{gbfield1}), the latter becomes:
 \bea
\label{gfield2}
R_{\mu\nu}&=& 8\,\pi\,G \Big[T_{\mu\nu}-\frac{1}{2}g_{\mu\nu}T+\frac{1}{2}\lambda\partial_\mu \phi \partial_\nu \phi+\frac{1}{2}g_{\mu\nu}V(\phi)+\nn\\& &+2(\nabla_\mu \nabla_\nu f(\phi)) R-g_{\mu\nu}(\Box f(\phi)) R-\nn\\& &-4(\nabla^\rho \nabla_{\mu} f(\phi))R_{\nu\rho}-4(\nabla^\rho \nabla_{\nu} f(\phi))R_{\mu\rho}+\nn\\& &+4(\Box f(\phi))R_{\mu\nu}+2g_{\mu\nu}(\nabla^\rho \nabla^\sigma f(\phi))R_{\rho\sigma}-\nn\\& &-4(\nabla^\rho \nabla^\sigma f(\phi))R_{\mu\rho\nu\sigma}\Big]
\eea

Following the standard approach for post-Newtonian expansions (see
\cite{willbook}), we choose a system of coordinates in which the
metric can be perturbatively expanded around Minkowski spacetime. We
write the metric and the scalar field as
 \be
\phi=\phi_{0}+\delta \phi,\qquad
g_{\mu\nu}=\eta_{\mu\nu}+h_{\mu\nu},
\ee
 where the value of $\phi_0$ is determined by the cosmological
solution. The perturbed field equations are
 \begin{align}
\label{perphi}
\lambda[\Box_{\rm flat}\delta\phi +(\delta\Box)\delta\phi] &-V^{\prime\prime}(\phi_0)\delta \phi-\frac12 V^{\prime\prime\prime}(\phi_0)(\delta \phi)^2\nn\\&+f^\prime(\phi_0) {\cal G}={\cal O}(\delta\phi^3,\delta\phi(h_{\mu\nu})^2,h_{\mu\nu}\dot\phi_0,h_{\mu\nu}\ddot\phi_0),\\
\label{00_perturbed_eq}
R_{00}= 8\,\pi\,G \Big\{T_{00}+\frac12T &-\frac12h_{00}T+\frac{1}{2}\lambda\partial_0 \delta\phi \partial_0 \delta\phi
+\frac12\lambda\,\dot\phi_0^2-\frac12V(\phi_0)\nn\\&+\frac12V'(\phi_0)\delta\phi\, (-1+h_{00})+f'(\phi_0)\Big[2(\partial_0\partial_0 \delta\phi) R\nn\\&+(\Box_{\rm flat} \delta\phi) R-8(\partial^\rho \partial_0 \delta\phi)R_{0\rho}
+4(\Box_{\rm flat}\delta\phi)R_{00}\nn\\&-2(\partial^\rho \partial^\sigma \delta\phi)R_{\rho\sigma}-4(\partial^\rho \partial^\sigma \delta\phi)R_{0\rho0\sigma}\Big]\Big\}\nn\\&+{\cal O}(\delta\phi^2h_{\mu\nu},\delta\phi^3,\dot\phi_0\delta\phi,\ddot\phi_0h_{\mu\nu},V(\phi_0)h_{00}),\\
\label{0i}
R_{0i}= 8\,\pi\,G  T_{0i} +{\cal O} (\delta\phi & h_{\mu\nu},\delta\phi^2,Th_{0i},\dot\phi_0\delta\phi,\ddot\phi_0h_{\mu\nu},V(\phi_0)h_{0i}),\\
\label{ij}
R_{ij}= 8\,\pi\,G \Big[T_{ij} +\frac12\delta_{ij}  & \left(-T+V'(\phi_0)\delta\phi+V(\phi_0)\right)\Big]\nn\\&+{\cal O}(\delta\phi h_{\mu\nu},\delta\phi^2,Th_{ij},\ddot\phi_0h_{\mu\nu},V(\phi_0)h_{ij}),
\end{align}
 where $\Box_{\rm flat}$ denotes the D'Alembertian of flat spacetime. 
Notice that, as usually done in scalar-tensor theories
\cite{wagoner,steinwill}, we have neglected all of the terms
containing derivatives of $\phi_0$ multiplying perturbed quantities
(e.g. $\dot\phi_0 \delta\phi$). This is due to the fact that $\phi_0$
changes on cosmological timescales and consequently one expects that
it remains practically constant during local experiments. Therefore
its time derivatives can be neglected as far as Solar System tests are
concerned. 

This can easily be verified by some order-of-magnitude analysis. Take
for instance Eq.~(\ref{00_perturbed_eq}): the terms containing a time
derivative of $\phi_0$ multiplying a perturbation are ${\cal
O}(\ddot{f}(\phi_0)h_{\mu\nu}/(r^2M_p^2))$ and ${\cal
O}(\dot\phi_0\delta\dot\phi/M_p^2)$, where $\dot\phi_0\sim H_0M_p$ and
$\ddot{f}\sim H_0^2$ ($M_p=(8\,\pi\,G)^{-1/2}$ is the Planck mass and
$H_0$ the present Hubble constant) and $h_{00}\sim h_{ij}\sim
r\delta\phi\sim h_{0i}/v\sim r^2 \delta\dot\phi/v\sim GM_\odot/r$
($r$ is the distance from the Sun, $M_\odot$ is the Solar mass and $v=\sqrt{GM_\odot/r}$). On the
other hand, the ${\cal O} (v^4)$ post-Newtonian correction to $R_{00}$
is $\sim (GM_\odot)^2/r^4\sim 10^{-55}{\cal
O}(\ddot{f}(\phi_0)h_{\mu\nu}/(r^2M_p^2),
\dot\phi_0\delta\dot\phi/M_p^2)$ even if $r$ is taken as large as
$1000$ AU. Therefore, the corrections coming from terms containing
time derivatives of $\phi_0$ multiplying perturbations are at least 55
orders of magnitude smaller than the post-Newtonian corrections, and
neglecting these terms cannot affect our results in any way. A similar
treatment applies to the terms containing the potential $V$
multiplying perturbed quantities (e.g.$V(\phi_0)h_{00}$): in order to
give a reasonable description of Cosmology, $V(\phi_0)$ should be of
the same order as the energy density of the cosmological constant and
these terms cannot therefore lead to any observable deviations at
Solar System scales. 

In the perturbed field equations, $V(\phi_0)$ and $\frac12\dot\phi^2$
are also present without multiplying perturbations. We will adopt a
different treatment for these simple $V(\phi_0)$ and
$\frac12\dot\phi^2$ terms: since they need to be of the same order as
the energy density of the cosmological constant, they will not lead to
any observational consequences as far as Solar System tests are
concerned (see \cite{sereno} and references therein). For the sake of
the argument, we will keep track of them but, due to their small
values, we can treat them as ${\cal O}(v^4)$ quantities following
\cite{sereno}. They will therefore not appear in the ${\cal O} (v^2)$
equations. As far as terms related to $V'(\phi_0)$ are concerned, we
intend to just keep track of them for the time being and discuss their
contribution later on. 

Up to now, we have just perturbed the field equations. The further
step needed to arrive at a post-Newtonian expansion is to expand the
perturbations of the metric and the scalar field in post-Newtonian
orders, {\em i.e.~}orders in the velocity $v$. The Parametrized
Post-Newtonian expansion requires that we expand $\phi$ and $h_{00}$
to ${\cal O} (v^4)$, $h_{ij}$ to ${\cal O} (v^2)$ and $h_{0i}$ to
${\cal O} (v^3)$. Therefore, we write
 \bea
\delta \phi&=&\two\delta \phi+\four\delta\phi\ldots\\
h_{00}&=&\two h_{00}+\four h_{00}\ldots\\
h_{ij}&=&\two h_{ij}+\ldots\\
h_{0i}&=&\three h_{0i}+\ldots
\eea
 where the subscript denotes the order in the velocity, {\em
i.e.~}quantities with a subscript $\two$ are ${\cal O}(v^2)$,
quantities with a subscript $\three$ are ${\cal O}(v^3)$, {\em etc}.

We can now write the field equations for each post-Newtonian order. 
To derive the parametrized post-Newtonian metric, we need to solve
these equations at each order and then use our results to solve to the
next order, and successively repeat the process. We start from the
field equation for the scalar, eq.~(\ref{perphi}). To order ${\cal
{O}}(v^2)$ this gives
 \be
\label{2phi}
\lambda\nabla^2(\two\delta\phi)-V^{\prime\prime}(\phi_0)\two\delta\phi=0\;:
\ee
 where $\nabla^2\equiv\delta_{ij}\partial_i\partial_j$. Note that,
since the metric is flat in the background, ${\cal G}={\cal O} (v^4)$.
This explains why we do not get any contribution from the coupling
with ${\cal G}$ in eq.~(\ref{2phi}). We want $\phi$ to take its
cosmological value at distances far away from the sources. This is
equivalent to saying that the perturbations due to the matter present
in the Solar System should vanish at cosmological distances, and this
can be achieved by imposing asymptotic flatness for the solution of
eq.~(\ref{2phi}), {\em i.e.~}$\two\delta\phi \to 0$ for $r \to
\infty$. This implies that
 \be
\two\delta\phi=0.
\ee

Now we turn our attention to eqs.~(\ref{00_perturbed_eq}), (\ref{0i})
and (\ref{ij}). To order ${\cal O}(v^2)$ for the components $00$ and
$ij$ and ${\cal O}(v^3)$ for the components $0i$, and after applying
the standard gauge conditions
 \be
\label{gauge1}
h^\mu_{i,\mu}-\frac{1}{2}h^\mu_{\mu,i}=0\;,\qquad
%\label{gauge2}
h^\mu_{0,\mu}-\frac{1}{2}h^\mu_{\mu,0}=\frac12 h^0_{0,0}\;,
\ee
 the field equations for the metric take the form
\bea
-\nabla^2{(\two h_{00})}&=& 8\,\pi\,G \rho\\
-\nabla^2{(\two h_{ij})}&=& 8\,\pi\,G \rho\delta_{ij}\\
{1\over 2}\left(\nabla^2{(\three h_{0i})}+\frac12(\two h_{00,j0})\right)&=& 8\,\pi\,G \rho v^i
\eea
 which, remarkably, is exactly the same as in General Relativity 
\cite{willbook}. The well-known solutions are
\bea
\two h_{00}&=&2U,\label{two_00}\\
\two h_{ij}&=&2U\delta_{ij}\label{two_ij},\\
\three h_{0i}&=&-\frac{7}{2}V_i-\frac{1}{2}W_i
\eea
where, following \cite{willbook}, we define the post-Newtonian 
potentials
\bea
U&=&G\int d^3 x'\frac{\rho(x',t)}{|x-x'|},\\
V_i&=&G\int d^3 x'\frac{\rho(x',t) v_i(x',t)}{|x-x'|},\\
W_i&=&G\int d^3 x'\frac{\rho(x',t) v^k(x',t)(x-x')_k(x-x')_i}{|x-x'|^3}.
\eea

We can already see that the theory has no deviations away from General
Relativity at order ${\cal O} (v^3)$: in particular it gives the
correct Newtonian limit. It is now easy to go one step further and
write down the perturbed equations that we need to ${\cal O}(v^4)$.
For the scalar field, using $\two\delta\phi=0$, we get
 \begin{equation}
\label{4phi}
\lambda\nabla^2(\four\delta\phi)-V^{\prime\prime}(\phi_0)\, \four\delta\phi+f^\prime(\phi_0)\, \four {\cal G}=0\;,
\end{equation}
with
\bea
\label{gbpert}
\four {\cal G}&=&(\two h_{00,ij})(\two h_{00,ij})-(\two h_{00,ii})(\two h_{00,jj})+(\two h_{ij,ij})^2+\nn\\& &+(\two h_{ij,kl})(\two h_{ij,kl})-(\two h_{ij,kk})(\two h_{ij,kk})-\nn\\& &-2 (\two h_{ij,kl})(\two h_{il,jk})+(\two h_{ij,kl})(\two h_{kl,ij})\;,
\eea
where we have again applied the gauge conditions (\ref{gauge1}).
Using eqs.~(\ref{two_00}) and~(\ref{two_ij}),  eq.~(\ref{gbpert}) becomes
\be
\label{4G}
\four {\cal G}=8\,U_{,kl}U_{,kl}-8\,(U_{,kk})^2\;.
\ee
The solution of eq.~(\ref{4phi}) is therefore
\be
\label{phisol}
\four\delta\phi=\frac{f^\prime(\phi_0)}{4\,\pi}\,
%\times\\&
\int d^3 x'\frac{\four {\cal G}(x',t)}{|x-x'|} e^{-\sqrt{V^{\prime\prime}(\phi_0)}|x-x'|}
\ee

The time-time component of the perturbed field equations for the
metric to ${\cal O}(v^4)$ is
 \bea
\label{4hoo}
\four R_{00}&=& 8\,\pi\,G \Big[(\four T_{00})+\frac12 (\four T)-\frac12(\two h_{00})(\two T)\nonumber\\
&&-\frac12V'(\phi_0)(\four\delta\phi)-\frac12V(\phi_0)+\frac12\lambda\dot\phi_0^2\Big],
\eea
 where we have already used the fact that $\two\delta\phi=0$. Note
also that no contribution coming from the coupling between $\phi$ and
the curvature terms in eq.~(\ref{gbfield1}) is present in the above
equations. This was to have been expected since in
eq.~(\ref{gbfield1}) these terms always have the structure of two
derivatives of $\phi$ times a curvature term, and so, due to the fact
that in the background the metric is flat and $\phi_0$ is slowly
varying, they can only contribute to orders higher than ${\cal
O}(v^4)$.

Let us discuss the contribution of the term proportional to
$V'(\phi_0)$. Using eqs. (\ref{phisol}) and (\ref{4G}), we can write
this term as an integral over the sources times a dimensionless
coefficient $ 8\,\pi\,G V'(\phi_0)f'(\phi_0)$. One can argue that
$V'(\phi)$ should be practically zero as far as the post-Newtonian
expansion is concerned \cite{wagoner,steinwill}. This is equivalent to
saying that the cosmological solution corresponds to a minimum of the
potential. Even though such assumptions are not exact, they are
accurate enough for our purposes. Note that even in cases where $V$
does not have a minimum, well-motivated models usually introduce
exponential forms for the potential and the coupling function,
{\textit i.e.~}$V=V_0 e^{-a\kappa\phi}$ and $f=f_0 e^{b\kappa\phi}$
where $\kappa^2=8\,\pi\,G$, $a$, $b$ and $f_0$ are of order unity
while $V_0$ is as small as the energy density of the cosmological
constant in order to guarantee that the theory will fit observations
related to the late-time cosmological expansion. This implies that,
since $ G \sim 1/M_p^2$, then $8\pi\,G V'(\phi_0)f'(\phi_0)$ is
dimensionless and of the order of the now renowned $10^{-123}$.
Therefore, we will not take the term proportional to $V'(\phi_0)$ into
account for what comes next. We will return to this issue shortly in
order to discuss how this choice affects the generality of our
results.

We can use the solutions for $\two h_{00}$ and $\two h_{ij}$, the
gauge conditions (\ref{gauge1}) and the standard post-Newtonian
parametrization for matter \cite{willbook} to write eq.~(\ref{4hoo})
as
 \begin{align}
-\nabla^2(\four h_{00}+2U^2-8{\boldsymbol\Phi_2})= 8\,\pi\,G  \Big[2\rho & \left(v^2-U+\frac{1}{2}\Pi-\frac{3p}{2\rho}\right)\nn\\&-V(\phi_0)+\frac12\lambda\dot\phi_0^2\Big],
\end{align}
 where $\Pi$ is the specific energy density (the ratio of the energy 
density to the rest-mass density) \cite{willbook} and
\be
\Phi_2=G\int d^3 x'\frac{\rho(x',t) U(x',t)}{|x-x'|}.
\ee
The solution to this equation is
\be
\four h_{00}=2U^2+4\Phi_1+4\Phi_2+2\Phi_3+6\Phi_4+\frac{ 8\,\pi\,G }{6}\left(V(\phi_0)-\frac12\lambda\dot\phi_0^2\right)|x|^2,
\ee
where
\bea
\Phi_1&=&G\int d^3 x'\frac{\rho(x',t) v(x',t)^2}{|x-x'|},\\
\Phi_3&=&G\int d^3 x'\frac{\rho(x',t)\Pi(x',t) v(x',t)^2}{|x-x'|},\\
\Phi_4&=&G\int d^3 x'\frac{p(x',t)}{|x-x'|}.
\eea
Therefore the metric, expanded in post-Newtonian orders, is
\bea
g_{00}&=&-1+2U-2U^2+4\Phi_1+4\Phi_2+\nn\\& &+2\Phi_3+6\Phi_4+\frac{ 8\,\pi\,G }{6}\left(V(\phi_0)-\frac12\lambda\dot\phi_0^2\right)|x|^2,\\
g_{0j}&=&-\frac{7}{2}V_i-\frac{1}{2}W_i,\\
g_{ij}&=&(1+2U)\delta_{ij},
\eea
 which, apart from the term related to
$V(\phi_0)-1/2\lambda\dot\phi_0^2$, is exactly the result that one
obtains for General Relativity. This term corresponds to the standard
correction normally arising from a cosmological constant and, since
$V(\phi_0)-1/2\lambda\dot\phi_0^2$ should indeed be of the same order
as the energy density of the cosmological constant, the contribution
of this term is negligible on Solar System scales. Since the metric is
written in the standard PPN gauge, one can read off the PPN parameters
\cite{willbook}. The only non-vanishing ones are $\gamma$ and $\beta$,
which are equal to $1$. Therefore, the theory discussed here seems to
be indistinguishable from General Relativity at the post-Newtonian
order.

The above implies that a gravitational theory with a scalar field
coupled to the Gauss--Bonnet invariant trivially satisfies the
constraints imposed on the post-Newtonian parameters by Solar System
tests, if the reasonable assumptions that we made for the values of
$V(\phi_0)$, $V'(\phi_0)$ and $f'(\phi_0)$ hold. This appears to be
due to the fact that the terms arising in the field equations for the
metric from the coupling between the scalar field and ${\cal G}$ in
the action always have the structure of two derivatives of $f$ times a
curvature term. Such terms do not contribute to the post-Newtonian
expansion to ${\cal O}(v^4)$. This is not the case for other possible
couplings of a scalar to a quadratic curvature term, such as $\phi
R^2$. Remarkably, the characteristic structure of such terms can be
traced back to the special nature of ${\cal G}$, {\em i.e.~}to the
fact that it is a topological invariant in four dimensions.

We now return to discuss how strongly our result depends on the
assumption that $V(\phi_0)$ and $V'(\phi_0)$ are reasonably small so
as to give a negligible contribution in the PPN expansion. This
assumption stems from the fact that $V(\phi_0)$ will play the role of
an effective cosmological constant if the theory is to account for the
late-time accelerated expansion of the universe and should therefore
be of the relevant order of magnitude. Additionally we expect that
$V'(\phi_0)$ will also be small enough so that its contribution can be
considered negligible, based on the fact that either the field
approaches a minimum at late times, or the potential is of the form
$V=V_0 e^{-a\kappa\phi}$, where $a$ is of order unity, and therefore
$V'(\phi_0)\sim\kappa V(\phi_0)$. The above should be true for all
models that lead to a reasonable cosmological phenomenology, once the
latter is attributed mainly to presence of the potential $V(\phi)$.

An alternative which one could consider is to attribute the
cosmological phenomenology to the coupling function $f(\phi)$.
However, it is important to stress that the values of $f'(\phi_0)$ and
$f''(\phi_0)$ should be suitable in order for the post-Newtonian
expansion to remain trustworthy. From eq.~(\ref{phisol}) we see that
non-trivial corrections will indeed be present at post-post-Newtonian
orders and, even though such corrections are normally subdominant, if
$f'(\phi_0)$ or $f''(\phi_0)$ is sufficiently large it can become
crucial for the viability of the theory. This was first observed in
\cite{farese} where the same theory, but without a potential $V$, was
confronted with Solar System observations, considering a nearly
Schwarzschild metric as an approximation. As mentioned before, the
potential plays the role of an effective cosmological constant if one
wants a theory that leads to a late-time accelerated expansion as in
\cite{nojodsas,nojodisam,carneu,neup,neup2,tsuji,tomigb1,tomigb2,cogneli}.
If this potential is not present, it is the coupling $f(\phi)$ between
the scalar field and the Gauss--Bonnet term that will have to account
for this phenomenology. In that case, it turns out that $f''(\phi_0)$
has to be of the same order as the inverse of the cosmological
constant, and this is enough to make the post-Newtonian approximation
break down. Fortunately, models with a potential do not suffer from
this problem and, in fact, $f$ is usually assumed to be of the form
$f=f_0 e^{b\kappa \phi}$ where both $f_0$ and $b$ are of order unity.
Therefore, as shown here and also predicted in \cite{farese},
reasonable models with a potential will pass the Solar System tests. 

There is yet another possibility: to consider models in which the
presence of both the potential and the coupling will somehow be
responsible for the cosmological phenomenology. In this sense, the
assumptions which we made here for $V(\phi_0)$, $V'(\phi_0)$ and
$f'(\phi_0)$ can become more loose and the resulting model would not
straightforwardly satisfy the Solar System constraints. This
possibility has very recently been considered in Ref.
\cite{Amendola:2007ni} and relevant constraints have been derived. In
any case, it is striking that, according to the analysis which we have
presented here, it is very easy to propose well-motivated models of
Gauss--Bonnet gravity which are practically indistinguishable from
General Relativity on Solar System scales. It is also worth commenting
that if the coupling function $f(\phi)$ is set to a constant, the
action (\ref{gbaction}) simply describes General Relativity with a
minimally coupled scalar field or, in other words, quintessence. This
implies that as long as the coupling is undetectable on Solar System
scales, the theory also cannot be distinguished from any successful
quintessence model on these scales.

Finally, let us discuss the possibility of including a second scalar
field in the action, coupled to the Gauss--Bonnet invariant, which
could, for example, be the dilaton. If this second scalar field is not
coupled to matter or to the Ricci scalar, then it can be treated using
the same approach as above. If the coupling functions with the
Gauss--Bonnet invariant and with the potential, if present, have
similar properties to those discussed above, we expect our result to
remain unaffected. Of course, there is also the possibility that the
dilaton is coupled to matter. This goes beyond the scope of the
present discussion since in this case the theory would be
phenomenologically different not only regarding Solar System tests,
but also for other aspects such as cosmological phenomenology,
covariant conservation of matter, the equivalence principle ({\em
e.g.} see ref.~\cite{taylor}) {\em etc}.

\section{Non-vacuum solutions in Palatini $f(R)$ gravity}
\label{nogo}

We have already established that, in vacuum, Palatini $f(R)$ gravity
reduces to General Relativity with an effective cosmological constant
and that, consequently, vacuum spherically symmetric solutions will be
Schwarzschild--de Sitter or Schwarz\-schild--anti-de Sitter. However, one
would like to go further than that and derive solutions in the
presence of matter as well. The first and simplest step in this
direction is to consider solutions with a high degree of symmetry.
Indeed spherically symmetric static solutions are quite realistic when
it comes to the description of stars and compact objects. 

In this section we will consider static spherically symmetric
solutions in Palatini $f(R)$ gravity, in the presence of matter.
Examining such solutions, apart from the usual interest in them as
descriptions of stars, will turn out to be crucial for the
understanding of the theory, as will become clear later on. In fact,
we will see that serious doubts concerning the viability of the theory
will be raised \cite{Barausse:2007pn}.

The standard procedure for arriving at a full solution describing the
spacetime inside and outside a static spherically symmetric object is
to separately find an exterior solution and an interior solution and
then match them together using appropriate junctions condition on the
matching surface (Israel junction conditions) \cite{gravitation}. In
General Relativity, in order to determine the interior solution one
needs, apart from the field equations, also the
Tolman-Oppenheimer-Volkoff (TOV) hydrostatic equilibrium equation (see
{\em e.g.~}\cite{schutz}). 

In \cite{TOV} a generalisation of the TOV equation for Palatini $f(R)$
gravity was derived. Let us briefly review the derivation of this
generalized TOV equation and then proceed along the lines of
\cite{Barausse:2007pn} to discuss the solutions. As shown in Section
\ref{manfield}, after suitable manipulations the field equations of
Palatini $f(R)$ gravity can be rewritten as a single one
(eq.~(\ref{eq:field}))
 \begin{align}
\label{eqfield2}
G_{\mu \nu} \!&=\! \frac{8\pi}{F}T_{\mu \nu}- \frac{1}{2}g_{\mu \nu}\! 
                        \left(\!{\cal R} - \frac{f}{F} \right)\! +\! \frac{1}{F} \left(
			\nabla_{\mu} \nabla_{\nu}
			\!- g_{\mu \nu} \Box
		\right)\! F-\nn\\
& \quad- \frac{3}{2}\frac{1}{F^2} \left(
			(\nabla_{\mu}F)(\nabla_{\nu}F)
			- \frac{1}{2}g_{\mu \nu} (\nabla F)^2
		\right),
\end{align}
 where $\nabla_{\mu}$ is the covariant derivative with respect to the
Levi--Civita connection of $g_{\mu\nu}$, $\Box\equiv
g^{\mu\nu}\nabla_{\mu}\nabla_{\nu}$ and $F=\partial f/\partial {\cal
R}$.

Using the static spherically symmetric ansatz
\begin{equation}
\label{metric}
	ds^2 \equiv -e^{A(r)}{\rm d}t^2 + e^{B(r)}{\rm d}r^2 + r^2{\rm d}\Omega^2
\end{equation}
 in eq.~\eqref{eqfield2}, considering perfect-fluid matter with
$T_{\mu\nu}=(\rho+p)u^\mu u^{\nu}+pg_{\mu\nu}$ (where $\rho$ is the
energy density, $p$ is the pressure and $u^\mu$ is the fluid
4-velocity) and representing $d/dr$ with a prime \footnote{In this
section we modify our standard notation and instead of using a prime
to denote differentiation with respect to the argument of the
function, we use it to denote differentiation with respect to the
radial coordinate. This significantly lightens the notation.}, one
arrives at the equations
 \bea\label{eq:Ap}
A' & = & \frac{-1}{1 + \gamma} \left(
	\frac{1 - e^B}{r} - \frac{e^B}{F}8\pi Grp
			+ \frac{\alpha}{r}
		\right), \\
	B' & = &  \frac{1}{1 + \gamma} \left(
			\frac{1 - e^B}{r} + \frac{e^B}{F}8\pi Gr\rho
			+ \frac{\alpha + \beta}{r}
		\right),\\
\alpha  &\equiv&  r^2 \left(
			\frac{3}{4}\left(\frac{F'}{F}\right)^2  + \frac{2F'}{rF}
			+ \frac{e^B}{2} \left( {\cal R} - \frac{f}{F} \right)
		\right), \\
	\beta  &\equiv&  r^2 \left(
			\frac{F''}{F} - \frac{3}{2}\left(\frac{F'}{F}\right)^2
		\right),\qquad
\gamma \equiv \frac{rF'}{2F}.\label{eq:abc}
\eea
 Making the definition $m_{\rm tot}(r)\equiv r(1 - e^{-B})/2$ and
using Euler's equation,
 \begin{equation}\label{eq:euler}
p'=-\frac{A'}{2}(p+\rho),
\end{equation}
one gets the generalised TOV equations~\cite{TOV}:
\begin{align}
\label{eq:OVB}
&p' = -\frac{1}{1 + \gamma} \frac{(\rho + p)}{r(r - 2m_{\rm tot})}\times\\
&\qquad\times\left( m_{\rm tot} + \frac{4\pi r^3 p}{F}  
          - \frac{\alpha}{2} (r - 2m_{\rm tot} ) \right),
\nn\\
\label{eq:mass}
m_{\rm tot}' &= \frac{1}{1 + \gamma} \bigg(\frac{4\pi r^2\rho}{F} + 
\frac{\alpha\!+\!\beta}{2}
- \frac{m_{\rm tot}}{r}(\alpha\! +\! \beta\! - \!\gamma) \bigg).
\end{align}

In order to determine the interior solution, one needs, besides
eqs.~(\ref{eq:OVB}) and (\ref{eq:mass}), to have information about the
microphysics of the matter configuration under investigation. In the
case of a perfect fluid this is effectively given by an equation of
state (EOS). A one-parameter EOS relates the pressure directly to the
energy density, {\em i.e.~}$p=p(\rho)$. This is the case which we will
consider here. 

Equations (\ref{eq:OVB}) and (\ref{eq:mass}) are implicit, their
right-hand sides effectively including through $F'$ and $F''$ both
first and second derivatives of the pressure, \textit{e.g.} $F'=
d/dr\,[F({\cal R}(T))]=(dF/d{\cal R})\,(d{\cal R}/dT)\,(dT/dp)\,p'$.
Therefore, they are difficult to solve so as to derive an interior
solution. We therefore first put them into an explicit form, which
allows us not only to solve them numerically, but also to study their
behaviour at the stellar surface where the matching with the exterior
solution occurs.

Multiplying eq.~\eqref{eq:OVB} by $dF/dp$ and using the definitions of
$\alpha$ and $\gamma$, we get a quadratic equation in $F'$ whose
solution is
  \be
\label{eq:F1}
  F'=\frac{-4 r F ({\cal C}-F) (r-2 m_{\rm tot})+D\sqrt{2\Delta}}{r^2
(3 {\cal C}-4 F) (r-2 m_{\rm tot})}
  \ee
where $D=\pm1$ and where we have defined 
\bea 
{\cal{C}}&=&\frac{dF}{dp}(p+\rho)=\frac{dF}{d\rho}\frac{d\rho}{dp}(p+\rho),
\label{eq:calC}\\
\Delta&=&F r^2 (r-2 m_{\rm tot}) \left[8 F ({\cal C}-F)^2 
(r-2 m_{\rm tot})\right.-\\
& &-\left.{\cal C} (4 F-3
     {\cal C}) \left((16 \pi  p-F {\cal R}+f) r^3+4 F m_{\rm tot}\right)\right].\nn
\eea
We will now focus on polytropic EOSs given by $p=k\rho_0^\Gamma$, where $\rho_0$ is the rest-mass density and $k$ and $\Gamma$ are constants. Note that the rest-mass density can be expressed in terms of the energy density $\rho$ and the internal energy $U$ as $\rho_0=\rho-U$. Assuming an adiabatic transformation and using the first law of thermodynamics one can express the internal energy in terms of the pressure, {\em i.e.} $U=p/(\Gamma-1)$. Therefore, the polytropic EOS can be rewritten as
\begin{equation}
\rho=\left(\frac{p}{k}\right)^{1/\Gamma}+\frac{p}{\Gamma-1},
\end{equation}
giving a direct link between $p$ and $\rho$. In eq.~\eqref{eq:calC}, we have written $\cal C$ in terms
of $dF/d\rho$ because this is finite at the stellar surface ($r =
r_{\rm out}$ where $p = \rho = 0$). In fact, $dF/d\rho=(dF/d{\cal
R})\,(d{\cal R}/dT)\,(3dp/d\rho-1)$, where $dF/d{\cal R}$ and $d{\cal
R}/dT$ are in general finite even when $T=3p-\rho$ goes to zero and
$dp/d\rho\to0$ for $p\to0$. This can be easily checked, for instance,
for the ${\cal R}^2$ or $1/{\cal R}$ models and it appears to be quite
a general characteristic that only very special models (and definitely
none of the cosmologically interesting ones) might be able to escape. 
Note also that while $d\rho/dp$ diverges when $p\to0$, the product
$(p+\rho)\,{d\rho}/{dp}$ goes to zero for $p\to0$ if $\Gamma<2$.
Therefore, for a polytrope with $\Gamma<2 $, ${\cal C}=0$ at the
surface. 

We now consider the matching between the interior and exterior
solutions. For the latter, the general solution is that of General
Relativity plus a cosmological constant. Here, the value of the
cosmological constant is equal to ${\cal R}_0/4$, where ${\cal R}_0$
is the vacuum value of ${\cal R}$ (see Section \ref{manfield}), {\em
i.e.~}
 \be
\exp(-B(r))=\ell\exp(A(r))=1-2m/r-{\cal R}_0 r^2/12,
\ee
 where $\ell$ and $m$ are integration constants to be fixed by
requiring continuity of the metric coefficients across the surface.
Using the definition of $m_{\rm tot}(r)$ this gives, in the exterior,
 \begin{equation} \label{eq:m_ext} 
m_{\rm tot}(r) = m+\frac{r^3}{24}{\cal R}_0\;. 
\end{equation} 
 Besides continuity of the metric, the junction conditions also
require continuity of $A'$, since $\rho=0$ at the matching surface
and, therefore, no surface layer approach can be adopted. For the
exterior, at the surface one has
\begin{gather}
\label{eq:requested_A_prime}
A'(r_{\rm out})=\frac{2 \left(r_{\rm out}^3 {\cal R}_0-12 m\right)}{r_{\rm out} 
\left({\cal R}_0 r_{\rm out}^3-12 r_{\rm out}+24 m\right)}\:.
 \end{gather}
 and this must be matched to the value of $A'(r_{\rm out})$ calculated
for the interior solution using eq.~\eqref{eq:Ap}. For this we need
$F'(r_{\rm out})$. Evaluating eq.~\eqref{eq:F1} at the surface, where
${\cal C}=p=0$ and $R$, $F$ and $f$ take their constant vacuum values
${\cal R}_0$, $F_0$ and $f_0=F_0 {\cal R}_0/2$, we get
\begin{equation} \label{eq:F1_surface}
F'(r_{\rm out})=-\frac{(1+\widetilde{D})F_0}{r_{\rm out}}\;,
\end{equation}
 where $\widetilde{D}=D\,{\rm sign}(r_{\rm out}-2 m_{\rm tot})$. Note
that, differently from GR, one cannot prove here that $r_{\rm out}>2
m_{\rm tot}$ from eq.~\eqref{eq:OVB} because $p'$ is not necessarily
positive, although one might expect $r_{\rm out}>2 m_{\rm tot}$ in
sensible solutions. 

In any case, it is easy to see that $\widetilde{D}=1$ does not give a
satisfactory solution, since it implies $\gamma=-1$ at the surface
[\textit{cf.} eq.~\eqref{eq:abc}] giving $A'\to\infty$ for $r\to
r_{\rm out}^{-}$ [see eq.~\eqref{eq:Ap}]. Since $F'$ has a
discontinuity for $\widetilde{D}=1$ ($F'\to-2F_0/r_{\rm out}$ when
$r\to r_{\rm out}^-$, $F'=0$ when $r> r_{\rm out}$) Dirac deltas will
appear on the right-hand side of eq.~\eqref{eq:field} due to the
presence of second derivatives of $F$ and one could hope that they
might cancel out with the Dirac deltas arising in the field equations
due to the discontinuity of $A'$. However, the discontinuity in $A'$
is an infinite one and therefore the Dirac deltas arising on the
left-hand side of eq.~\eqref{eq:field} can never be cancelled by those
on the right-hand side.  As already mentioned, one cannot attempt to
use here a surface layer approach to avoid discontinuities, because
$\rho=0$ on the surface for a polytrope. In addition, even if it were
possible to add a surface layer, the infinite discontinuity of $A'$
would require this layer to have an infinite surface density. We
conclude, then, that $\widetilde{D}=1$ cannot give a satisfactory
solution. For $\widetilde{D}=-1$, on the other hand, $F'(r_{\rm out})=
0$ for $r\to r_{\rm out}^-$ giving the correct interior value of $A'$
required for matching to eq.~\eqref{eq:requested_A_prime} and making
$F'$ continuous across the surface. We will concentrate only on this
case in the following.

In order to study the behaviour of $m_{\rm tot}$ at the surface, we
need first to derive an explicit expression for $F''$. If we take the
derivative of eq.~\eqref{eq:F1}, $F''$ appears on the left-hand side
and also on the right-hand side [through $m_{\rm tot}'$, calculated
from eq.~\eqref{eq:mass} and the definition of $\beta$,
eq.~\eqref{eq:abc}], giving a linear equation in $F''$. The solution
to this, evaluated at the surface, is
\be
\label{eq:F''}
 F''(r_{\rm out})=\frac {\left({\cal R}_0 r_{\rm out}^3-8 m_{\rm out}\right) 
{\cal C}'}{8 r_{\rm out} (r_{\rm out}-2 m_{\rm out})}
\ee
 Evaluating $\alpha$, $\beta$ and $\gamma$ at the surface using $F'=0$
and $F''$ given by eq.~\eqref{eq:F''}, and inserting into
eq.~\eqref{eq:mass} gives
\begin{equation}
\label{eq:mprime} 
 m_{\rm tot}'(r_{\rm out})=\frac{2 F_0 {\cal R}_0 r_{\rm out}^2+\left(r_{\rm 
out}^3 {\cal R}_0-8 m_{\rm tot}\right) {\cal C}'}{16 F_0}\;.
 \end{equation}

 For $1<\Gamma<3/2$, ${\cal C}'=d{\cal C}/dp\,p'\propto d{\cal
C}/dp\,(p+\rho)\to 0$ at the surface so that expression
\eqref{eq:mprime} is finite and it gives continuity of $m_{\rm tot}'$
across the surface [\textit{cf.} eq.~\eqref{eq:m_ext}]. However, for
$3/2<\Gamma<2$, ${\cal C}'\to\infty$ as the surface is approached,
provided that $dF/d{\cal R}({\cal R}_0)\neq0$ and $d{\cal
R}/dT(T_0)\neq0$ (note that these conditions are satisfied by generic
forms of $f({\cal R})$, {\em i.e.~}whenever an ${\cal R}^2$ term or a
term inversely proportional to ${\cal R}$ is present). While $m_{\rm
tot}$ keeps finite (as can be shown using the fact that $p'=0$ at the
surface), the divergence of $m_{\rm tot}'$ drives to infinity the
Riemann tensor of the metric, $R_{\mu\nu\sigma\lambda}$, and curvature
invariants, such as $R$ or
$R^{\mu\nu\sigma\lambda}R_{\mu\nu\sigma\lambda}$, as can easily be
checked\footnote{This seems to have been missed in
Ref.~\cite{barraco2}.}. This singular behaviour would give rise to
unphysical phenomena, such as infinite tidal forces at the surface
[\textit{cf.} the geodesic deviation equation] which would destroy
anything present there.

We can then conclude that no physically relevant solution exists for
any polytropic EOS with $3/2<\Gamma<2$. Certainly, it is clear that
polytropes give only simplified toy models for stars and one would
like to use a more accurate description of the interior structure. As
an example, we can consider neutron stars, in which case one has a
much more complicated dependence of pressure on density, taking
account of variations of composition (see, for example,
Ref.~\cite{haensel} and references therein). The behaviour of the EOS
in the outer layers would be critical for the behaviour of $m'_{\rm
tot}$ at the surface in the non-GR case.  However, while there are
indeed cases where a reasonable solution would be attainable (for
instance when $p\propto \rho_0$), one can argue that the viability of
a gravity theory should not depend on details such as this and that a
real difficulty has been identified.

\begin{figure}%[h]
\begin{center}
\includegraphics[width=13cm]{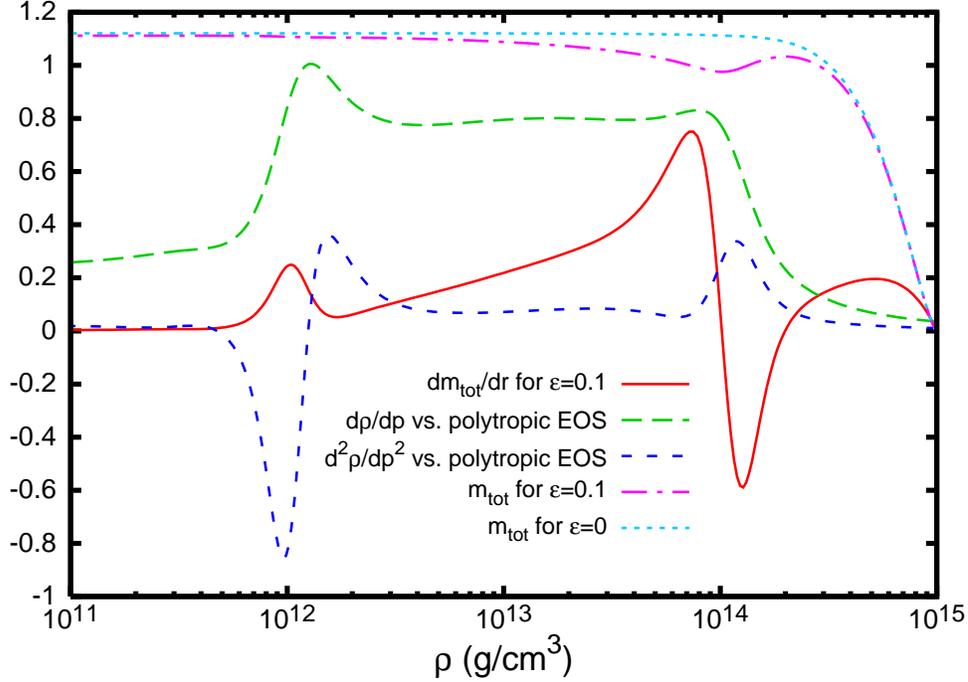}
 \caption{\label{fig}Profiles of $m_{\rm tot}$ (measured in units of
$M_\odot$) and of other associated quantities plotted against density
in the interior of neutron-star models with central density
$10^{15}{\rm g/cm^3}$ and $p'=0$ in the centre as required by local
flatness. We have used an analytic fit to the FPS EOS~\cite{haensel}
and the gravity theory given by $f({\cal R})={\cal R}+\epsilon {\cal
R}^2$. The dot-dashed purple line shows $m_{\rm tot}$ as calculated
with $\epsilon=0.1$ and the dotted cyan line shows the equivalent
curve in GR ($\epsilon=0$); the solid red line shows $dm_{\rm tot}/dr$
(in $M_\odot/{\rm km}$) for $\epsilon=0.1$. Note the bumps in the
$dm_{\rm tot}/dr$ curve which correspond to rapid composition changes
in the EOS (the corresponding features in the $m_{\rm tot}$ curve for
$\epsilon=0.1$ are less apparent but a noticeable dip is seen at
around $\rho=10^{14}{\rm g/cm^3}$). To make evident the influence of
composition changes, we also show comparisons between the FPS EOS and
a corresponding polytrope (with $\Gamma=4/3$ and $\kappa=10^{15}$
cgs): the green long-dashed curve and blue short-dashed curve show
$0.1\times(d\rho/dp)_{\rm FPS}/(d\rho/dp)_{\rm polytrope}$ and
$0.01\times(d^2\rho/dp^2)_{\rm FPS}/(d^2\rho/dp^2)_{\rm polytrope}$,
respectively.}
 \end{center}
%\vskip -0.5cm
\end{figure}

Setting aside the surface singularity issue, we next focus in neutron
stars in order to investigate the interior solution.  For such stars
we do have more physical descriptions of the interior than a
polytrope, a typical example being the FPS EOS, as given
in~\cite{haensel}, which we use here. As can be seen from
eq.~\eqref{eq:field}, the metric coefficients will be sensitive to
derivatives of the matter fields, since ${\cal R}$ is a function of
$T$~\footnote{The unusual behaviour of this class of theories has been
mentioned in a different context in Ref.~\cite{olmo3}. However, we
disagree with the claims made there about the violation of the
equivalence principle, because they seem to be based on an ill-posed
identification of the metric whose geodesics should coincide with
free-fall trajectories.}. This can be seen in Fig.~\ref{fig}: For
$f({\cal R})={\cal R}+\epsilon {\cal R}^2$, $m_{\rm tot}$, which in GR
has a smooth profile, now develops peculiar features when $d\rho/dp$
and $d^2\rho/dp^2$ change rapidly in going from the core to the inner
crust and from the inner crust to the outer crust. If $m_{\rm tot}$
were plotted against the radius, these features would look much more
abrupt, because they occur in a small range of radii close to the
surface. While $m_{\rm tot}$ does not represent a real mass in the
interior, such a strong dependence of the metric on the derivatives of
the matter field is not very plausible and could have dramatic
consequences. 

In our example for the neutron star interior we have chosen $f({\cal
R})={\cal R}+\epsilon {\cal R}^2$, even though most interesting
models, at least from a cosmological perspective, include a $1/{\cal
R}$ term. The reason for this is that, since ${\cal R}$ is
algebraically related to the energy density, a $1/{\cal R}$ term is
not going to produce large deviations from General Relativity in the
interiors of compact objects. Therefore, an ${\cal R}^2$ term was
definitely more suitable for the specific example considered here.  To
this one can add that an ${\cal R}^2$ term should generically be
present in the action if $f({\cal R})$ is taken to be some power
series representing the effective low-energy action of a more
fundamental theory, even if the $1/{\cal R}$ is greatly dominant at
cosmological scales. It should be stressed, in any case, that a
$1/{\cal R}$ term will have similar effects in the interior to those
for an ${\cal R}^2$ term but they will be more prominent in diffuse
objects, where the ${\cal R}^2$ term will be quite ineffective. This
can actually be even more critical, since the gravitational behaviour
of more diffuse objects is even more well established than that of
compact objects. 

In our attempt to determine and study non-vacuum static spherically
symmetric solutions, we have then found two unappealing
characteristics of Palatini $f(R)$ gravity as applied to stellar
models, each of which arises because of the dependence of the metric
on higher order derivatives of the matter field. First: whether or not
a regular matching can be made to the exterior solution depends
crucially on the microphysics, through the EOS, with polytropic EOSs
having $3/2<\Gamma<2$ being ruled out for generic $f({\cal R})$. 
Second: even if an EOS does allow for a regular solution at the
surface, the interior metric depends on the first and second
derivatives of the density with respect to the pressure, giving a
problematic behaviour. While polytropic EOSs are highly idealised, we
note that $\Gamma=5/3$, corresponding to an isentropic monatomic gas
or a degenerate non-relativistic particle gas, falls within the range
not giving a regular solution. The demonstration that the gravity
theory is unable to provide a consistent description for this
perfectly physical sort of matter configuration strongly suggests that
it is not suitable for being considered as a viable alternative to GR.

Since the problems discussed here arise due to the dependence of the
metric on higher order derivatives of the matter fields, one can
expect that they will also appear in other gravity theories having
these characteristics. Any theory having a representation in which the
field equations include second derivatives of the metric and higher
than first derivatives of the matter fields will face similar problems
because having a higher differential order in the metric than in the
matter field is what guarantees that the metric depends in a
cumulative way on the matter. If this is not the case then the metric
loses its immunity to rapid changes in matter gradients since it is
directly related to them instead of being an integral over them. This
is the same issue that was pointed out in Section \ref{postnewtpal},
where the post-Newtonian corrections to the metric were found to
depend directly on $T$ instead of being an integral over the sources
[eqs.~(\ref{olmo1s}) and (\ref{olmo2s}) and the related discussion
about the role of $\Omega(T)$] and in Section \ref{palinst}, where $R$
was found to be very sensitive to matter perturbations. 

 Such shortcomings should be expected for any theory which includes
fields other than the metric for describing the gravitational
interaction (\textit{e.g.} scalar fields) which are algebraically
related to matter rather than being dynamically coupled. In this case
one can always solve the field equations of the extra field and insert
the solution into the field equation for the metric, inducing a
dependence of the metric on higher derivatives of the matter fields. A
typical example of such a theory is a scalar-tensor theory with
Brans--Dicke parameter $\omega=-3/2$, which is anyway an equivalent
representation of Palatini $f(R)$ gravity (see Chapter
\ref{equivtheor}). One should mention that this problem could probably
be addressed in Palatini $f(R)$ gravity by adding higher order
curvature invariants in the action ({\it e.g.~}$f({\cal R}, {\cal
R}^{\mu\nu}{\cal R}_{\mu\nu})$), since this would introduce more
dynamics and break the non-dynamical coupling between matter and the
extra gravitational degrees of freedom.

The results presented in this section can be interpreted as a no-go
theorem for theories including higher order derivatives of the matter
fields in one of their possible representations, such as Palatini
$f(R)$ gravity or $\omega=-3/2$ scalar-tensor theory.

\section{Conclusions}

To summarise: in this chapter we have discussed viability constraints
related to the weak and strong gravity regimes for metric and Palatini
$f(R)$ gravity and for Gauss--Bonnet gravity. It has been shown that
such constraints can act in a complementary manner to the cosmological
constraints discussed in the previous chapter. Additionally, given
that most of the models considered in the literature are actually
tailored to fit cosmological observations, non-cosmological
constraints, such as those mentioned here, are crucial for
establishing the overall viability of alternative theories of gravity.

Specifically, we have seen that the post-Newtonian limit and stability
considerations severely constrain the parameter space of metric $f(R)$
gravity models. In the case of Palatini $f(R)$ gravity, even though
issues of stability similar to those present for metric $f(R)$ gravity
do not appear, the post-Newtonian limit provides serious indications
of non-viability for most models. However, the crucial problem with
this theory, its inability to give reasonable solutions for common
matter configurations, signalling an incompleteness, becomes apparent
when one considers non-vacuum solutions.  Finally, well motivated
models in Gauss--Bonnet gravity seem to be indistinguishable from
General Relativity as far as the Solar System tests are concerned. 
These last two results highlight, in different ways, the importance of
going beyond the standard weak-field-limit tests when trying to
constrain alternative theories of gravity.

\chapter{Future perspectives and conclusions}
\label{concl}

\section{Brief summary}

Before concluding this thesis or discussing future perspectives of the
work presented here, let us attempt to summarize in this section some
of the results presented so far. The motivation of this thesis has
been thoroughly discussed in Chapter \ref{intro} and a general
discussion about modifications of gravity was laid out in Chapter
\ref{foundations}. In Chapter \ref{modeltheo}, a number of specific
model theories were introduced and in Chapter \ref{equivtheor} the
relation between them was explored. Chapters \ref{cosmology} and
\ref{weakstrong} focused on the cosmological and astrophysical aspects
of these theories and on their viability.

As mentioned in the Introduction, these theories were introduced as
tools that could help us to examine how much and in which ways one can
deviate from General Relativity. Our intention was neither to tailor a
model within the framework of any of these theories that would fit the
data adequately nor to pick out a specific well-motivated low-energy
effective action from some fundamental theory and to confront it with
observations. The task which we undertook was to consider theories
that were easy to handle, each of them deviating from the framework of
General Relativity in a distinct way, and to exploit them in order to
get a deeper understanding of the difficulties and limitations of
modified gravity. In the light of this, it is probably preferable to
provide here a qualitative summary of our results which summarizes the
lessons learned from this procedure, instead of repeating in detail
the results already presented in the previous chapters.

Starting from the theoretical side, one of the clear outcomes of this
thesis is that generalizing the Einstein--Hilbert action to include
higher-order curvature invariants is not such a straightforward
procedure as it might seem. Even when considering the simplest of
generalisations: an $f(R)$ action as studied here, two distinct
classes of theory arise depending on the variational principle which
one decides to apply. The metric variational principle leads to fourth
order equations for the metric, whereas the Palatini variational
principle, which treats the connection as an independent variable,
leads to second order equations for the metric and an algebraic
equation relating the metric and the connections. Remarkably, both
approaches lead to General Relativity for the Einstein--Hilbert action.
Additionally, we saw that allowing the independent connection to
couple to the matter in order to restore its geometrical meaning ---
that of defining parallel transport and the covariant derivative ---
led again to a distinct class of theories: metric-affine $f(R)$
theories of gravity, which present an enriched phenomenology since the
independence of the connection allows for torsion and non-metricity.
In practice, metric-affine $f(R)$ gravity appears to comprise a very
general class of theories from which metric $f(R)$ gravity, Palatini
$f(R)$ gravity and General Relativity can come about after making a
number of simplifying assumptions.

As discussed in Chapter \ref{equivtheor} some of the theories
presented here can acquire different representations. For instance,
metric and Palatini $f(R)$ theories of gravity can be rewritten as
Brans--Dicke theories with Brans--Dicke parameters $\omega_0=0$ and
$\omega_0=-3/2$ respectively. This equivalence between theories has
proved fruitful for clarifying their characteristics. For example, the
equivalence between Palatini $f(R)$ gravity and $\omega_0=-3/2$
Brans--Dicke theory served as a straightforward demonstration of the
fact that even though the former theory has an independent connection,
it is intrinsically a metric theory of gravity. As we will see in the
next section, where we will resume the discussion of theories and
their representations, there is much more to be said about this issue.

The discussion about the cosmological and astrophysical aspects of the
theories examined here and the confrontation of the theories with
cosmological, astrophysical and Solar System observations hopefully
clarified that it is very difficult to construct a simple viable model
in an alternative theory of gravity. Mainly using metric and Palatini
$f(R)$ gravity as toy theories, it was demonstrated that observations
which are relevant to different scales provide different constraints
for the model examined and that simplistic models which provide an
adequate description of the phenomenology related to one scale are
easily ruled out when the experimental bounds related to a different
scale are taken into account. Solar System tests and bounds from Large
Scale Structure perturbations, appear to be very difficult to satisfy
with a single theory and, in most cases, constrain the parameter space
of the theory unnaturally close to the $\Lambda$CDM model.

One might ask how discouraging is the fact that simple models fail to
be viable? Indeed an Ockham's razor approach strongly disfavours very
complicated models. On the other hand, it should be stressed that in
order to explain with an alternative gravitation theory, phenomenology
that General Relativity cannot explain without the inclusion of new
mysterious matter components, one will inevitably have to add to this
theory more complexity and more dynamics. Even though simplicity
should not be given away lightheartedly, the best theory is always the simplest
one among those that do account for the observations.

Allowing for more dynamics in a gravitational theory, however, has
proved to be a far from easy task during the course of this work. Even
if the theory is tailored to fit cosmological observations and pass
Solar System tests, we saw that problems related to stability can very
easily appear. A typical example is the curvature scalar instability
in metric $f(R)$ gravity discussed in Section \ref{palinst}. On the
other hand, in Palatini $f(R)$ gravity, in which the equations are not
fourth order in the metric, this instability is not present.

As just mentioned, in order to account for the phenomenology remaining
unexplained by General Relativity, one inevitably needs to add more
dynamics to the theory. The fact that this dynamics was not added in
terms of the metric in Palatini $f(R)$ gravity, did help with issues
of stability and simplify the field equations, but this came at a very
high price as we saw in Section \ref{nogo}. The extra dynamics were
implicitly added in the matter part of the theory, even if this is not
at all obvious in the standard formulation, and this has dramatic
consequences for commulativity. This last example also highlights the
importance of going beyond applications to Cosmology and the Solar
System when testing alternative theories of gravity.

\section{What comes next?}
\subsection{Towards a theory of gravitation theories?}

Clearly this thesis is far from being an exhaustive study of the
theories considered: scalar-tensor theory, $f(R)$ gravity and
Gauss--Bonnet gravity. One could, therefore, list here a number of
proposals for future work on these theories, some of which have indeed
already been mentioned in the previous chapters. For instance,
metric-affine gravity is the least studied of the theories considered
here and several of its aspects are completely obscure, such as exact
solutions, post-Newtonian expansions and Solar System tests,
cosmological phenomenology, structure formation, {\em etc}. Exact solutions
have also not been studied in Gauss--Bonnet gravity and there is
definitely more to be said about this issue in metric and Palatini
$f(R)$ gravity as well. 

Instead of continuing this list, which indeed can get quite long, we
prefer to follow a different perspective here. We remind the reader
once more that all of the above theories should be viewed mainly as
toy or straw-man theories used to provide a better understanding of
gravity. A more elaborate plan could be, therefore, to go beyond such
approaches and this is what we would like to consider here.

Going beyond a trial-and-error approach to modified gravity has very
important advantages. From the theoretical side, one has to bear in
mind that what we are aiming for is really a better understanding of
the conceptual basis of gravity. Even though proposing an alternative
theory that violates one of the assumptions of General Relativity and
examining whether it is viable or not is a straightforward procedure,
it is definitely not the most efficient one. It is a complicated
procedure and in many cases it can be misleading, since more than one
characteristic of the specific theory can often influence the result.
On the other hand, we hope that the reader will be convinced by now
that there are already a very large number of alternative theories of
gravity in the literature (viable or not) and it is not always clear
how much we have managed to learn by studying them.

The benefits at the experimental level are even more clear. Past
experience has taught us that experiments test principles and not
theories ({\em e.g.} weak equivalence principle tests such as the
gravitational redshift tests \cite{PoundRebka}, which were initially
regarded as tests of General Relativity). This directly indicates that
the most efficient approach, from an experimental point of view, is to
boot-strap our way to a theory starting from the principles which it
should satisfy. This would save us a lot of the effort required in
bringing the theory to a form suitable for confrontation with
observations. The Parametrized Post-Newtonian expansion for $f(R)$
gravity, presented earlier, serves as an ideal example with all of its
complications. 

Of course, the above hardly constitutes an easy project. In some sense
what is being discussed here is essentially the need for an axiomatic
formulation of gravitation theories in general. Even in the simplest
of these theories, General Relativity, such an axiomatic formulation
is not yet available. One could, of course, ask how useful a
collection of axioms would be for a theory like General Relativity,
when we already know the field equations and the action. Indeed,
knowledge of either of these suffices to fully describe the dynamics
of the theory, at least at the classical level, which makes the
absence of an axiomatic formulation less significant as far as
practical purposes are concerned. However, as soon as one moves to even the
simplest generalizations of Einstein's theory, such as scalar-tensor
gravity for instance, the problem becomes acute as argued above. 

A set of axioms could help us to understand the theory in depth and
provide a better insight for finding solutions to long standing
problems, the most prominent being that of Quantum Gravity. For
example, it could help us to determine the fundamental classical
properties which one expects to recover in the classical limit and to
recognise which of them should break down at the quantum level.  Even
more, if emergent gravity scenarios are considered ({\em
i.e.}~scenarios in which the metric and the affine-connections are
collective variables and General Relativity would be a sort of
hydrodynamics emergent from more fundamental constituents) such a set
of axioms could provide much needed guidance for reconstructing the
microscopic system at the origin of classical gravitation, for example
by constraining its microscopic properties so as to reproduce the
emergent physical features encoded in these axioms.

However, with such a large number of alternative theories of gravity,
how can we characterise the way in which they differ from General
Relativity, group them, or obtain some insight into which of them are
preferable to others? Even if we are far from having a coherent and
strict axiomatic formulation, at least a set of principles would
definitely prove useful towards this end, as well as for analyzing
experimental results to assess the viability of alternative theories.

Already in the 1970s there were attempts to present a set of ground
rules, sometimes referred to as a ``theory of gravitation theories'',
which gravitation theories should satisfy in order to be considered
viable in principle and, therefore, interesting enough to deserve
further investigation. However, no real progress seems to have been
made in this direction over the last thirty years, even though the
subject of alternative gravity theories has been an active one. It is
important to understand the practical reasons for this lack of
progress if we wish to proceed beyond the trial-and-error approach
that is mostly being used in current research on modified gravity. 
Hopefully, this exploration, largely based on \cite{thovalste}, will
also give as a byproduct some interesting clarifications of some 
common misconceptions (regarding the WEP, equivalence of theories, 
{\em etc.}) and serve as a motivational point of reference for 
future work \footnote{In what follows purely classical physics will be considered. The 
issue of the compatibility between the Equivalence Principle(s) and 
quantum mechanics, although rich in facets and consequences (see {\em e.g.}~\cite{EPQM1, EPQM2, EPQM3, EPQM4, EPQM5, EPQM6}) is beyond the scope of this discussion.}.

\subsection{From principles to practice and vice-versa}

We have argued why it would be interesting to utilise some
theory-inde\-pend\-ent observations to enunciate general viability
criteria as a set of theoretical principles that can help us to
distinguish potentially viable theories from theories which are
ill-posed from the very beginning. As already mentioned, providing a strict axiomatic 
formulation is hardly an easy goal\footnote{See, however, Refs.~\cite{mathapproach1, mathapproach2, mathapproach3, mathapproach4, mathapproach5} for an attempt towards an axiomatic 
formulation of gravitational theories from a more 
mathematically-minded point of view.}, but one could hope to give at least some set of physical viability principles, even if the latter are not necessarilly at the level of axioms. It is clear that in order to be useful 
 such statements need to be formulated in a theory-independent 
 way and should be amenable to experimental tests so that we 
could  select at least among classes of gravitational theories 
by suitable  observations/experiments.The best example in this direction
so far is the Equivalence Principle in its various versions, {\em
i.e.} the Weak Equivalence Principle (WEP), the Einstein Equivalence
Principle (EEP) and the Strong Equivalence Principle (SEP)
\cite{willbook}. We have already discussed extensively in Section
\ref{ep} the three forms of the equivalence principle as well as their
implications for a gravitation theory, such as the existence of a
metric and of local Lorentz frames, the coupling of the metric to
matter fields {\em etc}. Therefore, let us just recall the following
important remarks and refer the reader back to Section \ref{ep} for
more details:

The WEP only says that there exist some preferred trajectories, the
free fall trajectories, that test particles will follow and that these
curves are the same independently of the mass and internal composition
of the particles that follow them (universality of free fall). The WEP
does not imply, by itself, that there exists a metric, geodesics, {\em
etc.} --- this comes about only through the EEP by combining the WEP
with the requirements of Local Lorentz invariance (LLI) and Local
Position Invariance (LPI). The same is true for the covariance of the
field equations. As far as the SEP is concerned, the main thrust
consists of extending the validity of the WEP to self-gravitating
bodies and the applicability of LLI and LPI to gravitational
experiments, in contrast to the EEP. As mentioned in Section \ref{ep},
even though there are experimental tests for all of the EPs, the most
stringent ones are those for the WEP and the EEP. 

Let us stress that there are at least three subtle points in relation
to the use and meaning of the EP formulations, the first one
concerning the relation between the SEP and General Relativity.  While
there are claims that the SEP holds only for General Relativity
\cite{willbook}, no proof of this statement has so far been given.
Indeed, it would be a crucial step forward to pinpoint a one-to-one
association between GR and the SEP but it is easy to realize that it
is difficult to relate directly and uniquely a qualitative statement,
such as the SEP, to a quantitative one, namely Einstein's equations.
The second subtle point is the reference to test particles in all of
the EP formulations. Clearly, no true test particles exist, hence the
question is: how do we know how ``small'' a particle should be in
order for it to be considered as a test particle ({\em i.e.} so that
its gravitational field can be neglected)?  The answer is likely to be
theory-dependent\footnote{See \cite{Geroch} and references therein for
the case of General Relativity.}, and there is no guarantee that a
theory cannot be concocted in which the WEP is valid in principle but,
in practice, experiments would show a violation because, within the
framework of the theory, a ``small'' particle is not close enough to
being a test particle. Of course, such a theory would not be viable
but this would not be obvious when we refer to the WEP only from a
theoretical perspective ({\em e.g.} if we calculate free fall
trajectories and compare with geodesics). A third subtlety, which we
shall come back to later, is related to the fact that sometimes the
same theory can appear to either satisfy or not satisfy some version
of the EP depending on which variables are used for describing it, an
example being the contrast between the Jordan and Einstein frames in
scalar-tensor theories of gravity.

Taking all of the above into consideration, it seems that the main
problem with all forms of the equivalence principle is that they are
of little practical value. As principles they are by definition
qualitative and not quantitative. However, quantitative statements are
what is needed in practice.

To this end, Thorne and Will \cite{thornewill} proposed the metric
theories postulates, which were presented in Section \ref{metpost}.
Essentially, the metric postulates require the existence of a metric
$g_{\mu\nu}$ and that the matter stress-energy tensor $T_{\mu\nu}$
should be divergence free with respect to the covariant derivative
defined with the Levi--Civita connection of this metric. We have
already thoroughly discussed in Sections \ref{ep} and \ref{metpost}
how the metric postulates encapsulate the validity of the EEP, a key
point being that $\nabla_\mu T^{\mu\nu}=0$ leads to geodesic motion
for test particles \cite{fock}. 

Appealing as they may seem, however, the metric postulates lack
clarity. As pointed out also by the authors of Ref.~\cite{thornewill},
any metric theory can perfectly well be given a representation that
appears to violate the metric postulates (recall, for instance, that
$g_{\mu\nu}$ is a member of a family of conformal metrics and that
there is no {\it a priori} reason why this particular metric should be
used to write down the field equations)\footnote{See also Ref.~\cite{Anderson} for an earlier criticism 
of the need for a metric and, indirectly, of the metric 
postulates.}. One of our goals here is to
demonstrate this problem, and also some other prominent ambiguities
that we have already very briefly stated in Section \ref{metpost}, and
to trace their roots. 

\subsubsection{What precisely is the definition of  stress-energy 
tensor?}
\label{set1}

In order to answer this question one could refer to an action.  This
would be a significant restriction to begin with though, since it
would add to the EEP the prerequisite that a reasonable theory has to
come from an action. Even so, this would not solve the problem: one
could claim that $T_{\mu\nu}\equiv -(2/\sqrt{-g})\delta S_M/\delta
g^{\mu\nu}$ but then how is the matter action $S_M$ defined?  Claiming
that it is the action from which the field equations for matter are
derived is not sufficient since it does not provide any insight about
the presence of the gravitational fields in $S_M$. Invoking a minimal
coupling argument, on the other hand, is strongly theory-dependent
(which coupling is really minimal in a theory with extra fields or an
independent connection? \cite{Sotiriou:2006qn}). Furthermore, whether
a matter field couples minimally or non-minimally to gravity or to
matter should be decided by experiments. Since a non-minimal coupling
could be present and yet evade experimental detection (as proposed in
string theories \cite{taylor}), it seems prudent to allow for it in
the action and in the theory.

Setting actions aside and resorting to the correspondence with the
stress-energy tensor of Special Relativity does not help either. There
is always more than one tensor that one can construct which will
reduce to the special-relativistic stress-energy tensor when gravity
is ``switched off'' and it is not clear what ``switched off'' exactly
means when extra fields describing gravity (scalar or vector) are
present in the theory together with the metric tensor.

Finally, mixing the two tentative definitions described above makes
the situation even worse: one can easily imagine theories in which
 \be T_{\mu\nu}\equiv -(2/\sqrt{-g})\delta S_M/\delta g^{\mu\nu}
\ee
 does not reduce to the special-relativistic stress-energy tensor in
some limit. Are these theories necessarily non-metric? This point
highlights also another important question: are the metric postulates
necessary or sufficient conditions for the validity of the EEP?
Concrete examples are provided in Sections \ref{ST}, \ref{fR} and \ref{ECSK}.

\subsubsection{What does ``non-gravitational field'' mean?}
\label{matgeom}
 There is no precise definition of ``gravitational'' and
``non-gravitational'' field. One could say that a field which is
non-minimally coupled to the metric is gravitational whereas all
others are matter fields. This definition does not appear to be
rigorous or sufficient though and it is shown in the following that it
strongly depends on the perspective and terminology that one chooses.

Consider, for example, a scalar field $\phi$ non-minimally coupled to
the Ricci curvature in $\lambda \phi^4 $ theory, as described by the
action
 \begin{equation} \label{NMCaction}
S=\int d^4x \, \sqrt{-g}\, \left[ \left( \frac{1}{16\,\pi\,G}-\xi 
\phi^2 \right) R-\frac{1}{2}\nabla^{\mu}\phi\nabla_{\mu}\phi  
-V(\phi) \right] \;.
\end{equation}
 If one begins with a classical scalar field minimally coupled to the
curvature ({\em i.e.} $\xi=0$) in the potential $V(\phi)=\lambda
\phi^4 $ and quantizes it, one finds that first loop corrections
prescribe a non-minimal coupling term ({\em i.e.} $\xi\neq 0$) if the
theory is to be renormalizable, thus obtaining the ``improved
energy-momentum tensor'' of Callan, Coleman, and Jackiw \cite{CCJ}
(see also \cite{ChernikovTagirov}). Does quantization change the
character of this scalar field from ``non-gravitational'' to
``gravitational''? Formally, the resulting theory is a scalar-tensor
theory according to every definition of such theories that one finds
in the literature ({\em e.g.} \cite{valeriobook,willbook,wagoner,
bergmann, nordvedt2, fuji}) but many authors consider $\phi$ to be a
non-gravitational field, and certainly this is the point of view of
the authors of Ref.~\cite{CCJ} (in which $\phi$ is regarded as a
matter field to be quantized) and of most particle physicists.

\subsection{Theories and representations}
\label{repres}

We have already discussed in Chapter \ref{equivtheor} the fact that
theories can acquire more than one representation. We used the term
``dynamical equivalence'' there, in order to refer to the fact that
two theories can describe the same dynamics. Within a classical
perspective, however, a gravitation theory is indeed a description of
the dynamics of a a gravitating system and in this sense, as also
mentioned in Chapter \ref{equivtheor}, when one refers to two
dynamically equivalent theories what is actually meant is two
different representations of the same theory. 

As will be demonstrated later, many misconceptions arise when a theory
is identified with one of its representations and other
representations are implicitly treated as different theories.  Even
though this might seem to be a very abstract point, to avoid
confusion, one would like to provide precise definitions of the words
``theory'' and ``representation''.  It is not trivial to do this,
however. For the term ``theory'', even if one looks at a popular
internet dictionary, a number of possible definitions can be found
\cite{wiktion}:
 \begin{enumerate} 

\item An unproven conjecture. 

\item An expectation of what should happen, barring unforeseen 
circumstances. 

\item A coherent statement or set of statements that attempts to
explain observed phenomena. 

\item A logical structure that enables one to deduce the possible
results of every experiment that falls within its purview. 

\item A field of  study attempting to exhaustively describe a 
particular class of constructs. 

\item A set of axioms together with all statements derivable from
them. 
 \end{enumerate}

Definitions (1) and (2) are not what is meant for scientific theories.
On the other hand, (3) and (4) seem to be complementary statements
describing the use of the word ``theory'' in natural sciences, whereas
(5) and (6) have mathematical and logical bases respectively. In a
loose sense, a more complete definition for the word ``theory'' in the
context of physics would probably come from a combination of (4) and
(6), in order to combine the reference to experiments in (4) and the
mathematical rigour of (6). An attempt in this direction could be:

\begin{defi} Theory: A coherent logical structure, which is preferably
expressed through a set of axioms together with all statements
derivable from them, plus a set of rules for their physical interpretation, that enables one to deduce the possible results
of every experiment that falls within its purview.\footnote{One might argue that when a theory is defined as a set of axioms, as suggested above, it is doomed to face the implications of G\"odel's incompleteness theorems. However, it is neither clear if  such theorems are applicable to physical theories, nor how  physically relevant they would be even if they were applicable \cite{Godel,Barrow:2006hi}.}
 \end{defi}

Note that no reference is made to whether there is agreement between
the predictions of the theory and actual experiments. This is a
further step which could be included in the characterization of a
theory. There could be criteria according to which the theory is
successful or not according to how large a class of observations is
explained by it and the level of accuracy obtained (see for example
\cite{hawkbft}).  Additionally, one could consider simplicity as a
merit and characterize a theory according to the number of assumptions
on which it is based (Ockham's razor).  However, all of the above
should not be included in the definition itself of the word
``theory''.

Physical theories should have a mathematical representation.  This
requires the introduction of physical variables (functions or fields)
in terms of which the axioms can be encoded in mathematical relations.
We attempt to give a definition:

\begin{defi} Representation (of a theory): A finite collection of
equations interrelating the physical variables which are used to
describe the elements of a theory and assimilate its axioms. 
 \end{defi}

The reference to equations can be restrictive, since one may claim
that in many cases a theory could be fully represented by an action.
At the same time it is obvious that any representation of a theory is
far from being unique. Therefore, one might prefer to modify the above
definition as follows:

\begin{defi} Representation (of a theory): A non-unique choice of
physical variables between which, in a prescribed way, one can form
inter-relational expressions that assimilate the axioms of the theory
and can be used in order to deduce derivable statements. 
 \end{defi}

It is worth stressing here that when choosing a representation for a theory it is 
essential to provide also a set of rules for the physical interpretation of the variables involved in it. This is needed for formulating the axioms ({\em i.e.} the physical statements) of the theory in terms of these variables.  It should also be noted that these rules come as extra information not {\em a priori} contained in the mathematical formalism. Furthermore, once they are consistently used to interpret the variables of the latter, they would allow to consistently predict the outcome of experiments in any alternative representation (we shall come back to this point and discuss an example later on in Section \ref{ST}).

All of the above definitions are, of course, tentative or even naive ones and others can be found that are more precise and comprehensive.
However, they are good enough to make the following point: the
arbitrariness that inevitably exists in choosing the physical
variables is bound to affect the representation. More specifically, it
will affect the clarity with which the axioms or principles of the
theory appear in each representation. Therefore, there will be some
representations in which it will be obvious that a certain principle
is satisfied and others in which it will be more intricate to see
that.  However, it is clear that the theory is one and the same and
that the axioms or principles are independent of the representation.
One may consider it a worthy goal to express theories in a
representation-invariant language. However, it should be borne in mind
that this is exactly what axiomatic formulation is all about and there
is probably no way to do this once reference to a set of physical
variables has been made.  In a sense, the loss of quantitative
statements is the price which one has to pay in order to avoid
representation dependence.

\subsection{Example no.~1: Scalar-tensor gravity}
\label{ST}

In order to make the discussion of the previous sections clearer, let
us use scalar-tensor theories of gravitation as an example.  As in
most current theories, scalar-tensor theories were not originally
introduced as collections of axioms but directly through a
representation. Instead of using the conventional notation found in
the literature, which we have also used when discussing scalar-tensor
theories in Section \ref{scalartensor} and in the rest of this thesis,
we will here write the action using the notation of
Ref.~\cite{Flanagan} (see also Ref.~\cite{shapiro}):
 \begin{equation} 
S=S^{(g)}+S^{(m)}\left[e^{2\alpha(\phi)}g_{ab}, 
\psi^{(m)}\right]\;,\label{eq:3.19}
\end{equation} 
where 
\begin{equation} \label{flaction}
S^{(g)}=\int 
d^{4}x\;\sqrt{-g}\left[\frac{A(\phi)}{16\pi 
G}R- \frac{B(\phi)}{2}g^{ab}\nabla_{a} 
\phi\nabla_{b}\phi-V(\phi)\right] \;
\end{equation}
 and $\psi^{(m)}$ collectively denotes the matter fields. Some of the
unspecified functions $A$, $B$, $V$, and $\alpha$ in this notation can
be fixed without loss of generality, {\em i.e.~}without choosing a
particular theory from within the class, and this is the way in which
one is led to the action of a scalar-tensor theory in the more
standard notation of Section \ref{scalartensor}. However, this would
come at the expense of fixing the representation, which is exactly
what we intend to analyse here.  Therefore, the present notation is
indeed the most convenient for our purposes.

 Let us first see how action (\ref{eq:3.19}) comes about from first
principles. As already discussed in Section \ref{ep}, following Will's
book \cite{willbook} one can argue that the EEP can only be satisfied
if some metric exists and the matter fields are coupled to it not
necessarily minimally but through a non-constant scalar, {\em i.e.}
they can be coupled to a quantity $\phi g_{\mu\nu}$, where $\phi$ is
some scalar. However, this coupling should be universal in the sense
that all fields should couple to $\phi$ in the same way \footnote{This
is not the case in supergravity and string theories, in which
gravivector and graviscalar fields can couple differently to particles
with different quark content \cite{Sherk, Gasperini}.}. Therefore, the
most general form of the matter action will have a dependence on $\phi
g_{\mu\nu}$. Of course, one can always choose to write $\phi$ as
$e^{2\alpha(\phi)}$, where $\phi$ is a dynamical field.

Now the rest of the action should depend on $\phi$, the metric and
their derivatives. No real principle leads directly to the action
above. However, one could impose that the resulting field equations
should be of second order both in the metric and in the scalar field
and utilize diffeomorphism invariance arguments to arrive at this
action. Then, (\ref{action}) is the most general scalar-tensor action
that one can write, once no fields other than $\phi$ and the metric
are considered, and no couplings other than a non-minimal coupling of
the scalar to the curvature is allowed.

We now return to the role of the four yet-to-be-defined functions
$A(\phi)$, $B(\phi)$, $V(\phi)$, $\alpha(\phi)$ and examine whether
there are redundancies.  As we have already said, the action
(\ref{action}) describes a {\em class} of theories, not a single
theory. Specifying some of the four functions will specialize it to a
specific theory within that class. However, one can already see that
this action is formally invariant under arbitrary conformal
transformations $\tilde{g}_{\mu\nu}=\Omega^2(\phi) g_{\mu\nu}$. In
fact, it can be recast into its initial form simply by redefining the
undetermined functions $A(\phi),B(\phi), V(\phi), \alpha(\phi)$ after
making the conformal transformation. This implies that any one of the
functions $A(\phi)$, $B(\phi)$, $V(\phi)$ and $e^{2\alpha(\phi)}$ can
be set to a (non-vanishing) constant by means of making a suitable
choice for $\Omega(\phi)$. Additionally, the scalar field $\phi$ can
be conveniently redefined so as to set yet another of these functions
to be a constant. Therefore, we conclude that setting two of these
functions to be constants (or just unity) is merely making a choice of
representation and has nothing to do with the content of the theory.
In fact, it does not even select a theory within the class.

This has a precise physical meaning: it demonstrates our ability to
choose our clocks and rods at will \cite{dicke}. One could decide not
to allow this in a theory (irrespectively of how natural that would
be). Therefore, it constitutes a very basic physical assumption which
can be described as an axiom.

Let us now turn our attention to the matter fields $\psi^{(m)}$: the
way in which we have written the action implies that we have already
chosen a representation for them. However, it should be clear that we
could always redefine the matter fields at will. For example, one
could set $\tilde{\psi}=\Omega^s \psi^{(m)}$ where $s$ is a
conveniently selected conformal weight \cite{wald} so that, after
making a conformal transformation, the matter action will be
 \begin{equation} 
S^{(m)}=S^{(m)}\left[\tilde{g}_{ab},\tilde{\psi}\right] \;. 
\end{equation} 
 The tilde is used here in order to distinguish between the physical
variables in the two representations. We can now make use of the
freedom discussed above to fix two of the four functions of the field
at will and set $A=B=1$. Then the action (\ref{action}) will formally
become that of General Relativity with a scalar field minimally
coupled to gravity. 

However this theory is not actually General Relativity, since now
$\tilde{\psi}=\tilde{\psi}(\phi)$ which essentially means that we have
allowed the masses of elementary particles and the coupling constants
to vary with $\phi$ and consequently with the position in spacetime.
From a physical perspective, this is translated into our ability to
choose whether it will be our clocks and rods that are unchanged in
time and space or instead the outcome of our measurements \cite{dicke}
(which, remember, can always be expressed as dimensionless constants
or dimensionless ratios, since even the measurement of a dimensional
quantity such as {\em e.g.}~a mass, is nothing more than a comparison
with a fixed standard unit having the same dimensions). We will return
to this issue again in Section \ref{ST}.

To summarize, we can in practice choose two of the four functions in
the action (\ref{eq:3.19}) without specifying the theory. In addition,
we can even fix a third function at the expense of allowing the matter
fields $\psi^{(m)}$ to depend explicitly on $\phi$, which leads to
varying fundamental units \cite{dicke}. Once either of these two
options is chosen, the representation is completely fixed and any
further fixing of the remaining function or functions leads to a
specific theory within the class. On the other hand, by choosing any
two functions and allowing for redefinitions of the metric and the
scalar field, it is possible to fully specify the theory and still
leave the representation completely arbitrary. 

It is now obvious that each representation might display different
characteristics of the theory and care should be taken in order not to
be misled into representation-biased conclusions, exactly as happens
with different coordinate systems. This highlights the importance of
distinguishing between different {\em theories} and different {\em
representations}.

This situation is very similar to a gauge theory in which one must be
careful to derive only gauge-independent results. Every gauge is an
admissible ``representation'' of the theory, but only gauge-invariant
quantities should be computed for comparison with experiment. In the
case of scalar-tensor gravity however, it is not clear what a
``gauge'' is and how one should identify the analogue of
``gauge-independent'' quantities.

\subsubsection{Alternative theories and alternative representations: \\Jordan and Einstein frames}
\label{JE}

Let us now go one step further and focus on specific scalar-tensor
theories. With $\psi^{(m)}$ representing the matter fields and
choosing $\alpha=0$ and $A(\phi)=\phi$, we fully fix the
representation. Let us now suppose that all of the other functions are
known. The action then takes the form
 \begin{equation} 
\label{rep1}
S=S^{(g)}+S^{(m)}\left[g_{ab},\psi^{(m)}\right]\;,
\end{equation} 
where 
\begin{equation}
 S^{(g)}=\int 
d^{4}x\;\sqrt{-g}\left[\frac{\phi}{16\pi G}R 
- \frac{B(\phi)}{2}g^{ab}\nabla_{a}\phi\nabla_{b} 
\phi-V(\phi)\right] \;,
\end{equation} 
 and it is apparent that $T_{\mu\nu}\equiv -(2/\sqrt{-g})\delta
S^{(m)}/\delta g^{\mu\nu}$ is divergence-free with respect to the
metric $g_{\mu\nu}$ and, therefore, the metric postulates are
satisfied.

Now we take a representation where $A=B=1$ and the action takes the form
 \begin{equation} 
\label{rep2}
S=S^{(g)}+S^{(m)}\left[e^{2\tilde{\alpha}(\phi)}\tilde{g}_{ab}, 
\psi^{(m)}\right]\;,
\end{equation} 
where 
\begin{equation} S^{(g)}=\int  
d^{4}x\;\sqrt{-\tilde{g}}\left[\frac{1}{16\pi G}\tilde{R} 
-\frac{1}{2}\tilde{g}^{ab}\tilde{\nabla}_{a} 
\phi\tilde{\nabla}_{b}\phi-\tilde{V}(\phi)\right] .
\end{equation} 
 As we have argued, for any (non-pathological) choice of $B$ and $V$
in the action (\ref{rep1}), there exists some conformal factor
$\Omega(\phi)$, relating $g_{\mu\nu}$ and $\tilde{g}_{\mu\nu}$, and
some suitable redefinition of the scalar $\phi$ to the scalar
$\tilde{\phi} $, which brings action (\ref{rep1}) into the form of
action (\ref{rep2}), therefore relating $B$ and $V$ with $\tilde{V}$
and $\tilde{\alpha}$. Actions (\ref{rep1}) and (\ref{rep2}) are just
different representations of the same theory after all, assuming that
$B$ and $V$ or $\tilde{V}$ and $\tilde{\alpha}$ are known. 

 According to the most frequently used terminology, the first
representation is called the {\em Jordan frame} and the second the
{\em Einstein frame} and the way in which we have just introduce them
should make it very clear that they are just alternative, but
physically equivalent, representations of the same theory.
(Furthermore, infinitely many conformal frames are possible,
corresponding to the freedom in choosing the conformal factor.)

Let us note, however, that if one defines the stress-energy tensor in
the Einstein frame as $\tilde{T}_{\mu\nu}\equiv -(2/
\sqrt{-\tilde{g}})\delta S^{(m)}/\delta \tilde{g}^{\mu\nu}$, one can
show that it is {\em not} divergence-free with respect to the
Levi--Civita connections of the metric $\tilde{g}_{\mu\nu}$. In fact,
the transformation property of the matter stress-energy tensor under
the conformal transformation $g_{\mu\nu}\rightarrow
\tilde{g}_{\mu\nu}=\Omega^2 \, g_{\mu\nu} $ is
$\tilde{T}_{\mu\nu}=\Omega^s \, T_{\mu\nu}$, where the appropriate
conformal weight in four spacetime dimensions is $s=-6$ \cite{wald}.
The Jordan frame covariant conservation equation
$\nabla^{\beta}T_{\alpha\beta}=0$ is therefore mapped into the
Einstein frame equation
 \begin{equation}\label{non-conservation}
\tilde{\nabla}_{\alpha}\tilde{T}^{\alpha\beta}=-\tilde{T}\, 
\, \frac{ \tilde{g}^{\alpha\beta} \tilde{\nabla}_{\alpha} 
\Omega}{\Omega} \;,
\end{equation}
 which highlights the fact that the Einstein frame energy-momentum
tensor for matter is not covariantly conserved unless it describes
conformally invariant matter with vanishing trace $T$ which is not, of
course, the general case.

In summary, we see that while the actions (\ref{rep1}) and
(\ref{rep2}) are just different representations of the same theory,
the metric postulates and the EEP are obviously satisfied in terms of
the variables of the Jordan frame, whereas, at least judging naively
from eq.~(\ref{non-conservation}), one could be led to the conclusion
that the the EEP is not satisfied by the variables of the Einstein
frame representation. However this is obviously paradoxical as we have
seen that the general form of the scalar-tensor action (\ref{eq:3.19})
can be derived from the EEP.

The point is that an experiment is not sensitive to the
representation, and hence in the case of the action (\ref{eq:3.19}) it
will not show any violation of the EEP. The EEP will {\em not} be
violated in {\em any} chosen representation of the theory. A common
misconception is that people speak about violation of the EEP or the
WEP in the Einstein frame simply implying that $\tilde{g}_{\mu\nu}$ is
not the metric whose geodesics coincide with free fall trajectories. 
Even though this is correct, it does not imply a violation of the WEP
or the EEP simply because all that these principles require is that
there should exist {\em some} metric whose geodesics coincide with
free fall trajectories, and indeed we do have one, namely
$g_{\mu\nu}$, the metric tensor of the Jordan frame. The fact of
whether or not one chooses to represent the theory with respect to
this metric is not relevant.

To go another step further, let us study free fall trajectories in the
Einstein frame. Considering a dust fluid with stress-energy tensor
$\tilde{T}_{\alpha\beta} = \tilde{\rho}\,\tilde{u}_{\alpha}
\tilde{u}_{\beta}$, eq.~(\ref{non-conservation}) becomes
 \begin{equation}
\tilde{\nabla}_{\alpha}\left( \tilde{\rho}\,\tilde{u}^{\alpha}
\tilde{u}^{\beta} \right)=
\tilde{\rho} \,
\frac{\tilde{g}^{\alpha\beta}\,\tilde{\nabla}_{\alpha} \Omega}{\Omega} \;.
\end{equation}
 By projecting this equation onto the 3-space orthogonal to
$\tilde{u}^{\mu}$ by means of the operator $\tilde{h}^{\mu}_{\nu}$
defined by $ \tilde{g}_{\mu\nu}=-\tilde{u}_{\mu}\tilde{u}_{\nu} +
\tilde{h}_{\mu\nu}$ and satisfying
$\tilde{h}^{\alpha}_{\beta}\,\tilde{u}^{\beta}=0$, one obtains
 \begin{equation}\label{correctedgeodesic}
\tilde{a}^{\gamma}\equiv 
 \tilde{h}^{\gamma}_{\beta}\tilde{u}^{\alpha} 
\tilde{\nabla}_{\alpha}\tilde{u}^{\beta}=\delta^{\gamma\alpha}
\frac{\partial_{\alpha} \Omega (\phi) }{\Omega (\phi) } \;.
\end{equation}
 The term on the right hand side of eq.~(\ref{correctedgeodesic}),
which would have been zero if the latter was the standard geodesic
equation, can be seen as arising due to the gradient of the scalar
field field $\phi$, or due to the variation of the particle mass
$\tilde{m}=\Omega^{-1} \, m $ along its trajectory, or due to the
variation with position in spacetime of the Einstein frame unit of
mass $\tilde{m}_u =\Omega^{-1}\, m_u $ (where $m_u$ is the constant
unit of mass in the Jordan frame) --- see Ref.~\cite{VFSN} for an
extensive discussion.

Massive particles in the Einstein frame are {\em always} subject to a
force proportional to $\nabla^{\mu}\phi$, hence there are no massive
test particles in this representation of the theory. From this
perspective, the formulation of the EEP ``(massive) test particles
follow (timelike) geodesics'' is neither satisfied nor violated: it is
simply empty. Clearly, the popular formulation of the EEP in terms of
the metric postulates is representation-dependent.

In this sense, the metric $g_{\mu\nu}$ certainly has a distinguished
status with respect to any other conformal metric, including
$\tilde{g}_{\mu\nu}$. However, it is a matter of taste and sometimes
misleading to call a representation physical or non-physical. The fact
that it is better highlighted in the Jordan frame that the theory
under discussion satisfies the EEP, does not make this frame
preferable, in the same sense that the Local Lorentz coordinate frame
is not a preferred one. The Einstein frame is much more suitable for
other applications, {\em e.g.} finding new exact solutions by using
mappings from the conformal frame, or computing the spectrum of
density perturbations during inflation in the early universe.

Let us now concentrate on the ambiguities related to the metric
postulates mentioned in Sections \ref{set} and \ref{matgeom}. One
should already be convinced that these postulates should be
generalized to include the phrase ``there exists a representation in
which''. But apart from that, there are additional problems. For
example, in the Jordan frame $\phi$ couples explicitly to the Ricci
scalar. One could, therefore, say that $\phi$ is a gravitational field
and not a matter field. In the Einstein frame, however, $\phi$ is not
coupled to the Ricci scalar---it is actually minimally coupled to
gravity and non-minimally coupled to matter. Can one then consider it
as being a matter field? If this is the case then maybe one should
define the stress-energy tensor differently from before and include
the $\phi$ terms in the matter action, {\em i.e.} define
 \begin{eqnarray} 
\label{rep3}
\bar{S}^{(m)}& = & \int  d^{4}x\; 
\sqrt{-\tilde{g}}\left[-\frac{1}{2} 
\tilde{g}^{ab}\tilde{\nabla}_{a}\tilde{\phi}\tilde{\nabla}_{b}\tilde{\phi}-\tilde{V}(\tilde{\phi})\right] \nonumber \\
&& \nonumber \\
&+&  S^{(m)}\left[e^{2\tilde{\alpha}(\tilde{\phi})}\tilde{g}_{ab},\psi^{(m)}\right] 
\end{eqnarray} 
and 
\be
\label{setbar}
\bar{T}_{\mu\nu}\equiv -(2/\sqrt{-\tilde{g}})\delta 
\bar{S}^{(m)}/\delta \tilde{g}^{\mu\nu}.
\ee

In this case though, $\bar{T}_{\mu\nu}$ will indeed be divergence-free
with respect to $\tilde{g}_{\mu\nu}$! The easiest way to see this is
to consider the field equations that one derives from the action
(\ref{rep2}) through a variation with respect to $\tilde{g}_{\mu\nu}$
with the redefinitions in eqs.~(\ref{rep2}) and (\ref{setbar}) taken
into account. This gives
 \be
\tilde{G}_{\mu\nu}=\kappa\bar{T}_{\mu\nu},
\ee
 where $\tilde{G}_{\mu\nu}$ is the Einstein tensor of the metric
$\tilde{g}_{\mu\nu}$. The contracted Bianchi identity
$\tilde{\nabla}_\mu \tilde{G}^{\mu\nu}=0$ directly implies that
$\tilde{\nabla}_\mu \bar{T}^{\mu\nu}=0$. 

Does this solve the problem, and was the fact that it was not apparent
that the EEP is not violated in the Einstein frame just due to a wrong
choice of the stress-energy tensor? Unfortunately, this is not the
case. First of all, $\tilde{g}^{\mu\nu}$ is still not the metric whose
geodesics coincide with free fall trajectories, as shown earlier.
Secondly, $\bar{T}_{\mu\nu}$ has the following form
 \be
\bar{T}_{\mu\nu}=\tilde{\nabla}_{\mu}\tilde{\phi}\tilde{\nabla}_{\nu}\tilde{\phi}-\frac{1}{2}\tilde{g}_{\mu\nu}\tilde{\nabla}^{\sigma}\tilde{\phi}\tilde{\nabla}_{\sigma}\tilde{\phi}-\tilde{g}_{\mu\nu}\, \tilde{V}(\tilde{\phi})+\tilde{T}_{\mu\nu},
\ee
 with $\tilde{T}_{\mu\nu}$ depending on $\tilde{\phi}$ as well as on
the matter, and it will not reduce to the special-relativistic
stress-energy tensor for the matter field $\psi^{(m)}$ if
$\tilde{g}_{\mu\nu}$ is taken to be flat. The same is true for the
action $\bar{S}^{(m)}$. Both of these features are due to the fact
that $\bar{T}_{\mu\nu}$ includes a non-minimal coupling between the
matter fields $ \psi^{(m)}$ and the scalar field $\phi$. Actually,
setting $\tilde{g}_{\mu\nu}$ equal to the Minkowski metric does not
correspond to choosing the Local Lorentz frame: that would be the one
in which $g_{\mu\nu}$ is flat to second order (see Section \ref{ep}).

The moral of this is that one can find quantities that indeed formally
satisfy the metric postulates but these quantities are not necessarily
physically meaningful. There are great ambiguities, as mentioned
before, in defining the stress-energy tensor or in judging whether a
field is gravitational or just a matter field, that in practice make
the metric postulates useless outside of a specific representation
(and how does one know, in general, when given an action, whether it
is in this representation, {\em i.e.}~whether the quantities of the
representation are the ones to be used directly to check the validity
of the metric postulates or whether a representation change is
necessary before doing this?).

\subsubsection{Matter or geometry? An ambiguity}
\label{vacuum}

We already saw that treating $\phi$ as a matter field merely because
it is minimally coupled to gravity and including it in the
stress-energy tensor did not help in clarifying the ambiguities of the
metric postulates. Since, however, this did not answer the question of
whether a field should be considered as gravitational (``geometric'')
or as non-gravitational (``matter''), let us try to get some further
insight into this.

Consider again, as an example, scalar-tensor gravity. Choosing
$A(\phi)=8\pi\,G\,\phi$ and $\alpha$ to be a constant, the action
(\ref{eq:3.19}) can be written as
 \begin{eqnarray}
S&=&\int d^4 x \sqrt{-g} \left[ \frac{\phi R}{2} 
-\frac{B(\phi)}{2}\, g^{\mu\nu} \nabla_{\mu}\phi 
\nabla_{\nu}\phi -V(\phi) \right. \nonumber \\
&& \nonumber \\
&& \left. +\alpha_{\psi}{\cal L}^{(\psi)}\left( 
g_{\mu\nu}, \psi^{(m)} \right) \right]\;,\label{Jframeaction}
\end{eqnarray}
 where $\alpha_{\psi}$ is the coupling constant between gravity and
the specific matter field $\psi^{(m)}$ described by the Lagrangian
density ${\cal L}^{(\psi)}$. This representation is in the Jordan
frame and it is no different from that of the action (\ref{rep1}),
apart from the fact that we have not specified the value of the
coupling constant to be $1$.

It is common practice to say that the Brans--Dicke scalar field $\phi$
is gravitational, {\em i.e.} that it describes gravity together with
the metric $g_{\mu\nu} $ \cite{willbook,wagoner, dicke, bransdicke}.
Indeed, $1/\phi$ plays the role of a (variable) gravitational
coupling. However, this interpretation only holds in the Jordan frame.
As discussed earlier, the conformal transformation to the Einstein
frame $g_{\alpha\beta} \rightarrow \tilde{g}_{\alpha\beta}=\Omega^2 \,
g_{\alpha\beta}$ with $\Omega=\sqrt{G\,\phi}$, together with the scalar
field redefinition
 \begin{equation}\label{scalarredefinition}
d\tilde{\phi}=\sqrt{ \frac{ 2\omega(\phi)+3}{16\pi G}}\, 
\frac{d\phi}{\phi}
\end{equation}
casts the action into the form
\begin{equation}\label{Eframeaction}
S=\int d^4 x \sqrt{-\tilde{g}} \left[ \frac{\tilde{R} }{2} 
-\frac{1}{2}\, \tilde{g}^{\mu\nu} 
\tilde{\nabla}_{\mu}\tilde{\phi} 
\tilde{\nabla}_{\nu}\tilde{\phi} 
-\tilde{V}\left( \tilde{\phi} \right)+\tilde{\alpha}_{\psi}{\cal 
L}^{(\psi)}\right] \;,
\end{equation}
where 
\begin{equation}
\tilde{V}\left( \tilde{\phi} \right)=\frac{ V\left[ \phi \left( 
\tilde{\phi} \right) \right] }{ \phi^2\left( \tilde{\phi} 
\right)}
\end{equation} 
and 
\begin{equation} 
\tilde{\alpha}_{\psi}\left( \tilde{\phi} 
\right)=\frac{ 
\alpha_{\psi} }{ \phi^2\left( \tilde{\phi} 
\right)} \;.
\end{equation} 
 The ``new'' scalar field $\tilde{\phi}$ is now minimally coupled to
the Einstein frame Ricci scalar $\tilde{R}$ and has canonical kinetic
energy: {\em a priori}, nothing forbids one to interpret
$\tilde{\phi}$ as being a ``matter field''. The only memory of its
gravitational origin as seen from the Jordan frame is in the fact that
now $\tilde{\phi}$ couples non-minimally to matter, as described by
the varying coupling $\tilde{\alpha}_{\psi}(\tilde{\phi})$.  However,
by itself this coupling only describes an interaction between
$\tilde{\phi}$ and the ``true'' matter field $\psi^{(m)}$. One could,
for example, take $\psi^{(m)}$ to be the Maxwell field and consider an
axion field that couples explicitly to it, obtaining an action similar
to (\ref{Eframeaction}) in which case it would not be possible to
discriminate between this axion field and a putative ``geometrical''
field on the basis of its non-minimal coupling. Even worse, this
``anomalous'' coupling of $\tilde{\phi}$ to matter is lost if one
considers only the gravitational sector of the theory by dropping
${\cal L}^{(\psi)}$ from the discussion.  This is the situation, for
example, if the scalar $\tilde{\phi}$ is taken to dominate the
dynamics of an early, inflationary, universe or a late,
quintessence-dominated, universe.

More generally, the distinction between gravity and matter
(``gravitational'' versus ``non-gravitational'') becomes blurred in
any change of representation involving a conformal transformation of
the metric $g_{\mu\nu}\rightarrow \tilde{g}_{\mu\nu}=\Omega^2 \,
g_{\mu\nu} $. The transformation property of the Ricci tensor is
\cite{wald,Synge}
 \bea\label{Riccitransform}
\tilde{R}_{\alpha\beta}&=& R_{\alpha\beta}
-2\nabla_{\alpha}\nabla_{\beta} \left( \ln \Omega \right) 
-g_{\alpha\beta} g^{\gamma\delta} \nabla_{\gamma}\nabla_{\delta} 
\left( \ln \Omega \right) \nonumber\\ & &
+2 \left( \nabla_{\alpha}\ln \Omega \right)
\left( \nabla_{\beta}\ln \Omega \right)
-2g_{\alpha\beta} g^{\gamma\delta} 
\left( \nabla_{\gamma}\ln \Omega \right)
\left( \nabla_{\delta}\ln \Omega \right) \;.
\eea
 The conformal transformation maps a vacuum solution in the Jordan
frame ({\em i.e.} one with $R_{\alpha\beta}=0$) into a non-vacuum
solution in the Einstein frame ($\tilde{R}_{\alpha\beta}\neq 0$). The
conformal factor $\Omega$, which was a purely ``geometrical'' field in
the Jordan frame, is now playing the role of a form of ``matter'' in
the Einstein frame. 

A possible way of keeping track of the gravitational nature of
$\Omega$ is by remembering that the Einstein frame units of time,
length, and mass are not constant but scale according to
$\tilde{t}_u=\Omega\, t_u$, $\tilde{l}_u=\Omega\, l_u$, and
$\tilde{m}_u=\Omega^{-1} \, m_u$, respectively (where $t_u, l_u$, and
$m_u$ are the corresponding constant units in the Jordan frame)
\cite{dicke}. However, one would not know this prescription by looking
only at the Einstein frame action (\ref{Eframeaction}) unless the
prescription for the units is made part of the theory ({\em i.e.}~by
carrying extra information additional to that given by the action!).
In practice, even when the action (\ref{Eframeaction}) is explicitly
obtained from the Jordan frame representation, the variation of the
units with $\Omega$ (and therefore with the spacetime location) is
most often forgotten in the literature \cite{VFSN} hence leading to the study of a different theory with respect to that expressed by the action (\ref{Jframeaction}).

Going back to the distinction between material and 
gravitational fields, an 
alternative possibility to distinguish between ``matter'' and 
``geometry'' would seem to arise by labeling  as ``matter 
fields'' only those described by a stress-energy tensor that 
satisfies some energy condition. In fact, a  conformally 
transformed field that originates from Jordan frame geometry 
does not, in general, satisfy any energy condition. The 
``effective stress-energy tensor'' of the field $\Omega$ derived 
from eq.~(\ref{Riccitransform}) does not have the canonical 
structure quadratic in the first derivatives of the field but 
contains instead terms that are linear in the second 
derivatives. Because of this structure, the stress-energy tensor $\Omega$ violates the 
energy conditions. While it would seem that labelling as ``matter 
fields'' those that satisfy the weak or null energy condition 
could eliminate the ambiguity, this is not the case. As we have previously seen,  one can always redefine the scalar field in such a way that it is minimally coupled to 
gravity and has canonical kinetic energy (this is precisely the 
purpose of the field redefinition~(\ref{scalarredefinition})). 
Keeping track of the 
transformation of units in what amounts to a full specification 
of the representation adopted  (action plus information on 
how the units scale with the scalar field) could help making the 
property of satisfying energy conditions frame-invariant, but at the cost of extra ``structure'' in defining a given theory.

As  a conclusion, the concept of vacuum versus non-vacuum, or of 
``matter field'' versus ``gravitational field'' is 
representation-dependent. One might be prepared to accept {\em a priori} and without any real physical justification that one representation should be chosen in which the fields are to be characterized as gravitational or non-gravitational and might be willing to carry this extra ``baggage" in any other representation in the way described above. However, even if such a compromise is made, the problem discussed here is  just ``tackled'' instead of actually being solved in a clean and tidy way.

These considerations, as well those discussed at the previous sub-section, elucidate a more general point: it is not 
only the mathematical formalism associated with a theory that is 
important, but the theory must also include a set of rules to 
interpret physically the mathematical laws. As an example 
from the classical mechanics of point particles, consider two 
coupled harmonic oscillators described by the Lagrangian
\begin{equation}\label{Lq}
L=\frac{\dot{q}_1^2}{2}+ \frac{\dot{q}_2^2}{2}-
\frac{q_1^2}{2}-\frac{q_2^2}{2}+\alpha \, q_1 q_2 \;.
\end{equation}
A different representation of this physical system is obtained 
 by using normal coordinates $ Q_1\left( q_1, q_2 \right) , 
Q_2\left( q_1, q_2 \right) $, in terms of which the Lagrangian 
(\ref{Lq}) becomes
\begin{equation}\label{LQ}
L=\frac{\dot{Q}_1^2}{2}+ \frac{\dot{Q}_2^2}{2}-
\frac{Q_1^2}{2}-\frac{Q_2^2}{2} \;.
\end{equation}
Taken at face value, this Lagrangian describes a different 
physical system, but we know that the mathematical expression 
(\ref{LQ}) is not all there is to the theory: the {\em 
interpretation} of $q_1$ and $q_2$ as the degrees of freedom of 
the two original oscillators prevents viewing $Q_1$ and 
$Q_2$  as the physically measurable quantities. In addition to the equations of 
motion, a set of interpretive rules constitutes a fundamental 
part of a theory. Without such rules it is not only impossible to connect the results derived through the mathematical formalism to a physical phenomenology but one would not even be able to distinguish alternative theories from alternative representations of the same theory.  Note however, that once the interpretative rules are assigned to the variables in a given representation they do allow to predict the outcome of experiments in  any other given representation of the theory (if consistently applied), hence assuring the physical equivalence of the possible representations.

While the above comments hold in general for any physical theory,
it must however be stressed that gravitation theories are one of those cases in which the problem is more acute. 
In fact, while the physical interpretation of the variables is clear in simple systems, such as the example of the two coupled oscillators discussed above, the physical content of  complex theories (like quantum mechanics or gravitation theories) is far less intuitive. Indeed, for what regards gravity, what we actually  
know more about is the phenomelogy of the system instead of the system itself. Therefore, it is often difficult, or even arbitrary, to formulate explicit interpretive rules, which should nevertheless be provided in order to completely specify the theory.

\subsection{Example no.~2: $f(R)$ gravity}
\label{fR}
To highlight further the ambiguity concerning whether a field is a
gravitational or matter field, s well as to demostrate how the problems discussed here can actually go beyond representations that just involve conformal redefinitions of the metric, let us examine one further example:
that of $f(R)$ gravity in the metric and Palatini formalisms. We have
already extensively discussed these theories and in Chapter
\ref{equivtheor} we have established that they can acquire the
representation of a Brans--Dicke theory. Metric $f(R)$ gravity can be
re-written as Brans--Dicke theory with Brans--Dicke parameter
$\omega_0=0$. In terms of the action \ref{eq:3.19}, this corresponds
to the choice $A=\phi$, $B=0$, $\alpha=0$. Palatini $f(R)$ gravity, on
the other hand, can be re-written as an $\omega_0=-3/2$ Brans--Dicke
theory, corresponding to the choice $A=\phi$, $B=-3/2$, $\alpha=0$
when one refers to the action (\ref{eq:3.19})

Note that the general representation used in the
action~(\ref{eq:3.19}) is actually not as general as one might expect,
since we have just shown that theories described by
this action can even, with suitable choices of the parameters, acquire
completely different non-conformal representations. One can, in
principle, add at will auxiliary fields, such as the scalar field used
above, in order to change the representation of a theory and these
fields need not necessarily be scalar fields. Therefore, all of the
problems described so far are not specific to conformal
representations. In this $f(R)$ representation, the scalar $\phi$ is
not even there, so how can one decide whether it is a gravitational
field or a matter field? For the case of metric $f(R)$ gravity, the
scalar field was eliminated without introducing any other field and
the metric became the only field describing gravity. On the other
hand, in the Palatini formalism the outcome is even more surprising if
one considers that the scalar field was replaced with an independent
connection which, theoretically speaking, could have forty degrees of
freedom assuming that it is symmetric but in practice has only one! 

\subsection{Example no.~3: Einstein--Cartan--Sciama--Kibble theory}
\label{ECSK}

 Our final example is Einstein--Cartan--Sciama--Kibble theory. In this
theory, one starts with a metric and an independent connection which
is not symmetric but has zero non-metricity. We will not present
extensive calculations and details here but, instead, address the
reader to Ref.~\cite{hehlrev} for a thorough review. What we would
like to focus on is the fact that, since the theory has an independent
connection, one usually arrives at the field equations through
independent variations with respect to the metric and the connections. 
Additionally, since the matter action depends on both the metric and
the connections, its variation will lead to two objects describing the
matter fields: the stress-energy tensor $T_{\mu\nu}$, which comes from
varying the matter action with respect to the metric as usual, and the
hypermomentum $\Delta^\lambda_{\phantom{a}\mu\nu}$, which comes from
varying the matter action with respect to the independent connections.

 In this theory, $T_{\mu\nu}$ is not divergence-free with respect to
either the covariant derivative defined with the Levi--Civita
connection or with respect to the one defined with the
independent connection. It also does not reduce to the
special-relativistic stress-energy tensor in the suitable limit.
However, it can be shown that a suitable non-trivial combination of
$T_{\mu\nu}$ and $\Delta^\lambda_{\phantom{a}\mu\nu}$ does lead to a
tensor that indeed has the latter property \cite{hehlrev}. What is
more, a third connection can be defined which leads to a covariant
derivative with respect to which this tensor is divergence-free
\cite{hehlrev}! This is sufficient to guarantee that the EEP is
satisfied. Does this make Einstein--Cartan theory a metric theory? And
how useful are the metric postulates for discussing violations of the
EEP if, in order to show that they are satisfied, one will already
have demonstrated geodesic motion or LLI on the way?

\subsection{Discussion}
\label{discussion}

We have attempted to shed some light on the differences between
different theories and different representations of the same theory
and to reveal the important role played by a representation in our
understanding of a theory. For doing this, several examples have been
presented which hopefully highlight this issue. It has been argued
that certain conclusions about a theory which may be drawn in a
straightforward manner in one representation, might require serious
effort when a different representation is used and vice-versa.
Additionally, care should be taken as certain representations may be
completely inconvenient or even misleading for specific applications.

It is worth commenting at this point, that the literature is seriously
biased towards particular representations and this bias is not always
a result of the convenience of certain representations in a specific
application, but often is a mere outcome of habit. It is common, for
instance, to bring alternative theories of gravity into a
General-Relativity-like representation due to its familiar form, even
if this might be misleading when it comes to getting a deeper
understanding of the theory.

This seemingly inevitable representation-dependent formulation of our
gravitation theories has already been the cause of several
misconceptions. What is more, one can very easily recognise a
representation bias in the definition of commonly used quantities,
such as the stress-energy tensor. Notions such as that of vacuum and
the possibility of distinguishing between gravitational fields and
matter fields are also representation-dependent. This is often
overlooked due to the fact that one is very accustomed to the
representation-dependent definitions given in the literature. On the
other hand, representation-free definitions do not exist.

Note that even though the relevant literature focuses almost
completely on conformal frames, the problems discussed here are not
restricted to conformal representations. Even if conformally invariant
theories were considered, nothing forbids the existence of other
non-conformal representations of these theories in which the action or
the field equations will not, of course, be invariant. This might
imply that creating conformally invariant theories is not the answer
to this issue. After all, even though measurable quantities are
always dimensionless ratios and are therefore conformally invariant,
matter is not generically conformally invariant and, therefore,
neither can (classical) physics be conformally invariant, at least
when its laws are written in terms of the fields representing this
matter.

The issue discussed here seems to have its roots in a more fundamental
problem: the fact that in order to describe a theory in mathematical
terms, a non-unique set of variables has to be chosen. Such variables
will always correspond to just one of the possible representations of
the theory. Therefore, even though {\em abstract statements such as
the EEP are representation-independent}, attempts to turn such
statements into {\em quantitative mathematical relations that are of
practical use, such as the metric postulate, turn out to be severely
representation-dependent}. 

The comparison between a choice of representation and a choice of
coordinate system is practically unavoidable.  Indeed, consider
classical mechanics: one can choose a coordinate system in order to
write down an action describing some system. However, such an action
can be written in a coordinate invariant way. In classical field
theory one has to choose a set of fields --- a representation --- in
order to write down the action. From a certain viewpoint, these fields
can be considered as generalized coordinates. Therefore, one could
expect that there should be some representation-independent way to
describe the theory. However, up to this point no real progress has
been made on this issue.

The representation dependence of quantitative statements acts in such
a way that, instead of merely selecting viable theories for us, they
actually predispose us to choose theories which, in a specific
representation, appear more physically meaningful than others
irrespective of whether this is indeed the case. The same problem is
bound to appear if one attempts to generalise a theory but is biased
towards a specific representation, since certain generalisations might
falsely appear as being more ``physical'' than others in this
representation. This effectively answers the question of why most of
our current theories of gravitation eventually turn out to be just
different representations of the same theory or class of theories. 
Scalar-tensor theories and theories which include higher order
curvature invariants, such as $f(R)$ gravity or fourth order gravity,
are typical examples.

Even though this discussion might at some level appear to be purely
philosophical, the practical implications of representation dependence
should not be underestimated. For instance, how can we formulate
theories that relate matter/energy and gravity if we do not have a
clear distinction between the two, or if we cannot even conclude
whether such a distinction should be made? Should we then aim to avoid
any statement based on a sharp separation between the matter and
gravity sectors?

\section{Concluding remarks}

To conclude, even though some significant progress has been made with
developing alternative gravitation theories, one cannot help but
notice that it is still unclear how to relate principles and
experiments in practice, in order to form simple theoretical viability
criteria which are expressed mathematically. Our inability to express
these criteria and also several of our very basic definitions in a
representation-invariant way seems to have played a crucial role in
the lack of development of a theory of gravitation theories. 
 This is a critical obstacle to overcome if we want to go beyond a
trial-and-error approach in developing alternative gravitation
theories.

It is the author's opinion that such an approach should be one of the
main future goals in the field of modified gravity. This is not to
say, of course, that efforts to propose or use individual theories,
such as $f(R)$ gravity or Gauss--Bonnet gravity, in order to deepen our
understanding about the gravitational interaction should be abandoned
or have less attention paid to them. Such theories have proved to be
excellent tools for this cause so far, and there are still a lot of
unexplored corners of the theories mentioned in this thesis, as well
as in other alternative theories of gravity. 

The motivation for modified gravity coming from High Energy Physics,
Cosmology and Astrophysics is definitely strong. Even though modifying
gravity might not be the only way to address the problems mentioned in
Chapter \ref{intro}, it is our hope that the reader is by now
convinced that it should at least be considered very seriously as one
of the possible solutions and, therefore, given appropriate attention.
The path to the final answer is probably long. However, this has never
been a good enough reason for scientists to be discouraged.

\vspace*{1cm}
\begin{quote}

\em
\raggedleft
 If I have ever made any valuable discoveries, it has been owing more 
to patient attention, than to any other talent.

   {\bf Isaac Newton}
\end{quote}

%%%%%%%%%%%%%%%%%%%%%%
%%%%%%%%%%%%%%%%%%%%%%
\backmatter
%%%%%%%%%%%%%%%%%%%%%%
%%%%%%%%%%%%%%%%%%%%%%

%%%%%%%%%%%%%%%%%%%%%%%%%%%%%%%%%%%%%%%%%%%%
% Thomas Sotiriou- PhD Thesis Bibliography %
%%%%%%%%%%%%%%%%%%%%%%%%%%%%%%%%%%%%%%%%%%%%

%%%%%%%%%%%%%%%%%%%%%%
%%%%%%%%%%%%%%%%%%%%%%
%%%%%%%%%%%%%%%%%%%%%%
%%%%%%%%%%%%%%%%%%%%%%
%%%%%%%%%%%%%%%%%%%%%%
\end{document}